\documentclass[usenatbib]{mn2e}
\usepackage{times}
\usepackage{epsfig}
\usepackage{float}
\newcommand{\be}{\begin{equation}}
\newcommand{\ee}{\end{equation}}

\title[Luminosity Functions of XMM-LSS C1 Galaxy Clusters]
{Luminosity Functions of XMM-LSS C1 Galaxy Clusters}
\author[A. Alshino et al.]
  {Abdulmonem ~Alshino,$^1$\thanks{E-mail: alshino@star.sr.bham.ac.uk}
  Habib ~Khosroshahi,$^2$ Trevor ~Ponman,$^1$ Jon ~Willis,$^3$
  \newauthor
  Marguerite ~Pierre$^4$, Florian Pacaud$^5$ and Graham P. Smith$^1$ \\
  $^1$School of Physics and Astronomy, The University of
  Birmingham, Birmingham B15 2TT, UK. \\
  $^2$ School of Astronomy, Institute for Research in Fundamental Sciences (IPM), P. O. Box 19395-5531, Tehran, Iran. \\
  $^3$Department of Physics and Astronomy, University of Victoria, Elliot Building, 3800 Finnerty Road, Victoria, BC, V8P 1A1
  Canada.\\
  $^4$DAPNIA/SAp CEA Saclay, 91191 Gif sur Yvette, France.\\
  $^5$Argelander-Institut f\"ur Astronomie, University of Bonn, Auf dem H\"ugel 71, 53121 Bonn, Germany.
  }
\begin{document}

\date{Accepted XXXX Xxxxx XX. Received XXXX Xxxxx XX; in original from XXXX Xxxxx XX}

\pagerange{\pageref{firstpage}--\pageref{lastpage}} \pubyear{2009}

\maketitle

\label{firstpage}

\begin{abstract}

CFHTLS optical photometry has been used to study the galaxy
luminosity functions of 14 X-ray selected clusters from the XMM-LSS survey. 
These are mostly groups and poor clusters, with
masses ($M_{500}$) in the range 0.6 to 19 $\times 10 ^{13}
M_{\odot}$ and redshifts $0.05\leq z \leq 0.61$. 
Hence these are some of the highest redshift X-ray selected groups to have
been studied. Lower and upper colour
cuts were used to determine cluster members. We derive individual luminosity
functions (LFs) for all
clusters as well as redshift-stacked and temperature-stacked LFs in three
filters, $g^\prime$, $r^\prime$ and $z^\prime$, down to $M=-14.5$.  All LFs
were fitted by Schechter functions which constrained the faint-end
slope, $\alpha$, but did not always fit well to the bright end.
Derived values of $\alpha$ ranged from
$-1.03$ to as steep as $-2.1$. We find no evidence for upturns at faint
magnitudes. Evolution in $\alpha$ was apparent in all bands: it
becomes shallower with increasing redshift; for example, in the
$z^\prime$ band it flattened
from -1.75 at low redshift to -1.22 in the redshift range
$z=$0.43-0.61.  
Eight of our systems lie at $z\sim 0.3$, and we combine these
to generate a galaxy LF in three colours for X-ray selected groups and poor
clusters at redshift 0.3. We find that at $z\sim 0.3$ $\alpha$ 
is steeper (-1.67) in the
green ($g^\prime$) band than it is (-1.30) in the red ($z^\prime$) band.
This colour trend disappears at low redshift, which we attribute to
reddening of faint blue galaxies from $z\sim 0.3$ to $z\sim 0$. We also
calculated the total optical luminosity and found it to correlate
strongly with X-ray luminosity ($L_{X}\propto L_{OPT}^{2.1}$), and also
with ICM temperature ($L_{OPT}\propto T^{1.62}$), consistent with
expectations for self-similar clusters with constant mass-to-light ratio.
We did not find any convincing correlation of Schechter
parameters with mean cluster temperature.

\end{abstract}

\begin{keywords}
galaxies: clusters: general - galaxies: evolution - galaxies:
luminosity function - galaxies: structure.
\end{keywords}

\section{Introduction}

Most of our knowledge of galaxies is based on observations of the local universe,
although distant universe observations have also provided a
wealth of information. Statistical studies of galaxies at high
redshift are mostly limited to rich galaxy clusters mainly due to
observational limitations.  Galaxy clusters are important
cosmological environments where key galaxy transformation
such as stripping and strangulation occur. However, in the hierarchical
formation of structure rich clusters are the latest structures to be formed.
Lower mass systems or galaxy groups may have been
the place where galaxies experience a substantial degree of evolution through
processes such as mergers and tidal interaction, as a result of the higher
efficiency of these processes in the lower velocity dispersion environment of groups.

The Galaxy luminosity function (LF) -- the number of galaxies per unit
volume in the luminosity interval $L$ to $L+dL$ -- has been widely used to
study the formation of galaxies and the evolution of galaxy populations with
redshift. It is also an excellent statistical tool for
describing how different environments influence the properties of galaxies.

Both the bright end (\cite{Bower2006}, \cite{Naab2007}) and the faint end
(\cite{Marzke1994}, \cite{Khochfar2007}) of the LF have been the subject of
in-depth studies, as they offer strong observational constraints for 
models of galaxy formation and evolution.
While the bright end of the LF is affected by AGN
feedback (\cite{Bower2006}), the faint-end slope is predominantly
influenced by feedback from supernovae (\cite{Dekel1986}), and provide a direct
indicator of the significance of dwarf galaxies, which are expected to
behave differently in rich and poor clusters. Multi-colour LFs, in
particular, probe the history of the faint galaxy
population, including its star formation history -- see for example,
\cite{Adami07}.

The vast majority of studies of the galaxy LF give faint-end slopes in the
range $\sim -1$ to $\sim -2$. Most of these have limited magnitude depth
($M > -16$) and recent deep studies are mostly confined to rich local
clusters (See Table 1 in \cite{Popesso2005a} and Table A.1 in \cite{Boue08}
and references therein). These studies not only disagree on the value of
the faint-end slope, but they also disagree on the exact form of it, as
some studies (e.g. \cite{gonzalez2006}) found upturns; a single
Schechter function was not an adequate fit to the faint end, and a double
Schechter function was required to give a reasonable fit.  The existence of these upturns 
is very sensitive to the method used to determine galaxy membership, with some 
approaches including spurious galaxies or excluding genuine
cluster members due to their low surface brightness.

The evolution of the faint-end slope is hard to study, mainly because the
number of faint galaxies detected decreases sharply with increasing
redshift.  \cite{Liu08} found that the faint-end slope of a field galaxy
population became shallower with increasing redshift (up to $z=0.5$) for
all galaxy spectral types. However, to account for the photometric redshift
errors of the galaxies, they weighted the galaxies as probability-smoothed
luminosity distribution at the redshift at which they were measured. This
places an important caveat on the interpretation of their data, and hence on their
results. On the other hand, simulations by \cite{Khochfar2007} show a
measurable dependence of the faint-end slope of the galaxy luminosity
function on redshift. However, most of this dependence is seen over a
relatively large redshift range,  $\Delta z \ge2$. Furthermore, it is
hard to discriminate galaxy environments in such studies.

X-ray surveys remain one of the most popular methods of finding galaxy
systems. Due to the strong density dependence of X-ray emissivity, X-ray 
cluster selection is much less vulnerable to contamination
along the line-of-sight than optical methods. The XMM-Large Scale Survey
(XMM-LSS) (\cite{Pierre2004}), a contiguous X-ray survey, has a well-defined selection function 
which is used to produce a sample of galaxy groups to study their intracluster medium and
galaxy properties at medium to high redshift. \citet{pacaud07} have
presented a study of a sample of 29 galaxy systems from the XMM-LSS
survey, drawn from an area of 5 deg$^2$ out to a redshift of z = 1.05. The
cluster distribution peaks around z = 0.3 and T=1.5 keV, half of the
objects being groups with a temperature below 2 keV.

In this paper, we use the XMM-LSS optical follow-up observations to study
the evolution of the galaxy luminosity function in galaxy groups and poor
clusters since $z\approx 0.6$. Given the observational biases --
distant groups are more massive and hotter -- we study whether the redshift
dependencies are weaker or stronger when the intrinsic properties of the
systems, for instance, intracluster medium temperature, are taken into
account.

By using a deep ($m_{g^\prime}=24$) optical survey of X-ray selected
galaxy clusters up to redshift of $z=0.61$, we aim in this
paper to clarify the debate on the faint-end slope of the LFs of low-mass
($M_{500} \leq 20 \times 10 ^{13} M_{\odot}$) galaxy clusters (or groups),
and to explore the existence of any dips, or upturns at the faint end,
and to establish whether the slope shows trends with redshift or intracluster
medium temperature. Comparison with previous results can help to
elucidate the universality of galaxy cluster LFs. Furthermore, the scaling
relation of total optical luminosity with temperature and X-ray luminosity
for our cluster sample can shed light on the mass-to-light ratios of
low-mass systems when compared to rich clusters.

The paper is constructed as follows: In section 2, we describe the optical catalogue used to calculate 
the LFs. Then, we describe the data reduction and the method used to construct the colour-magnitude 
diagrams (CMD) and the subsequent LFs, and the technique adopted for the background 
subtraction. In section 3, we describe our results, starting with the individual cluster LFs, and then
the redshift-stacked clusters and temperature-stacked clusters. In section 4, we discuss our results 
and compare them with other studies. Finally, in section 5, we summarise our conclusions.

Throughout this article, we adopt  the cosmological
parameters estimated by \cite{Spergel2007}, namely:
$H_0=73$~km~s$^{-1}$~Mpc$^{-1}$, $\Omega_m=0.24$, $\Omega_\Lambda=0.76$.

\section{Data}

\subsection{Observations}

Optical photometry of the XMM-LSS survey was obtained from the
Canada-France-Hawaii Telescope Wide Synoptic Legacy
Survey\footnote{See http://www.cfht.hawaii.edu/Science/CFHLS/},
referred to as the CFHTLS Wide survey. Data were obtained in five
passbands ($u^\ast$, $g^\prime$, $r^\prime$,$i^\prime$, $z^\prime$) down to a
nominal magnitude limit of $i^\prime=24.5$. Of the 19 deg$^2$ of
CFHTLS Wide data available in the W1 survey area, 4 deg$^2$ overlap
with the X-ray selected cluster catalogue presented by \cite{pacaud07}.
Hence our photometric data are drawn from four 
$ 1^{\circ} \times 1^{\circ} $ catalogues derived from the survey data.

The data used in this paper are based upon the reduction procedure
outlined in \cite{Hoekstra2006}. Source extraction and photometry
were performed using {\tt SExtractor v2.5.0} (Bertin and Arnouts
1996). Zero point information for sources detected in the CFHTLS Wide
field survey W1 area was extrapolated from common sources detected in the
Sloan Digital Sky Survey equatorial patch which overlaps the southern
edge of the W1 area.

XMM-LSS Class 1 (C1) clusters are a well-controlled X-ray selected and
spectroscopically confirmed cluster sample. The criteria used to construct
the sample guarantee negligible contamination by point-like sources. The
observations of the clusters were performed in a homogeneous way (10-20 ks
exposures). For full details of the C1 sample, see \cite{pacaud07}. The
main properties of the sample are shown in Table \ref{C1usters}. Detailed
information on the C1 selection process can be found in \cite{pacaud06}. 
17 out of the 29 XMM-LSS C1 clusters are covered by the CFHTLS Wide field
survey. In this paper, we study the luminosity functions of 14 of 
these 17 clusters --
dropping the three with the highest redshifts (clusters with XLSSC 
numbers 2,29, and 1) because their photometric data is too poor
to allow useful constraints to be obtained.

\subsection{Analysis}

Galaxies were detected by SExtractor (\cite{Bertin1996}). 
Luminosity functions (LFs) were produced in three of the five
CFHTLS ($u^\ast$, $g^\prime$, $r^\prime$,$i^\prime$, $z^\prime$) filter
bands, namely, $g^\prime$, $r^\prime$ and $z^\prime$. To determine the 
completeness of the LFs, we took into account the limiting 
apparent magnitude in each field. 
The completeness limits for each filter was determined using the 
apparent magnitude LFs of all data (down to the 
faintest magnitudes available) for each C1 cluster individually.
Variations in seeing and exposure time across the CFHTLS fields
used here are small, and 
it was found that for each filter there was a common 
completeness limit at which all LFs started to drop below the 
faint end power law slope. Note that the LF turn-up reported
by some authors (see section 4.2), which could potentially introduce
an error into this method for estimating completeness, falls beyond the
faint limit of our LFs (e.g. at -16 in $g^\prime$ band), except in our 
three closest clusters, and hence cannot seriously bias our estimates
of the completeness limits. The completeness
threshold magnitudes for the three filters $g^\prime$, $r^\prime$ 
and  $z^\prime$ were found to be 24, 23.5 and 23, respectively. 
These values are also consistent with results based on comparison of the 
number counts per field to deeper data from the CFHTLS Deep Field and 
CCCP Megacam observations (\cite{Urquhart2009}).

Each entry in the catalogues is associated with a FLAG value which
indicates the degree of reliability of the data.  Flag is a short integer,
and a value of $0$ denotes good data. The more unreliable the data is, the
higher the FLAG value becomes. We included all catalogue entries with
FLAG $\leq 3$, which includes sources with very close and bright neighbours
or some bad pixels and sources which are originally blended with other
sources. This may admit some problematic galaxies 
but this is better than excluding many genuine cluster members, because many
clusters contain significant number of blended sources. Factors that may
raise the FLAG to $>3$ include sources with saturated pixels, truncated
sources, incomplete or corrupted data and sources with memory overflow
during deblending or extraction. Catalogue entries with FLAG $>3$ constitute 
only $\simeq 5\%$ of the total number of entries, and were all removed. 
Many of the removed entries are fainter than the threshold magnitude 
and hence would have been removed anyway.

Each entry in the catalogue also includes a stellarity class
value, STAR, with values ranging from 0 to 1. The lower its value,
the more likely the detected object is a galaxy. Data points with
different STAR values were checked by IRAF and their radial
profiles were examined to see if they matched the typical profile
of a star or a galaxy. Typically, objects with STAR $>$ 0.85 were found to be
stars, whilst those with $<$ 0.85 were galaxies. 
Therefore, only catalogue entries with
STAR values of 0.85 or less were included when constructing the LFs. 

Spectral temperatures of the XMM-LSS clusters, and the resulting
values of $R_{500}$ (which allow for the evolution in critical
density with redshift), were measured by the Saclay team (\cite{pacaud07}). 
To construct colour-magnitude diagrams, we selected all galaxies within a 
circle of radius, $R^{\ast} =1.5 \times R_{500}$ of  the clusters.
This radius limit,  $R^{\ast}$, represents an estimate of $R_{200}$.
Colour-magnitude diagrams were produced for all 14 clusters for
colour bands: $g^\prime$, $r^\prime$ and $z^\prime$. The factor, 1.5 
does not have a large effect on the fitted parameters of the Schechter 
function; we compared the results of $1.0 \times R_{500}$ to 
$1.8 \times R_{500}$ and found that the faint-end slope, $\alpha$, 
was only changed within its 1 $\sigma$ errors.

The CMD were used to colour select galaxies which might be cluster
members, hence reducing the background due to interlopers.
The colours used for this were $u^{\ast}_{2}-g^{\prime}_{2}$ versus
$g^{\prime}_{\rm kron}$, $g^{\prime}_{2}-r^{\prime}_{2}$ versus
$r^{\prime}_{\rm kron}$, and $i^{\prime}_{2}-z^{\prime}_{2}$ versus
$z^{\prime}_{\rm kron}$ for the three filters respectively, where the subscript
\textit{2} refers to the $2$ arcsec aperture used in the magnitude
measurements.  To define and select cluster members in the CMD, we defined
upper and lower colour cuts and only galaxies between these two lines were
used to produce the LF, as galaxies outside these
limits were most likely not cluster members. To define these two colour
cuts, we first defined the red sequence line in the CMD and then pushed
this line up and down to allow for statistical errors, and for the 
likely range of galaxy colours.

To define the red sequence line, we first defined the slope and then
its Y-intercept. We checked that the BCGs lay at the centre of the 
X-ray emission in all clusters, and then
calculated the red sequence slope in each case
by fitting a straight line to the bright galaxies in the
CMD. Bright galaxies are defined as those with magnitude ranging from
that of the brightest cluster galaxy, $m_{\rm{BCG}}$, to a magnitude of
$m_{\rm{BCG}}+3$, inclusive. We found that the slope of the red sequence
line for the 14 C1 clusters showed a mild trend with redshift: it was
steeper for high-redshift clusters. A similar trend was observed
by \citet{gilbank2008} and attributed to a deficit of faint red galaxies
at high redshifts, consistent with the galactic downsizing
picture. 

We divided our cluster sample into two redshift ranges:
low-redshift clusters ($z < 0.2$) and intermediate-redshift clusters ($0.2
\leq z \leq 0.61$). Clusters from each group share a common red sequence line
slope with a small variation. The common slope for the low-redshift range
was -0.007 and -0.025 for the second range.  Instead of using a different
red sequence slope for each cluster, we used the common slope of the
redshift groups for all clusters belonging to that redshift group.

The Y-intercepts of the red sequence lines were different for each galaxy
cluster and depended on the average colour of the bright galaxies as
defined above.  To fix the value of the intercept for each cluster, the red sequence line
was normalised so as to pass through the point in the CMD which has a magnitude of
$m_{\rm{BCG}}+1.5$ and colour equal to the average colours of the bright
galaxies. This point and
the value of the slope completes the definition of the red sequence line.

\begin{figure}

\epsfig{file=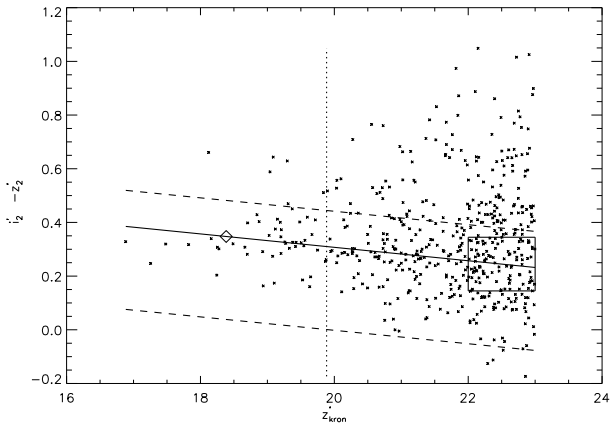,width=8.5cm,height=6cm}

\caption{Colour-magnitude diagram of cluster 25 (redshift=0.26). All galaxies (crosses) on the left of the vertical dotted line are the bright galaxies with magnitude $\leq m_{\rm{BCG}}+3$. The red sequence line (the solid line) is defined by the point (diamond) with magnitude of $m_{\rm{BCG}}+1.5$ and colour equal to the average colours of all bright galaxies and by the slope of -0.025. The statistical scatter, $\sigma$, is estimated by the average colour errors of the galaxies within the $1.0 \times 0.1$ box on the faint end of the red sequence line. In the case of cluster 25, $\sigma =0.07$. The dashed lines are the upper and lower colour cuts. The upper colour cut is the red sequence line pushed upward by $2\sigma$ and the lower colour cut is the red sequence line pushed downward by $2\sigma + \bigtriangleup$, where $\bigtriangleup = 0.175$ for redshift of 0.26. See text for definition of $\bigtriangleup$.}
\label{color_cuts}
\end{figure}

Both upper and lower colour cuts have the same slope as the red sequence
line. In order to define the upper colour cut, we have to determine the
upper (red) limit to the cluster red sequence. We took into account
the statistical scatter of the colours of the faintest galaxies on the red
sequence. These galaxies are defined as those inside a $1.0 \times 0.1$
(magnitude by colour units) box in the CMD centred on the faint end of the
red sequence line (see figure \ref{color_cuts} ). The size of this box was 
chosen to include the faintest
galaxies most probably belonging to the red sequence after studying the CMD
of the C1 sample. The expected scatter of these galaxies, $\sigma$, is calculated by
averaging their colour errors, that is the Y-axis errors in the CMD. The
upper colour cut, is then taken to be the red sequence line pushed upward by $2\sigma$.
By taking into account this scatter, we ensure that
almost all genuine cluster red sequence galaxies should fall beneath the red
cut, since the statistical error on the brighter galaxies will be smaller.

Similarly, the lower (blue) colour cut is the red sequence line pushed
downward in the CMD. In this case, the shift has to account for both
statistical scatter, and for the fact that late-type cluster galaxies are
intrinsically bluer than red sequence galaxies. The shift was therefore
taken to be $-(2\sigma+\bigtriangleup)$. Where $\bigtriangleup$ is the
theoretical colour difference between ellipticals and spirals. This was
estimated using a simple model which calculates what colour late-type
galaxies would have when redshifted by different amounts, as described in
\citet{King1985}. $\bigtriangleup$ is a function of redshift only and the
redshift of the galaxy cluster was used to determine its value. 
This method of estimating $\bigtriangleup$ ignores any intrinsic 
evolution in the colour offset between red and blue cluster galaxies.
However, the detailed COMBO-17 study of \cite{Bell2004} 
(see their Figure 1) shows that
the colour difference between blue and red sequence cluster
galaxies changes little over the redshift range (0-0.7) spanned
by our clusters. Figure \ref{color_cuts} shows an example of our
use of colour cuts for selection of cluster galaxies.


Of course, background and foreground galaxies will still contaminate the
sample after the colour cut has been applied, and this contamination must
be estimated and removed statistically. For this purpose we used all data
in the catalogue to which a given galaxy cluster belonged. 
In addition to simple Poisson fluctuations, uncertainties in removing
background and foreground galaxies arise from large scale structure.
To quantify the extra fluctuations arising from this, we proceeded
as follows. The whole catalogue $1^{\circ} \times 1^{\circ}$ area 
was divided into smaller blocks with
areas comparable to that of the cluster in question. Any of these blocks
covered mostly (60\% or more, by area) by a galaxy cluster, were considered
to be dominated by a cluster and hence were discarded from the background
calculation. Blocks covered by clusters to an extent less than 60\%, were
not discarded but the portion covered by the $R^{\ast}$ circle of any
galaxy cluster was removed, so the final blocks used have somewhat
different areas.

For each background block, an LF was produced in just the same way as for
the cluster itself. The same values of the upper and lower red sequence
limits of the galaxy cluster in question, were applied to all its
background block areas, so galaxies beyond those limits were removed. The
application of colour cuts to both source and background fields reduces the
noise level in both of them, and hence in the final background-subtracted
LF.

We then divided each background block LF by its area,  added them and
normalised the resulting single LF to the area of the galaxy cluster
in question.  
The error bars on the averaged background LF were calculated from the
scatter of the individual block LFs contributing to it.  This method of
estimating the background has some advantages over the more conventional
background estimation method using an outer annulus around the galaxy
cluster, since it uses a large background region, and the error estimate
allows for the variance arising from the large scale structure. Finally,
for each cluster we subtracted its composite background LF from the cluster
LF, and propagated the errors.

Apparent magnitudes were converted to absolute magnitudes, using the distance
for each cluster, and applying K-corrections calculated from
Table 3 (for Hubble type E) in \cite{Frei1994}. 
The use of early-type K-corrections is common in cluster studies, and
justified by the dominance of early-type galaxies in clusters.
However, if there were a systematic trend in early-type fraction
with magnitude, then this could lead to some distortion
of the LF slope. To quantify the maximum possible effect, we note that, 
using the tables in \cite{Frei1994}, 
the K-correction for Hubble type E at z=0.6 for $z^\prime$ is 0.37 while 
it is 0.05 for Hubble type Im). Assuming (very conservatively) a systematic
change from a 100\% early-type to 100\% late-type population across the
faint end slope of our LFs, the impact of a differential error of
0.3 magnitudes on our determination of $\alpha$ would still only
amount to $\Delta \alpha \approx 0.04$, which is small compared to
the trends in alpha which represent some of our main results.
The tables in \cite{Frei1994} apply to SDSS filters, which differ
slightly from the corresponding MegaCam filters. The resulting
differences in K-corrections 
are much smaller than the differences between early-type 
and late-type galaxies (about 0.03 at z=0.1 and 0.06 at z=0.6), and
will have  negligible effect on our derived LF slopes.

Finally, the data were binned into bins of width 0.5 magnitude
(experiments showed that this bin size was a good choice
in terms of fit quality and parameter confidence regions), and
the resulting LFs were then fitted by a Schechter  function model 
(\cite{Scechter1976}),
$$\phi(M)dM=0.4 ln(10) \phi^* e^{-X} X^{1+\alpha} dM ,$$ 
where $X=10^{-0.4(M-M^*)}$, $M^*$ is the characteristic magnitude, $\phi^*$ is the 
characteristic number density and $\alpha$ is the faint-end slope, \cite{Lin1996}. 
Contour plots of the $1\sigma$, $2\sigma$ and $3\sigma$ confidence levels of 
 $\alpha$ and $M^*$ were also produced. The errors in the text and tables refer to the
$1\sigma$ errors. We also calculated the total optical luminosity $L_{OPT}$ of 
 each cluster by integrating the fitted Schechter function from $5 \times M^*$ to -16. 

In addition to
single LFs for each  galaxy cluster in our sample, we produced
stacked luminosity functions.  The radius used to determine the volume is the $R^{\ast}$ of 
the cluster. Before stacking different clusters together, to correct for the evolution in the critical 
density of the universe, we multiply the LF by 
$$\frac{\rho_{c}(z=0)}{\rho_{c}(z=z_{cl})} ,$$ 
where $z$ is the redshift, $\rho_{c}$ is the  critical density of the universe, a function of $z$,
and $z_{cl}$ is the redshift of the cluster. This correction is necessary for high-redshift clusters 
if stacked with low-redshift clusters to scale the galaxy density in each cluster to the density at 
redshift=0.  The faintest magnitude bin is not necessarily the same for each cluster and to account 
for this, we divided the total number of galaxies
in each magnitude bin by the summed volume of galaxy clusters that
contributed to that bin only. The stacked LFs should enable us to
study the evolution of the LF with redshift and to explore any differences between
clusters of different temperature.


\begin{table*}
\begin{center}
\begin{tabular}{cccccccccccccccc}
\hline

 XLSSC  &  R.A.  &  Dec  &  Redshift  &  T  &  L$_{X}$  &  r$_{500}$  \\ 
 number  &  (J2000)  &  (J2000)  &  &  ($keV$)  &  10$^{43}  erg  s^{2}$  &  (Mpc)  \\ 
 \hline 
11 & 36.54 & -4.97 & 0.05 & 0.64 & 0.11 & 0.29 \\ 
21 & 36.23 & -5.13 & 0.08 & 0.68 & 0.11 & 0.3 \\ 
41 & 36.38 & -4.24 & 0.14 & 1.34 & 2.4 & 0.44 \\ 
25 & 36.35 & -4.68 & 0.26 & 2.0 & 4.6 & 0.53 \\ 
44 & 36.14 & -4.23 & 0.26 & 1.3 & 1.2 & 0.4 \\ 
22 & 36.92 & -4.86 & 0.29 & 1.7 & 6.2 & 0.47 \\ 
27 & 37.01 & -4.85 & 0.29 & 2.8 & 4.8 & 0.65 \\ 
8 & 36.34 & -3.8 & 0.3 & 1.3 & 1.2 & 0.4 \\ 
13 & 36.86 & -4.54 & 0.31 & 1.0 & 1.3 & 0.34 \\ 
40 & 35.52 & -4.55 & 0.32 & 1.6 & 1.6 & 0.44 \\ 
18 & 36.01 & -5.09 & 0.32 & 2.0 & 1.3 & 0.52 \\ 
6 & 35.44 & -3.77 & 0.43 & 4.8 & 60.3 & 0.84 \\ 
49 & 35.99 & -4.59 & 0.49 & 2.2 & 4.3 & 0.49 \\ 
1 & 36.24 & -3.82 & 0.61 & 3.2 & 33.2 & 0.58 \\ 
2 & 36.38 & -3.92 & 0.77 & 2.8 & 19.6 & 0.49 \\ 
29 & 36.02 & -4.23 & 1.05 & 4.1 & 48.3 & 0.52 \\ 
5 & 36.79 & -4.3 & 1.05 & 3.7 & 17.1 & 0.49 \\

\hline
\end{tabular}
\caption{List of the 17 C1 galaxy clusters covered by CFHTLS optical
survey and their properties sorted according to their redshifts (\protect \cite{pacaud07}). The three highest redshift
clusters (2,29 and 2) though covered by the survey, were not included in our analysis because their data were too poor
to yield useful fits.} \label{C1usters}
\end{center}
\end{table*}

\section{Results}

\subsection{Individual cluster luminosity functions}
The fitted values of the three Schechter parameters $\alpha$ and $M^*$ and
$L_{OPT}$ of the individual C1 clusters are presented in Table
\ref{IndividualResults}. The LF plots with the associated $1\sigma$,
$2\sigma$ and $3\sigma$ contours in the $M^*$-$\alpha$ plane for passbands
$r^\prime$ and $z^\prime$ are shown in Figures \ref{Ind_r} and \ref{Ind_z}
respectively. For some of the C1 clusters, the fitting failed to constrain
some of the parameters, $M^*$ in particular, due to poor statistics or the
lack of any well-defined turnover in the LF at the bright end. For these
clusters the LF and best fit are presented without any accompanying
confidence contour plot. These LFs are placed at the bottom of the
figures. For clusters with unconstrained $M^*$, $L_{OPT}$ was also not
constrained, because its value depends on both $\alpha$ and
$M^*$. Therefore, we excluded these clusters in the part of the analysis
related to $L_{OPT}$.

The average values of $\alpha$ for our sample of clusters are
$$-1.70\pm0.10, -1.64\pm0.04 \; \mathrm{and} -1.43\pm0.03$$ for the $g^\prime$,
$r^\prime$ and $z^\prime$ passbands respectively. The correlations between
$L_{OPT}$, $\alpha$ and $M^*$ and redshift, temperature ($T$) and the X-ray
luminosity ($L_{X}$) taken from \cite{pacaud07}, were tested
using Pearson's correlation coefficient. These coefficients are
computed from the ratio of the covariance of the tested variables, X and Y, to the
square root of the product of the variances of these variables, i.e.
$$r=\frac{COV(X,Y)}{\sqrt{VAR(X)*VAR(Y)}}.$$

This correlation coefficient measures the linear correlation, if it is
1 or -1 then the two variables are perfectly positively or negatively linearly
correlated, respectively. To compute the upper and lower 1 $\sigma$ errors
on the correlation coefficient r, we used Fisher's Z
transformation: $Z={\rm tan}^{-1} r$. The strongest correlations
found are those between $L_{OPT}$ and $T$ and between $L_{OPT}$ and $L_{X}$, both of
which are expected from the scaling relations of galaxy
clusters. In our sample, they both have a correlation coefficient of at
least 0.9, see Table \ref{Corr_Coeff}.

Because higher redshift clusters are more difficult to detect than nearby
ones, they will tend to be more massive and hence hotter than typical
nearby clusters, see Figure 3 in \cite{pacaud07}. This ({\it Malmquist})
selection effect is present in any deep
cluster survey. To account for the $T-z$ correlation arising from this
selection effect in the C1 sample, for each correlation coefficient between
a quantity and $T$ or $z$, we have also calculated the partial correlation
coefficient between the same two quantities, which attempts to remove any
part of the correlation which arises due to the intrinsic trend in $T$ with
$z$ within our sample. For this we used an Interactive Data Language
($\textit{IDL}$) routine, $\textit{p\_correlate.pro}$ to compute the
partial correlation coefficient. This uses the following method, which to
be concrete we explain using the example of the correlation between
$\alpha$ and redshift.
Let $\alpha$ and redshift $z$ are the variables of primary
interest, whilst temperature $T$ is a third variable whose effects we
wish to remove. First, the routine calculates the
residuals after regressing
$\alpha$ on $T$; these are the parts of $\alpha$ that cannot be
predicted by $T$.
Likewise, it calculates the residuals after regressing $z$ on $T$.
Finally, the partial correlation
coefficient between $\alpha$ and $z$, adjusted for $T$, is the
correlation between these two sets of residuals.

The results of our correlation analysis for the unstacked clusters, are
tabulated in the top section of Table \ref{Corr_Coeff}.  The correlation
coefficients between the faint-end slope, $\alpha$ of the individual
clusters and redshift are 0.44$\pm$0.27 for the $r^\prime$ band and
0.54$\pm$0.25 for the $z^\prime$ band. These coefficients, including the
coefficient for the $g^\prime$ band, get stronger after the application of
the partial analysis and the errors on the coefficients become
smaller. This strongly suggests evolution of $\alpha$ with redshift in our
sample. We will further scrutinise this possibility in the section of
redshift-stacked clusters, because stacking LFs of clusters with similar
redshifts should lower scatter in the data and provide a means to probe
possible trends. $M^*$ also shows a negative correlation with redshift and
with temperature but these correlations become insignificant in a partial
correlation analysis.

\subsection{Global scaling relations}
The relationships between the global cluster properties, $L_{OPT}$, $L_X$ and
$T$ provide a probe of cluster self-similarity. $L_{OPT}$ is strongly
correlated to the temperature of our clusters --
the correlation coefficients between $L_{OPT}$ and $T$ are  0.95$\pm$0.06, 0.96$\pm$0.04 and 0.97$\pm$0.03 for the
$g^\prime$, $r^\prime$ and $z^\prime$ bands respectively, whilst the partial
correlation coefficients  for the same quantities, factoring out the effects 
of $z$, are 0.87$\pm$0.16, 0.89$\pm$0.11
and 0.92$\pm$0.06, see third row in Table \ref{Corr_Coeff}. The removal of the $z$ effects has lowered the values of the
coefficients but they are still high and significant. Correlation between $L_{OPT}$ and $L_{X}$
is also quite strong: 0.92$\pm$0.11 ($g^\prime$ band), 0.93$\pm$0.07 ($r^\prime$ band) and 0.90$\pm$0.08 ($z^\prime$
band). 

\begin{figure}

\epsfig{file=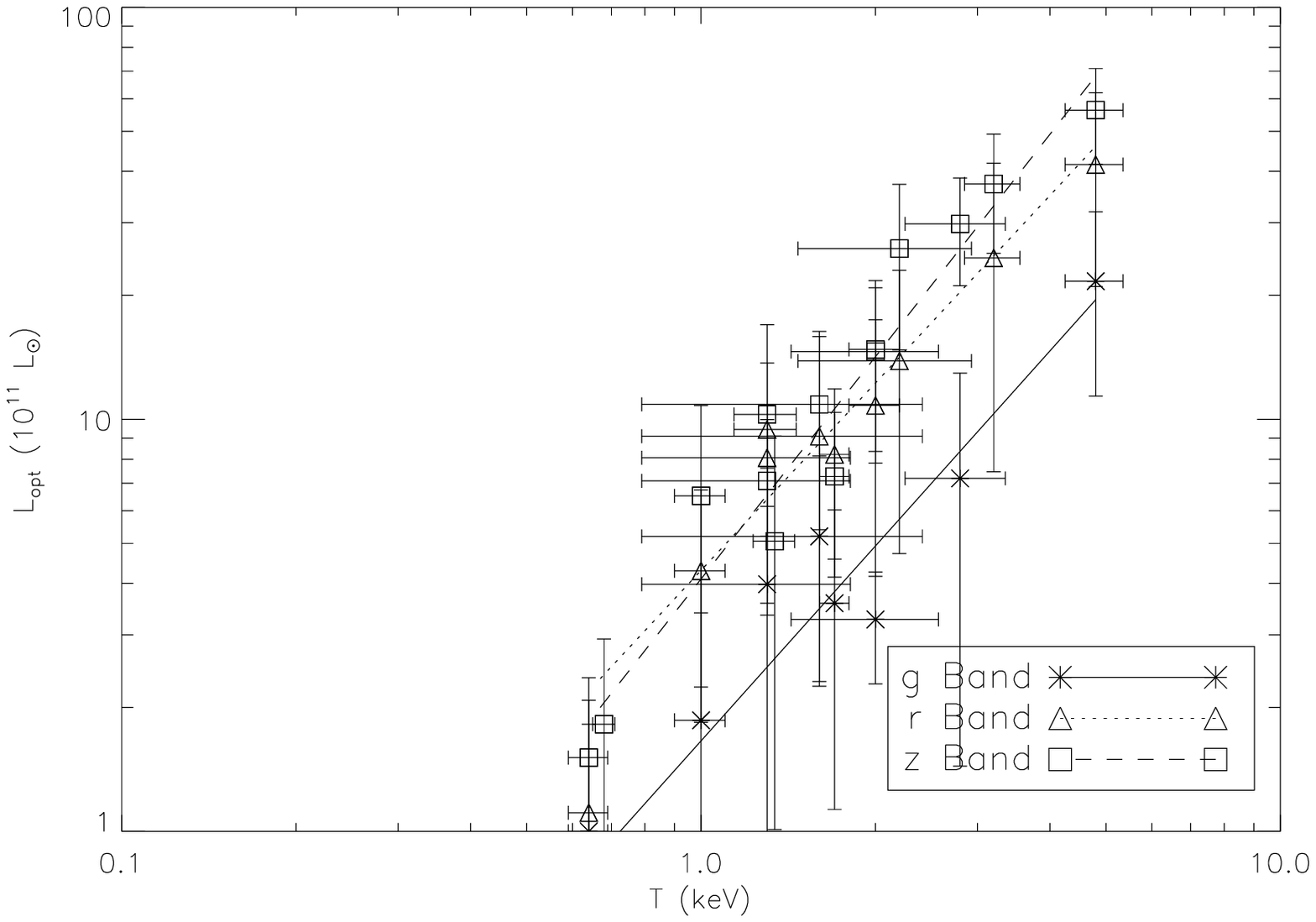,width=9cm,height=7cm}
\epsfig{file=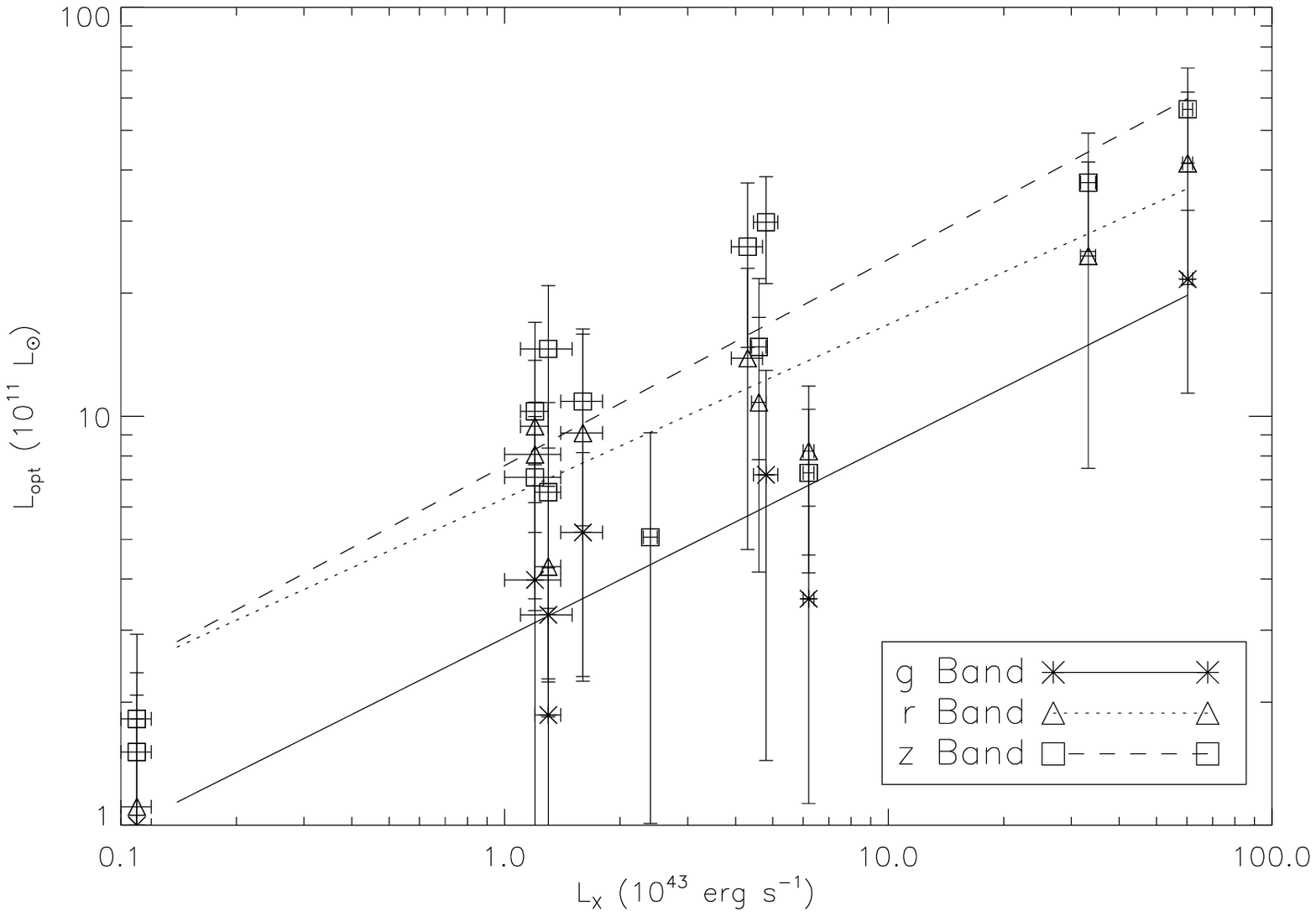,width=9cm,height=7cm}

\caption{Correlation diagrams of $L_{OPT}$ versus X-ray gas temperature, T (top panel) and $L_{OPT}$ versus X-ray luminosity, $L_{X}$ (bottom panel) of C1 clusters for passbands $g^\prime$ (stars),$r^\prime$ (triangles) and $z^\prime$ (squares). Clusters with unconstrained $M^*$ and hence unconstrained $L_{OPT}$ were excluded.}
\label{Corr_plots}
\end{figure}

In Figure \ref{Corr_plots},  we plot $L_{OPT}$ versus $T$ (top panel) and $L_{OPT}$ versus $L_{X}$ (bottom panel). We
calculate the slopes for these plots using the  $\textit{Fortran}$ package
$\textit{ODRPACK}$ (\cite{Akritas1996}), which uses numerical orthogonal distance regression method to minimise
perpendicular distances between points and  the fitted line. One advantage of this is that the slope value will not
change if the quantities in question switch axes. In addition, $\textit{ODRPACK}$ takes into account errors on
both X-values and Y-values which are available for $L_{OPT}$, $T$ and  $L_{X}$.

The logarithmic slopes for the $L_{OPT}-T$ relation for the three filters
$g^\prime$, $r^\prime$ and $z^\prime$, respectively are $1.57\pm 0.17,
1.51\pm 0.17 \; \mathrm{and} \; 1.79\pm 0.12 $, giving an average value of
$1.62\pm0.11$. For the $L_{OPT}-L_{X}$ relation, the slopes are $0.47\pm
0.07, 0.43\pm 0.08 \; \mathrm{and} \; 0.50\pm 0.07$, and the average value
is $0.47\pm0.05$. Note that the slopes do not differ significantly for the
three filters, except the slope of $L_{OPT}$ versus $T$ in the $z^\prime$
filter. Such relations between $L_{OPT}$ on one hand, and $L_{X}$ and the
gas temperature on the other, are expected because richer and hence more
luminous clusters have deeper gravitational potential wells which in turn
raise the ICM temperature and its X-ray output by adiabatic compression and
shocks generated by supersonic motion. We will discuss this further in
Sec. 4.7.

The correlation coefficients between $L_{OPT}$ and redshift are high
(all above 0.8), but when the effects of temperature are removed
they become insignificant in at least two of the filter set, therefore,
this correlation is most likely due to selection effects, $\textit{Malmquist
effect}$, and does not reflect any genuine relationship between $L_{OPT}$ and
$z$. Correlations between $\alpha$ and $M^*$ and $T$, $z$ and $L_{X}$
were also computed, but none of those showed significantly
high values.

Following the above analysis of trends in the properties of individual
clusters, we now perform a stacking analysis, grouping clusters
first by redshift, and then by temperature. This provides LFs of higher
statistical quality, enabling the behaviour to be examined in greater
detail.

\begin{table}
\begin{center}
\begin{tabular}{llllllllll}
\hline

 XLSSC    &  g  Band  &  r  Band  &  z  Band  \\ 
 number  &  &  &  \\ 
 \hline 
 &  &  $\alpha$  &  \\ 
 \hline 
1 &  -1.94$\pm$0.23  &  -1.59$\pm$0.2  &  -1.06$\pm$0.17  \\ 
6 &  -1.61$\pm$0.16  &  -1.7$\pm$0.09  &  -1.31$\pm$0.09  \\ 
8 &  -1.53$\pm$0.37  &  -1.39$\pm$0.2  &  -1.15$\pm$0.16  \\ 
11 &  -1.67$\pm$0.09  &  -1.8$\pm$0.05  &  -1.71$\pm$0.04  \\ 
13 &  -1.63$\pm$0.63  &  -1.5$\pm$0.07  &  -1.51$\pm$0.08  \\ 
18 &  -1.21$\pm$0.88  &  -1.76$\pm$0.13  &  -1.53$\pm$0.12  \\ 
21 &  -2.01$\pm$0.11  &  -1.89$\pm$0.06  &  -1.77$\pm$0.06  \\ 
22 &  -1.62$\pm$0.26  &  -1.19$\pm$0.19  &  -1.16$\pm$0.15  \\ 
25 &  -2.1$\pm$0.12  &  -1.73$\pm$0.09  &  -1.57$\pm$0.08  \\ 
27 &  -1.78$\pm$0.14  &  -1.85$\pm$0.12  &  -1.56$\pm$0.1  \\ 
40 &  -1.03$\pm$0.3  &  -1.55$\pm$0.13  &  -1.27$\pm$0.09  \\ 
41 &  -1.84$\pm$0.07  &  -1.86$\pm$0.09  &  -1.67$\pm$0.08  \\ 
44 &  -1.75$\pm$0.12  &  -1.47$\pm$0.07  &  -1.44$\pm$0.09  \\ 
49 &  -1.99$\pm$0.38  &  -1.65$\pm$0.16  &  -1.36$\pm$0.12  \\ 
 \hline 
 &  &  $M*$  &  \\ 
 \hline 
1 &  -34.65$\pm$***  &  -23.7$\pm$0.91  &  -23.47$\pm$0.32  \\ 
6 &  -20.96$\pm$0.43  &  -23.24$\pm$0.5  &  -22.98$\pm$0.23  \\ 
8 &  -21.42$\pm$2.11  &  -21.79$\pm$0.83  &  -22.55$\pm$0.69  \\ 
11 &  -20.61$\pm$1.5  &  -21.81$\pm$1.84  &  -21.13$\pm$0.73  \\ 
13 &  -19.78$\pm$1.22  &  -22.19$\pm$0.37  &  -24.31$\pm$0.73  \\ 
18 &  -19.66$\pm$1.43  &  -31.09$\pm$***  &  -22.23$\pm$0.41  \\ 
21 &  -30.21$\pm$***  &  -29.02$\pm$***  &  -21.16$\pm$0.8  \\ 
22 &  -20.26$\pm$0.84  &  -20.62$\pm$0.39  &  -22.15$\pm$0.5  \\ 
25 &  -29.29$\pm$***  &  -22.66$\pm$0.78  &  -22.98$\pm$0.52  \\ 
27 &  -22.22$\pm$1.32  &  -33.02$\pm$***  &  -23.52$\pm$0.76  \\ 
40 &  -21.3$\pm$0.78  &  -22.95$\pm$0.96  &  -23.21$\pm$0.5  \\ 
41 &  -30.57$\pm$***  &  -32.2$\pm$***  &  -23.38$\pm$1.5  \\ 
44 &  -33.23$\pm$***  &  -22.79$\pm$0.53  &  -23.42$\pm$0.62  \\ 
49 &  -31.66$\pm$***  &  -23.06$\pm$0.82  &  -23.72$\pm$0.46  \\ 
 \hline 
 &  &  $L_{OPT}$  10$^{11}$  $L_{\sun}$  &  \\ 
 \hline 
1 &  57.51$\pm$***  &  24.66$\pm$17.2  &  37.27$\pm$11.98  \\ 
6 &  21.64$\pm$10.25  &  41.53$\pm$20.52  &  56.29$\pm$14.72  \\ 
8 &  3.98$\pm$3.63  &  3.07$\pm$1.92  &  7.09$\pm$3.74  \\ 
11 &  1$\pm$0.82  &  1.11$\pm$0.97  &  1.51$\pm$0.85  \\ 
13 &  1.86$\pm$1.53  &  12.93$\pm$4.75  &  23.17$\pm$12.84  \\ 
18 &  1.27$\pm$0.99  &  107.34$\pm$***  &  14.61$\pm$6.25  \\ 
21 &  0.94$\pm$***  &  4.19$\pm$***  &  1.82$\pm$1.11  \\ 
22 &  3.58$\pm$2.45  &  4.22$\pm$1.64  &  7.27$\pm$3.13  \\ 
25 &  10.95$\pm$***  &  10.81$\pm$6.65  &  14.78$\pm$6.95  \\ 
27 &  7.19$\pm$5.75  &  42.62$\pm$***  &  14.81$\pm$8.72  \\ 
40 &  5.2$\pm$2.95  &  10.1$\pm$6.79  &  12.87$\pm$5.47  \\ 
41 &  12.99$\pm$***  &  12.21$\pm$***  &  5.06$\pm$4.05  \\ 
44 &  95.01$\pm$***  &  9.45$\pm$4.25  &  13.28$\pm$6.7  \\ 
49 &  17.41$\pm$***  &  13.86$\pm$9.14  &  25.97$\pm$11.23  \\

\hline
\end{tabular}
\caption{Results of the Schechter function fitting of the LFs of the 
14 C1 galaxy clusters. For some clusters, the $M*$ values were not 
constrained by the fitting program and the errors of these 
unconstrained $M*$ are starred. Also, the errors of the corresponding 
$L_{OPT}$ values are starred, since the computation of $L_{OPT}$ 
depends on both $\alpha$ and $M*$.} \label{IndividualResults}
\end{center}
\end{table}

\begin{table}
\begin{center}
\begin{tabular}{llllllll}
\hline

 & $g^\prime$ Band & $r^\prime$ Band & $z^\prime$ Band \\
Quantities & P.C. Coeff. & P.C. Coeff. & P.C. Coeff. \\
\hline       
\multicolumn{4}{|c|}{Individual non-stacked C1 clusters} \\       
\hline       
L$_{OPT}$,L$_{X}$ & 0.92$\pm$0.11 & 0.93$\pm$0.07 & 0.90$\pm$0.08 \\
L$_{OPT}$,T & 0.95$\pm$0.06 & 0.96$\pm$0.04 & 0.97$\pm$0.03 \\
L$_{OPT}$,T,z (Partial) & 0.87$\pm$0.16 & 0.89$\pm$0.11 & 0.92$\pm$0.06 \\
L$_{OPT}$,z & 0.82$\pm$0.21 & 0.83$\pm$0.16 & 0.86$\pm$0.10 \\
L$_{OPT}$,z,T (Partial) & 0.28$\pm$0.44 & 0.36$\pm$0.36 & 0.61$\pm$0.22 \\
$\alpha$,L$_{X}$ & -0.08$\pm$0.30 & 0.23$\pm$0.30 & 0.64$\pm$0.21 \\
$\alpha$,T & 0.01$\pm$0.29 & 0.05$\pm$0.30 & 0.54$\pm$0.25 \\
$\alpha$,T,z (Partial) & 0.00$\pm$0.29 & -0.37$\pm$0.28 & -0.17$\pm$0.30 \\
$\alpha$,z & 0.01$\pm$0.29 & 0.44$\pm$0.27 & 0.54$\pm$0.25 \\
$\alpha$,z,T (Partial) & 0.20$\pm$0.30 & 0.67$\pm$0.20 & 0.65$\pm$0.21 \\
$M^*$,T & -0.32$\pm$0.43 & -0.57$\pm$0.31 & -0.48$\pm$0.26 \\
$M^*$,T,z (Partial) & -0.47$\pm$0.41 & -0.15$\pm$0.37 & -0.02$\pm$0.29 \\
$M^*$,z & -0.06$\pm$0.43 & -0.57$\pm$0.31 & -0.48$\pm$0.26 \\
$M^*$,z,T (Partial) & 0.25$\pm$0.44 & 0.00$\pm$0.36 & 0.00$\pm$0.29 \\
\hline       
\multicolumn{4}{|c|}{Redshift-stacked} \\       
\hline       
$\alpha$,z & 0.97$\pm$0.10 & 0.88$\pm$0.29 & 0.89$\pm$0.29 \\
$\alpha$,z,T (Partial) & 0.91$\pm$0.24 & 0.65$\pm$0.58 & 0.51$\pm$0.65 \\
$M^*$,z & -0.86$\pm$0.33 & -0.87$\pm$0.31 & -0.71$\pm$0.53 \\
$M^*$,z,T (Partial) & -0.21$\pm$0.67 & -0.58$\pm$0.62 & -0.33$\pm$0.68 \\
\hline       
\multicolumn{4}{|c|}{Temperature-stacked} \\       
\hline       
$\alpha$,T & 0.10$\pm$0.64 & 0.31$\pm$0.68 & 0.75$\pm$0.49 \\
$\alpha$,T,z (Partial) & -0.44$\pm$0.67 & -0.85$\pm$0.35 & -0.45$\pm$0.67 \\
$M^*$,T & -0.46$\pm$0.67 & -0.26$\pm$0.67 & 0.67$\pm$0.56 \\
$M^*$,T,z (Partial) & 0.20$\pm$0.67 & -0.46$\pm$0.67 & 0.25$\pm$0.67 \\

\hline
\end{tabular}
\caption{Pearson's correlation coefficients (P.C. Coeff.) of individual 
C1 clusters, redshift-stacked clusters and temperature-stacked clusters for
the three-filter set ($g^\prime$, $r^\prime$,$z^\prime$).  'X,Y,Z
(Partial)' denotes partial correlation coefficient of quantities X and Y
with effects of quantity Z removed, to be compared with the line directly 
above it, where correlation coefficient of the same quantities X and Y is
presented without partial analysis. } \label{Corr_Coeff}
\end{center}
\end{table}

\subsection{Redshift-stacked clusters}
The 14 C1 clusters span a redshift range 0.05 to 0.61. This range was
divided into five redshift bins: 0.05-0.14, 0.26-0.26, 0.29-0.29, 0.30-0.32
and 0.43-0.61. The number of clusters in each bin ranges from two to
four. The redshift ranges of these bins were chosen according to two
criteria: first, the redshift range of the combined clusters was not too
large, and second we required adequate data quality in each bin, to allow a
well-constrained Schechter function fit.  We kept the number of bins to at
least five because a smaller number of bins increases the errors on the
correlation coefficients. Plots of the redshift-stacked data with fitted
Schechter functions for the three photometric bands are shown in Figures
\ref{z_stacked_g}, \ref{z_stacked_r} and \ref{z_stacked_z}, and results of
the fits are given in Table~\ref{Stacked_Results}.

The faint-end slope, $\alpha$ of the Schechter function of the stacked data
shows an evolutionary trend, becoming less steep with
increasing redshift. Three of the redshift bins (0.26-0.26, 0.29-0.29,
0.30-0.32) have very similar redshifts and in general the $\alpha$ values 
for these three bins agree within their errors.

The Pearson and partial correlation coefficients were calculated for
$\alpha$ and $z$, see Table \ref{Corr_Coeff}. The coefficients are high
($\geq0.88$) but with relatively large errors, mainly due to the small 
number of
bins. The partial correlation analysis lowered the values of the
coefficients and enlarged the errors. Evidence for evolution in $\alpha$ is 
seen in all three bands, arising primarily from the fact that the
faint-end slope is steeper ($\alpha=-1.75$ to -1.8) in the low $z$ bin 
than in the higher redshift bins.

One obvious concern in probing evolutionary trends in the Schechter
function fits is that the fitted magnitude range decreases systematically
with redshift, due to the apparent magnitude limit of our data.
A second effect which might bias $\alpha$ is that within a given
redshift bin, the contributing clusters are probed down to different
absolute magnitudes, according to their distance. Hence at the faint end, 
clusters may progressively drop out of the stacked LF. This is
especially the case for the lowest and highest redshift bin, which
are both much broader that the three bins at $z\approx 0.3$.
To show the scale of this latter effect, we
have drawn a vertical dotted line on each of the stacked LF plots
(\ref{Dip_tstacked_zband}, \ref{z_stacked_g}, \ref{z_stacked_r},
\ref{z_stacked_z}, \ref{t_gband_plot}, \ref{t_rband_plot} and
\ref{t_zband_plot}) to show the faintest magnitude to which {\it all}
clusters in the bin contribute.
To the right (fainter side) of this line, one or more of the
clusters in the redshift bin drop out of the stacked data. 

To check whether the trend of $\alpha$ with redshift is robust against
these two effects, we carried out tests on the stacked data, by
progressively removing the faintest magnitude bin in the stacked LFs and
re-fitting. In general, we found no significant change in the fitted
values of $\alpha$ (which changed only within their errors),
or in the $\alpha$-$z$ correlation when the LFs 
were truncated at the vertical dashed line, or when the LFs for all
redshift bins were fitted to the same limiting absolute magnitude
(which is set by the most distant systems). 
There was one exception
to this. The three clusters in the highest redshift band (clusters 1, 6 
and 49) all have $\alpha$ values (albeit with large errors)
steeper that the shallow slope of -1.31 which fits to the stacked data
in the $g^\prime$ band
for this high redshift bin. As the faintest bins, to the right of
the dashed line in the bottom plot of Figure \ref{z_stacked_g}, are
progressively removed, the fitted slope steepens. Hence the flat slope
of -1.31 must be regarded as unsafe, and the very high $\alpha$-$z$ 
correlation in $g^\prime$ band, given in Table~\ref{Corr_Coeff}, is
probably overestimated. Rather, we have a situation in all three
photometric bands, where the faint-end slope is steeper at $z<0.2$
than it is at higher redshift.

\begin{figure}
\center

\epsfig{file=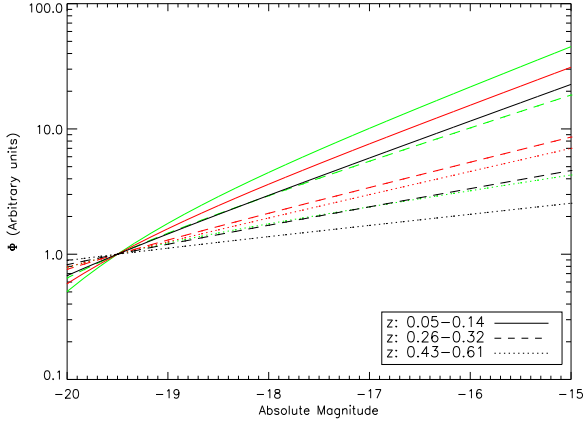,width=8cm,height=6.cm}

\caption{The faint end of the fitted LFs of the C1 sample grouped into three
redshift bins: low (0.05-0.14, solid lines), intermediate 
(0.26-0.32, dashed lines) and high (0.43-0.61, dotted lines). Colours 
represent the filter bands: green for $g^\prime$, red for $r^\prime$
and black for $z^\prime$. All LFs were normalised to have $\phi=1$ at
M=-19.5 for easy comparison. The faint end slopes become shallower 
with increasing redshifts. Also, at
intermediate redshift (dashed lines), the slope shows a trend with 
colour, becoming steeper towards the blue. This colour trend largely
vanishes at low redshifts (solid lines).}
\label{alpha_evolution}
\end{figure}

To visualise the behaviour of the faint-end slope in terms of both
redshift and colour, we plot the faint end of the fitted luminosity
functions for the three bands in Figure
\ref{alpha_evolution} using green for $g^\prime$ band,
red for $r^\prime$ band and black for $z^\prime$ band. For this
plot, we have divided the
sample into three redshift bins: low (0.05-0.14), intermediate
(0.26-0.32) and high (0.43-0.61), denoted by different line styles.
All LFs have been renormalised to have $\phi=1$ at M=-19.5.

The Figure shows how the faint end slope steepens towards low $z$.
It also illustrates colour trends in $\alpha$.
At low redshift (solid lines), the slopes are very similar (though the
curves are separated due to their different values of $M*$),
whilst at intermediate redshift (dashed lines), the slope shows a
strong trend with colour.

\begin{figure}
\center

\epsfig{file=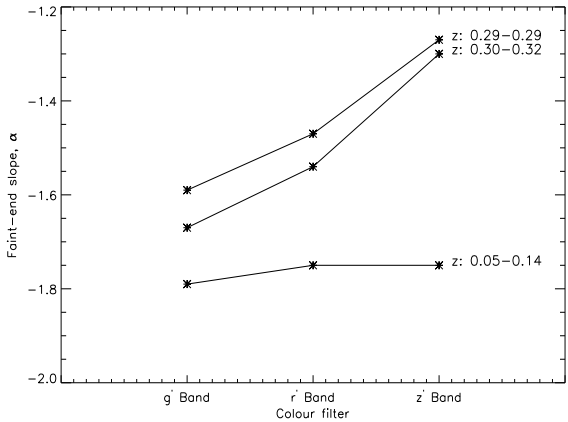,width=8cm,height=6.cm}

\caption{The faint-end slope of the fitted Schechter function in 
$g^\prime$, $r^\prime$ and $z^\prime$ bands for local clusters with 
redshift 0.05 to 0.14 and for intermediate redshift 
($0.29\leq z \leq 0.32$) clusters.}
\label{alpha_colour}
\end{figure}

The values of $\alpha$ in Table \ref{Stacked_Results} also show a trend
with colour. The faint-end slope of $z$-stacked clusters becomes steeper as
we move from $z^\prime$ (red side) band to $g^\prime$ (blue
side). This trend is very obvious in the second, third and fourth redshift
bins ($0.29\leq z
\leq 0.32$) and much less obvious and maybe absent (within the errors) in
the first bin($z\leq 0.14$), see Figure \ref{alpha_colour} in which we 
plotted the values of $\alpha$ for the three bands for the lower- and 
intermediate redshift bins. The increase in the faint-end slope of the
Schechter function in the bluer bands means that at the faint side of the
colour-magnitude diagram the blue galaxies outnumber red ones.

\begin{figure}

\epsfig{file=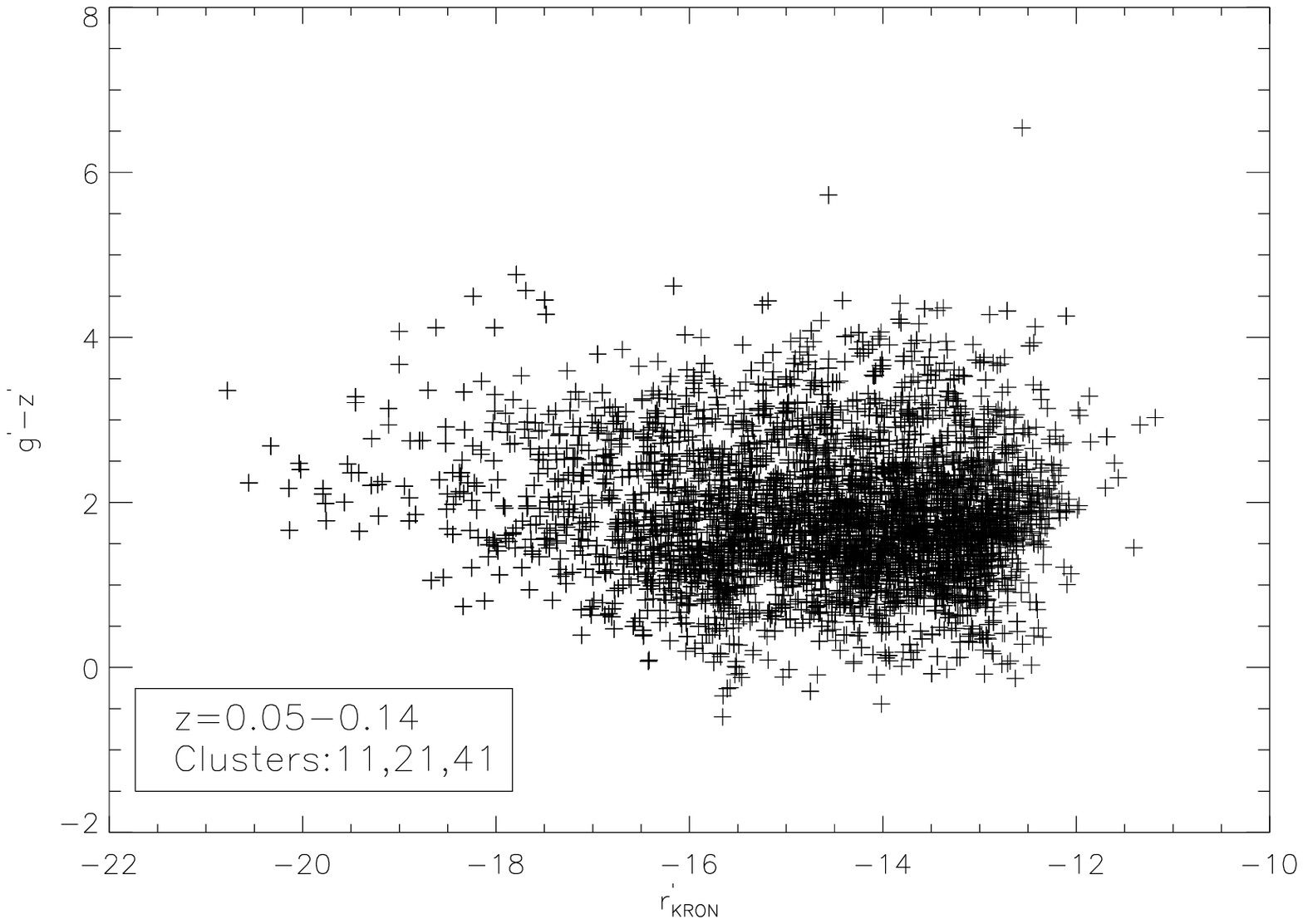,width=8cm,height=5cm}
\epsfig{file=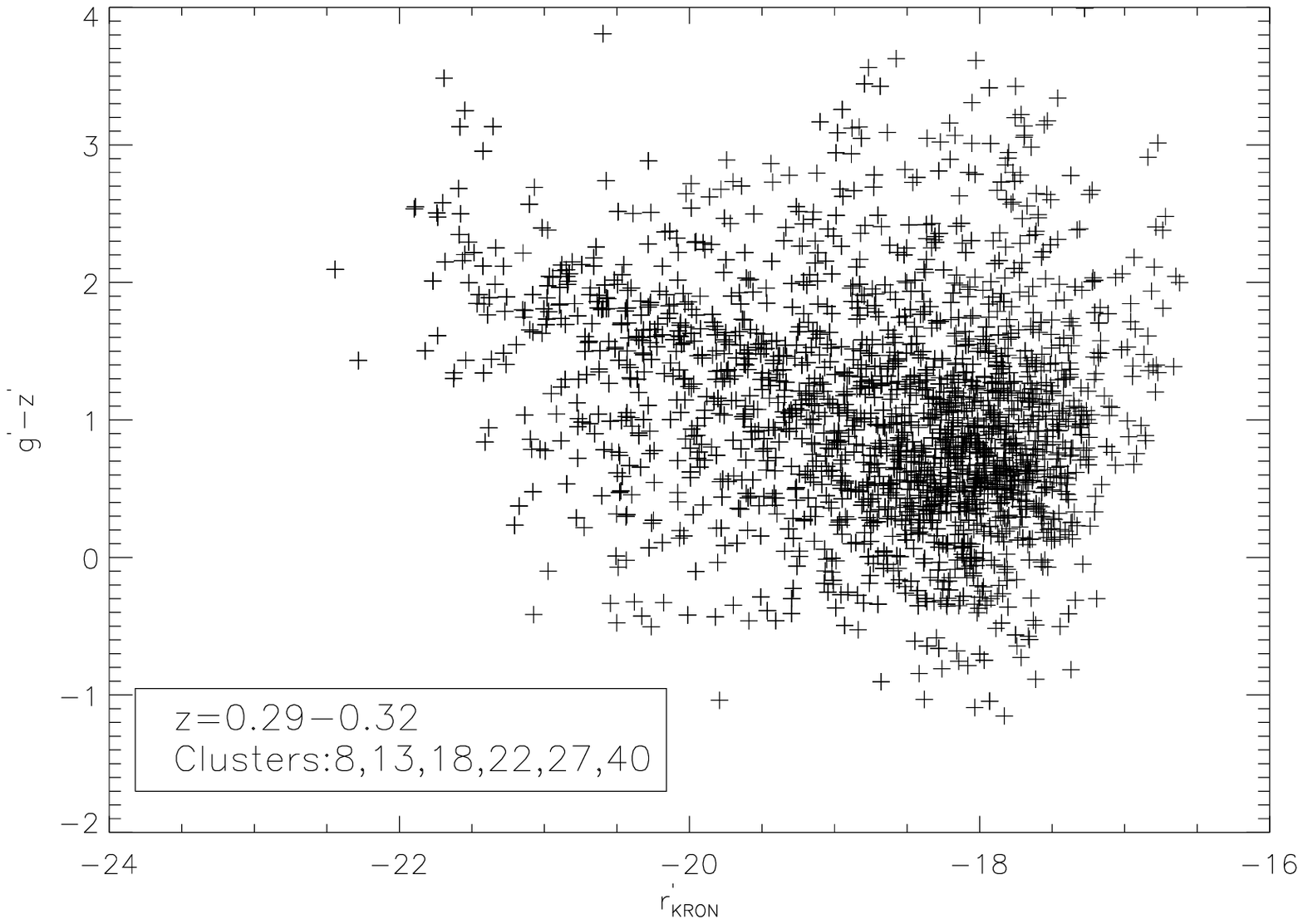,width=8cm,height=5cm}

\caption{Colour-magnitude diagram: $g^\prime - z^\prime$ versus $r^\prime$ 
(K-corrected) for low-redshift (z=0.05-0.14)
clusters and intermediate-redshift (z=0.29-0.32) clusters.}
\label{Colour_trend}
\end{figure}

To explore this we produced K-corrected colour-magnitude diagram
(Figure \ref{Colour_trend}) of $g^\prime_{2}-z^\prime_{2}$ versus absolute
$r_{\rm {kron}}$ magnitude for $0.29\leq z \leq0.32$ (six clusters: 
8,13,18,22,27 and 40) 
in which this trend is most obvious, and the same plot for the first
redshift bin, $0.05\leq z \leq0.14$ in which no such trend is apparent.
Figure \ref{Colour_trend} clearly demonstrate how the distribution
of cluster galaxy colours changes from $z\sim 0$ to $z\sim 0.3$.  
In the $g^\prime$ band the evolution of $\alpha$ is much stronger, 
especially after
removing the effects of the temperature (partial correlation). 
These trends in $\alpha$ show
that the fraction of blue faint galaxies at $z \sim 0.3$ was larger
than it is now, and suggests that these galaxies have reddened and moved


The Schechter function characteristic magnitude $M^*$ in the redshift-stacked
clusters showed a negative correlation with redshift. The correlation
coefficients between $M^*$ and redshift are high but less significant than
those between $\alpha$ and redshift.  However, when the partial
calculations were carried out, these coefficients dropped and became
consistent with zero. Hence the trend in $M^*$ with $z$ appears to be due
to a selection effect: hotter clusters are more luminous, and so
are more easily detected at high redshift, and these brighter clusters 
also tend to have brighter $M^*$ (\cite{zandivarez}).

\subsection{Luminosity functions of $z\sim 0.3$ clusters}

\begin{figure}
\center

\epsfig{file=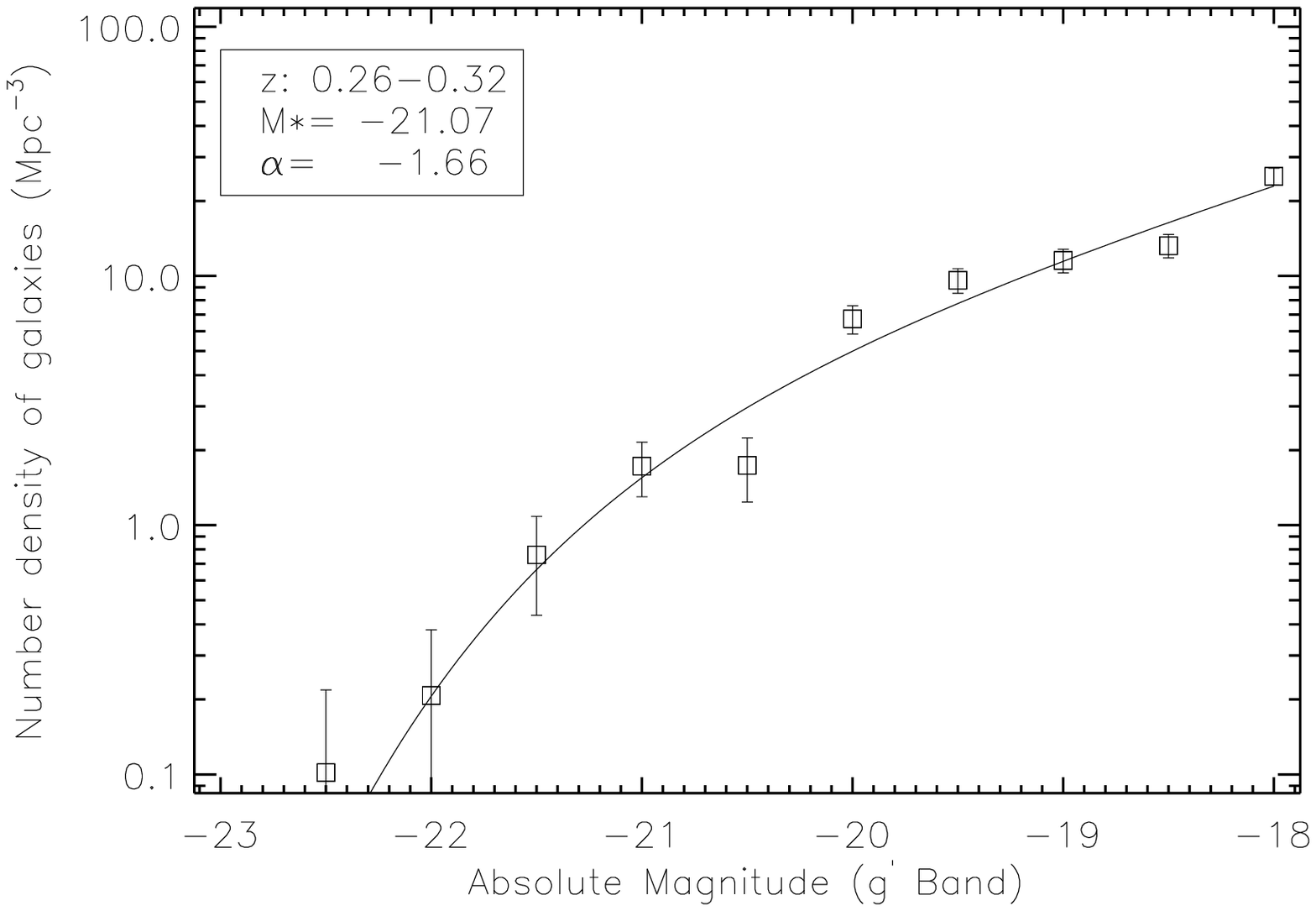,width=8.cm,height=6.cm}
\epsfig{file=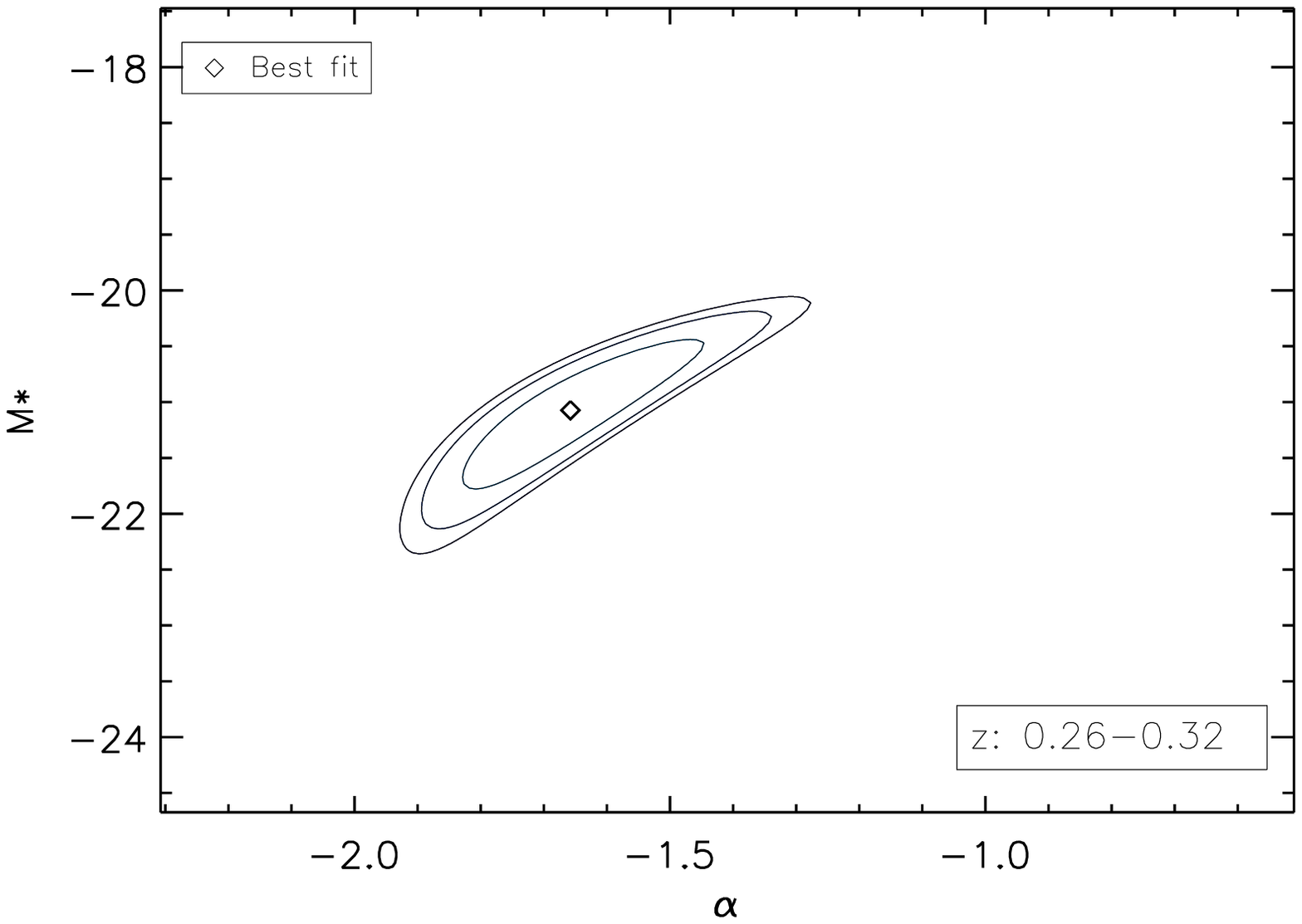,width=8.cm,height=6.cm}

\caption{LFs of the  8 stacked C1 clusters with redshift 0.2 to 0.4 and their associated $1\sigma$, $2\sigma$ and $3\sigma$ contours of confidence levels for  $\alpha$ and $M^*$ in the $g^\prime$ band.}
\label{0.2-0.4clustersLFs1}
\end{figure}

\begin{figure}
\center

\epsfig{file=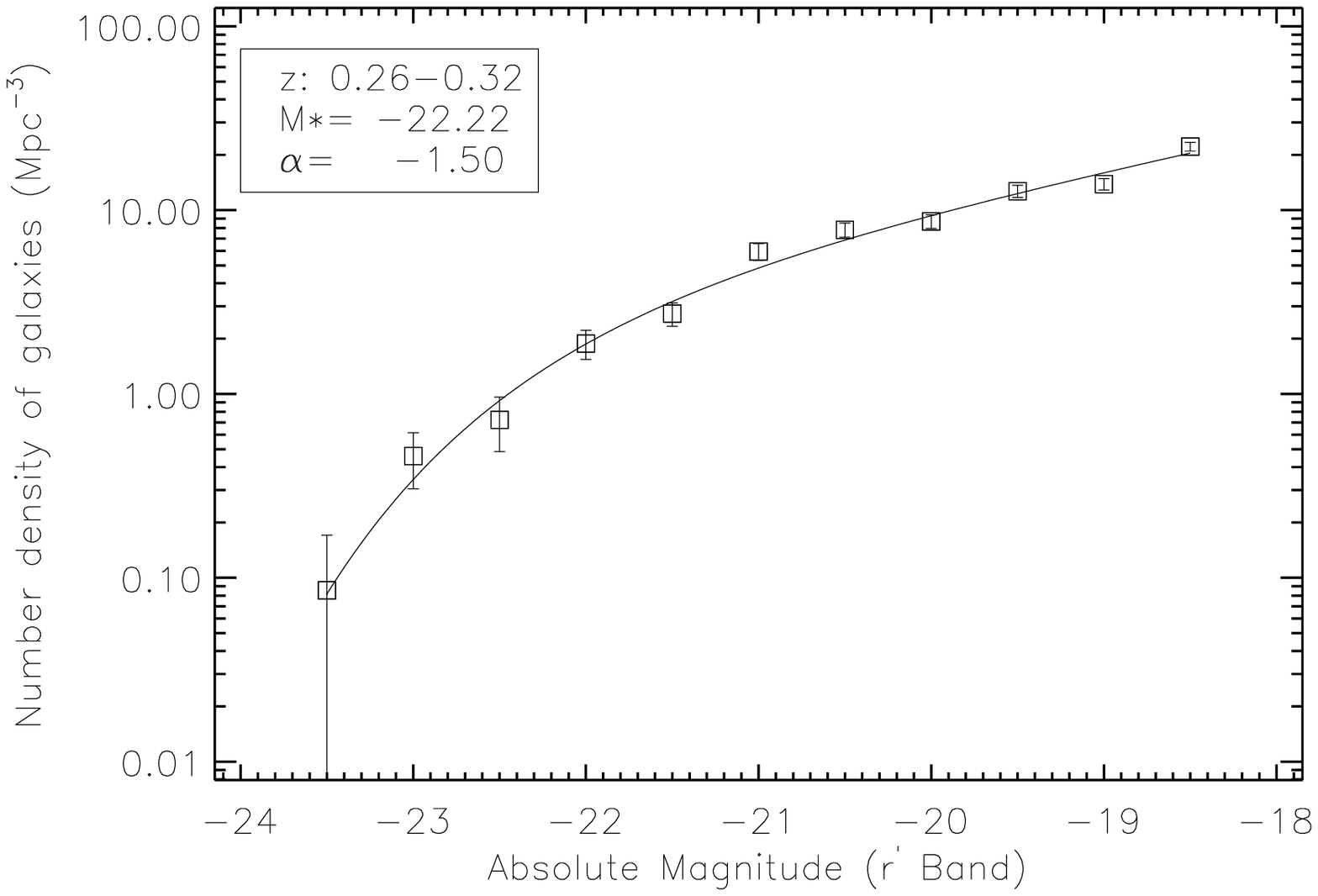,width=8.cm,height=6.cm}
\epsfig{file=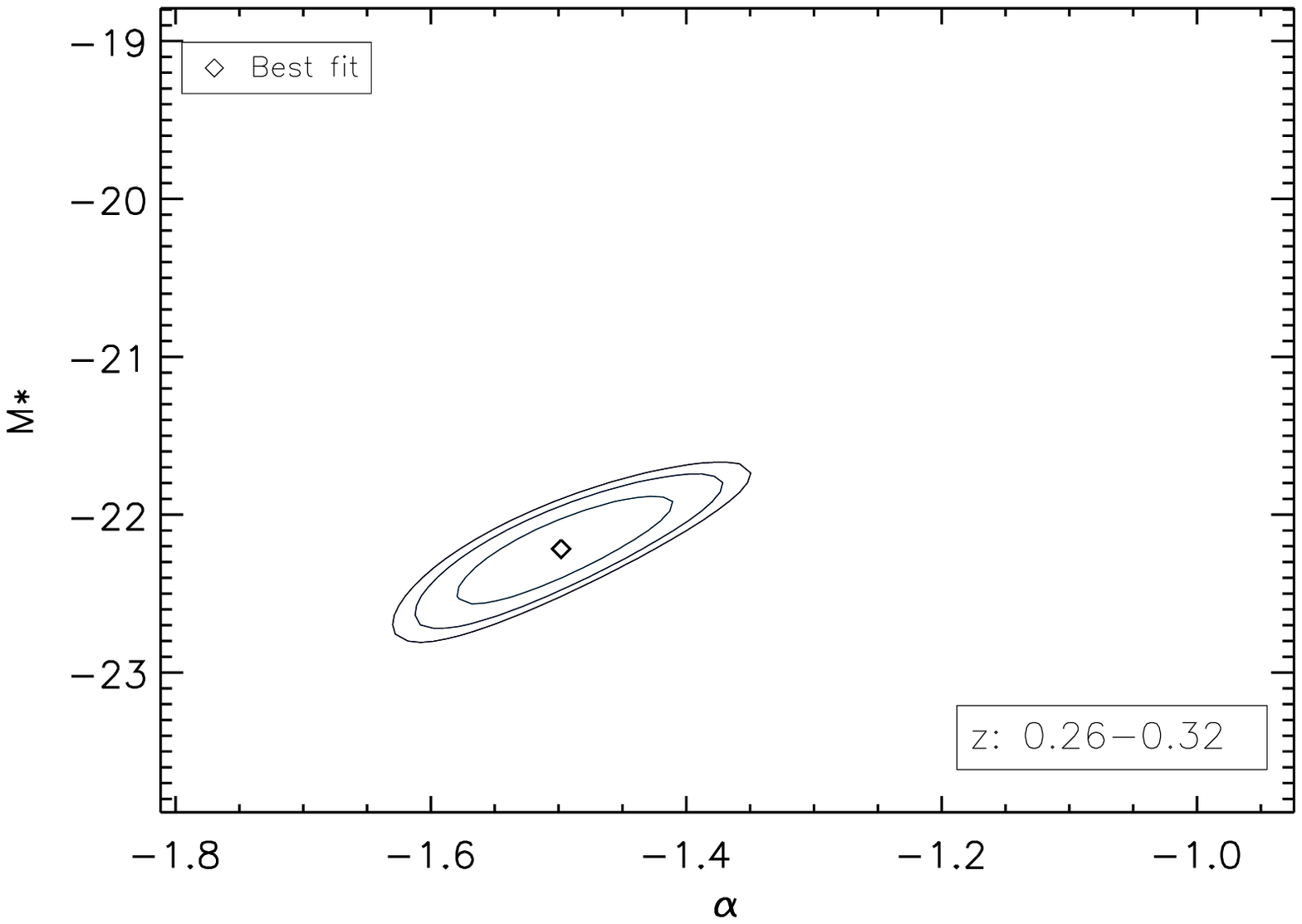,width=8.cm,height=6.cm}

\caption{LFs of the  8 stacked C1 clusters with redshift 0.2 to 0.4 and their associated $1\sigma$, $2\sigma$ and $3\sigma$ contours of confidence levels for  $\alpha$ and $M^*$ in the $r^\prime$ band.}
\label{0.2-0.4clustersLFs2}
\end{figure}

\begin{figure}
\center

\epsfig{file=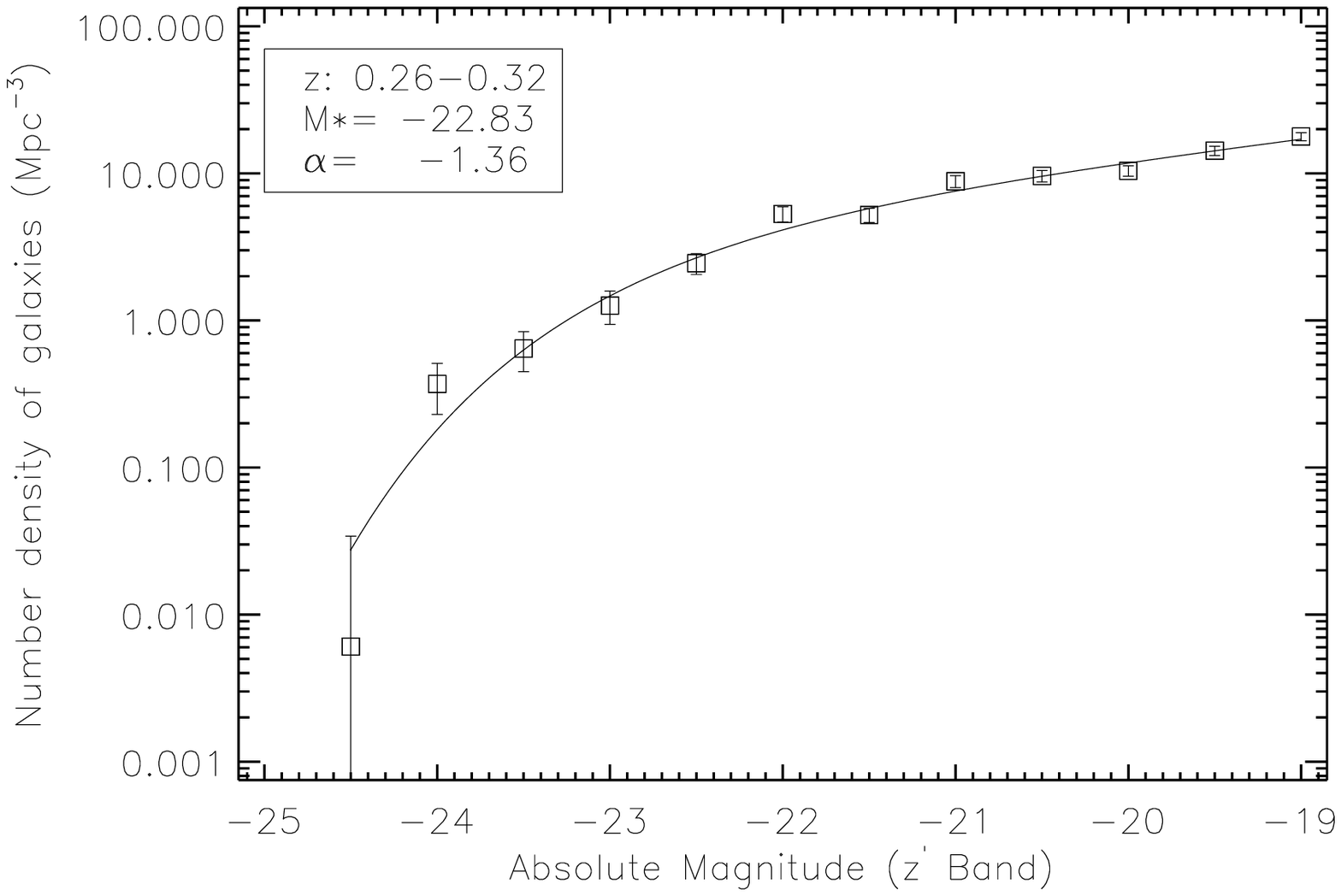,width=8.cm,height=6.cm}
\epsfig{file=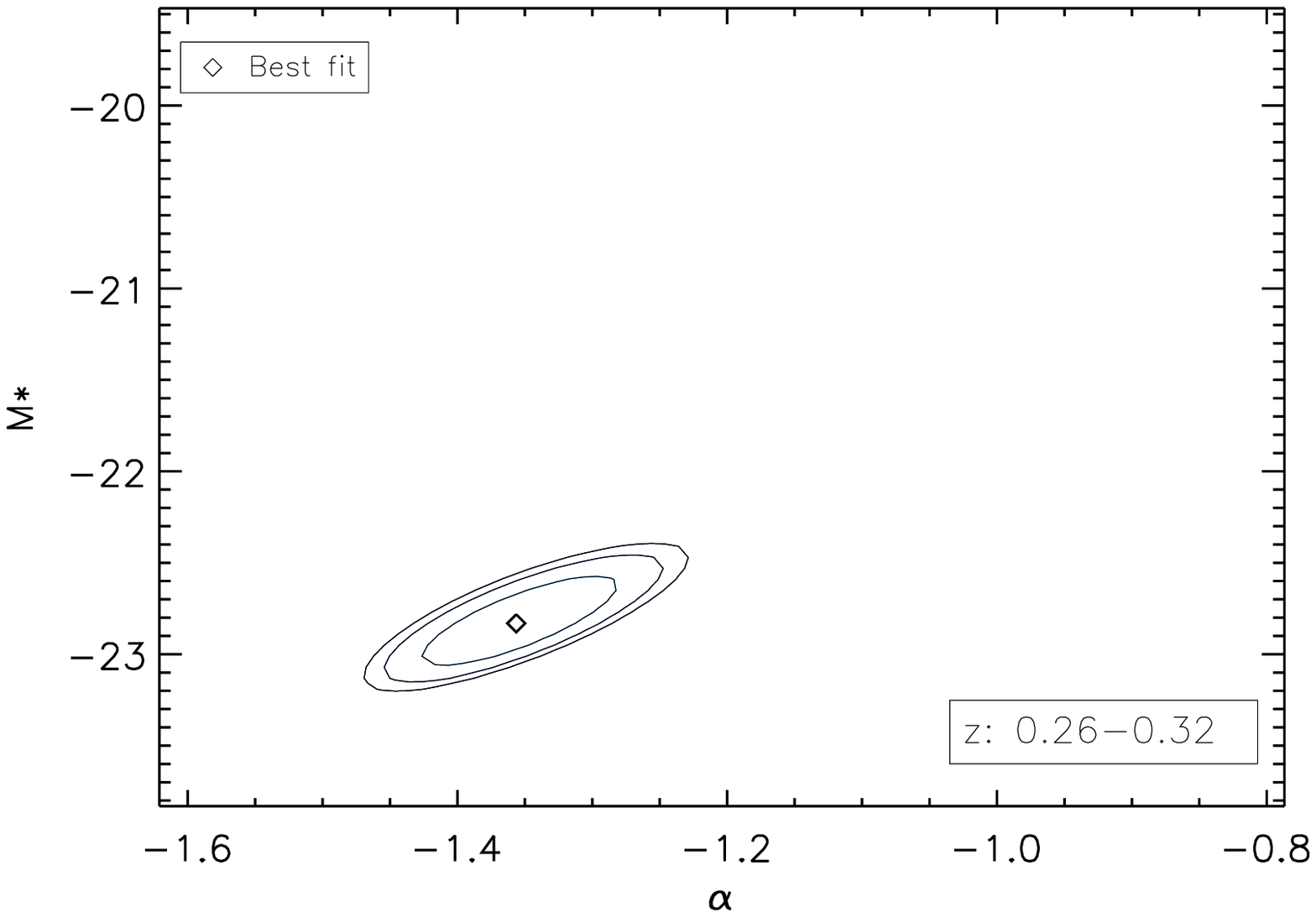,width=8.cm,height=6.cm}

\caption{LFs of the  8 stacked C1 clusters with redshift 0.2 to 0.4 and their associated $1\sigma$, $2\sigma$ and $3\sigma$ contours of confidence levels for  $\alpha$ and $M^*$ in the $z^\prime$ band.}
\label{0.2-0.4clustersLFs3}
\end{figure}

Eight amongst the 14 C1 clusters, more than half of our
sample, lie within the narrow redshift range 0.26 to 0.32. These 
clusters are representative of low-mass clusters at intermediate 
redshifts - a population which dominates the XMM-LSS cluster dataset. 
Stacking these clusters together provides the best available composite 
LF for X-ray selected poor
clusters at $z\sim 0.3$, which should be valuable for
future comparative studies. The LFs and their associated error 
contours are shown in 
Figures \ref{0.2-0.4clustersLFs1}, \ref{0.2-0.4clustersLFs2} and 
\ref{0.2-0.4clustersLFs3}.

These clusters range in temperature from 1.3 to 2.8 keV. 
Schechter fits give $\alpha$ values  -1.66$\pm$0.11, -1.50$\pm$0.05 and
$-1.36\pm0.05$, and  $M^*$ values -21.07$\pm$0.38, -22.21$\pm$0.22 and
$-22.83\pm0.17$, for the  $g^\prime$, $r^\prime$  and $z^\prime$ bands
respectively. Their faint-end slopes are shallower than the local clusters 
($z \leq 0.14$) but steeper than higher-redshift ($z \geq 0.43$) ones. 
The colour trend of $\alpha$ is very obvious and seems 
to be a characteristic of  $z\sim 0.3$ clusters compared to other clusters
in other redshift bins, as mentioned above. 
\cite{wilman2005a} studied a sample of poor 
clusters at redshift $z\sim 0.4$ selected from the CNOC2 galaxy redshift 
survey. Comparing this optically selected sample 
with  nearby clusters, they found that the fraction of passive 
galaxies declines with redshift, which is consistent with our finding
of  larger population of faint blue galaxies at $z\sim 0.3$. However, these
authors did not study the LF of their intermediate redshift groups.

In a recent study, \cite{harsono2009} presented composite LFs of 
six rich ($T\sim 7$-9~keV)
clusters with redshifts ranging from 0.14 to 0.40 (averaging to 0.246)
in the B,g,V,r,i and z bands. The LFs
were well fitted by a single Schechter function with $\alpha$  values
for $g$, $r$ and $z$ bands as follows: $-1.31\pm0.04$, $-1.33\pm0.03$ and 
$-1.45\pm0.02$ and the corresponding $M^*$ values were $-20.94\pm0.17$, 
$-21.95\pm0.29$ and $-22.26\pm0.30$. Their $M^*$ values are in reasonable
agreement with 
ours, but their slopes are shallower, and show no trend with colour. 
However, their data were limited to 20-40\% of the area within $r_{200}$, 
and they suggest
that the lack of any upturn in the slope at faint magnitudes may be related
to this -- the extra faint galaxies responsible for the upturn being
associated with a population infalling into clusters. In contrast,
our data extend to $1.5r_{500}$, which is approximately equal to $r_{200}$.

\begin{table*}
\begin{center}
\begin{tabular}{llllllllll}
\hline

\multicolumn{9}{|c|}{Redshift-Stacked} \\
\hline
z-Range & Average T & Clusters- & $\alpha$ ($g^\prime$ Band) & $\alpha$ ($r^\prime$ Band) & $\alpha$ ($z^\prime$ Band) & $M^*$ ($g^\prime$ Band) & $M^*$ ($r^\prime$ Band) & $M^*$ ($z^\prime$ Band) \\
 & (keV) & Stacked &  &  &  &  &  &  \\
\hline
0.05-0.14 & 0.89 & 11,21,41 & -1.79$\pm$0.05 & -1.75$\pm$0.03 & -1.75$\pm$0.02 & -20.15$\pm$0.51 & -20.69$\pm$0.48 & -22.03$\pm$0.54 \\
0.26-0.26 & 1.65 & 25,44 & -1.66$\pm$0.10 & -1.55$\pm$0.07 & -1.48$\pm$0.07 & -21.05$\pm$0.36 & -22.86$\pm$0.48 & -23.31$\pm$0.46 \\
0.29-0.29 & 2.25 & 22,27 & -1.67$\pm$0.13 & -1.54$\pm$0.10 & -1.30$\pm$0.10 & -20.76$\pm$0.47 & -21.90$\pm$0.46 & -22.25$\pm$0.35 \\
0.30-0.32 & 1.48 & 8,13,18,40 & -1.59$\pm$0.22 & -1.47$\pm$0.07 & -1.27$\pm$0.07 & -20.56$\pm$0.58 & -21.99$\pm$0.25 & -22.54$\pm$0.21 \\
0.43-0.61 & 3.40 & 1,6,49 & -1.31$\pm$0.09 & -1.46$\pm$0.08 & -1.22$\pm$0.06 & -21.40$\pm$0.26 & -22.99$\pm$0.31 & -23.49$\pm$0.18 \\
\hline
\hline
\multicolumn{9}{|c|}{Temperature-Stacked} \\
\hline
T-Range & Average z & Clusters- & $\alpha$ ($g^\prime$ Band) & $\alpha$ ($r^\prime$ Band) & $\alpha$ ($z^\prime$ Band) & $M^*$ ($g^\prime$ Band) & $M^*$ ($r^\prime$ Band) & $M^*$ ($z^\prime$ Band) \\
(keV) &  & Stacked &  &  &  &  &  &  \\
\hline
0.64-1.00 & 0.15 & 11,13,21 & -1.67$\pm$0.05 & -1.80$\pm$0.02 & -1.79$\pm$0.02 & -20.11$\pm$0.45 & -22.97$\pm$0.58 & -25.68$\pm$2.13 \\
1.30-1.34 & 0.23 & 8,41,44 & -1.57$\pm$0.05 & -1.54$\pm$0.05 & -1.42$\pm$0.04 & -21.12$\pm$0.30 & -22.32$\pm$0.31 & -23.05$\pm$0.28 \\
1.60-2.20 & 0.34 & 18,22,25,40,49 & -0.90$\pm$0.10 & -1.26$\pm$0.05 & -1.18$\pm$0.04 & -19.64$\pm$0.20 & -21.71$\pm$0.18 & -22.85$\pm$0.17 \\
2.80-3.20 & 0.45 & 1,27 & -1.45$\pm$0.08 & -1.22$\pm$0.07 & -1.15$\pm$0.05 & -22.24$\pm$0.44 & -22.78$\pm$0.28 & -23.58$\pm$0.20 \\
4.80-4.80 & 0.43 & 6 & -1.61$\pm$0.16 & -1.70$\pm$0.09 & -1.31$\pm$0.09 & -20.96$\pm$0.43 & -23.24$\pm$0.50 & -22.98$\pm$0.23 \\

\hline
\end{tabular}
\caption{Results of the Schechter function fitting of the redshift-stacked 
and temperature-stacked clusters for the three-filter set ($g^\prime$,
$r^\prime$,$z^\prime$).} \label{Stacked_Results}
\end{center}
\end{table*}

\subsection{Temperature-stacked clusters}

The 14 C1 clusters span a temperature range of 0.64 to 4.80 keV This
was divided into five subranges: 0.64-1.00, 1.30-1.34, 1.60-2.20, 2.80-3.20
and 4.80-4.80, using the same criteria, discussed
above, which were used for stacking into redshift bins.
The highest temperature
bin consists of only one cluster because after removing the three
high-redshift clusters (2,5,29), the temperature difference between the two
highest temperature clusters was too large to stack them together,
and the LF of the highest temperature cluster, cluster 6 (T=4.80 keV)
was of sufficient quality that it provides useful constraints on its own.
The second highest temperature bin consists of two clusters and the
rest have at least three clusters.

The correlation coefficients of $\alpha$ with temperature are not high
enough to establish any trend, especially when we take into consideration the
reverse in sign of the coefficients after the partial correlation
calculation, see Table \ref{Corr_Coeff}. But in Table \ref{Stacked_Results}
the highest temperature bin contains only one cluster (cluster 6) and the
other bins show some indication that $\alpha$ increases (slope decreases)
with temperature, especially in bands $r^\prime$ and $z^\prime$. Further
investigation is needed to arrive at more conclusive results about the
$\alpha$ tend with temperature.  As to $M^*$, the stacked results do
not show any trend with temperature. 
The LFs of the temperature stacked data are shown in Figures
\ref{t_gband_plot}, \ref{t_rband_plot} and \ref{t_zband_plot}.

\begin{figure}
\center

\epsfig{file=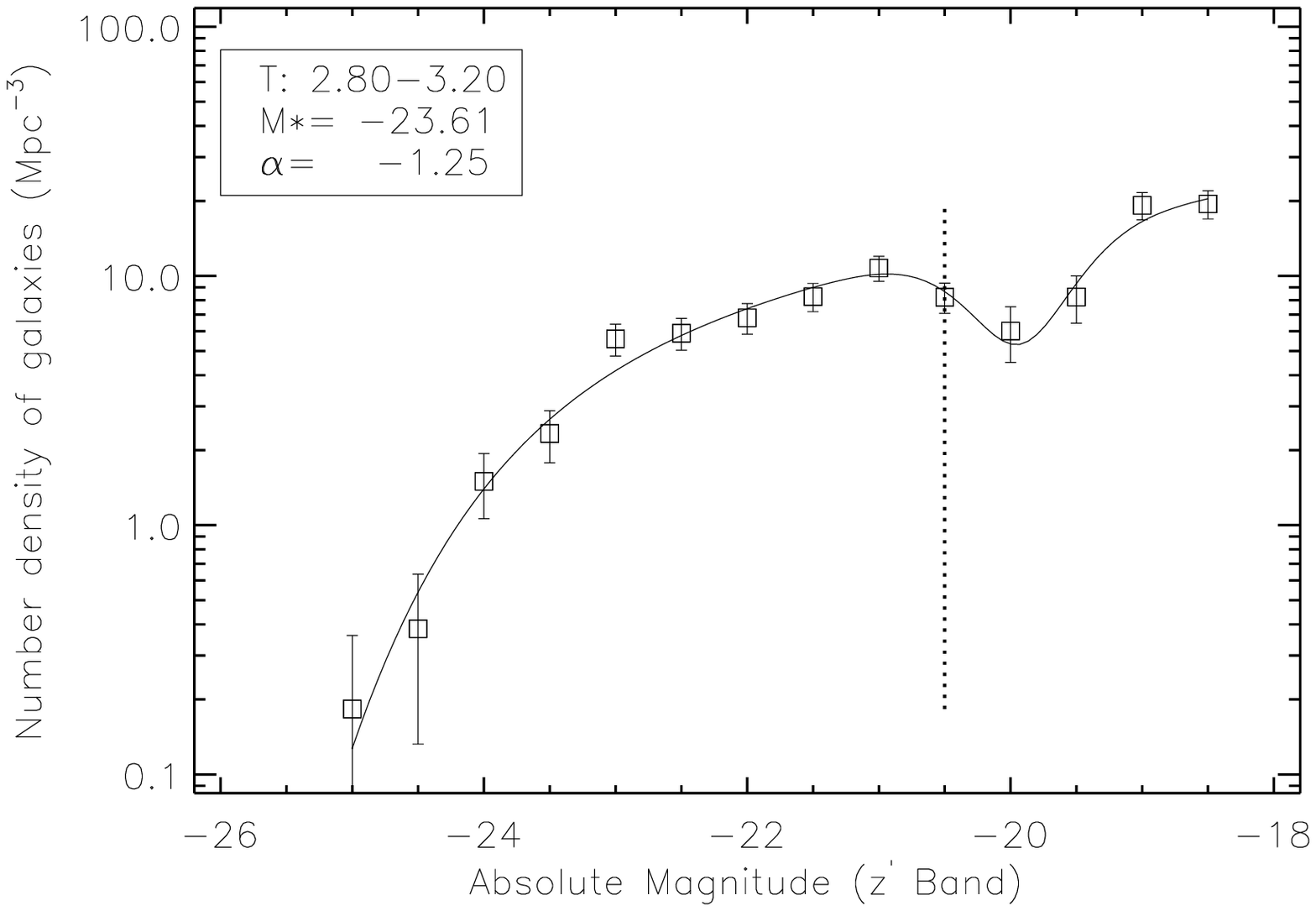,width=8cm,height=5cm}
\epsfig{file=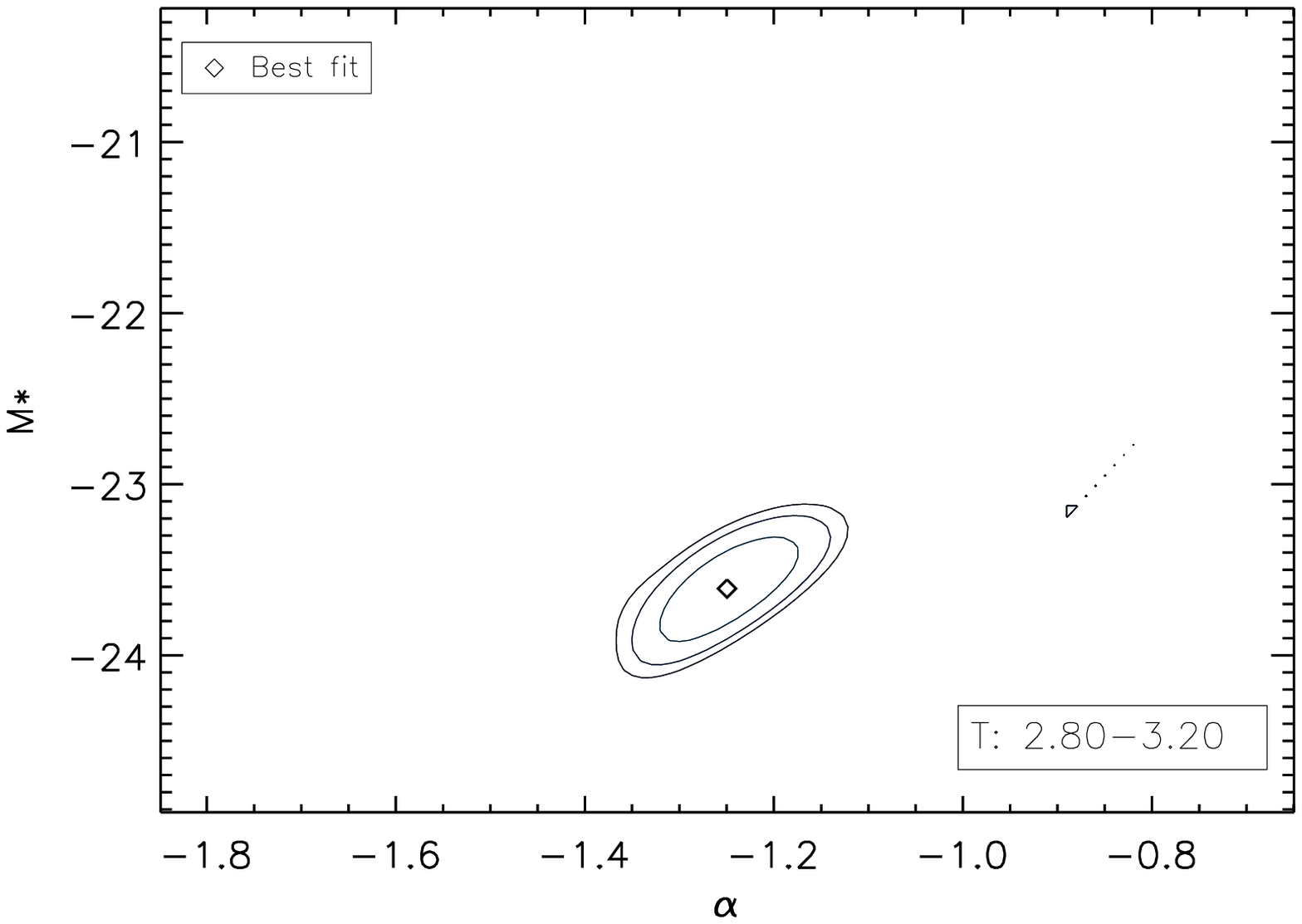,width=8cm,height=5cm}

\caption{LF of the stacked clusters 1 and 27 (fourth temperature bin: 
$2.80\leq T
\leq3.20$) with plot of the $1\sigma$, $2\sigma$ and
$3\sigma$ confidence contours in the $\alpha$-$M^*$ plane. The line is the fit
of a Schechter function plus a Gaussian dip. The fitted dip
position is $-19.9\pm0.1$ ($z^\prime$ filter). The vertical dotted
line marks the faintest magnitude at which both stacked
clusters contribute.}
\label{Dip_tstacked_zband}
\end{figure}


Some previous studies (see for example, \cite{Miles2004}) found that
galaxy clusters with low X-ray luminosity (comparable to the coolest
clusters in our C1 sample) exhibit dips in their LFs. In our
data, some of the
temperature stacked LF plots (\ref{t_gband_plot}, \ref{t_rband_plot}
and \ref{t_zband_plot}), especially those with high temperature ($\geq
2.8$ keV) showed signs of dips in the faint end of the LF. It can be hard to
distinguish between scatter of the data points and a genuine dip in
the LF. 

To test the genuineness of these dips we fitted a Schechter
function minus a Gaussian function defined by three parameters (central 
magnitude, width and depth) to these temperature
stacked LFs.  The two fits, with and without the Gaussian were
statistically compared using their $\chi^2$ values, and an F-test applied
to assess the significance of the improvement resulting from
inclusion of the dip. In some cases the 
dip improved the fit at
a confidence level of more than 90\%. See, for
example Figure \ref{Dip_tstacked_zband}. 

However, careful examination of
the stacked LF and the individual clusters LFs in these cases suggested
that the dip is
produced by the stacking of clusters with different
faintest magnitude limits, rather than lying within 
the magnitude range shared by all the combined clusters. This was found to be 
true for all stacked LFs that showed a
statistical improvement in fit on inclusion of a Gaussian dip.
We therefore conclude that our data show no evidence for real
dips in the optical LFs of the C1 clusters.

\section{Discussion}

\subsection{Faint-end slope of the luminosity function}

In this work we have studied the LFs of the individual clusters in the
C1 sample from XMM-LSS.
A Schechter function provided a reasonable fit across most of the LF for
most clusters, especially in the $z^\prime$ band. But the bright end
was poorly-fitted for nearly half of the sample (6 out of 14) and $M^*$
values were often not well constrained. The faint-end slope ranges are $-1.03
\leq \alpha \leq -2.1$, $-1.19 \leq \alpha \leq -1.89$ and $-1.06 \leq
\alpha \leq -1.77$ with averages $-1.70\pm0.10$, $-1.64\pm0.04$ and
$-1.43\pm0.03 $ for the $g^\prime$, $r^\prime$ and $z^\prime$ passbands
respectively. The mean faint-end slope, averaging over all the three filters, 
is $\alpha_{avg}=-1.59\pm0.05$. 

Comparison of fitted Schechter parameters
from different studies should take into
account the passband, cluster redshift, and the procedure used in
constructing the LF, including the methods used to determine the cluster
membership and the background subtraction, since all of these factors may
affect the results and therefore the accuracy of comparison.

Previous studies of galaxy cluster LFs have found a wide range 
for $\alpha$,
from $\alpha \sim -1$ (\cite{Paolillo01}) to $\alpha \sim -2$
(\cite{Popesso2006}), but generally, LFs of clusters (both high-mass and
low-mass systems) are found to have steeper slopes than field galaxy 
LFs, which usually span values
$\alpha \sim -0.7$ (\cite{Lin1996}) to -1 (\cite{Loveday1995}). 
The values we obtain for $\alpha$ fall
into the cluster LF range. The mass ($M_{500}$) range of the
C1 clusters is 0.6-19 $\times 10^{13} M_{\sun}$ (\cite{pacaud07}) and this
puts the C1 sample in the lower-mass class of galaxy clusters
(poor clusters and groups). This indicates that low-mass systems have 
almost the same range of faint-end LF slopes as more massive systems.

Moreover, \cite{gonzalez2006} studied LFs of galaxy clusters with a virial
mass range $0.01-20 \times 10^{13} M_{\sun}$ and redshift $0.03<z<0.06$
and found slopes of $-1.9< \alpha < -1.6$ at the faint end ($M_{r}
\geq -18$). This is consistent with our result for clusters with comparable
redshift; clusters 11 and 21, which have estimated masses of 0.6 and 0.9
$\times 10^{13} M_{\sun}$ and redshifts 0.05 and 0.08, show faint-end
slopes of $-1.80\pm 0.05$ and $-1.89\pm 0.06$ in the $r^\prime$ band, with
magnitude limits of -14.5 and -15 respectively.

The study of \cite{Popesso2005a} on X-ray selected rich clusters with
$z \leq 0.25$ also gave a steep faint-end ($M_{g} \geq -16$)
slope: $-2.1 \leq \alpha \leq -1.6$ in the SDSS $g$ band. C1 clusters
with redshifts $\leq 0.26$, namely clusters 11,21,25,41 and 44, have
a $g^\prime$ band slope range of $-2.1 \leq \alpha \leq -1.67$, which
agrees well with \cite{Popesso2005a}. The redshift-stacked
clusters with redshift $z \leq 0.32$ (the first four redshift bins)
gave a slope range of $-1.79 \leq \alpha \leq -1.59$ in the $g^\prime$
band which is also consistent with \cite{Popesso2005a}. 
The C1 clusters are low-mass systems, whilst the \cite{Popesso2005a} systems 
are rich clusters. The steep faint end slopes seen in both 
indicate a larger fraction of dwarf galaxies in both groups and 
clusters, compared with the shallower LF slopes usually found in the field.

However, as is the case with richer clusters, the results from different 
studies
of the luminosity function of group galaxies arrive at different results.
For instance, \cite{Miles2004} derived a very flat ($\alpha \sim -1$) Schechter
slope for X-ray bright groups -- though they found a faint upturn in X-ray
dim systems -- and \cite{zandivarez} derived similarly low faint end slopes
for SDSS groups.
\cite{Miles2004} used photometric data of X-ray selected systems
and used all galaxies with B-R $<$ 1.7 from the regions outside a 
radius of $R_{500}$ from the centre of the group as 
the background for subtraction, whilst \cite{zandivarez}
used spectroscopic data for membership determination for their 
friends-of-friends selected clusters. \cite{robotham06} extracted LFs
for 2PIGG groups, derived from the 2dF galaxy redshift survey, and
obtained good fits with Schechter functions, with faint end slopes which
increased from $\alpha \sim -1$ for red galaxies to $\alpha \sim -1.5$
for blue galaxies. These discrepancies in the faint-end slope 
from different studies could arise from a variety of causes:
different cluster selection methods (X-ray selected clusters in our case),
spectroscopic or photometric selection of cluster galaxies, 
different galaxy background
subtraction techniques (see discussion in section 4.3), and possibly
because different clusters have different faint-end slopes depending
on their large-scale environment, which will affect the incidence of
infalling galaxies.

\subsection{The absence of the upturn in the faint end of LFs}

Both \cite{Popesso2005a} and \cite{gonzalez2006} reported an upturn at the
faint end of their stacked LF, and required
a sum of two Schechter functions, rather than a single Schechter, to
obtain reasonable fits.
\cite{Popesso2005a} located the upturn at -16 in the
$g^\prime$ band, and -18.5 in $z^\prime$; the upturn of \cite{gonzalez2006} 
was found at a similar magnitude: -18 in the $r^\prime$ band.
In our sample, only the LFs for clusters 11,21 and 41 extend
to the faint magnitudes in which \cite{Popesso2005a} and
\cite{gonzalez2006} found their upturns. The composite LF for these
systems is the first in the redshift-stacked LFs, see 
Figures \ref{z_stacked_g}, 
\ref{z_stacked_r} and \ref{z_stacked_z}. Although our results agree with 
\cite{Popesso2005a} and \cite{gonzalez2006} regarding the steep values of 
the faint-end slope, we do not find any evidence for a departure from a
simple power law at the faint end.

Other studies gave steep slopes at the faint end of cluster LFs but 
without evidence of sudden upturns, see for example
\cite{Durret02}. 
 \cite{Garilli1999} studied composite LFs of 65 clusters ranging in redshift 
from 0.05 to 0.25 and did not find upturns in their composite LFs. 
\cite{Popesso2005a} argued that \cite{Garilli1999} did not see this upturn
 in their stacked LF because they used a weighting for the
 individual LFs which depends strongly on the cluster magnitude limit,
 such that clusters with fainter magnitude limits, which
 contribute to the faint magnitude bins of the stacked LF, were heavily
 down-weighted.  In our sample, only the LFs for clusters 11,21 and 41 extend
 to the faint magnitude region in which \cite{Popesso2005a} and
 \cite{gonzalez2006} found their upturn. The composite LF for these
 systems is the first in the redshift-stacked LFs, see Figures \ref{z_stacked_g}, \ref{z_stacked_r} and
 \ref{z_stacked_z}. We did not apply any weighting method that depends on
 the magnitude limit and although faint-end slopes are steep in all three
 bands for the stacked LF of clusters 11,21 and 41, they lack any upturn
 at the locations found by \cite{Popesso2005a} and \cite{gonzalez2006}. 
 Furthermore, individual LFs of these three clusters do not show any
 obvious upturn in the faint-end part of the LF that can be distinguished 
 from the scatter
 of the data relative to the fitted Schechter function.

 \cite{Popesso2006} decomposed their LF by galaxy type and showed that the
 late-type galaxies LF was well fitted by a single Schechter function with
 a steep slope ($\alpha = −2.0$ ), while the early-type galaxies LF
 could not be fitted by a single Schechter function, and a
 composite of two Schechter functions was needed, such that the faint-end
 upturn of the global cluster LF was due to the early-type cluster
 galaxies. This suggests one way of reconciling our results with those
 of \cite{Popesso2006}. If in our poorer clusters late-type galaxies 
 outnumber early-types
 in the intermediate and faint magnitude ranges then the LF would be
 steep and without any upturns. This needs to be further investigated by
 studying the early-type and the late-type LF separately.
 Another possibility for the difference between our results and those
 of \cite{Popesso2006} lies in the techniques used to remove
 non-cluster galaxies, as we discuss in the next section.

\subsection{Membership determination methods: Effects on $\alpha$}

The steepness of the faint end of the luminosity function reflects the
number of dwarf galaxies within a cluster. Estimates of this number
are very sensitive to the method used to estimate and remove the
contribution of background and foreground galaxies before constructing
the cluster luminosity function.

\cite{Rines08} compared methods of membership determination based on 
spectroscopic data and on photometric data 
(which we used) with regard to the resulting
LF. They highlighted the advantages of spectroscopic identification
of cluster members. Where automated photometric methods are used, they
found, for example, that many large galaxies, especially
those with low surface brightnesses, may be detected as many small separate
objects, and warned that if these pieces of galaxies are not removed, they
can produce an artificial excess of faint galaxies in cluster fields.

However, we have to emphasise that although spectroscopic data can give precise
information on the cluster membership, their use
to study cluster LFs is limited to relatively
nearby clusters, since for higher-redshift clusters, spectroscopy is 
feasible only for the bright
cluster galaxies. \cite{Boue08} used deep multicolour photometry to
study the LF of A496, using colour selection to reduce contamination
by red background galaxies, and did not find the large fraction of dwarf
galaxies ($\alpha = −2.0$) inferred by some other authors, including
\cite{Popesso2006}. They suggested that this excess of dwarf galaxies
in some studies might arise from inadequate removal of background, due to
use of inadequate (or no) colour
cuts. They claimed that the red sequence used by \cite{Popesso2006} was
polluted by field galaxies because they used $u^{\ast}-r^\prime$ vs
$i^\prime$ in their CMD which \cite{Boue08} showed was not
efficient in rejecting background galaxies. In our study, we did not use
$u^{\ast}-r^\prime$ vs $i^\prime$ to define the colour cuts. Instead, we
used $u^{\ast}-g^\prime$ vs $g^\prime$ for the $g^\prime$ band,
$g^\prime-r^\prime$ vs $r^\prime$ for the $r^\prime$ band and
$i^\prime-z^\prime$ vs $z^\prime$ for the $z^\prime$ band. Moreover, our
method of field LF subtraction is based on global background LF constructed
by using the whole $1^{\circ} \times 1^{\circ}$ field of the
cluster. Therefore, we don't see any obvious reason why we might have
contaminated the red sequence with field galaxies in such away as to give a
false steep faint-end slope.

\subsection{Origin of the faint galaxies}

Our results indicate that larger numbers of faint galaxies exist in
cluster environments than in the field. It is not straightforward
to understand this result, since
various dynamical processes
that can destroy dwarfs act more effectively in dense environments.
Several ideas have been proposed to explain the excess of dwarfs in
clusters. \cite{Babul92} argued that a primordial population of dwarf
galaxies is preserved in high-pressure environments, whilst it fades
away in low-pressure regions. Alternatively, dwarfs could be formed by
galaxies that fell into clusters from the surrounding field and were
morphologically transformed. The transformation mechanism could be
tidal fragmentation or so-called harassment of infalling late-type
spiral galaxies by the cluster potential or by close encounters
(\cite{Moore96}) or ram pressure stripping of dwarf irregular galaxies
(e.g., \cite{vanZee04}).

\cite{Boselli08} showed that both simulations and observations are consistent
regarding the scenario of recent accretion and transformation of 
low-luminosity star-forming galaxies in the Virgo cluster into quiescent
dwarfs due to ram pressure gas stripping and galaxy
starvation. They also showed that this process of transformation results
in galaxies with structural and spectrophotometric properties similar to
those of dwarf ellipticals. If the whole star-forming dwarf galaxy
population dominating the faint end of the field luminosity function were
accreted, it could be totally transformed by the cluster environment into
dwarf ellipticals on timescales as short as 2 Gyr. These vigorous forces
acting in cluster environments may explain the steepness of LFs faint-end
slopes in nearby clusters.

\subsection{The evolution of $\alpha$}

Our results show an evolutionary trend of the faint-end slope,
$\alpha$, in all bands used: $g^\prime$, $r^\prime$ and
$z^\prime$. 
\cite{Liu08} examined the faint-end slope of the V-band LF of {\it field}
galaxies with redshifts $z < 0.5$ and found that it
becomes shallower with increasing redshift: their $\alpha$ changed from
$-1.24$ for the lowest redshift bin $0.02\leq z<0.1$ to $-1.12$ for the
highest redshift bin $0.4\leq z<0.5$. In clusters, a recent study by 
\cite{Lu2009} of an optically selected cluster sample found
steepening of the faint end with decreasing 
redshift since $z \sim 0.2$, and that the relative number of red-sequence 
dwarf galaxies had increased by a factor of $\sim 3$.

It is possible that this LF slope trend with redshift is linked to the
finding of \cite{Harsono07} that the `upturn' in the LF faint end
(i.e. the excess of galaxies above a single Schechter function) is
found only in low redshift clusters. They attributed this to the
recent infall of star-forming field galaxies or the whittling down of
formerly more massive objects. The impact of recent infall of galaxies
into clusters is also supported by the work of
\cite{Lisker07}, who showed that dEs in the Virgo cluster fall into two major 
morphological subclasses: a)  dEs with blue centres, thick disks or features 
reminsiscent of late-type galaxies, such as spiral arms or bars; 
this class showed no central clustering, suggesting that they are an 
unrelaxed population formed from infalling galaxies. The second subclass is
b) nucleated dEs -- a fairly relaxed population of 
spheroidal galaxies indicating that they have resided in the cluster 
for a long time, or were formed along with it. \cite{Lisker07} also pointed to 
other studies deriving similar results (see
references therein), indicating that this subclassification is not
specific to the Virgo cluster.

\subsection{Colour trends}

The faint-end slopes, $\alpha$, of the redshift-stacked groups are
steeper in bluer bands in almost all redshift
bins. However, this trend is significant ($> 1\sigma$) only for the
redshift range $0.29 \leq z \leq 0.32$. The redshift bin in which this
effect seems to be absent is the first bin: $0.05 \leq z \leq 0.14$. This
suggests that the fraction of faint blue galaxies in clusters of redshift
$z \sim 0.3 $ are higher than in local systems. Figure \ref{Colour_trend}
further illustrates this and it also shows that these blue faint galaxies
appear to have reddened and moved upwards in the colour-magnitude diagram. 
This is consistent with the findings of \cite{wilman2005a} who compared the
fractions of passive (red and quiescent) and blue star-forming galaxies in
cluster at $0.3 \leq z \leq 0.55$ with nearby ($z \simeq 0$)
clusters. They found that the fraction of passive galaxies declined
strongly with redshift to at least $z \simeq 0.45$. These results are also
consistent with the well-known Butcher-Oemler effect in clusters and
support the idea that dense environments are responsible for galaxy
transformation from blue to red because these trends are less obvious in
field environments, see \cite{wilman2005b}. 

Our result is also consistent with \cite{yee2005} who studied the colours
of galaxies as a function of luminosity and environment using the Red Sequence
Cluster Survey and the SDSS. They found a higher incidence of faint to moderate
luminosity galaxies in high density environments at $z>0.2$ compared to
lower redshifts and lower density environments.
They interpreted this as arising from the shut-down of star formation
in low mass galaxies within clusters at $z<0.3$, in contrast to the situation
in the field (c.f. \cite{balogh2004}).

The fact that such transformations are
observed in low-mass clusters like our C1 sample, as well as in richer
clusters, favours mechanisms for suppression of star formation
which operate in shallower potential wells,
such as strangulation, tidal interactions and galaxy mergers, rather 
than ram pressure stripping, which is effective mostly in rich environments
with high velocity dispersions.

\subsection{Correlation between global properties of clusters}

The optical luminosity is a good indicator of cluster richness, and hence
should be closely related to cluster mass, velocity dispersion and temperature
(\cite{Popesso05b}). Assuming that cluster mass is directly proportional to
the optical light (i.e. $M/L_{OPT}$ is constant), that the ICM is 
in hydrostatic equilibrium and that 
X-ray luminosity $L_{X}$ scales with gas temperature T as
$L_{X} \propto T^{3}$ (\cite{Xue00}), it is expected that $L_{OPT} \propto
T^{1.5}$, and that $L_{OPT} \propto L_{X}^{0.5}$.

Our scaling results for $L_{OPT}$ with $L_{X}$ and $T$ mostly agree well with
these expectations. For the $L_{OPT}-L_{X}$ relation, the logarithmic
slopes are $0.47\pm 0.07$ ($g^\prime$ band), $0.43\pm 0.08$
($r^\prime$ band) and $0.50\pm 0.07$ ($z^\prime$ band). While for the
$L_{OPT}-T$ relation, we have $1.57\pm 0.17$ ($g^\prime$ band),
$1.51\pm 0.17$ ($r^\prime$ band) and $1.79\pm 0.12$ ($z^\prime$ band).

\cite{Popesso04} found $0.38\pm 0.02$ for the $L_{OPT}-L_{X}$ relation 
and $1.12\pm 0.08$ for the $L_{OPT}-T$ relation in
the $z$ SDSS band within a cluster radius of 0.5 Mpc (chosen to
minimise the scatter in their scaling relations). The systems they used 
for their analysis, the RASS-SDSS sample,
were X-ray selected, and ranged from low-mass systems of $10^{12.5}
M_{\odot}$ to massive clusters of $10^{15} M_{\odot}$, over a redshift
range from 0.002 to 0.45. Their logarithmic slope value for the
$L_{OPT}-L_{X}$ relation is not inconsistent with our value (within
the errors), however, their $L_{OPT}-T$ value is lower than ours. They
attributed the departure of their results from the expected values to
a breakdown in the assumption of constant mass-to-light ratio. More
precisely, they argued that if the assumption of hydrostatic equilibrium
was retained, their results would be in a good agreement with $M/L \propto
L^{0.33\pm 0.03}$, as found by \cite{Girardi02}. However, we note that
extracting $L_{OPT}$ within a fixed {\it metric} radius, will include
a smaller fraction of the virial radius for higher mass clusters.
Hence it should be no surprise if the $L_{OPT}-T$ relation is flattened
below the expected slope of 1.5 for self-similar clusters.
Using clusters from
the RASS-SDSS sample, \cite{Popesso05b} calculated $L_{OPT}$ within
$r_{500}$ and $r_{200}$. Their $r_{200}$ results were (we used
$1.5\times r_{500}$): $0.57\pm 0.03$ ($g$ band), $0.58\pm 0.03$ ($r$
band) and $0.58\pm 0.03$ ($z$ band) for the $L_{OPT}-L_{X}$ relation and
$1.62\pm 0.10$ ($g^\prime$ band), $1.64\pm 0.09$ ($r^\prime$ band) and
$1.62\pm 0.10$ ($z^\prime$ band) for $L_{OPT}-T$ relation. These are
in better agreement with our results than \cite{Popesso04}, and
demonstrate the importance
of the radius used to estimate the $L_{OPT}$.

\section{Conclusions}

We have studied the luminosity functions of 14 Class 1 (C1) XMM-LSS
galaxy clusters in three CFHTLS MegaCam bands: $g^\prime$, $r^\prime$
and $z^\prime$. The X-ray selected clusters have masses ranging from 0.6 to
19 $\times 10 ^{13} M_{\odot}$, a redshift range of 0.05 to 0.61, and ICM
temperature range of 0.64 to 4.80 keV. We used colour-magnitude
lower and upper cuts to reduce contamination by cluster 
non-members, and performed
background subtraction using the $1^\circ \times 1^\circ$ field of
view in which the cluster lies. K-corrected luminosity functions 
of galaxies within $1.5\times r_{200}$ were constructed for each
cluster and fitted with a Schechter function. Total optical luminosities
of the individual clusters were also computed by integrating over the
fitted Schechter functions.  The individual LFs were also 
stacked together into 5 different redshift and temperature bins. The main
findings are:

\begin{itemize}
 \item A Schechter function provides a good fit across most of the LF  
 for the majority of clusters in our sample. The value of $\alpha$ range
 from $-1.03$ to $-2.1$, but no evidence is found for upturns 
 at the faint end of the Schechter function, even in the lowest
redshift systems, for which our LFs extend well into the dwarf regime.
 
\item $M^{*}$ ranges from $-19.66$ to $-24.31$. However, for many 
(nearly a third) of the
clusters' $M^{*}$  values are not well-constrained.
 
\item The redshift-stacked LFs confirm that $\alpha$ becomes shallower with 
increasing redshift. The value of $\alpha$ is $-1.75\pm 0.02$ at 
low redshift (0.05-0.14), flattening to $-1.22\pm 0.06$ 
at high redshift (0.43-0.61) in the $z^\prime$ band. Similar trends are 
present in the other two bands.

\item  $\alpha$, also steepens significantly from the 
red ($z^\prime$) to the blue ($g^\prime$) band for clusters at
redshift $\sim 0.3$. This effect is not present in our local clusters
($z\sim 0$), suggesting reddening of the faint
blue galaxies from $z\sim 0.3$ to $z\sim 0$.


\item After removing the effects of redshift (correcting for the 
Malmquist effect), the temperature-stacked LFs
do not exhibit any strong evidence for trends of the 
Schechter parameters with ICM temperature. 

\item Total optical luminosities for our sample range from 1.0 to 
56.3 $\times 10^{11} L_{\odot}$, and correlate strongly with X-ray
luminosity. The logarithmic slopes of the $L_{OPT}-L_{X}$ relation are
$0.47\pm 0.07$, $0.43\pm 0.08$ and $0.50\pm 0.07$ for the $g^\prime$,
$r^\prime$ and $z^\prime$ bands respectively.

\item Also, $L_{OPT}$ correlate strongly with the X-ray gas temperature, T. 
The logarithmic slopes of the $L_{OPT}-T$ relation are $1.57\pm 0.17$,
$1.51\pm 0.17$ and $1.79\pm 0.12$ for the $g^\prime$, $r^\prime$ and
$z^\prime$ bands respectively.

\item The slopes  of the $L_{OPT}-L_{X}$ and $L_{OPT}-T$ relations are 
consistent with the established, non-self-similar, cluster $L_X-T$ relation 
and constant mass-to-light ratio, except for the $z^\prime$ band value of the
$L_{OPT}-T$ relation which is higher than the expected value (1.5) by
$\sim 0.3$.

\item Some of our stacked LFs show dips, but these appear to be artefacts
arising where clusters with different faintest magnitude limits are stacked
together. We therefore we conclude there is no evidence for real dips 
in the optical LFs of the C1 clusters.

\end{itemize}

\section{Acknowledgements}

The results presented in this paper are based on observations obtained
with MegaPrime/MegaCam, a joint project of CFHT and CEA/DAPNIA, at the
Canada-France-Hawaii Telescope (CFHT) which is operated by the
National Research Council (NRC) of Canada, the Institut National des
Science de l'Univers of the Centre National de la Recherche
Scientifique (CNRS) of France, and the University of Hawaii. This work
is based in part on data products produced at TERAPIX and the Canadian
Astronomy Data Centre as part of the Canada-France-Hawaii Telescope
Legacy Survey, a collaborative project of NRC and CNRS.

\onecolumn
\begin{figure}
\center

\epsfig{file=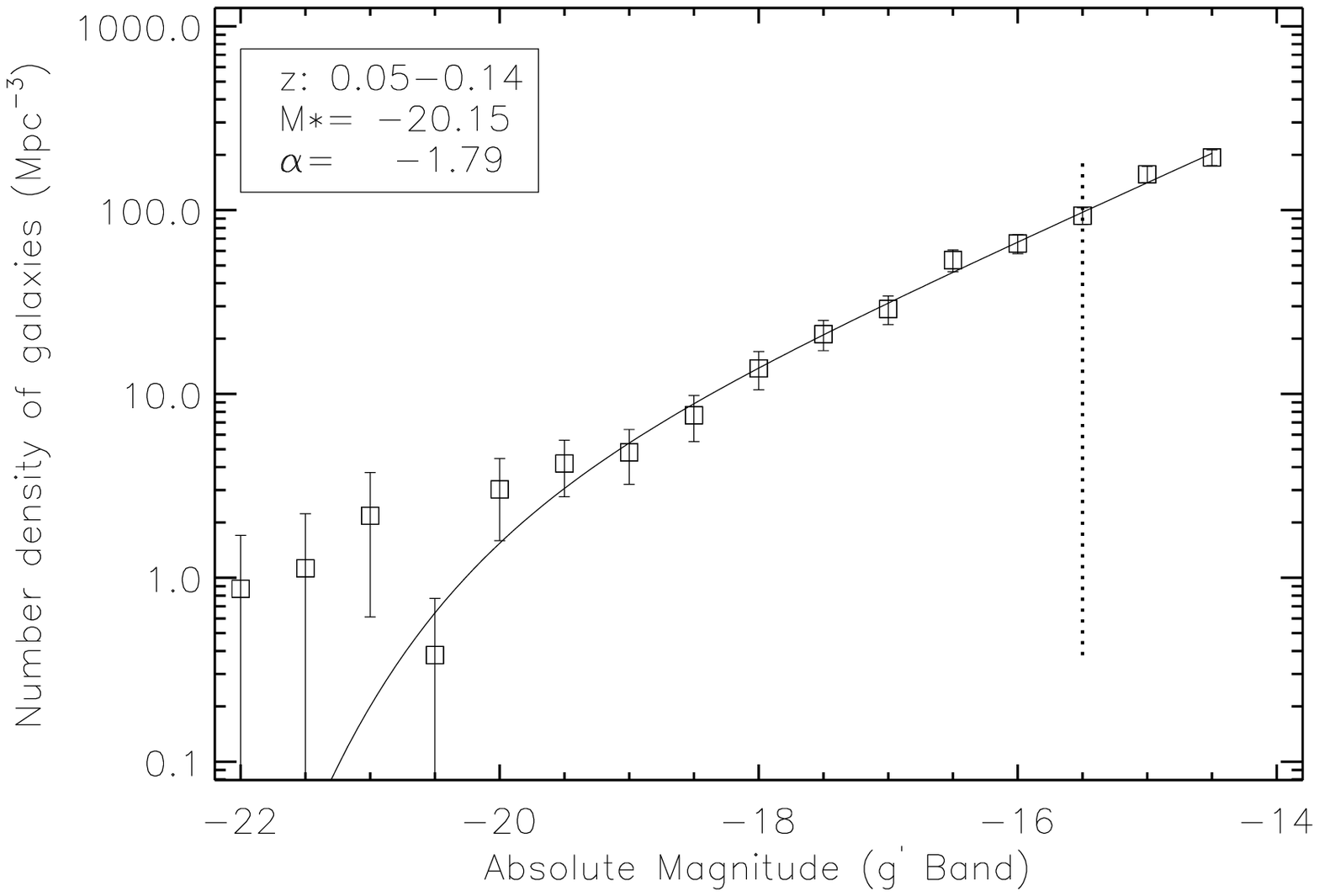,width=7.5cm,height=4.5cm}
\epsfig{file=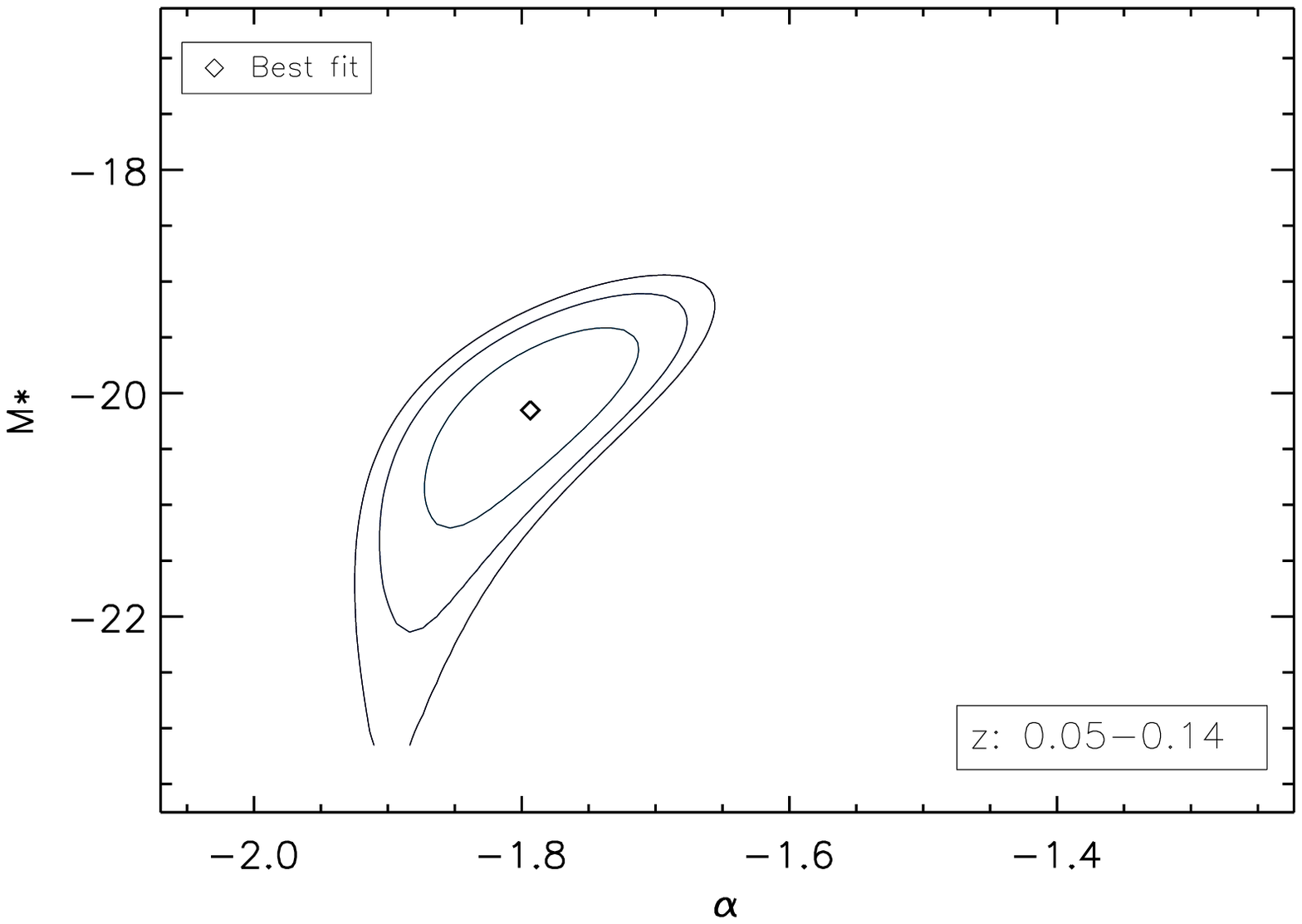,width=7.5cm,height=4.5cm}
\epsfig{file=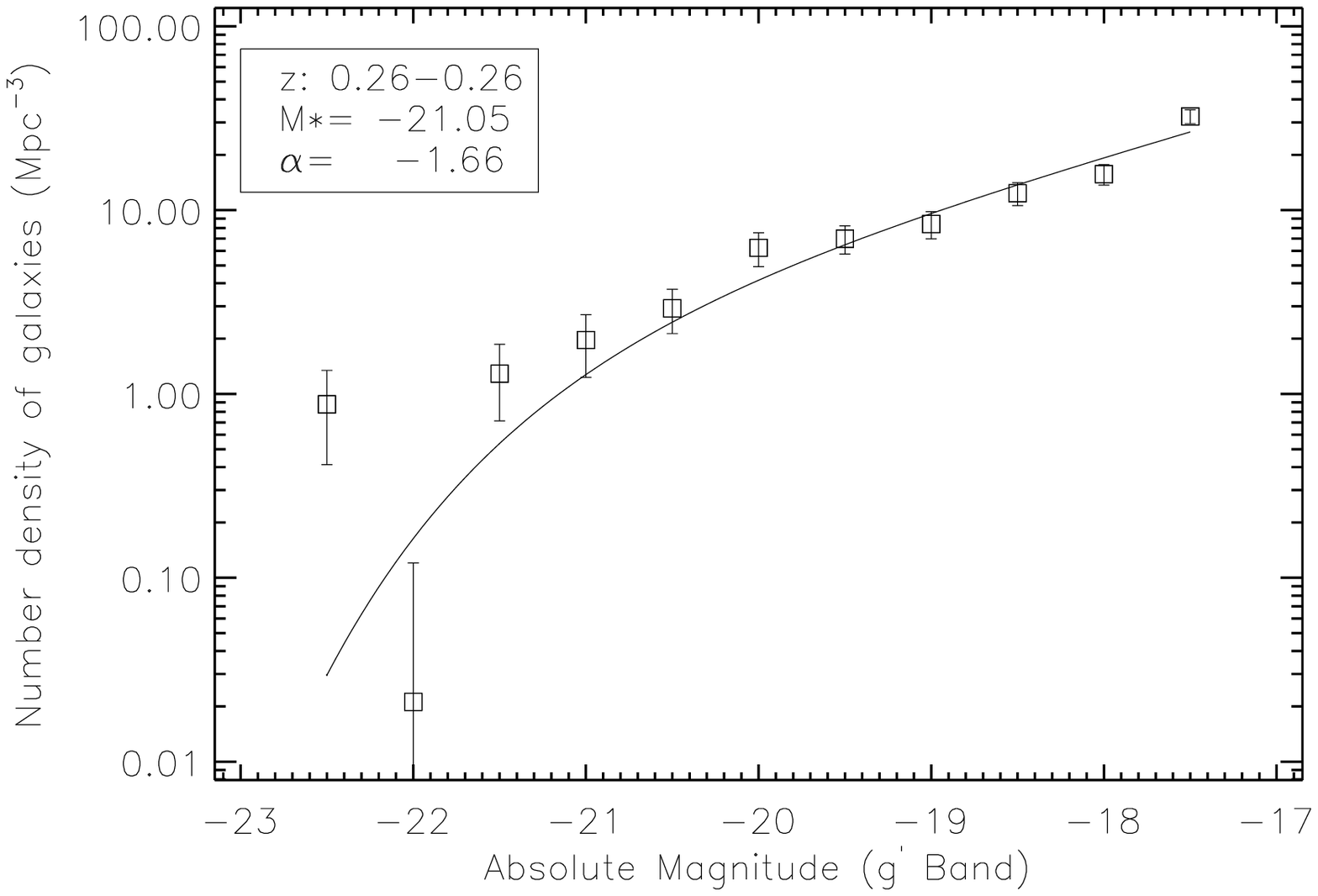,width=7.5cm,height=4.5cm}
\epsfig{file=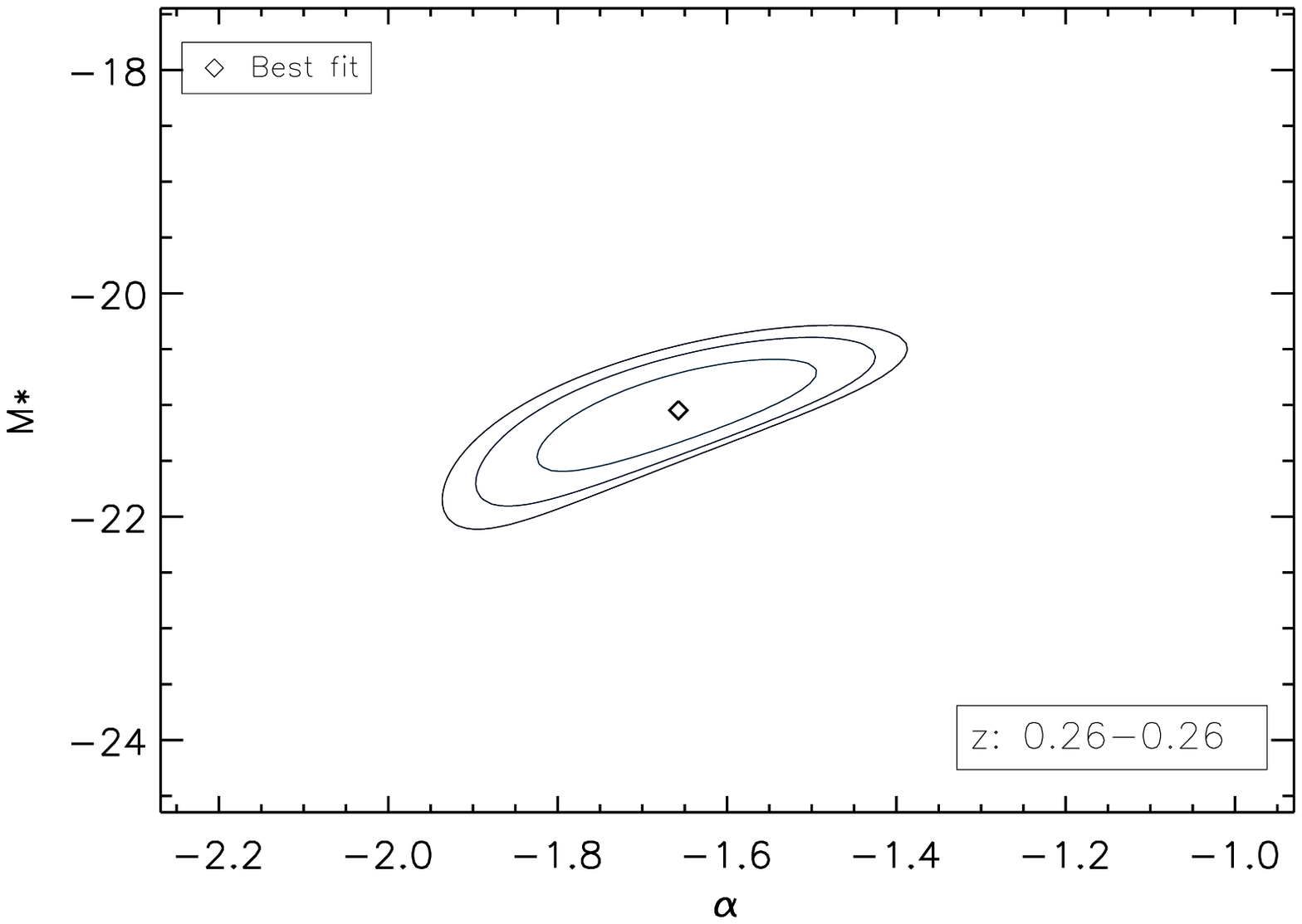,width=7.5cm,height=4.5cm}
\epsfig{file=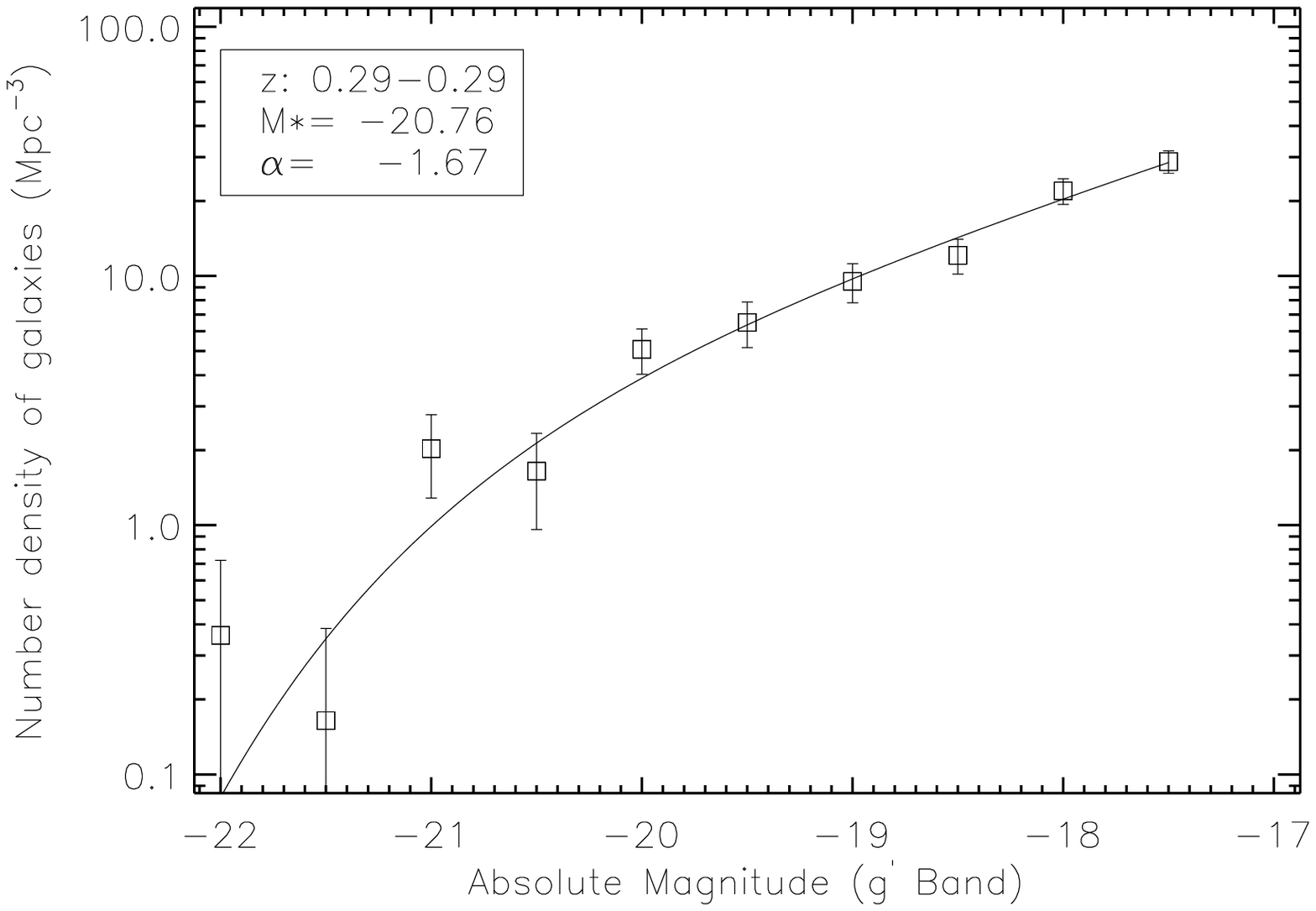,width=7.5cm,height=4.5cm}
\epsfig{file=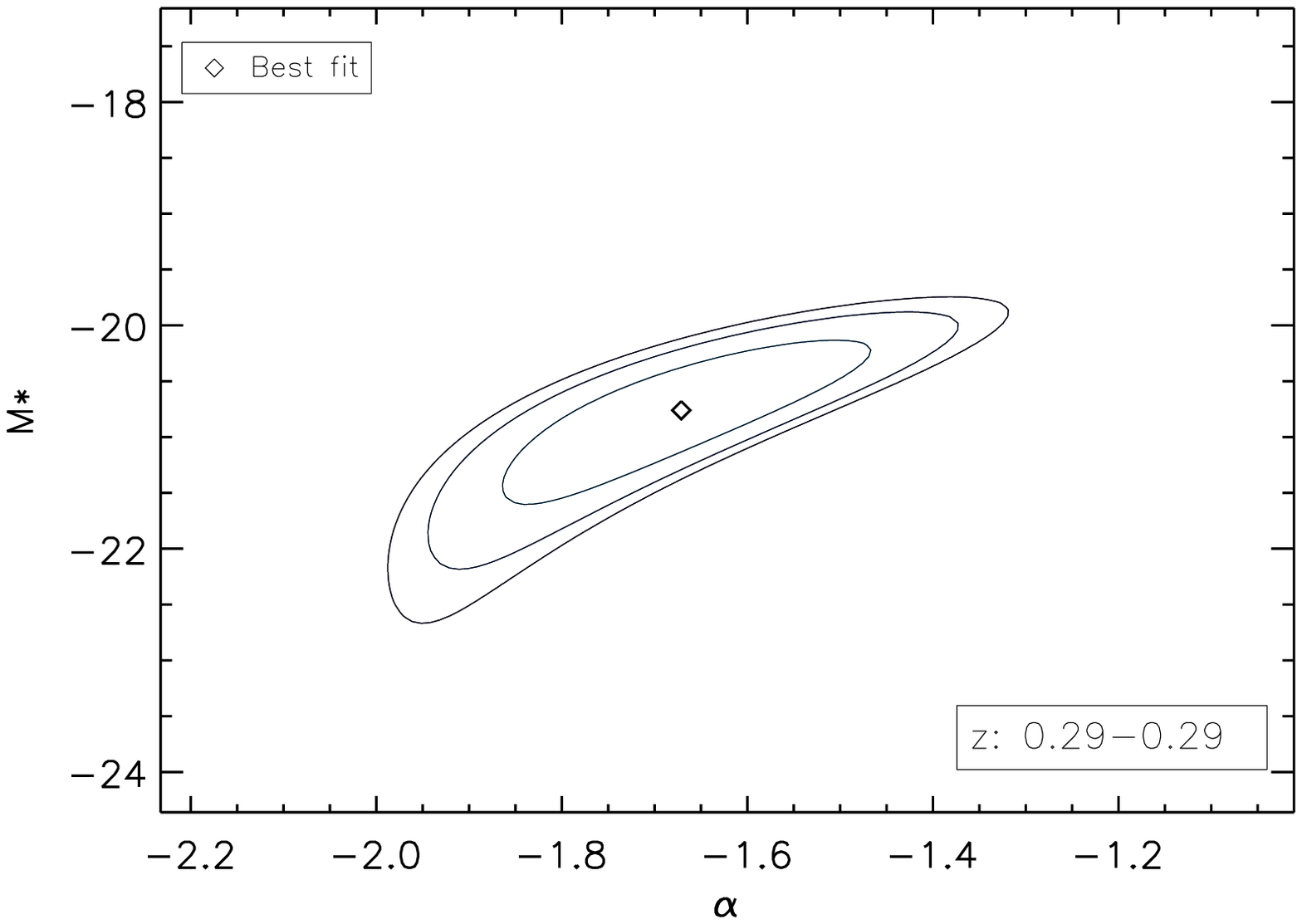,width=7.5cm,height=4.5cm}
\epsfig{file=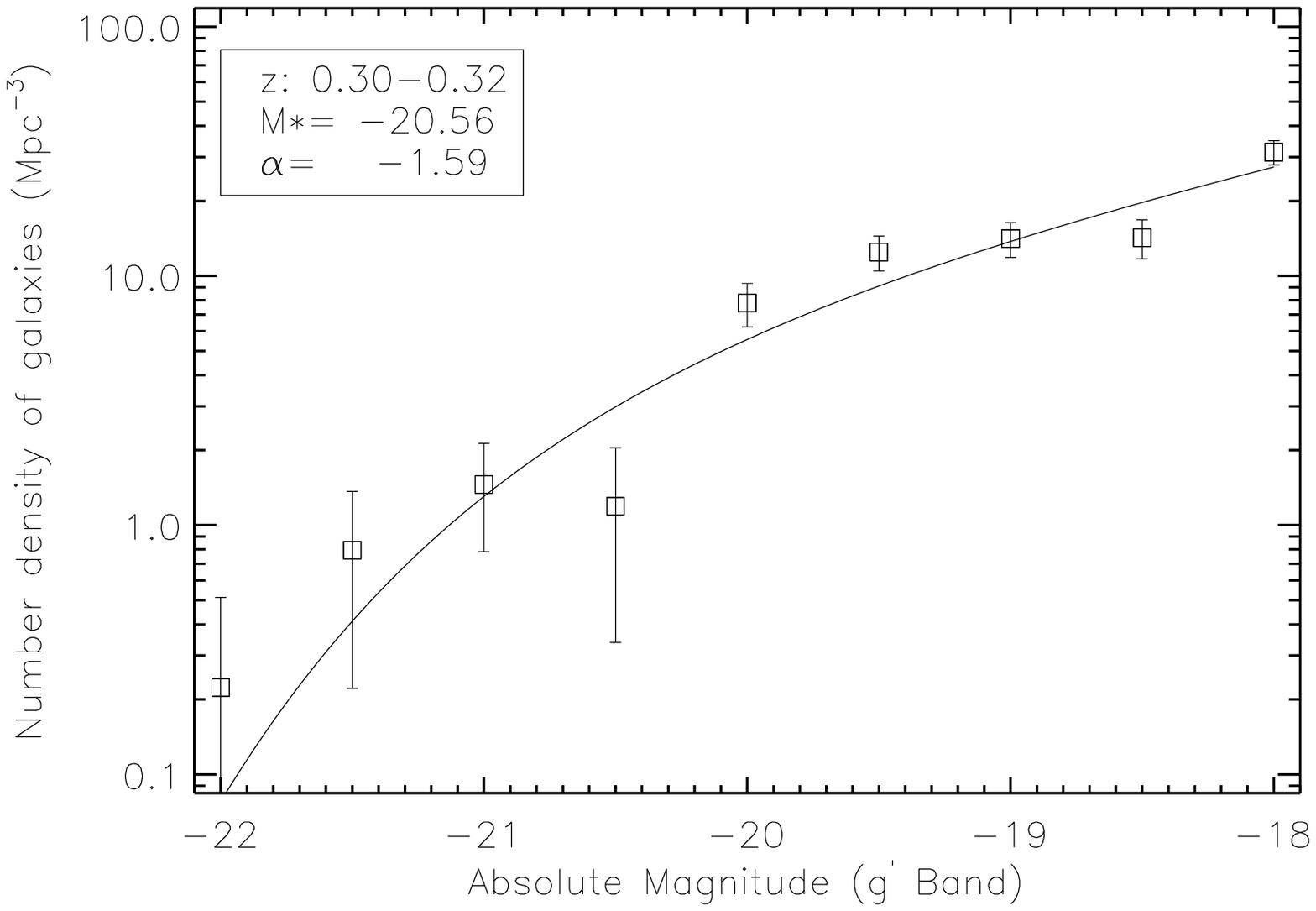,width=7.5cm,height=4.5cm}
\epsfig{file=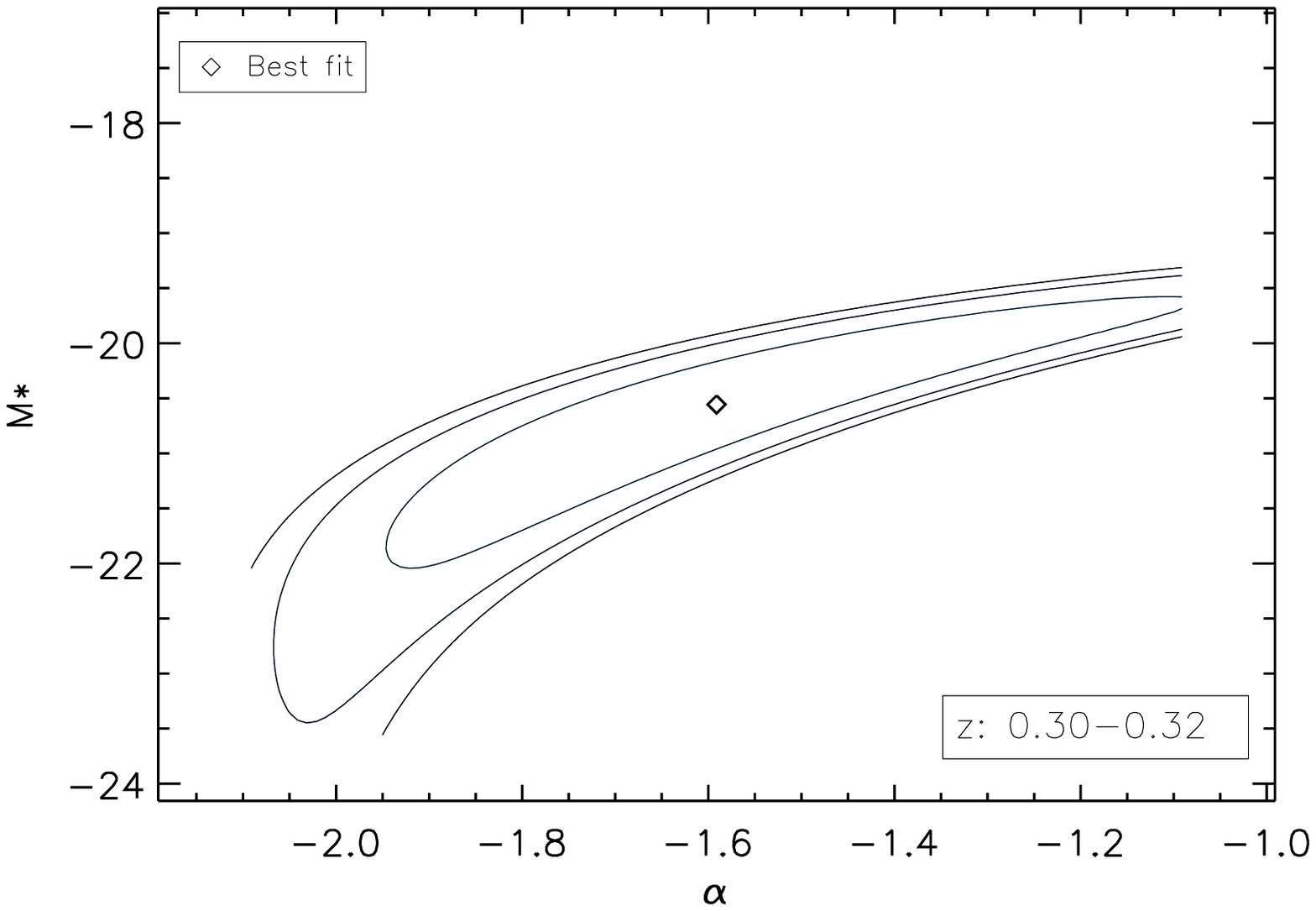,width=7.5cm,height=4.5cm}
\epsfig{file=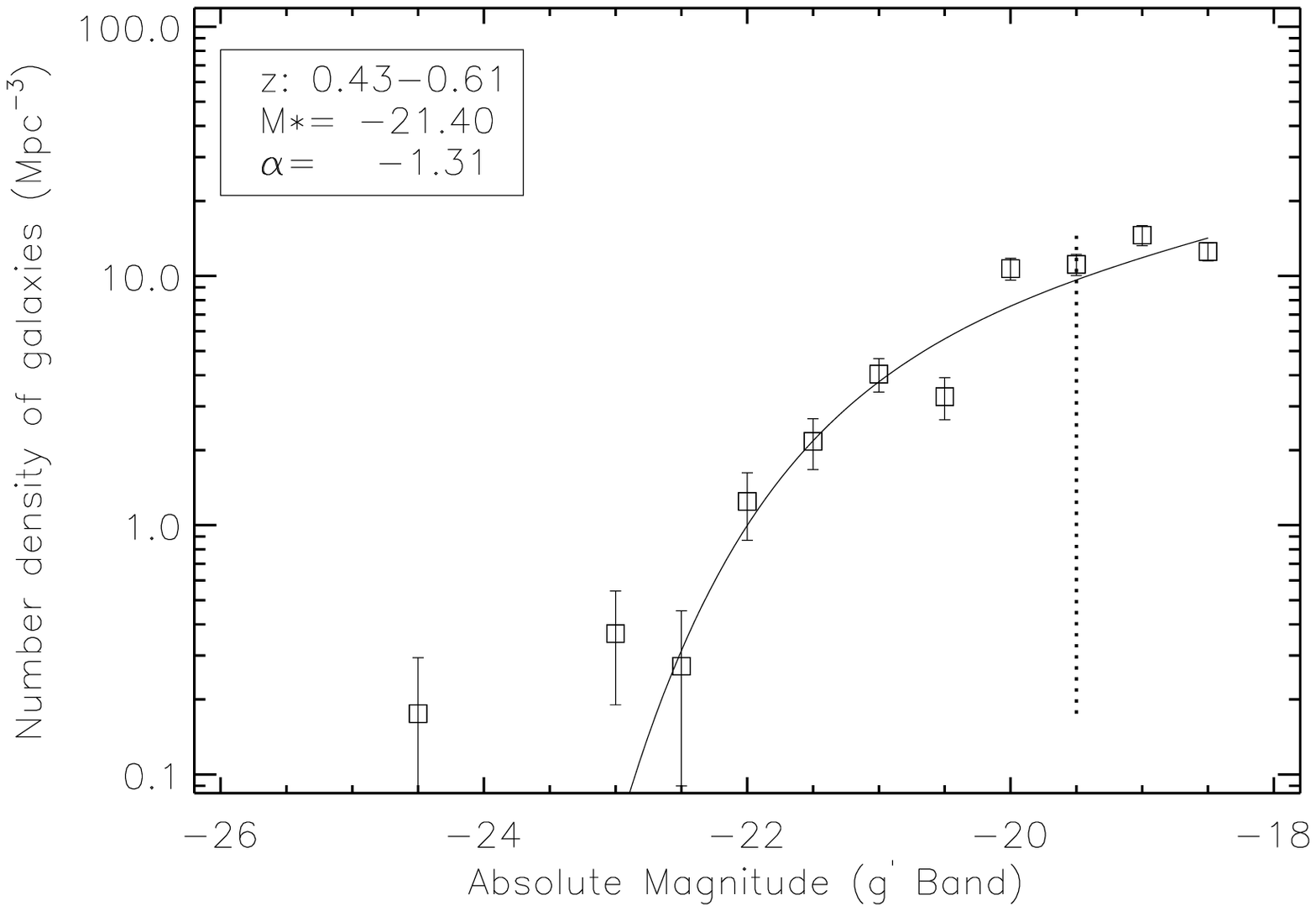,width=7.5cm,height=4.5cm}
\epsfig{file=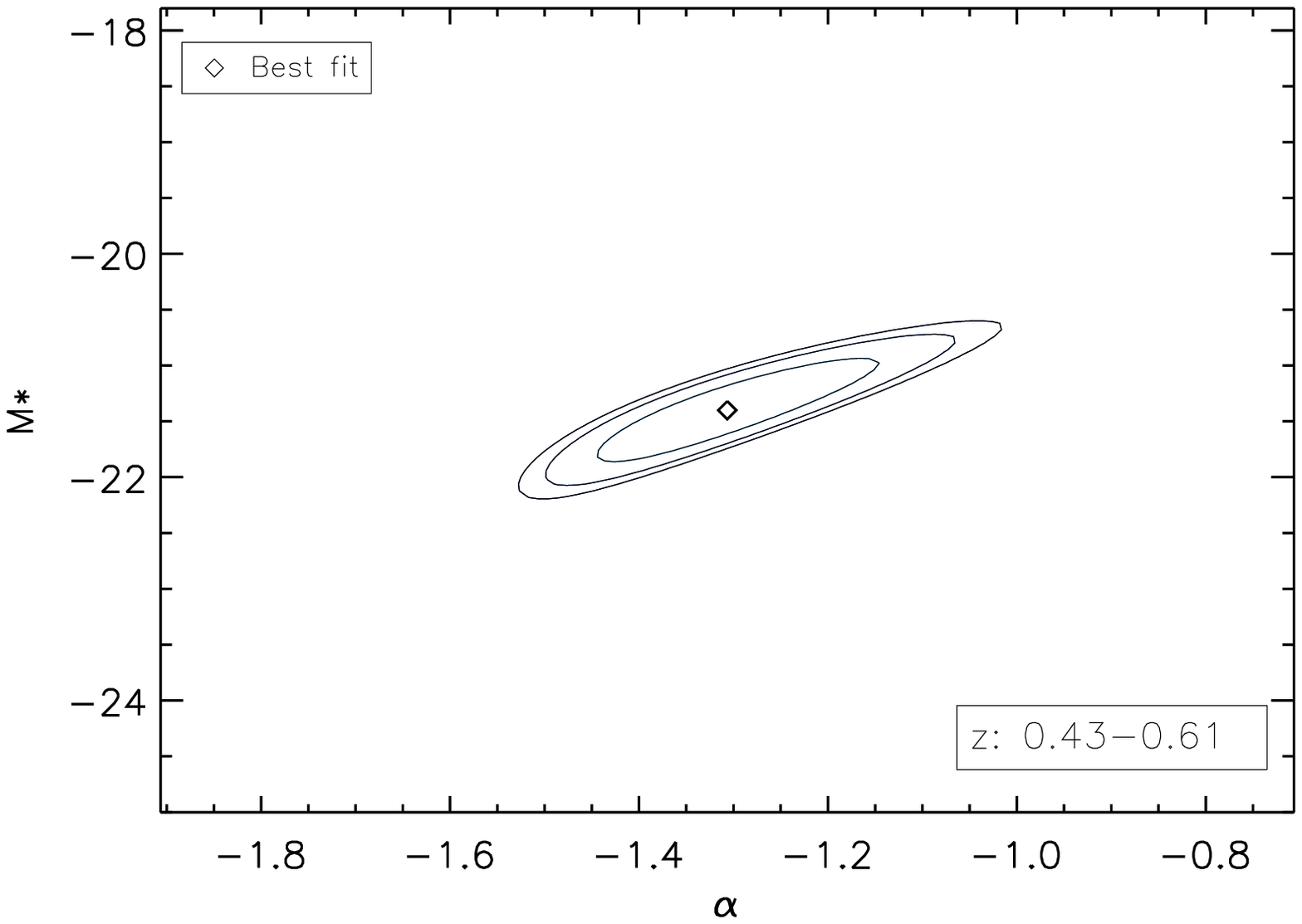,width=7.5cm,height=4.5cm}

\caption{LFs of the stacked clusters for 5 redshift ranges and the associated $1\sigma$, $2\sigma$ and $3\sigma$ contours for $g^\prime$ band. All  stacked clusters contributed to all magnitude bins are at the left side (brighter side) of the vertical dotted line which is at the faintest common magnitude bin of the clusters. Whereas at the right side (fainter side) of it, some clusters did not have data in some magnitude bins because they already reached their faintest magnitude limit.}
\label{z_stacked_g}
\end{figure}

\onecolumn
\begin{figure}
\center

\epsfig{file=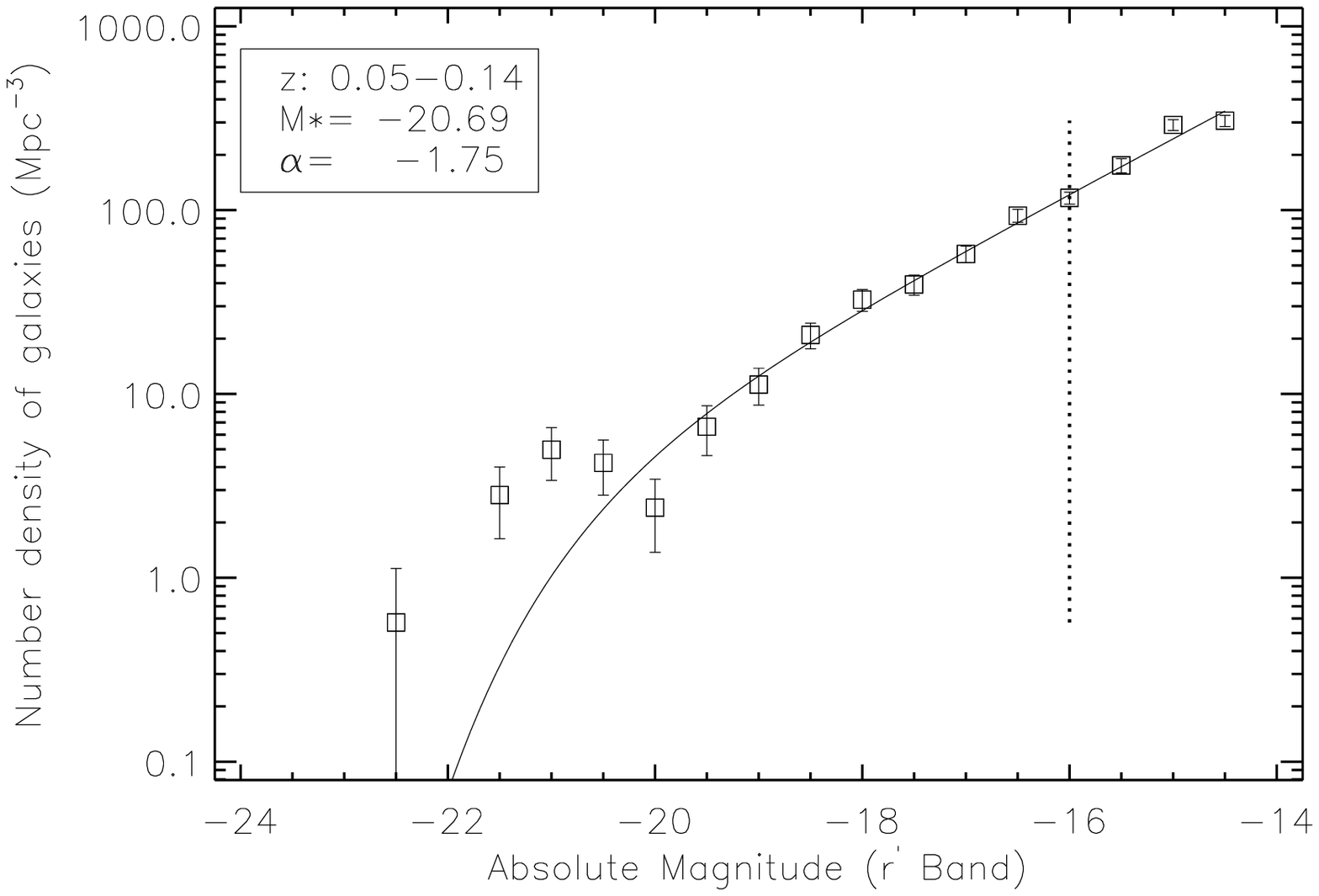,width=7.5cm,height=4.5cm}
\epsfig{file=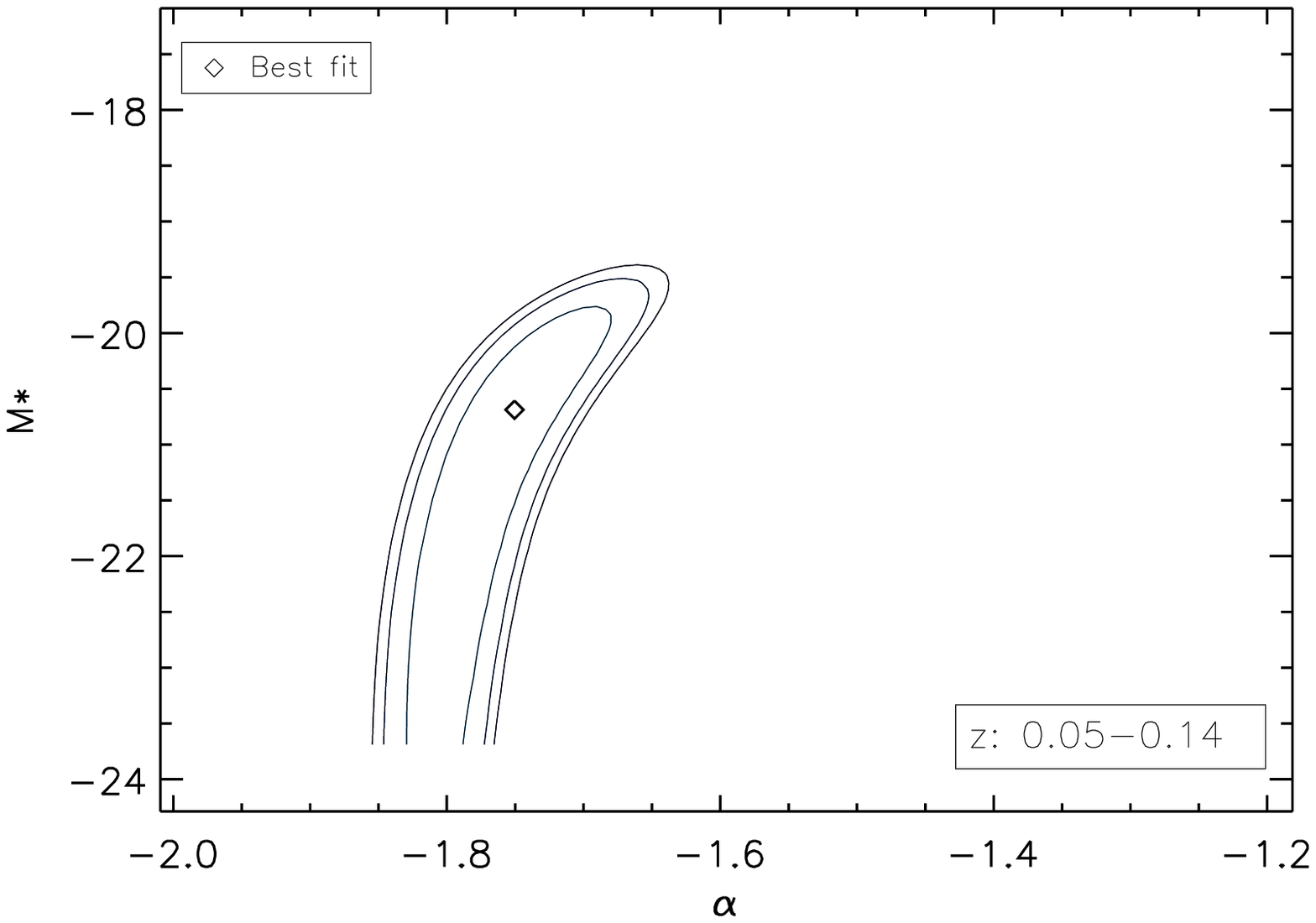,width=7.5cm,height=4.5cm}
\epsfig{file=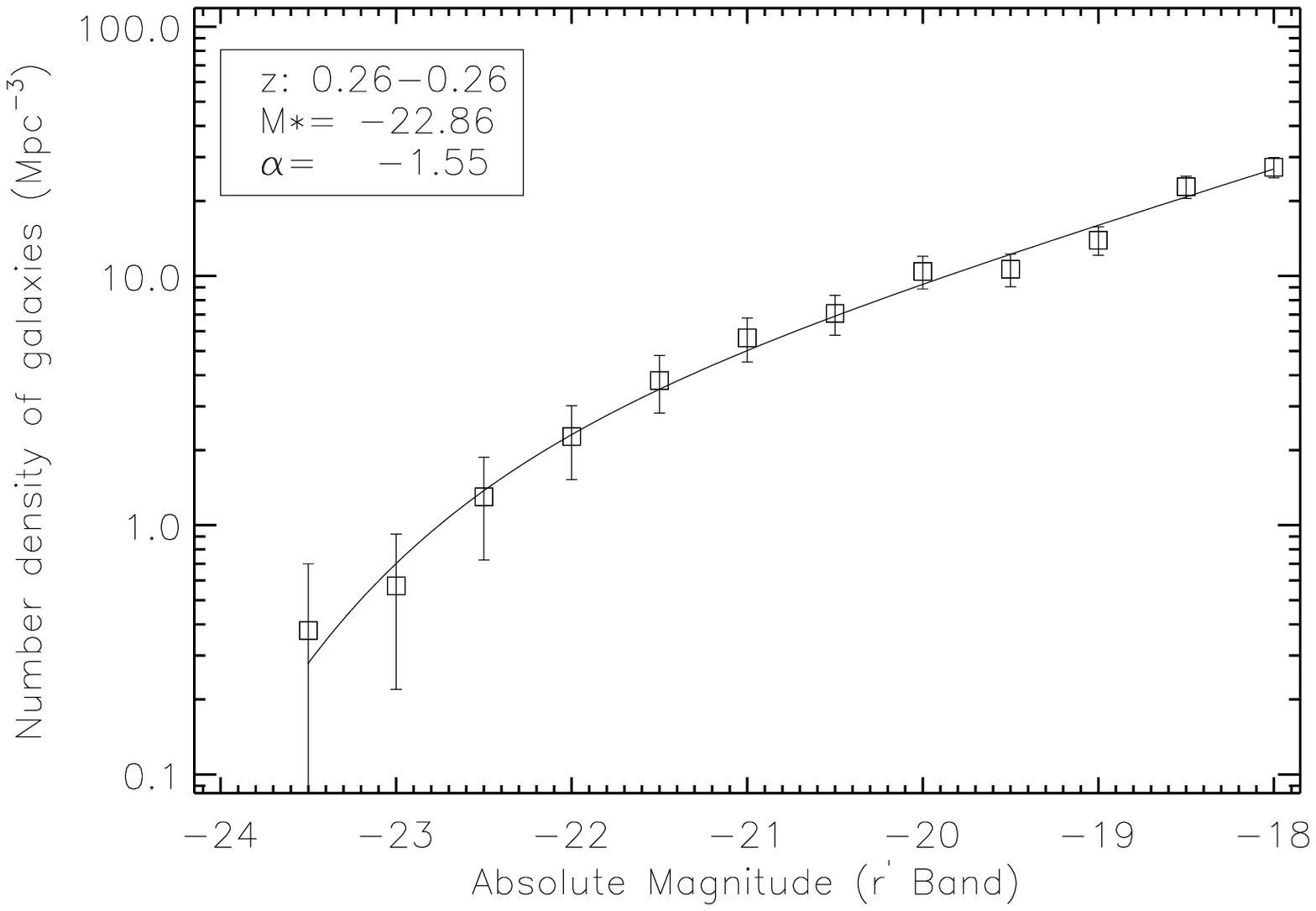,width=7.5cm,height=4.5cm}
\epsfig{file=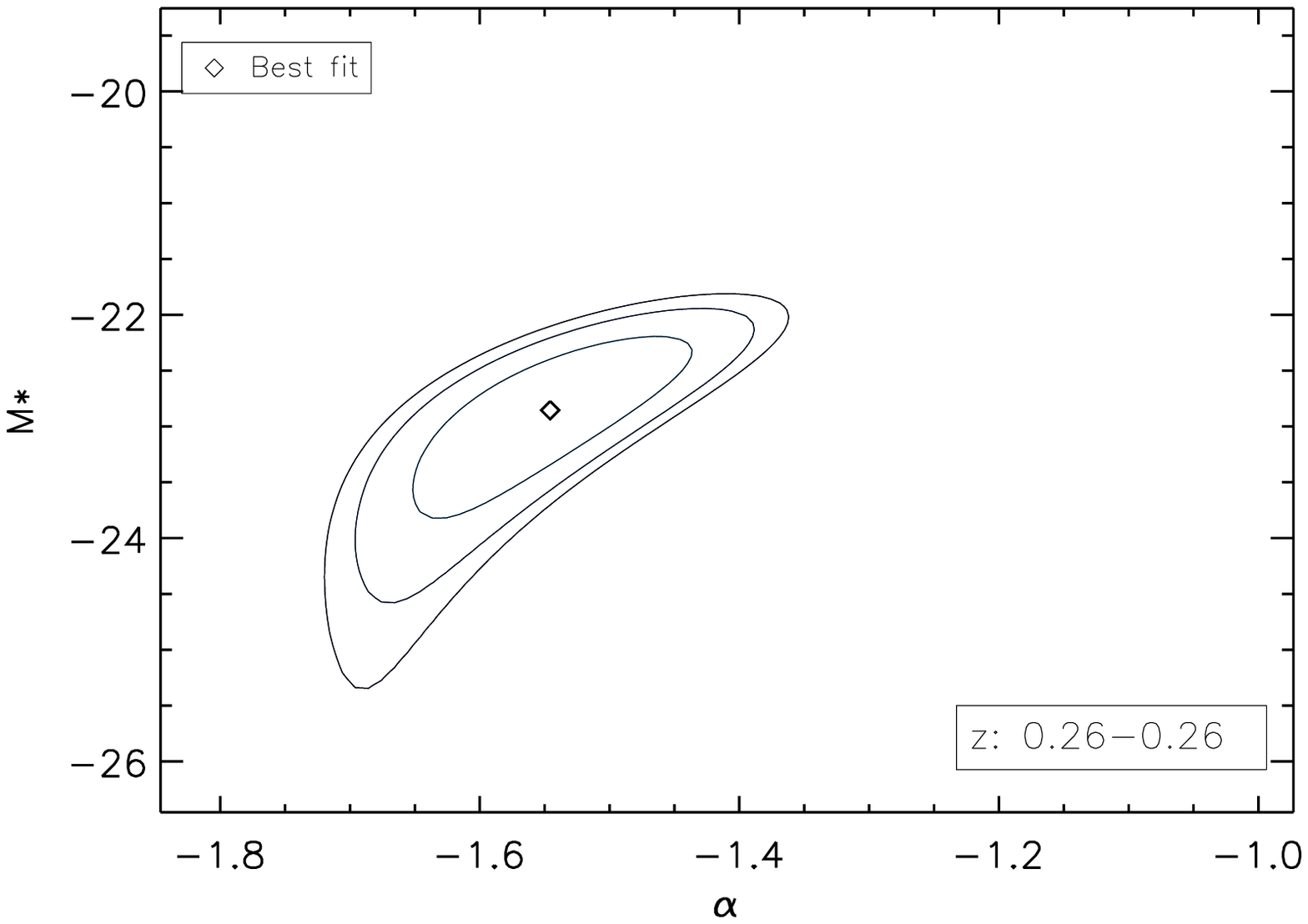,width=7.5cm,height=4.5cm}
\epsfig{file=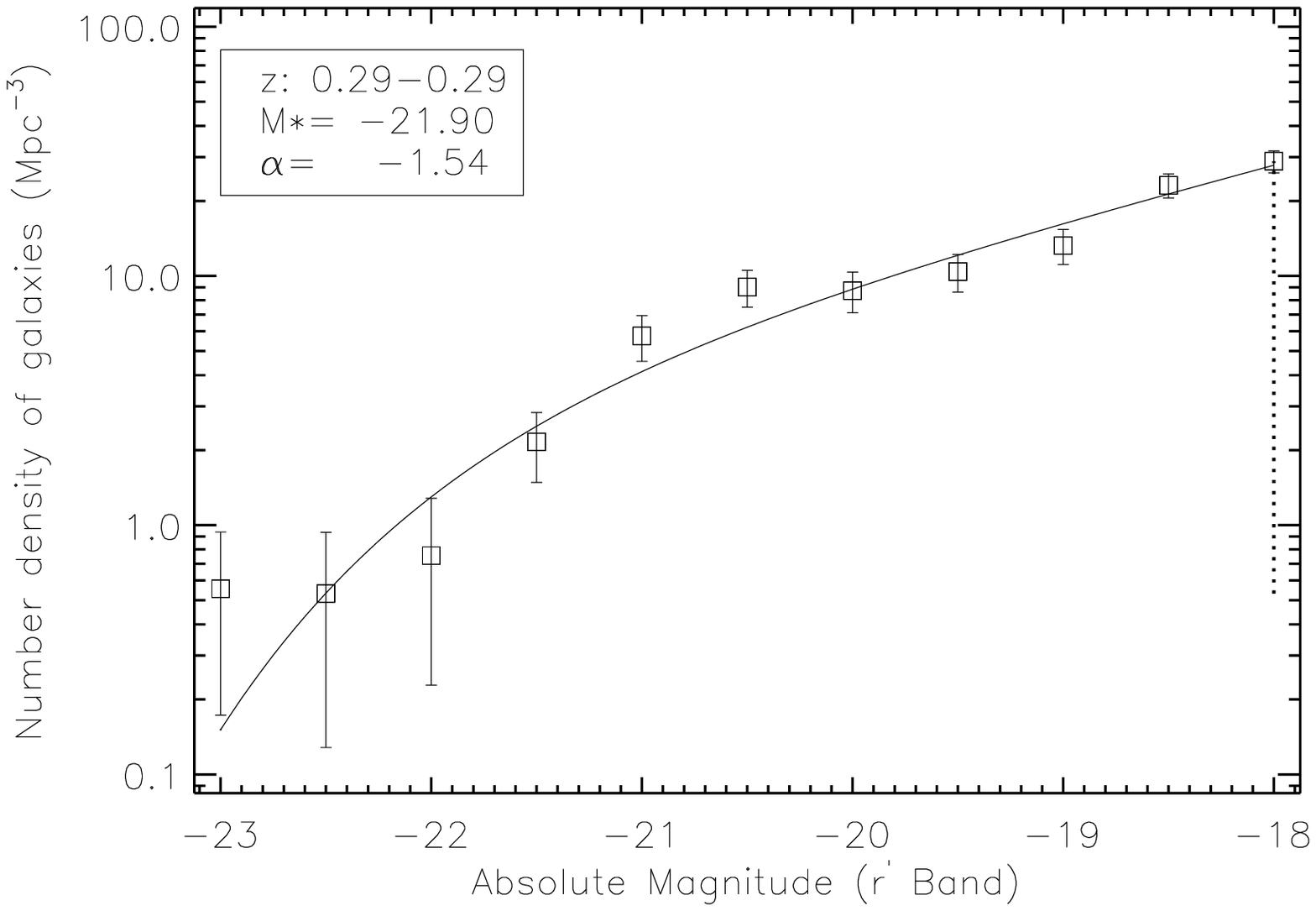,width=7.5cm,height=4.5cm}
\epsfig{file=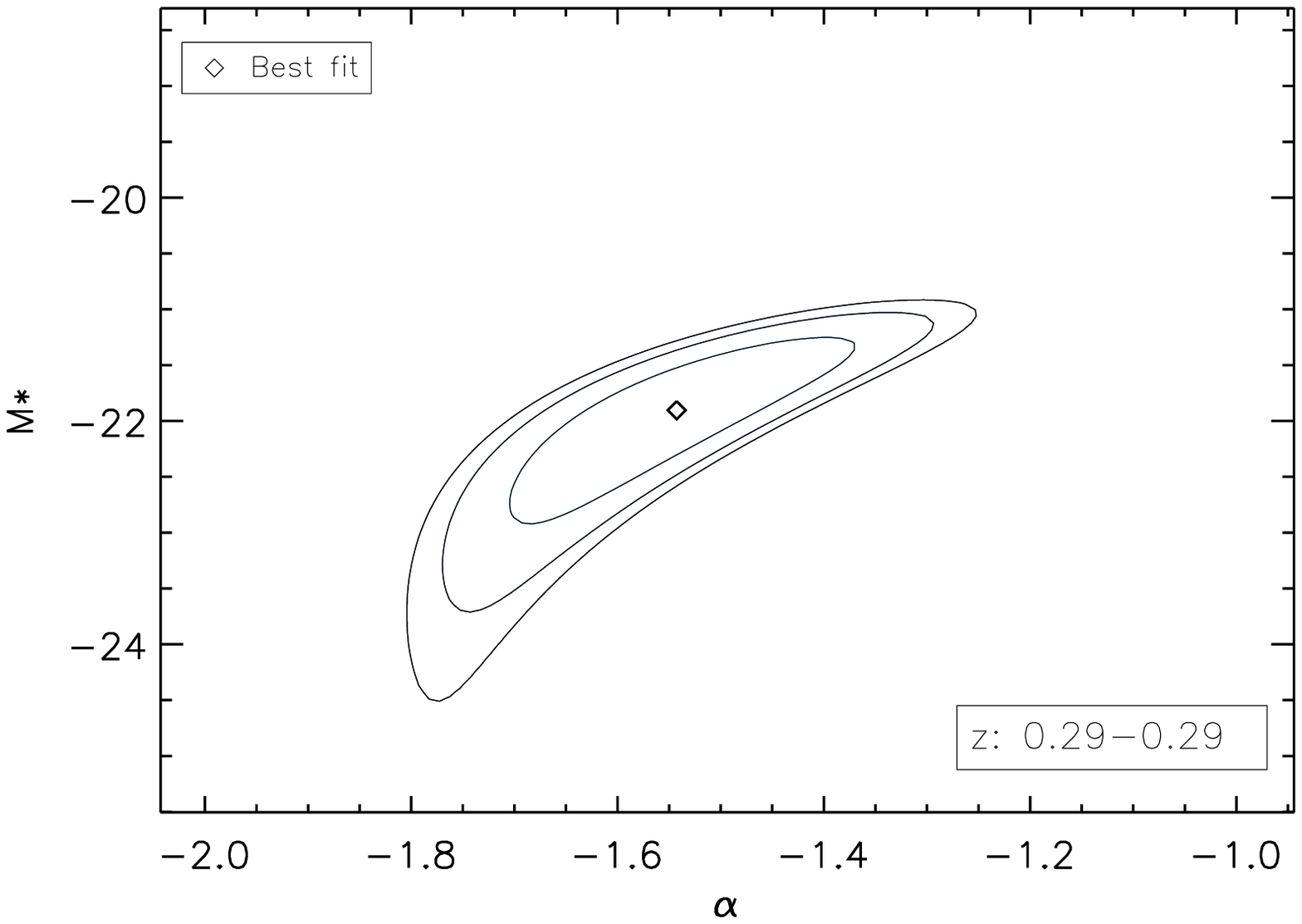,width=7.5cm,height=4.5cm}
\epsfig{file=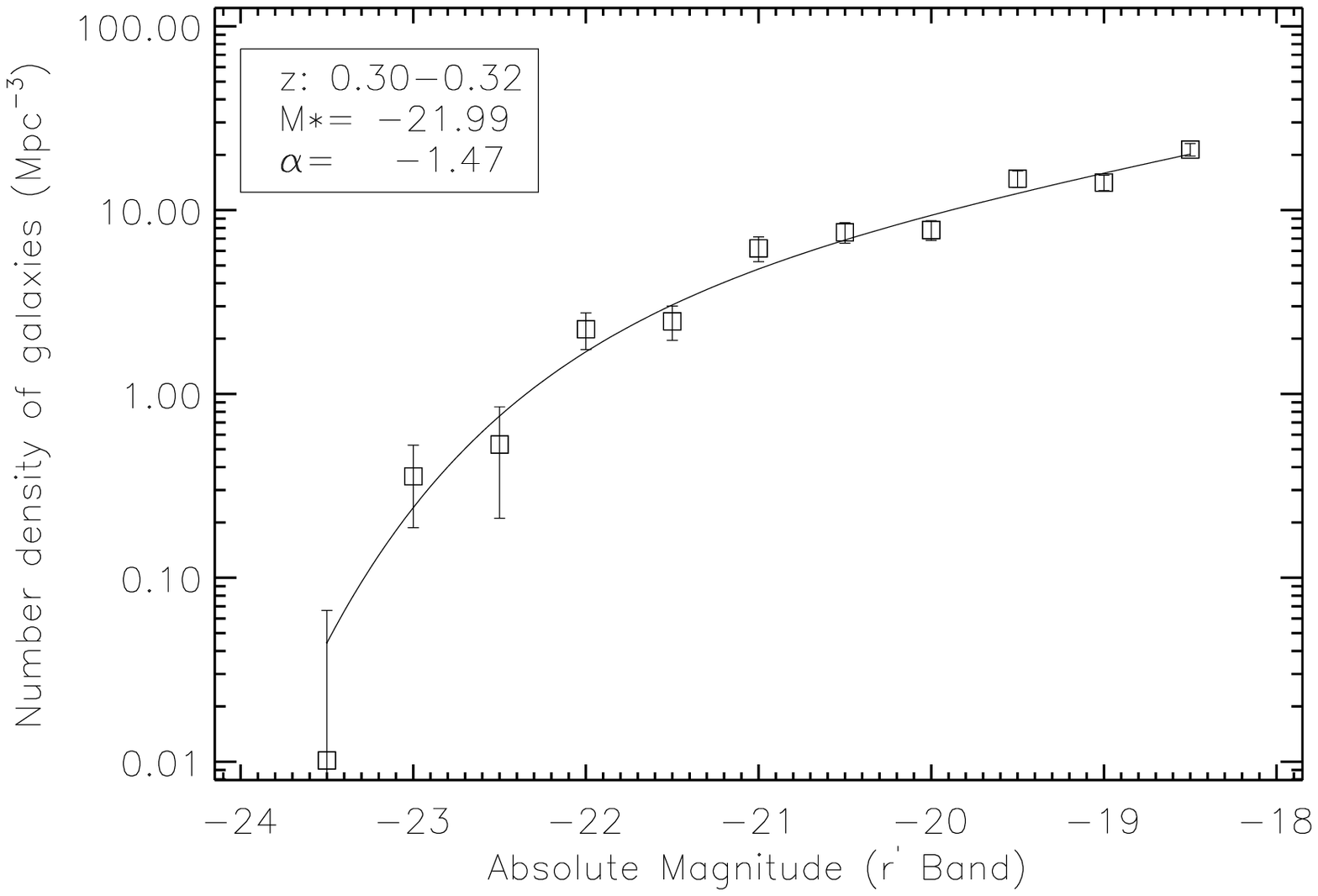,width=7.5cm,height=4.5cm}
\epsfig{file=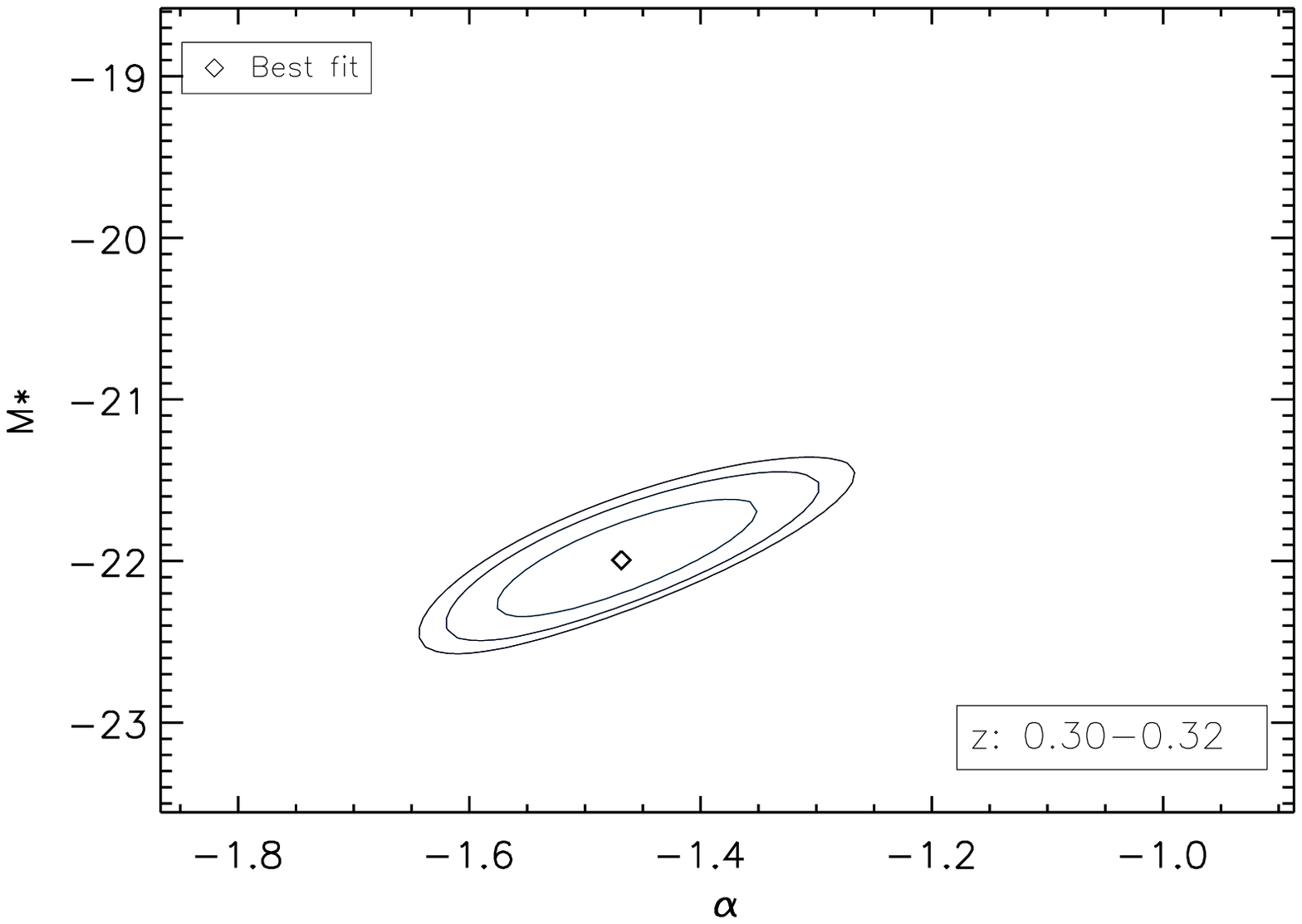,width=7.5cm,height=4.5cm}
\epsfig{file=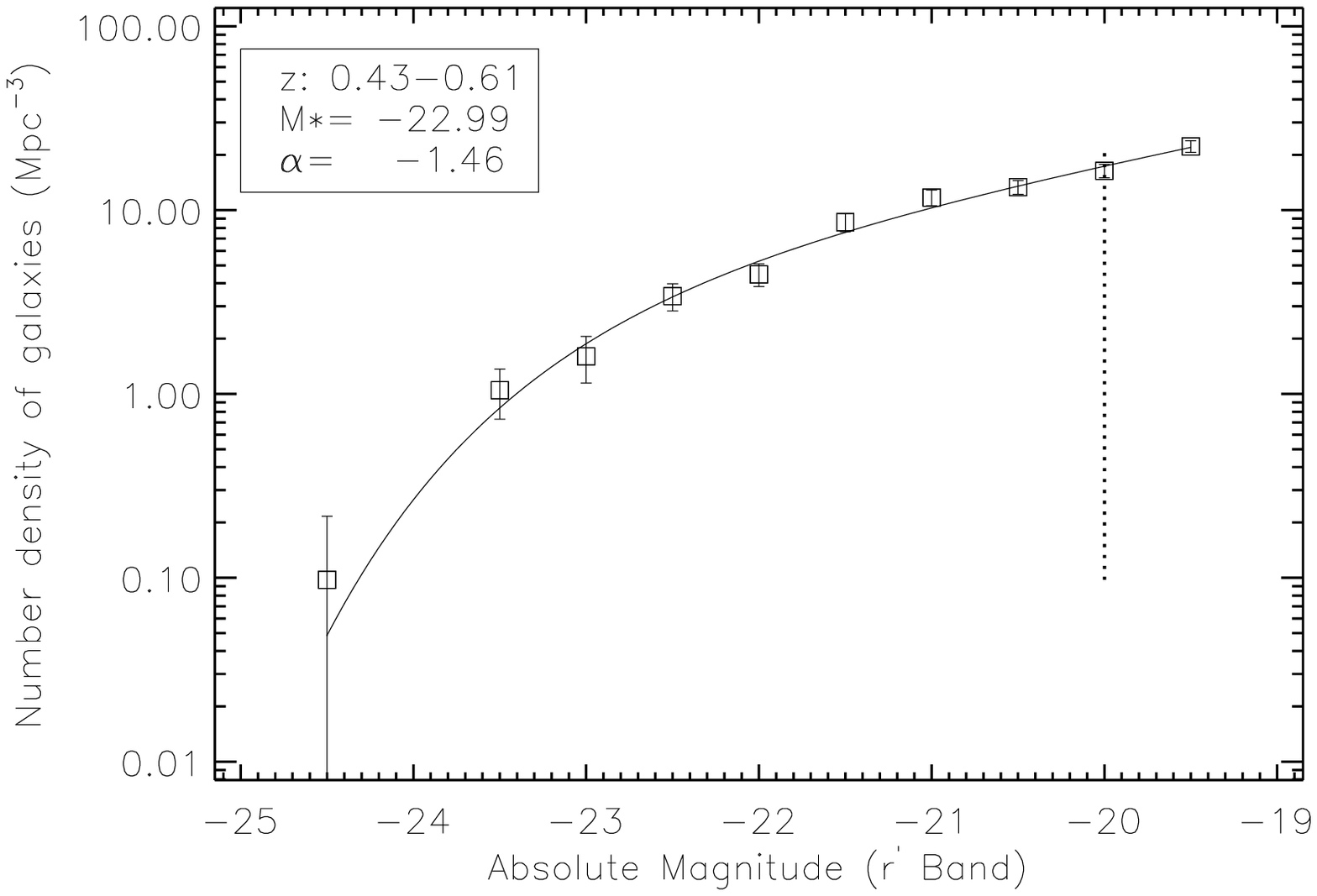,width=7.5cm,height=4.5cm}
\epsfig{file=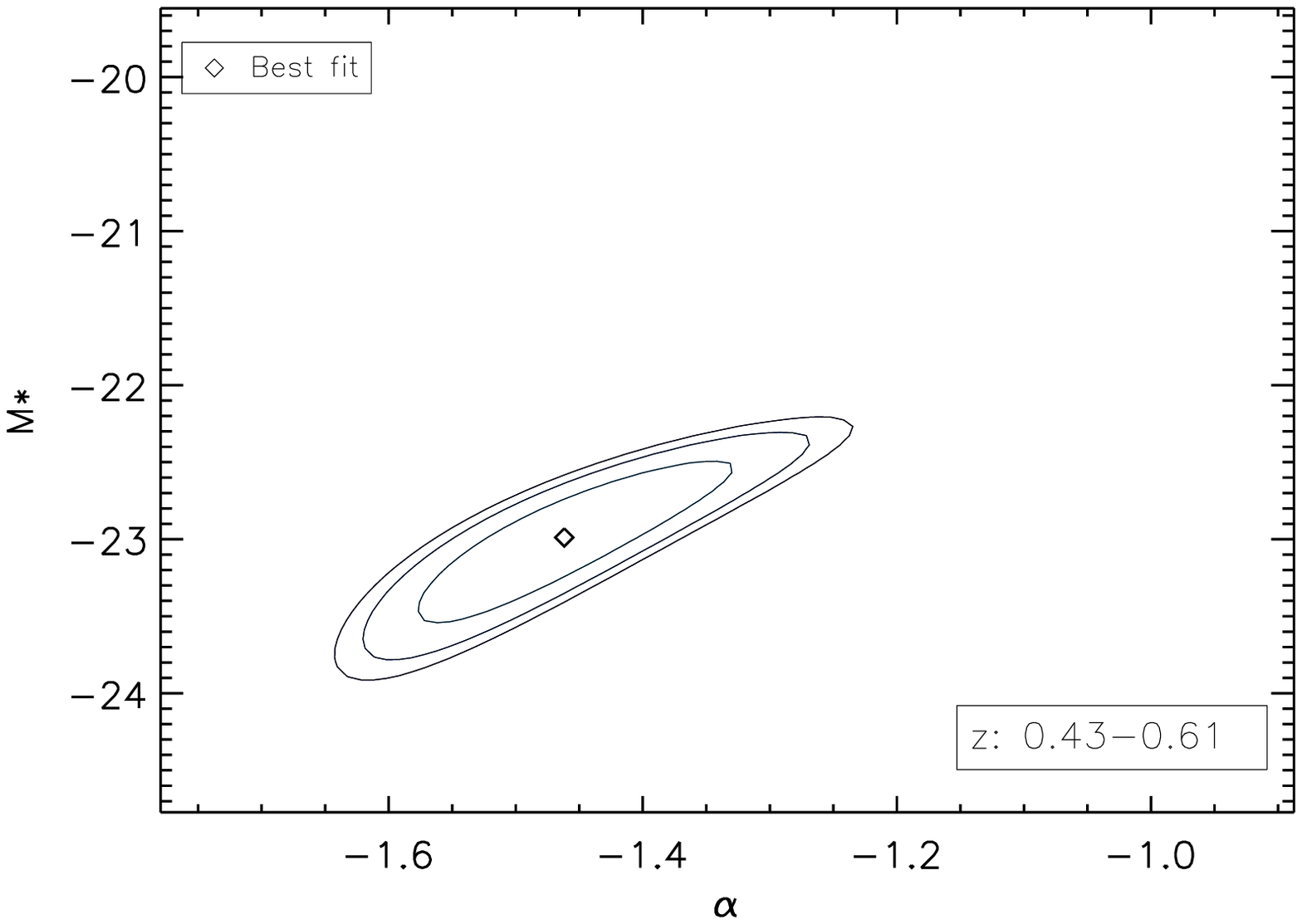,width=7.5cm,height=4.5cm}

\caption{LFs of the stacked clusters for 5 redshift ranges and the associated $1\sigma$, $2\sigma$ and $3\sigma$ contours for the $r^\prime$ band. The vertical dotted line is at the faintest common magnitude value of all stacked clusters.}
\label{z_stacked_r}
\end{figure}

\onecolumn
\begin{figure}
\center

\epsfig{file=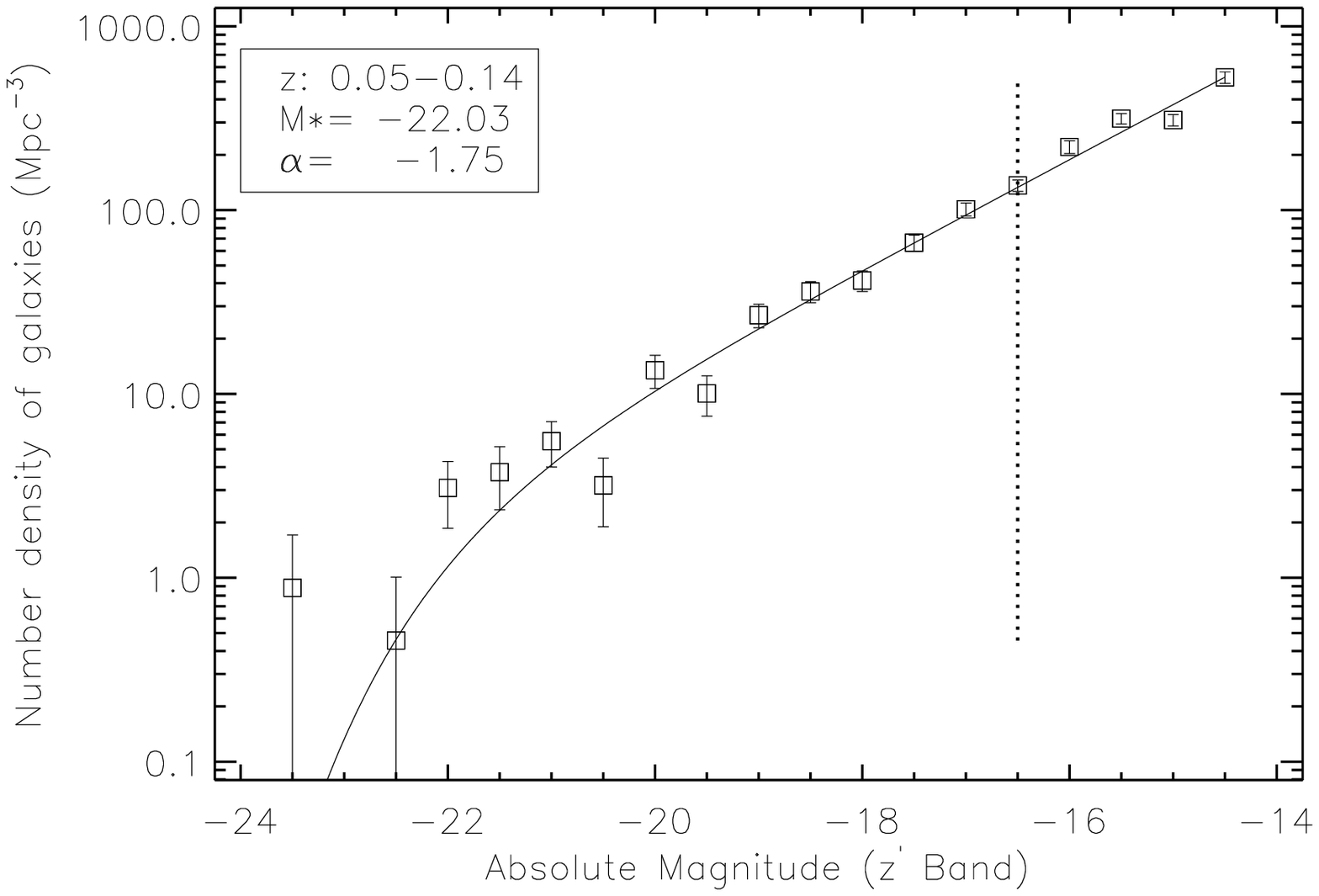,width=7.5cm,height=4.5cm}
\epsfig{file=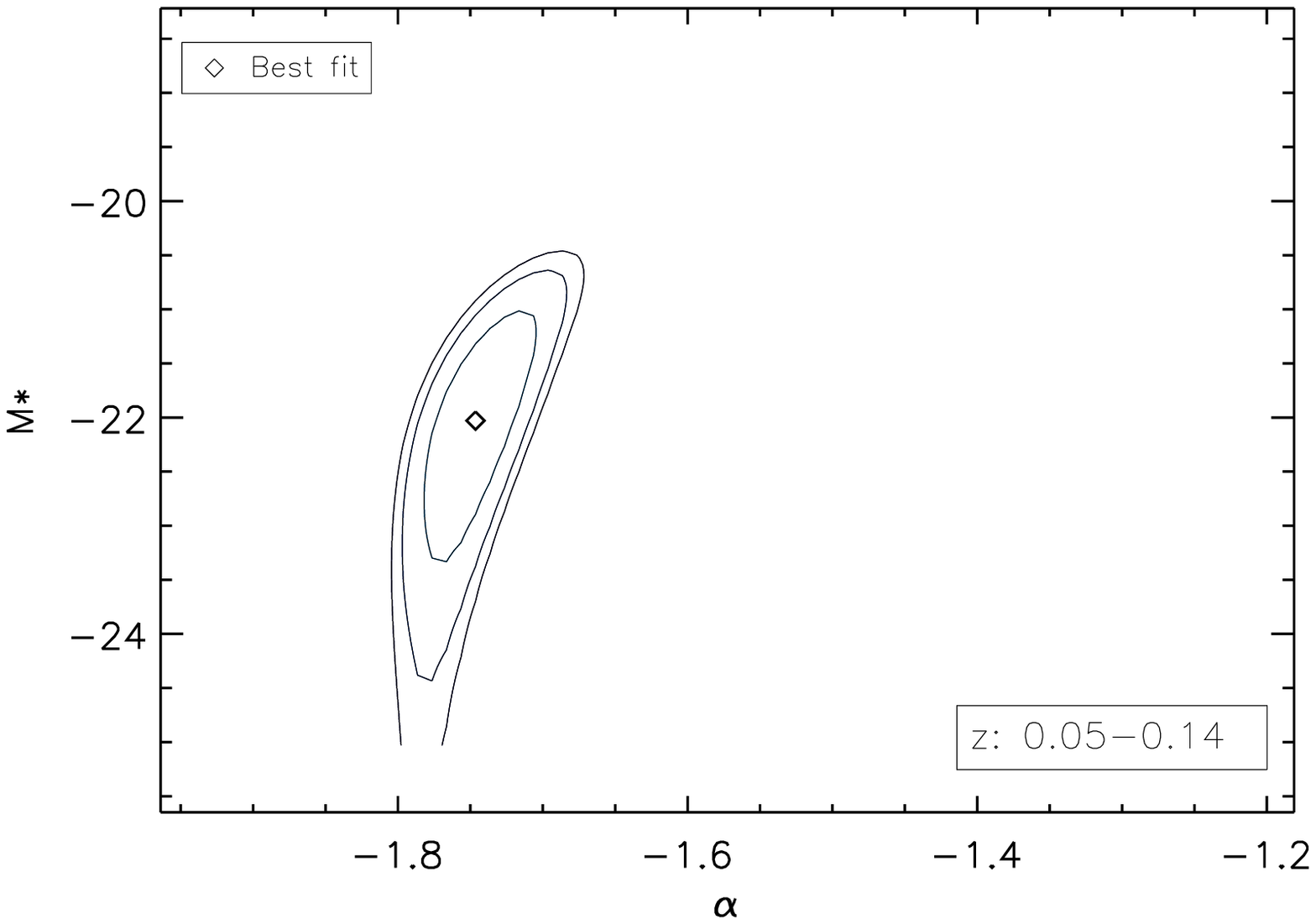,width=7.5cm,height=4.5cm}
\epsfig{file=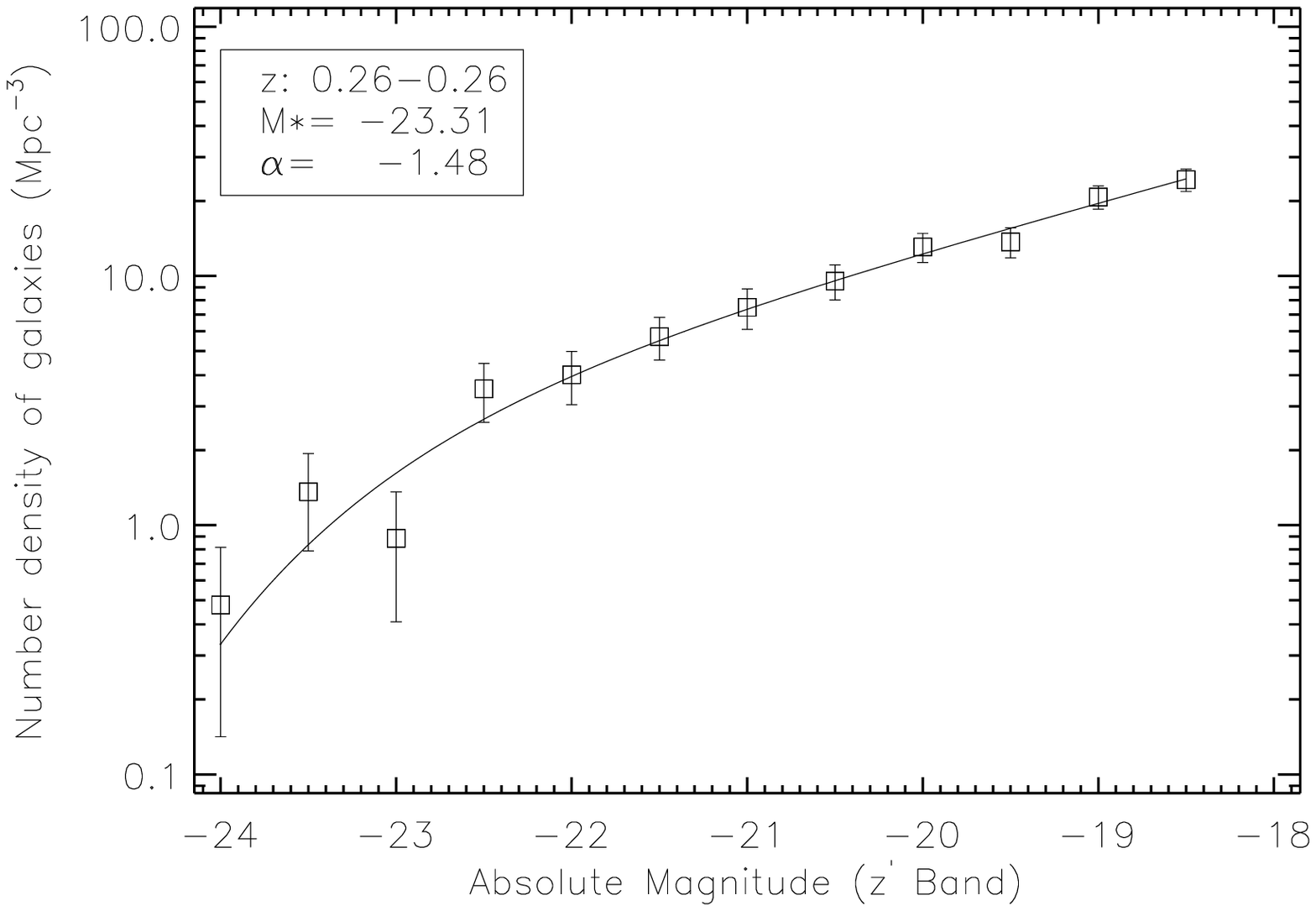,width=7.5cm,height=4.5cm}
\epsfig{file=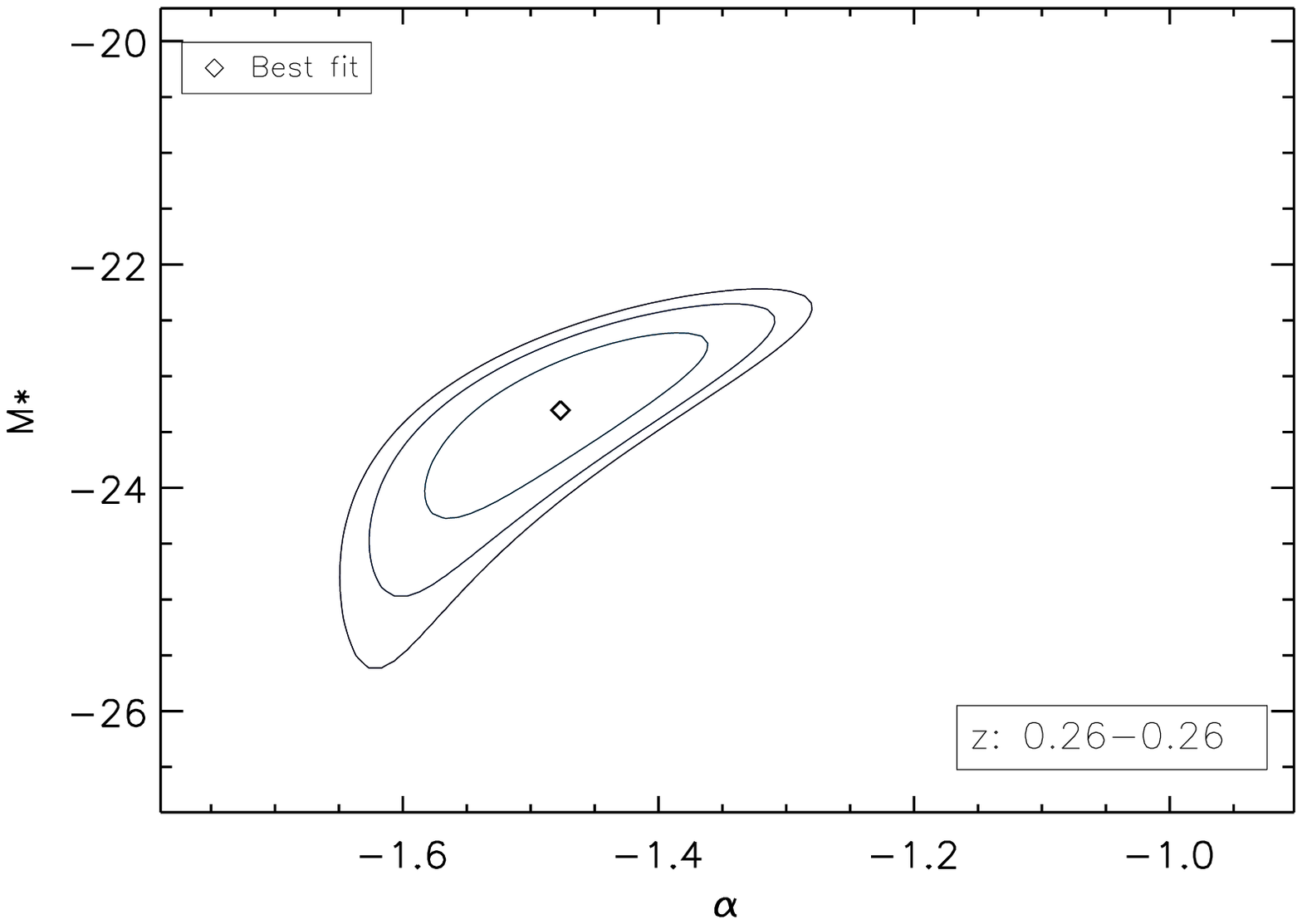,width=7.5cm,height=4.5cm}
\epsfig{file=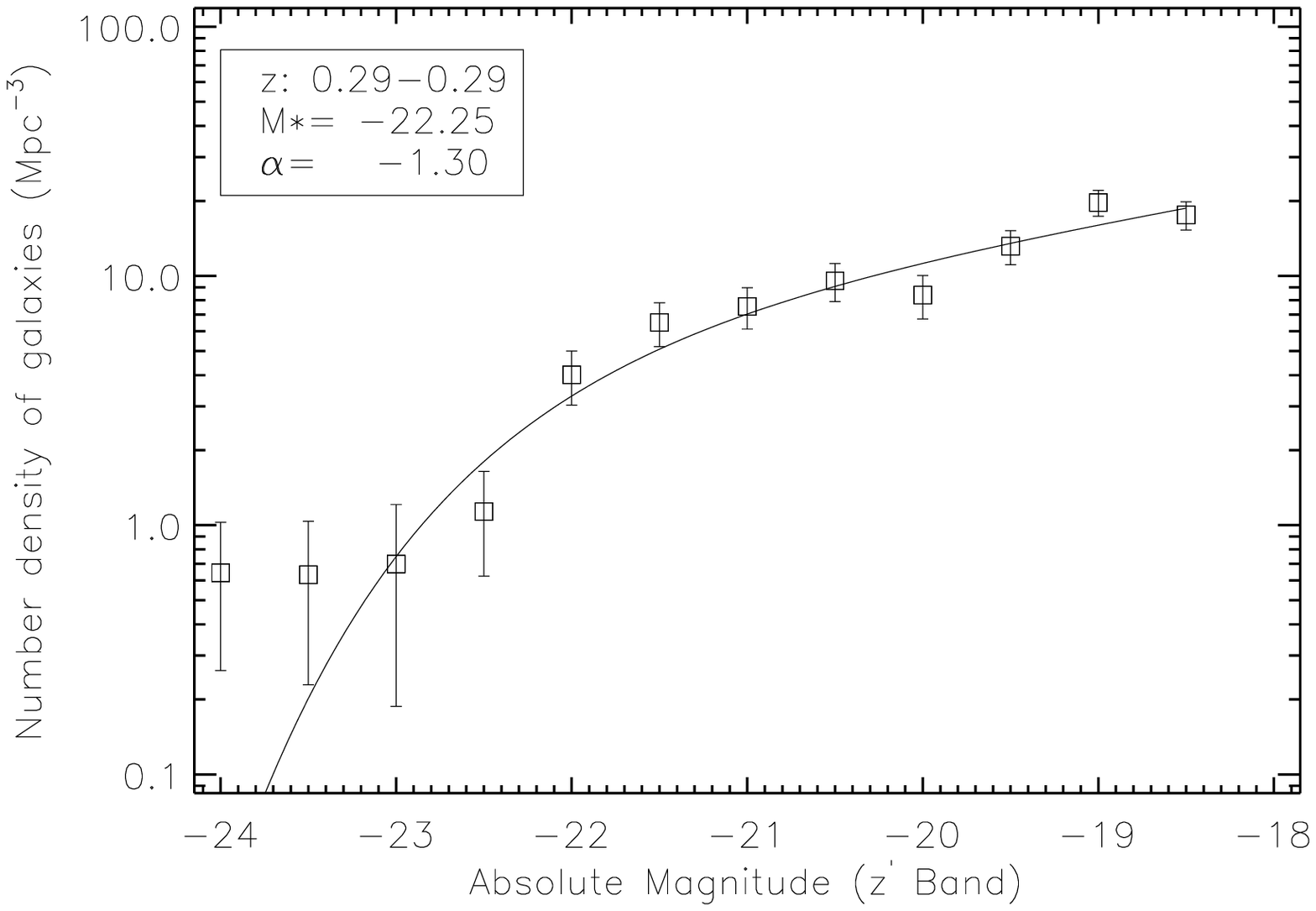,width=7.5cm,height=4.5cm}
\epsfig{file=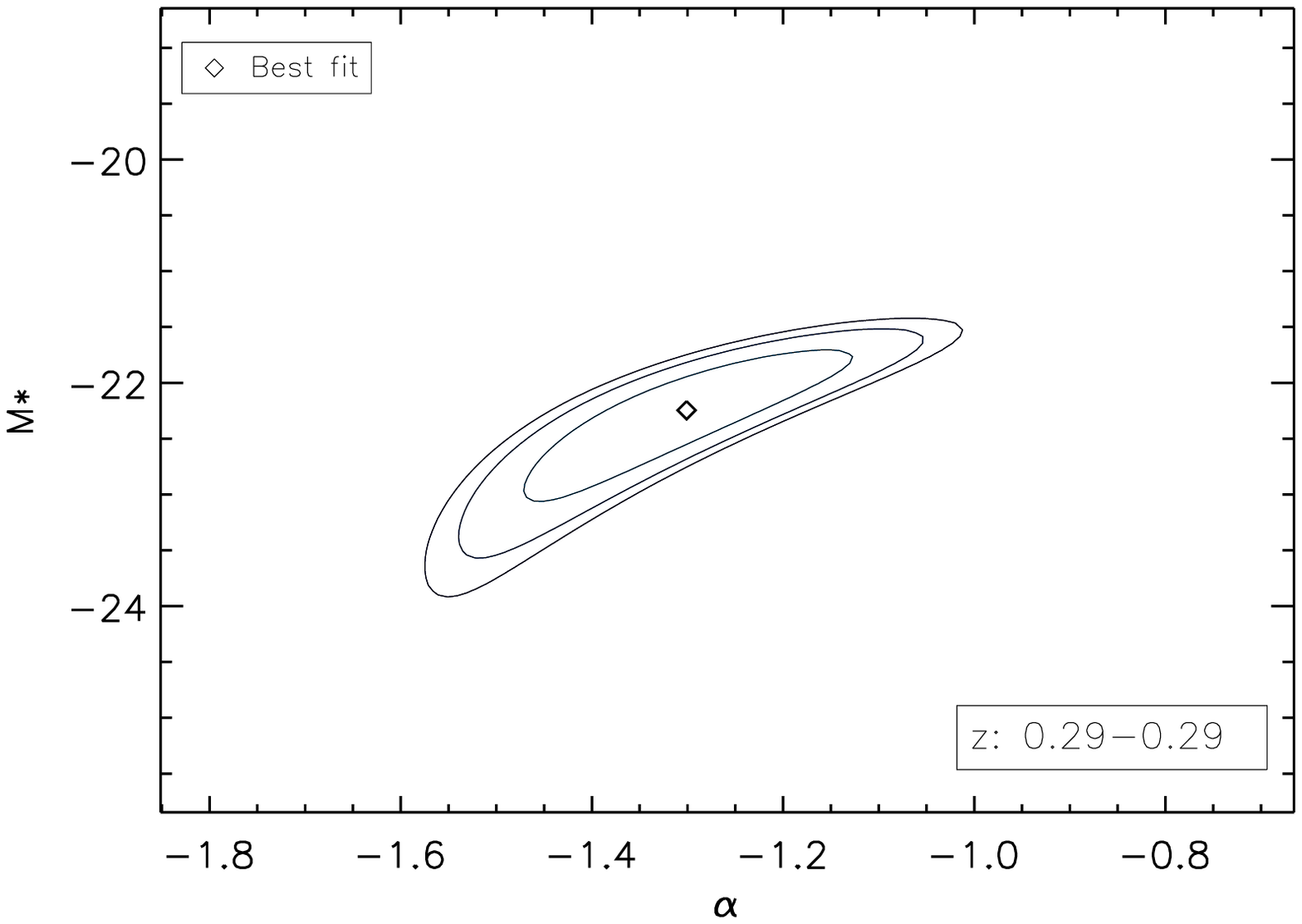,width=7.5cm,height=4.5cm}
\epsfig{file=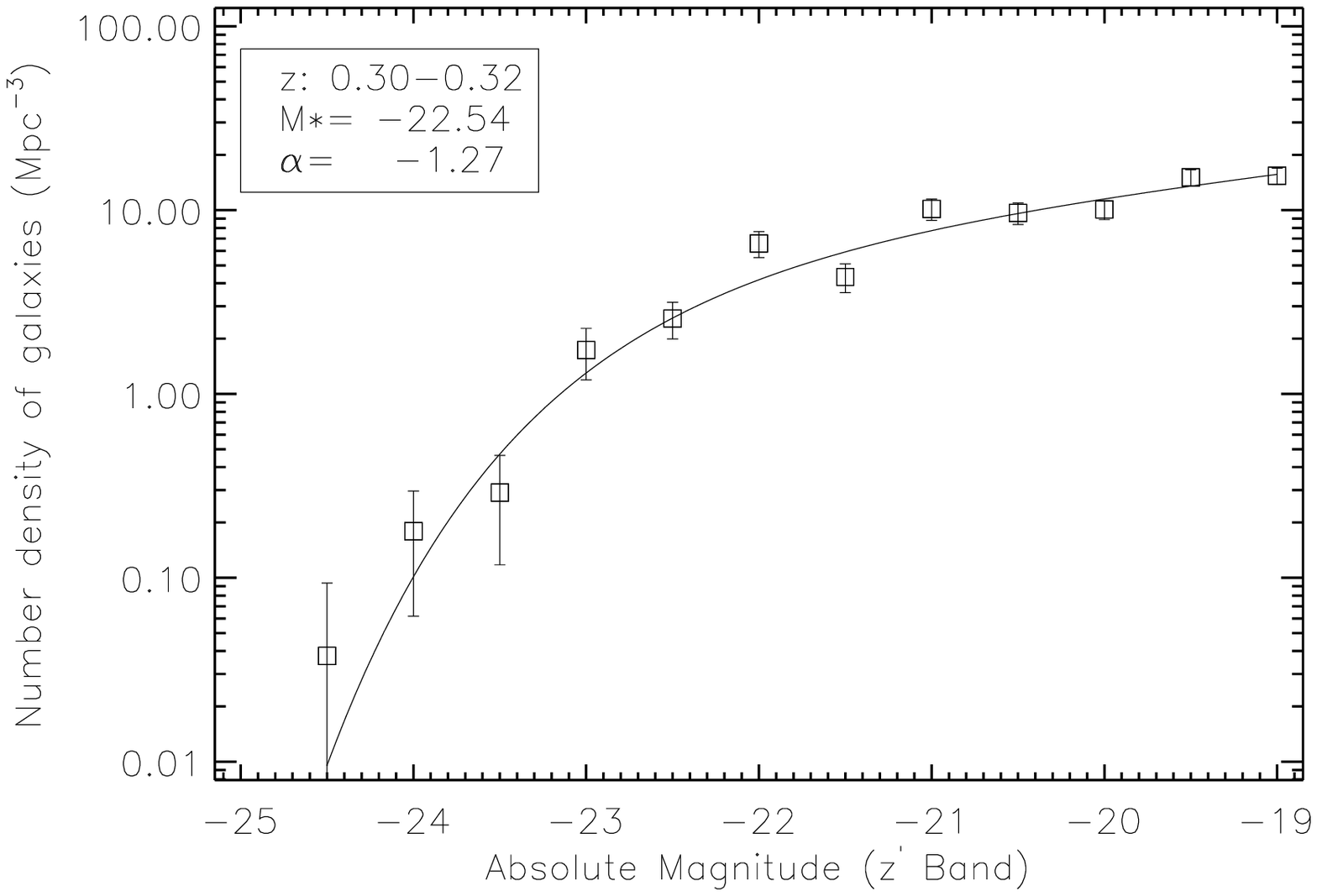,width=7.5cm,height=4.5cm}
\epsfig{file=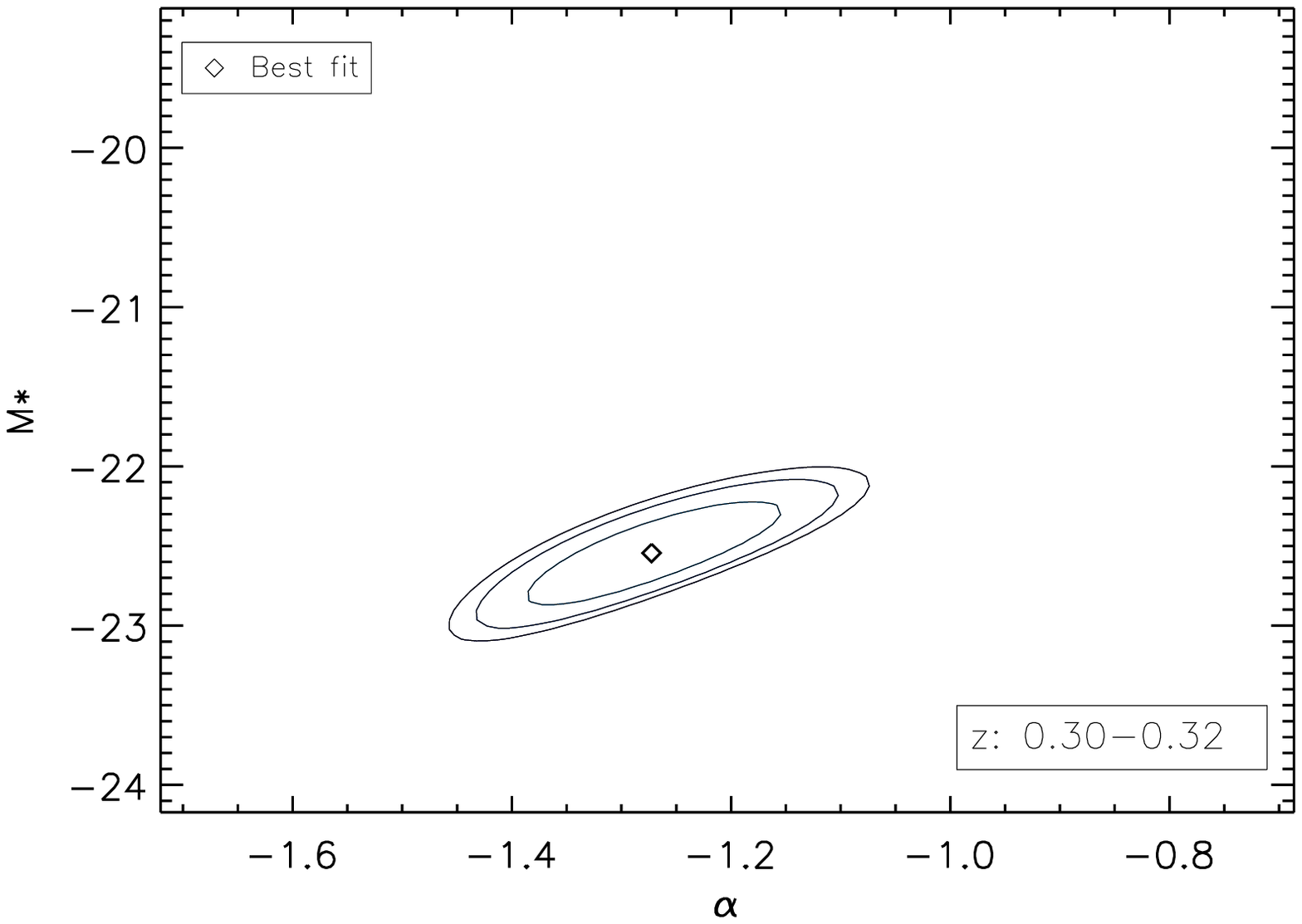,width=7.5cm,height=4.5cm}
\epsfig{file=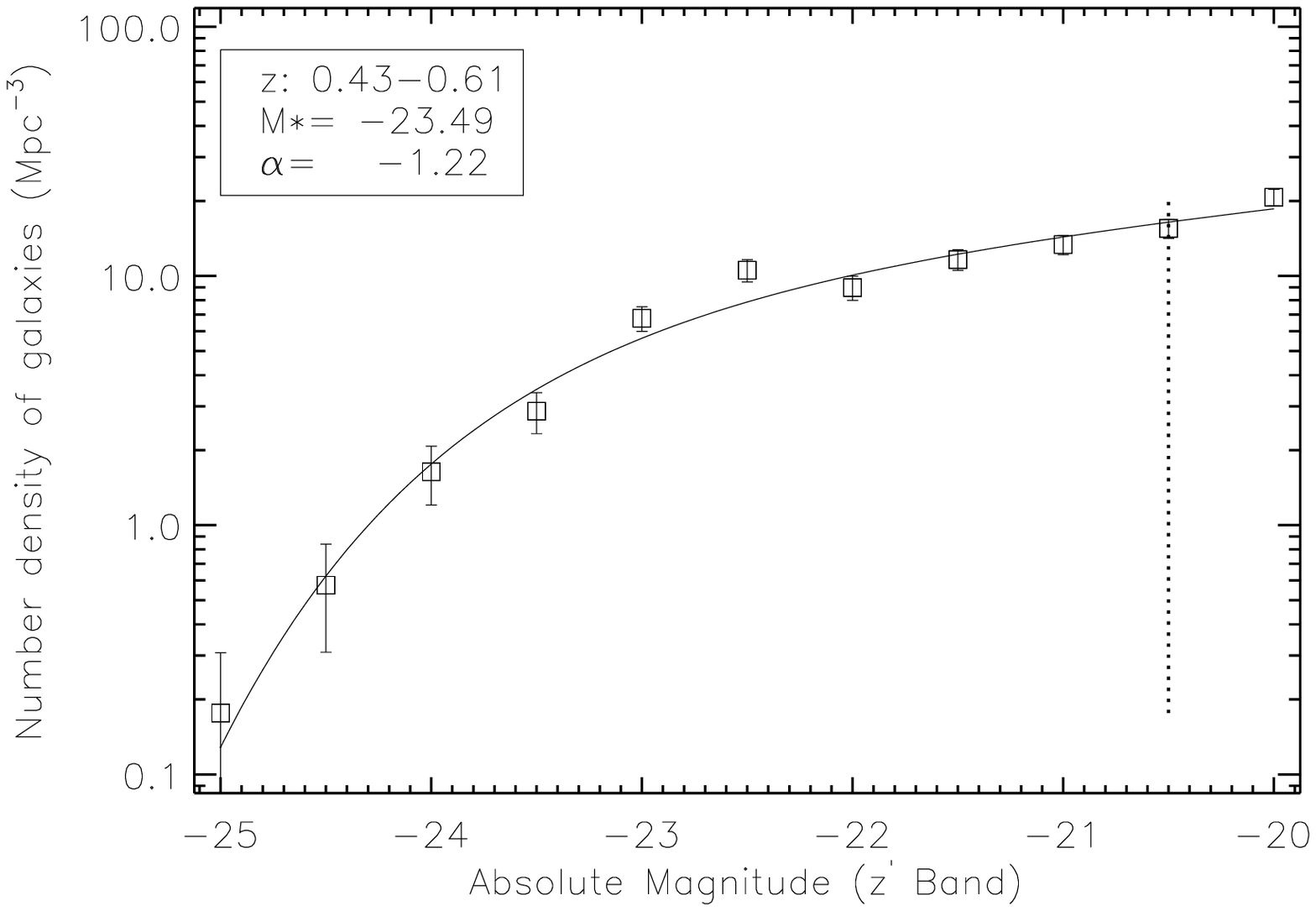,width=7.5cm,height=4.5cm}
\epsfig{file=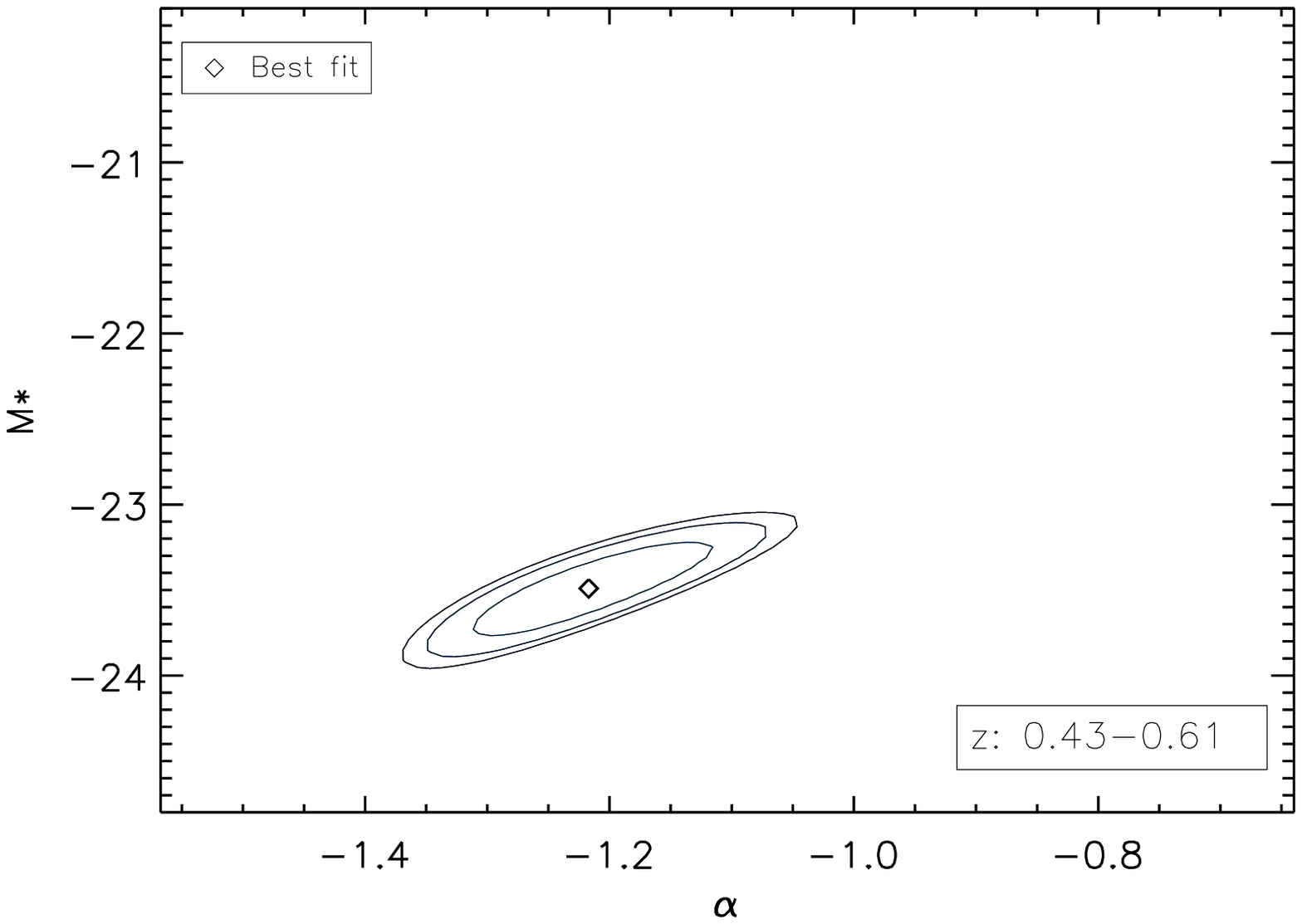,width=7.5cm,height=4.5cm}

\caption{LFs of the stacked clusters for 5 redshift ranges and the associated $1\sigma$, $2\sigma$ and $3\sigma$ contours for the $z^\prime$ band. The vertical dotted line is at the faintest common magnitude value of all stacked clusters.}
\label{z_stacked_z}
\end{figure}


\onecolumn
\begin{figure}
\center

\epsfig{file=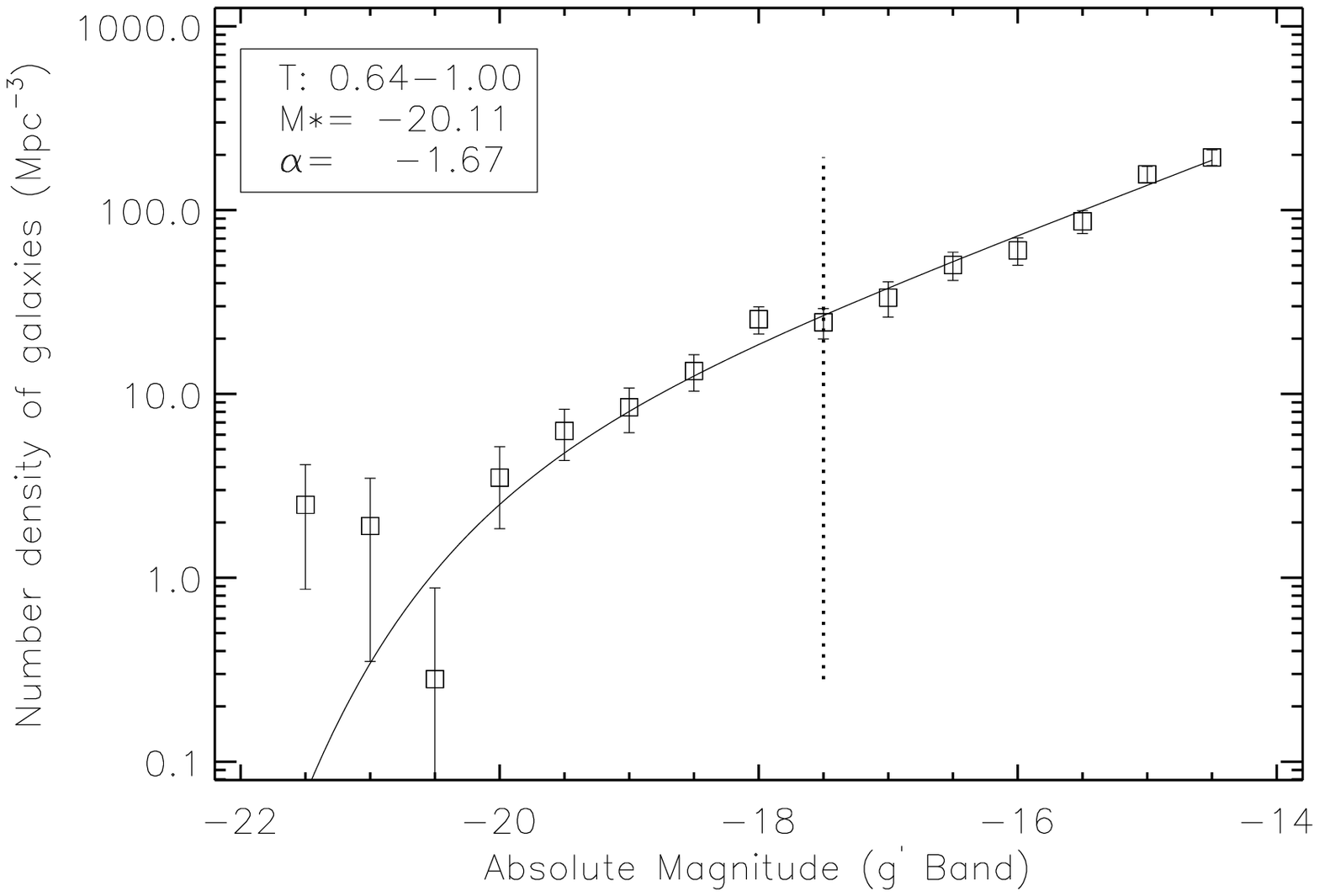,width=7.5cm,height=4.5cm}
\epsfig{file=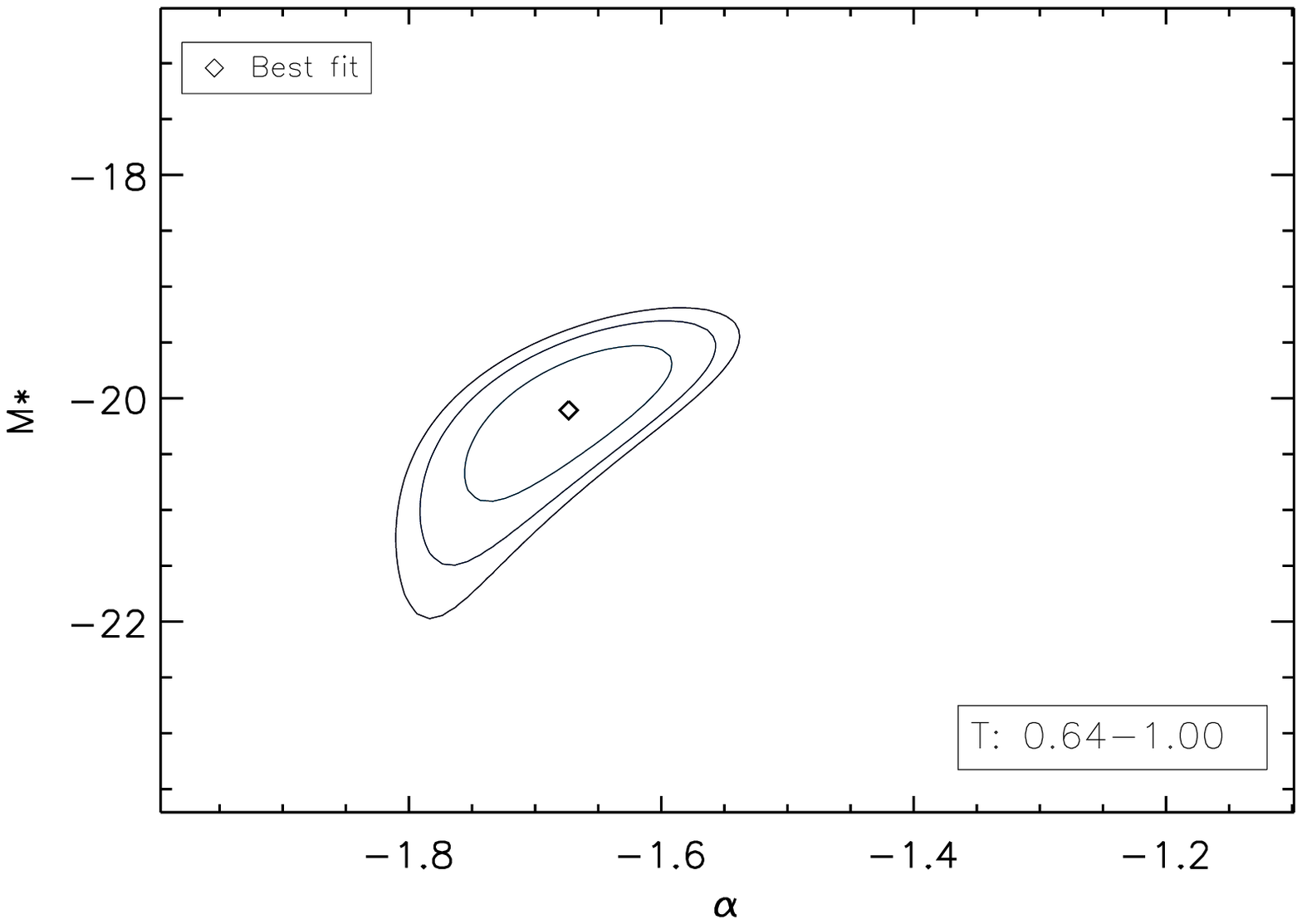,width=7.5cm,height=4.5cm}
\epsfig{file=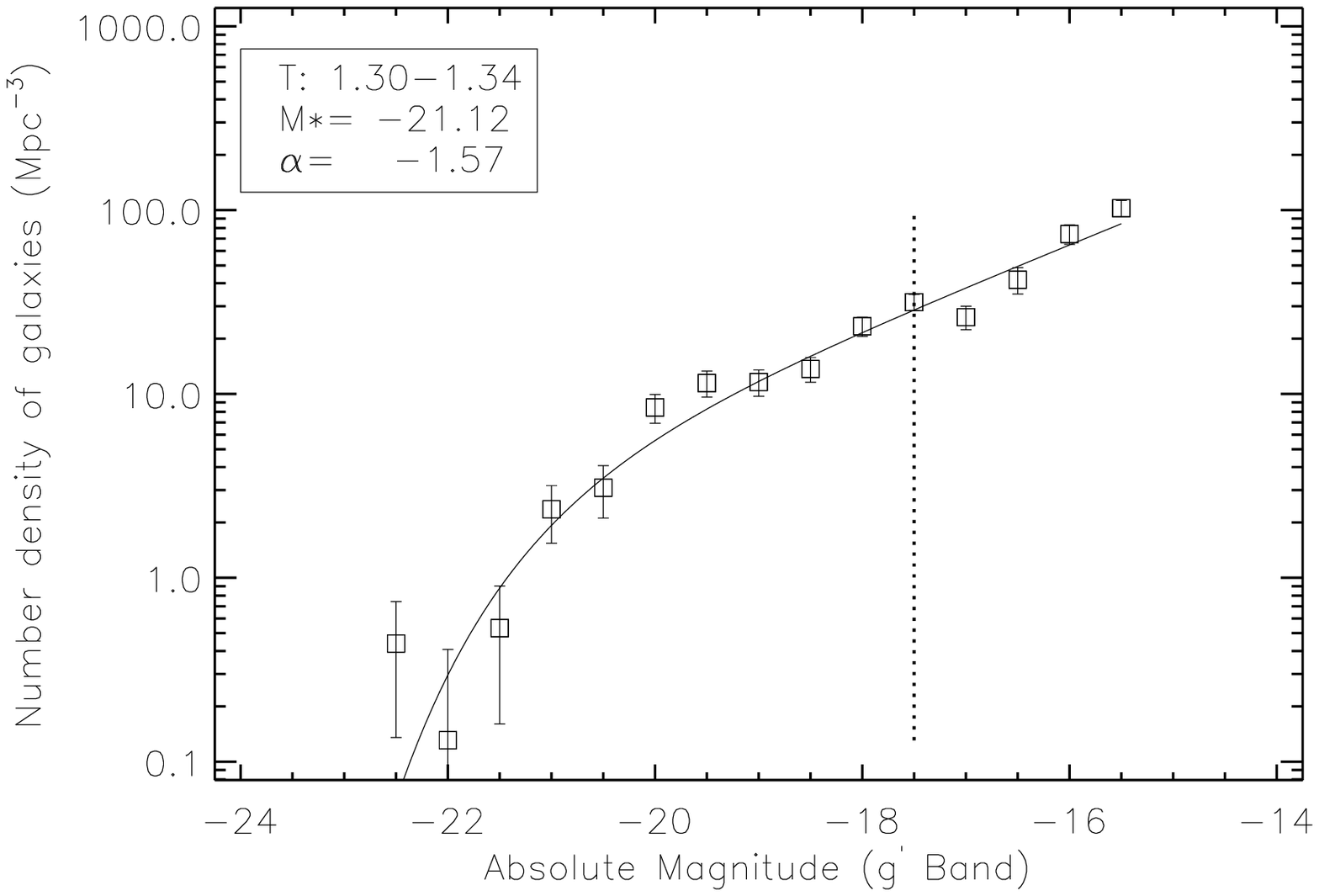,width=7.5cm,height=4.5cm}
\epsfig{file=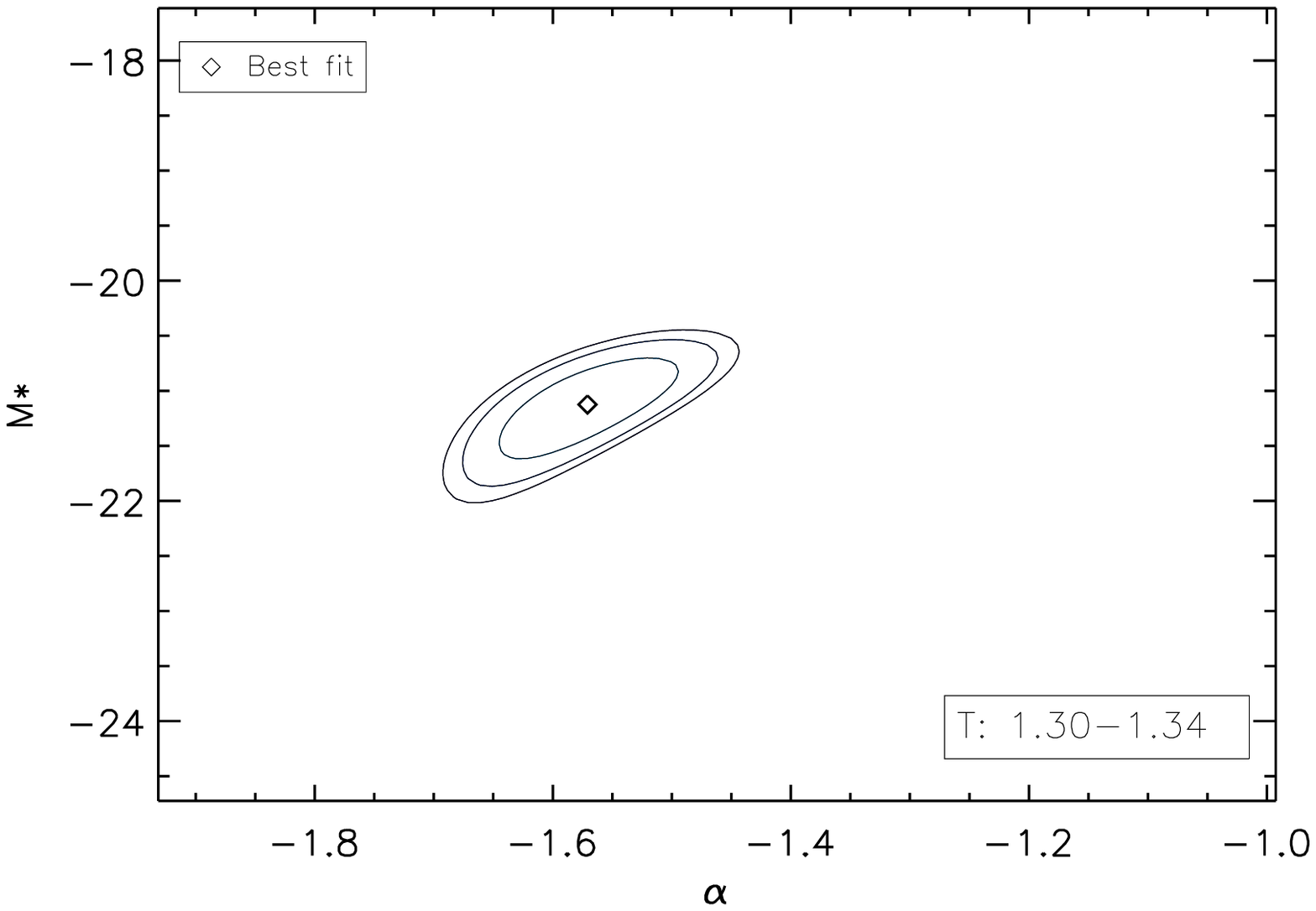,width=7.5cm,height=4.5cm}
\epsfig{file=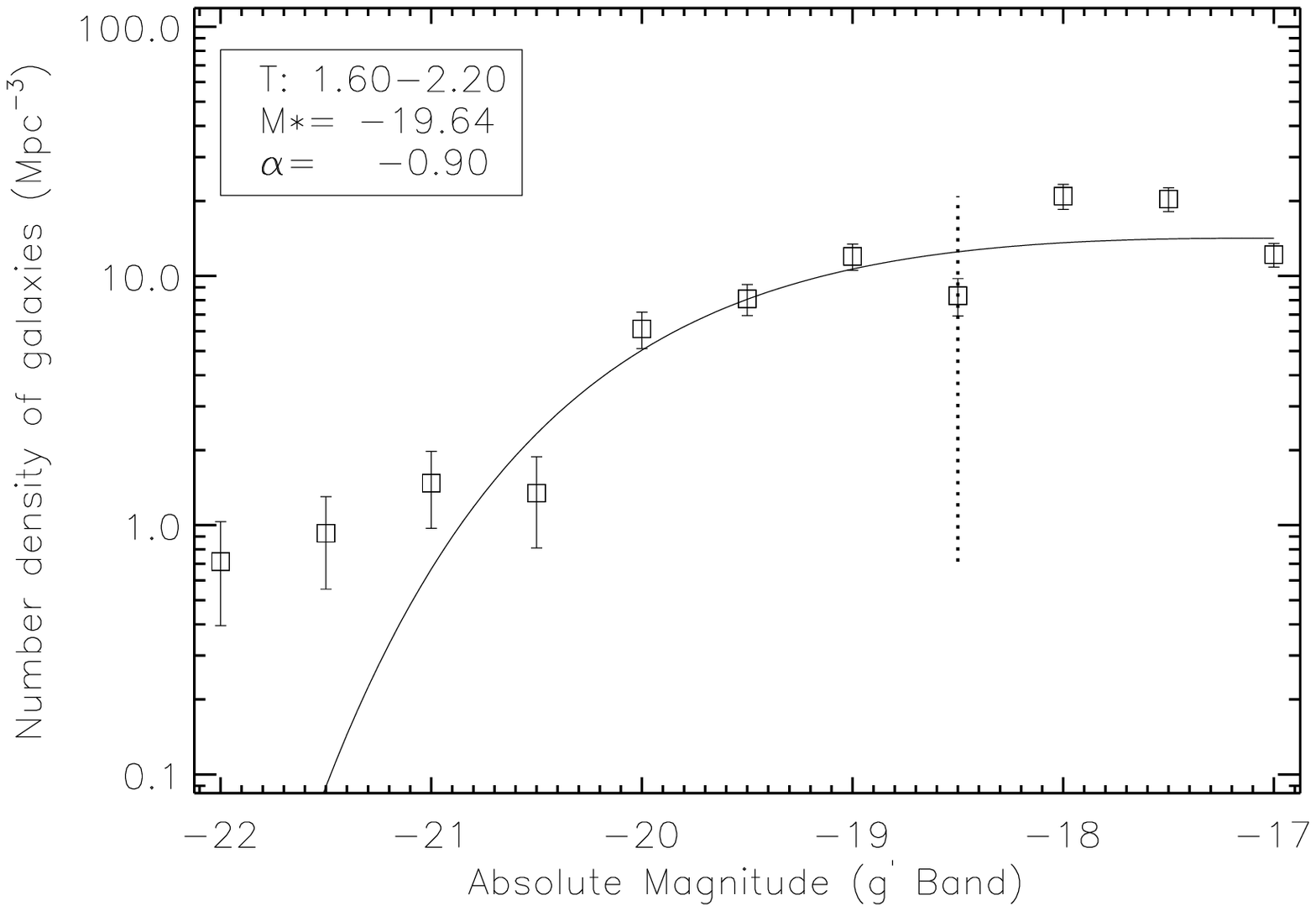,width=7.5cm,height=4.5cm}
\epsfig{file=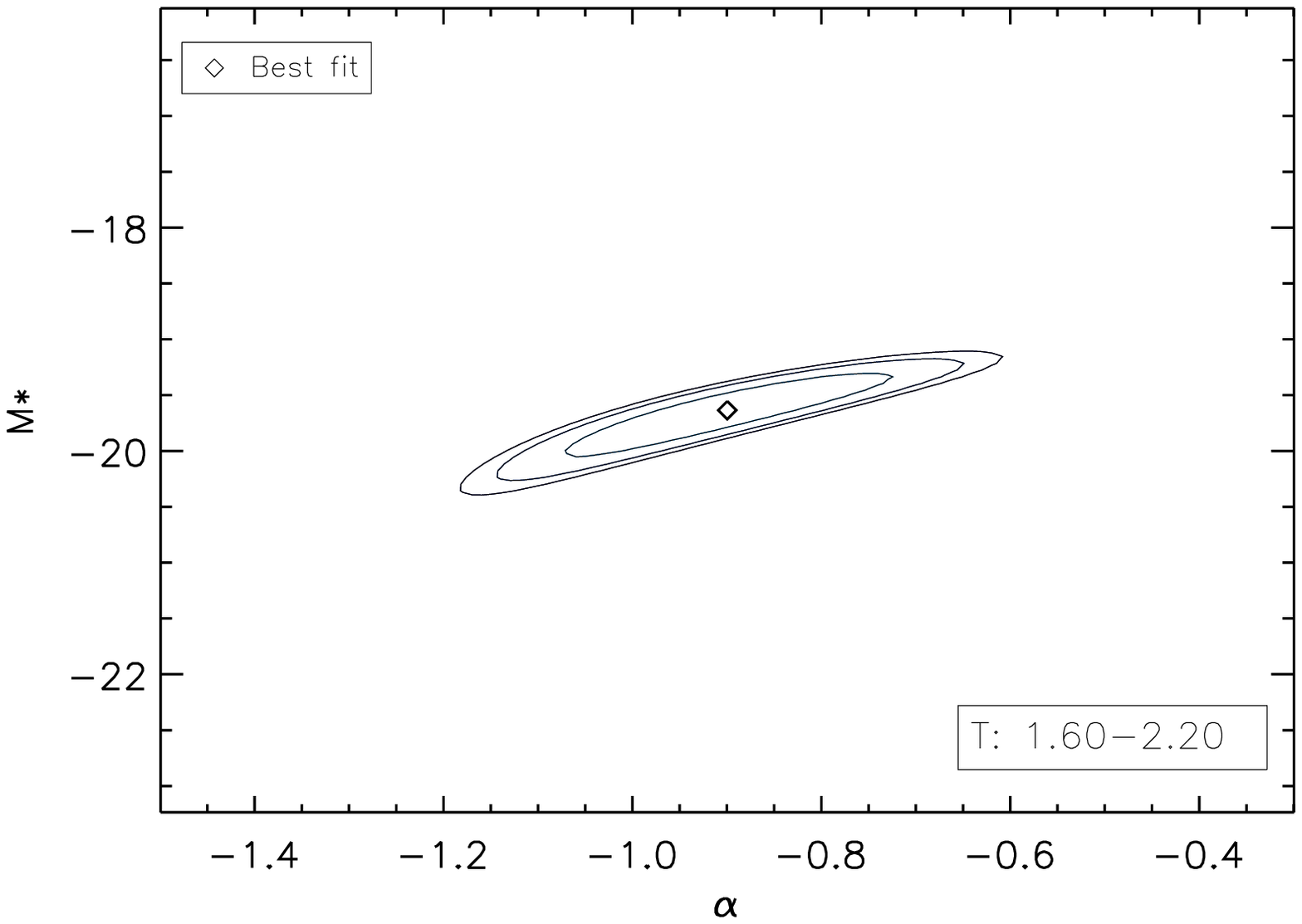,width=7.5cm,height=4.5cm}
\epsfig{file=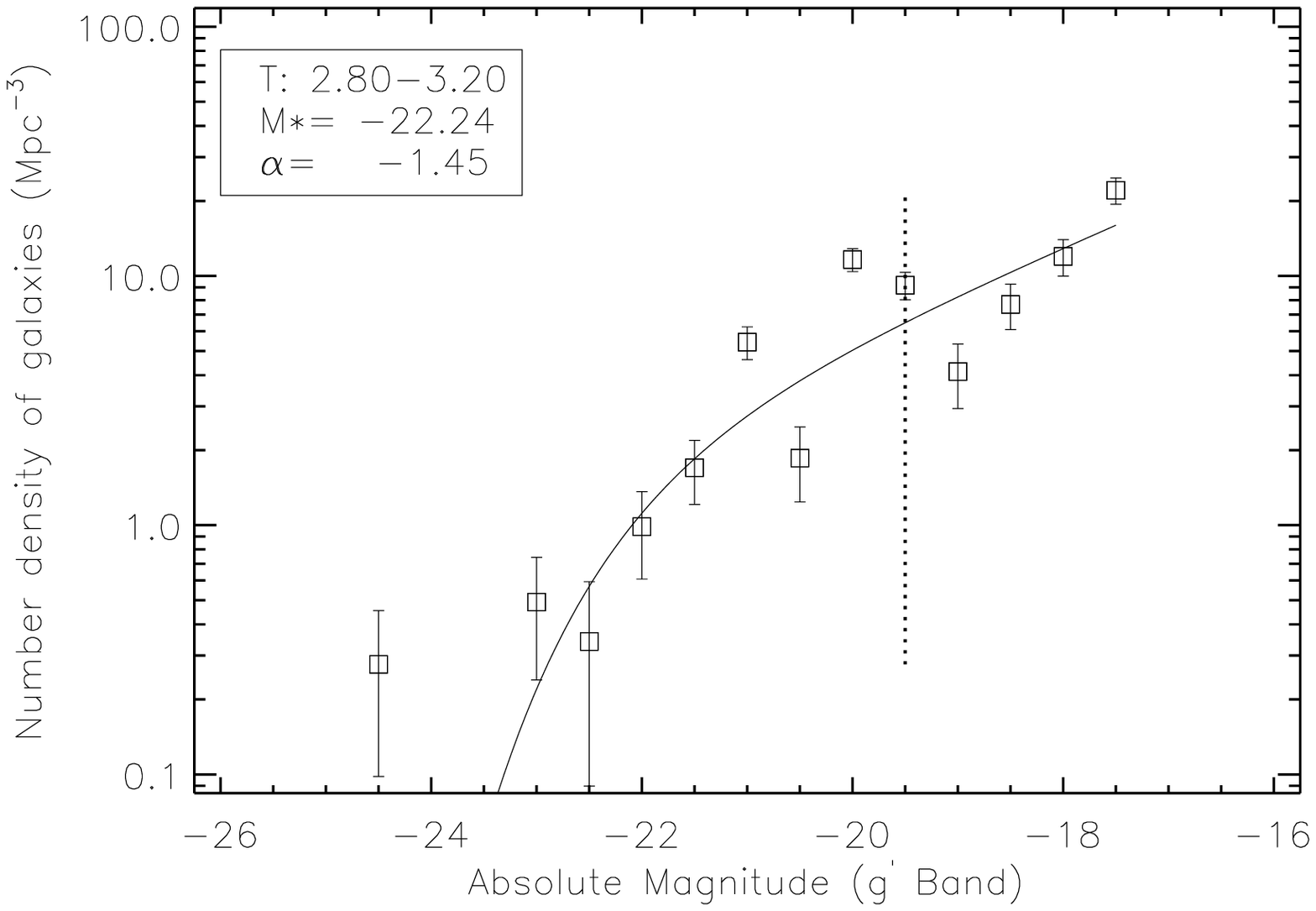,width=7.5cm,height=4.5cm}
\epsfig{file=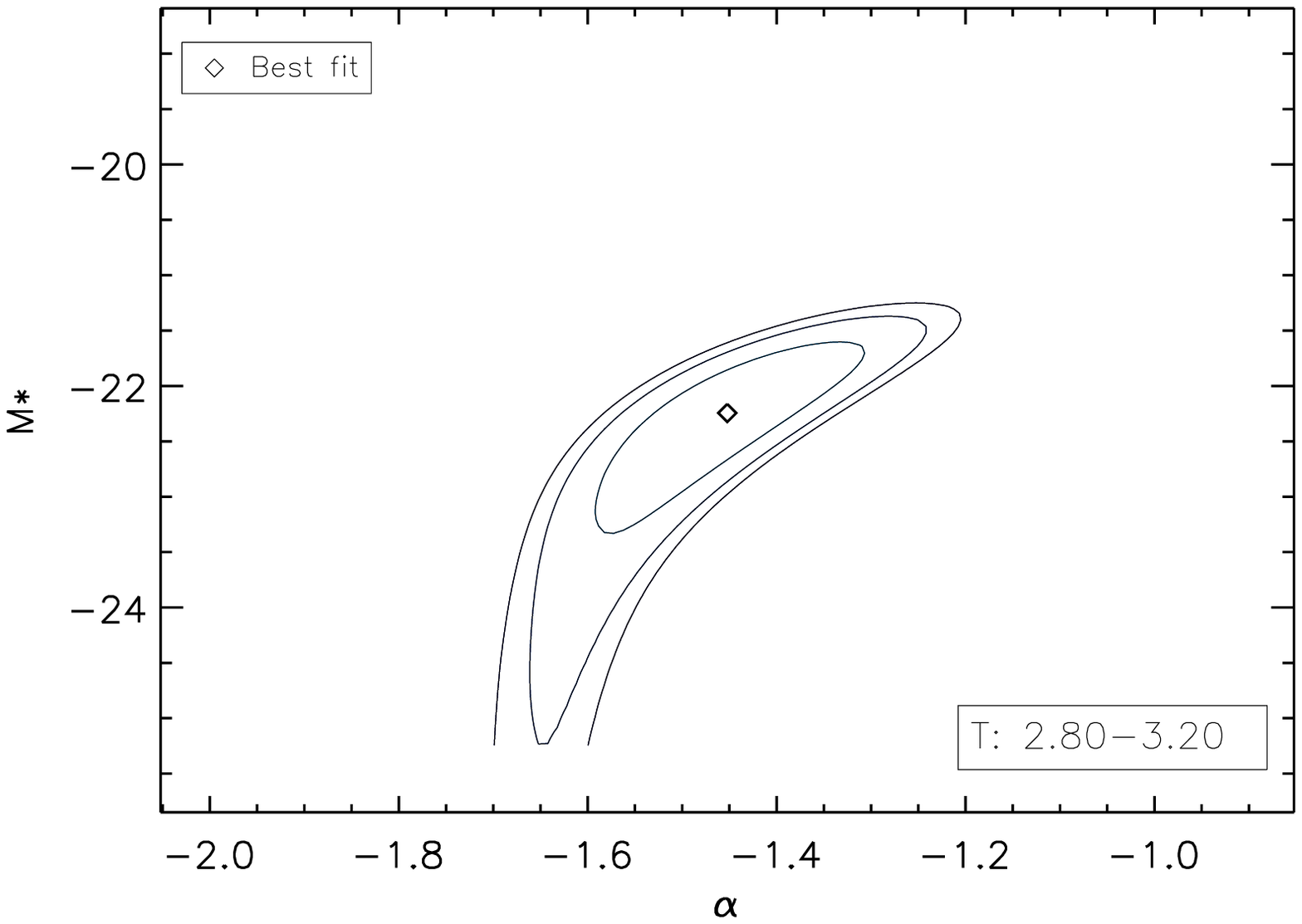,width=7.5cm,height=4.5cm}
\epsfig{file=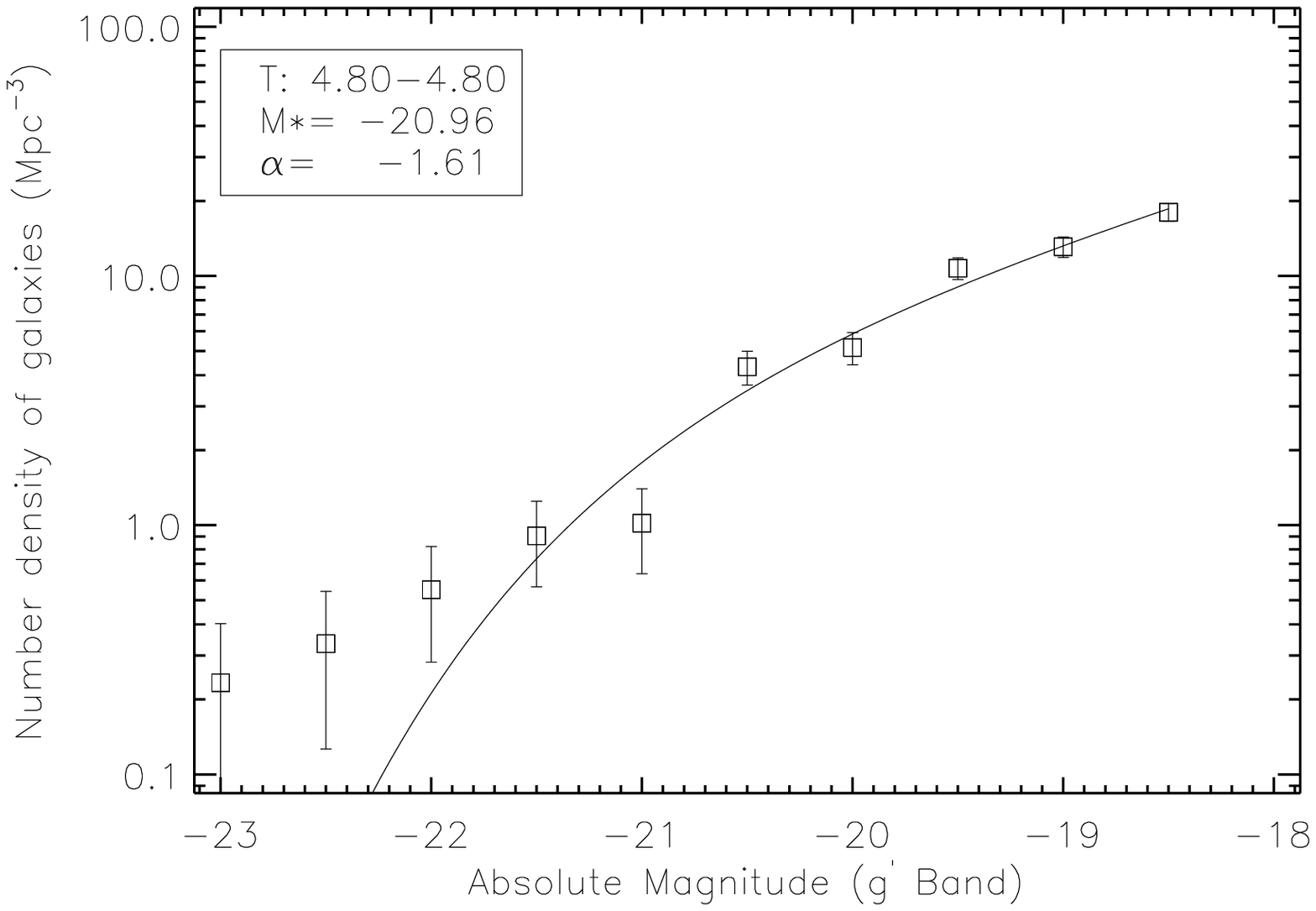,width=7.5cm,height=4.5cm}
\epsfig{file=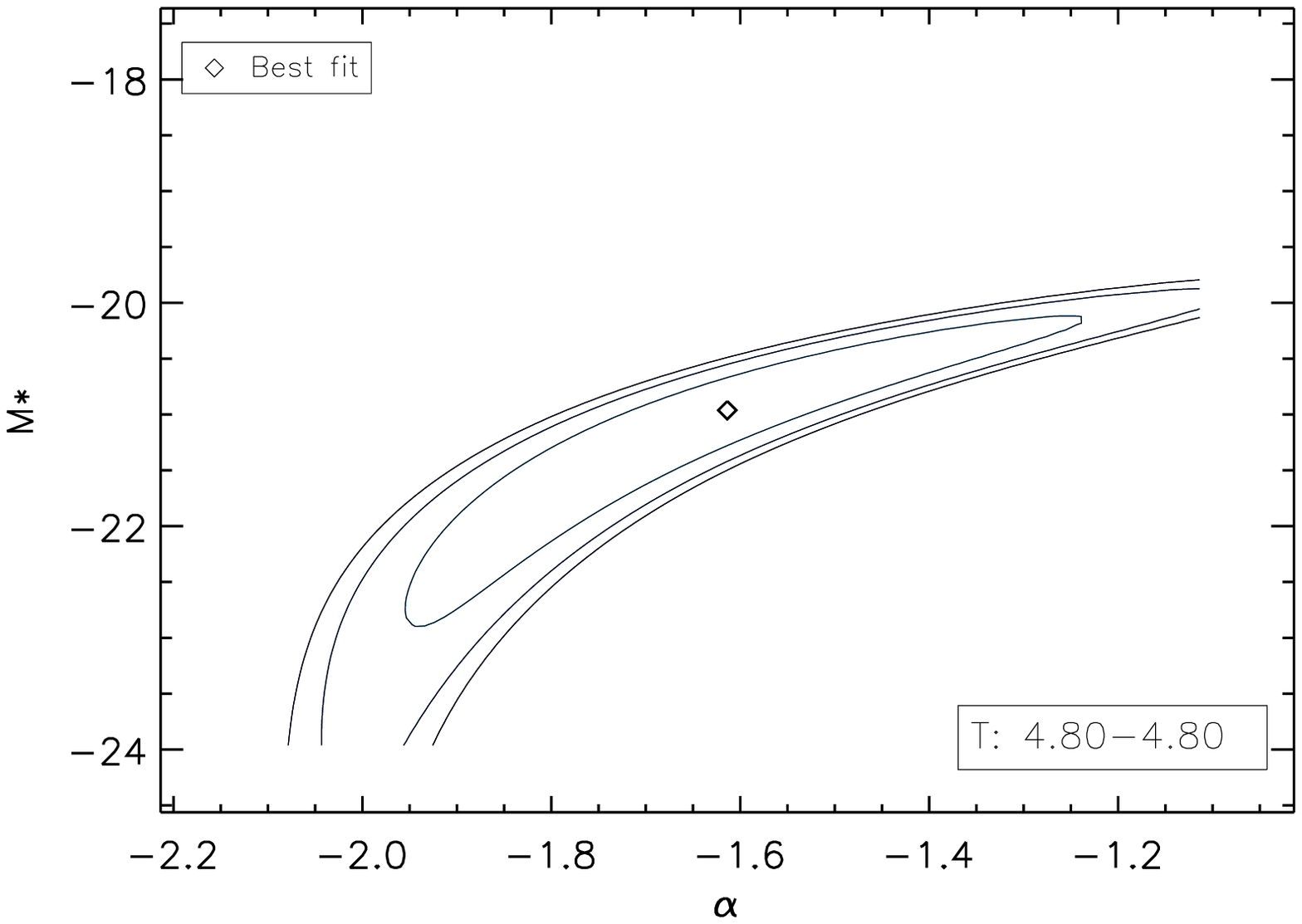,width=7.5cm,height=4.5cm}

\caption{LFs of the stacked clusters for 5 temperature ranges and the associated $1\sigma$, $2\sigma$ and $3\sigma$ contours for the $g^\prime$ band. The vertical dotted line is at the faintest common magnitude value of all stacked clusters.}
\label{t_gband_plot}
\end{figure}


\onecolumn
\begin{figure}
\center

\epsfig{file=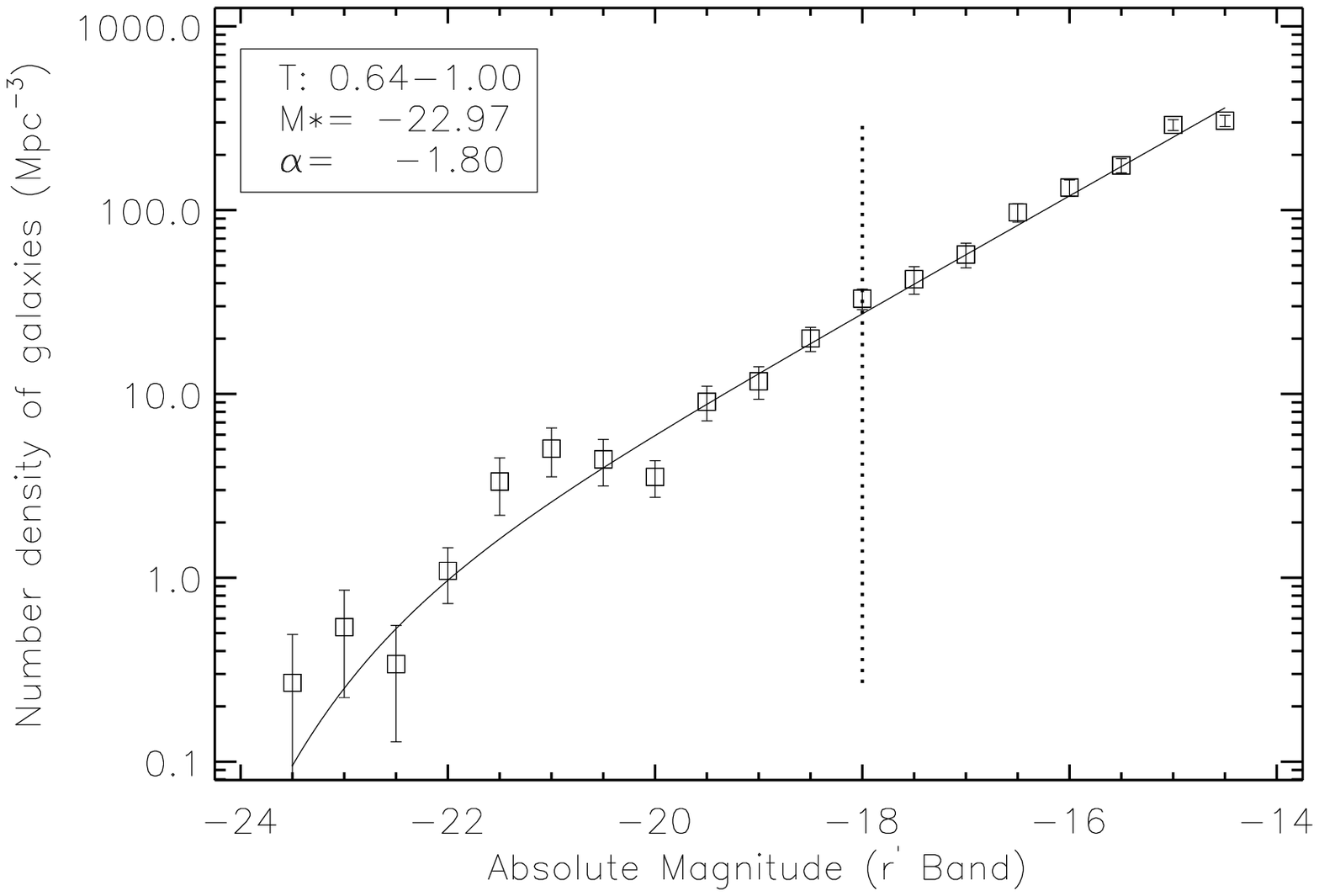,width=7.5cm,height=4.5cm}
\epsfig{file=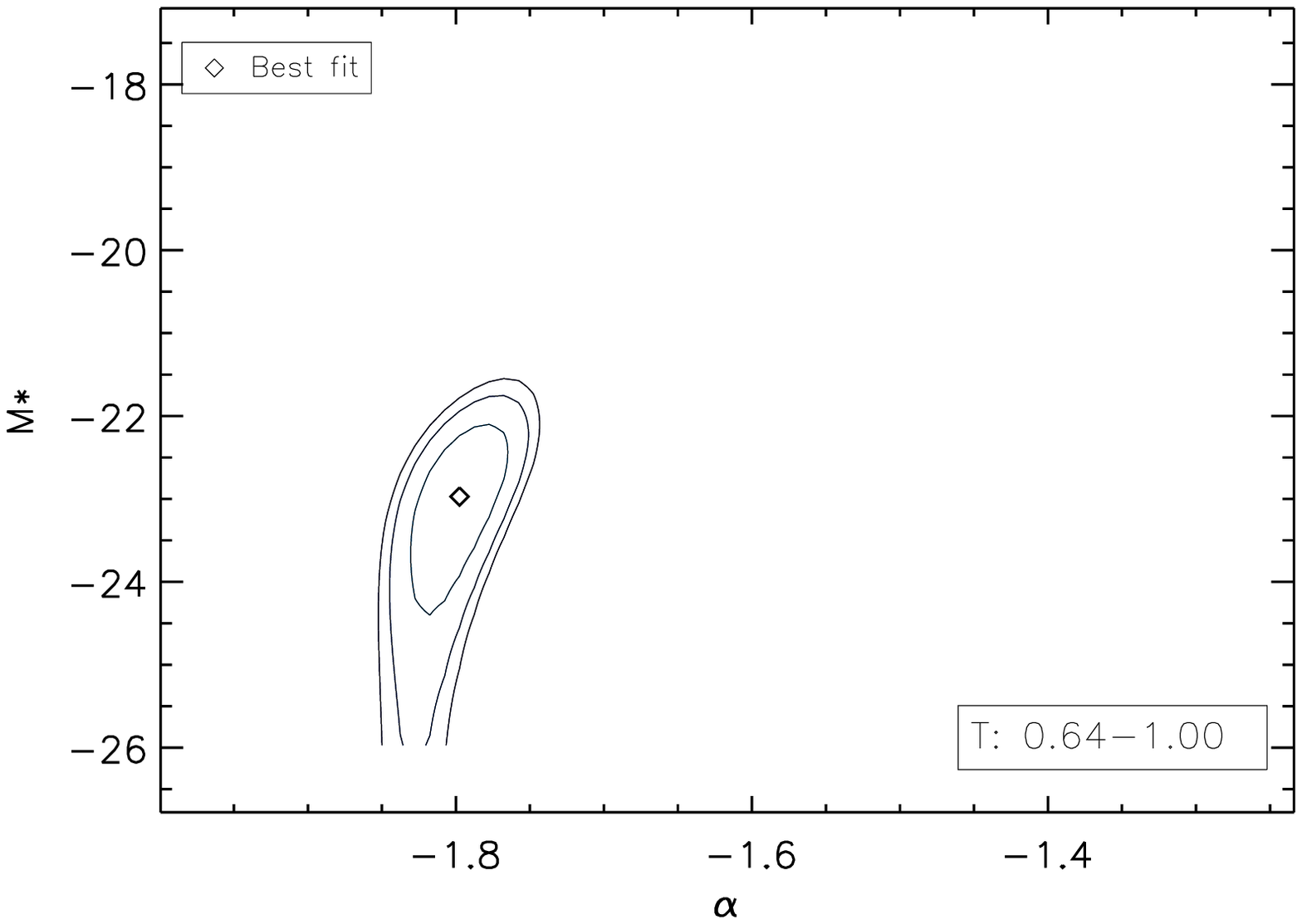,width=7.5cm,height=4.5cm}
\epsfig{file=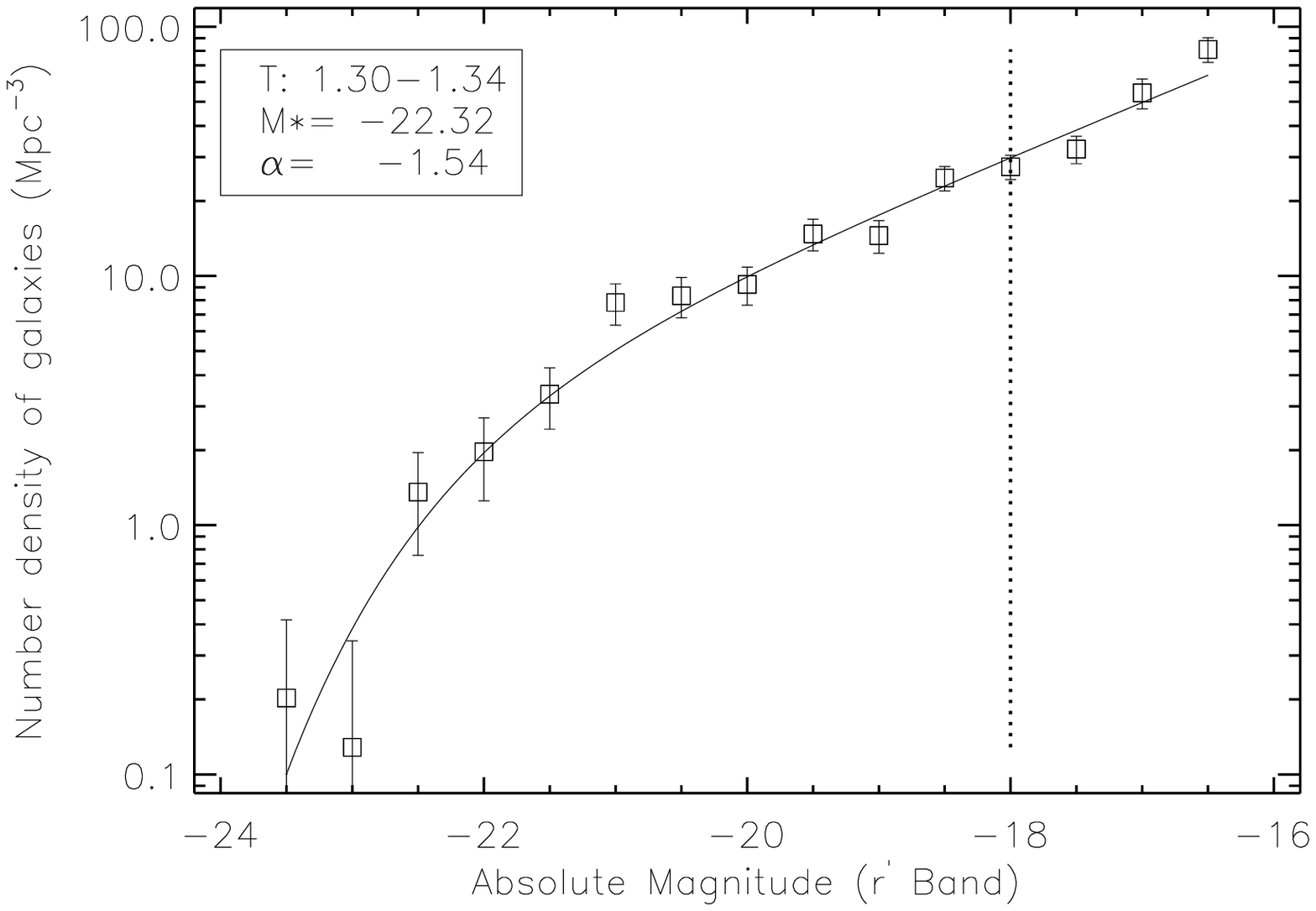,width=7.5cm,height=4.5cm}
\epsfig{file=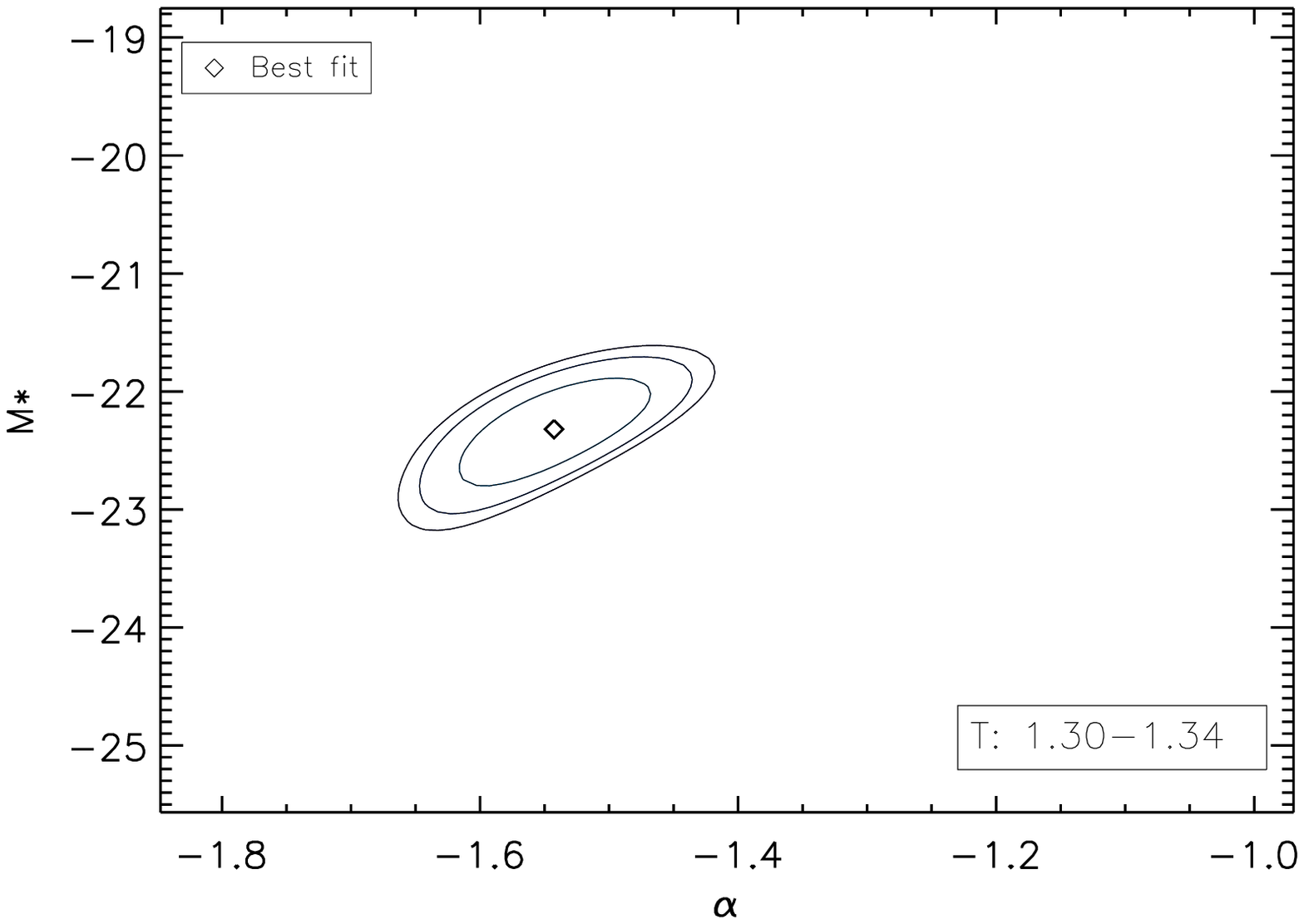,width=7.5cm,height=4.5cm}
\epsfig{file=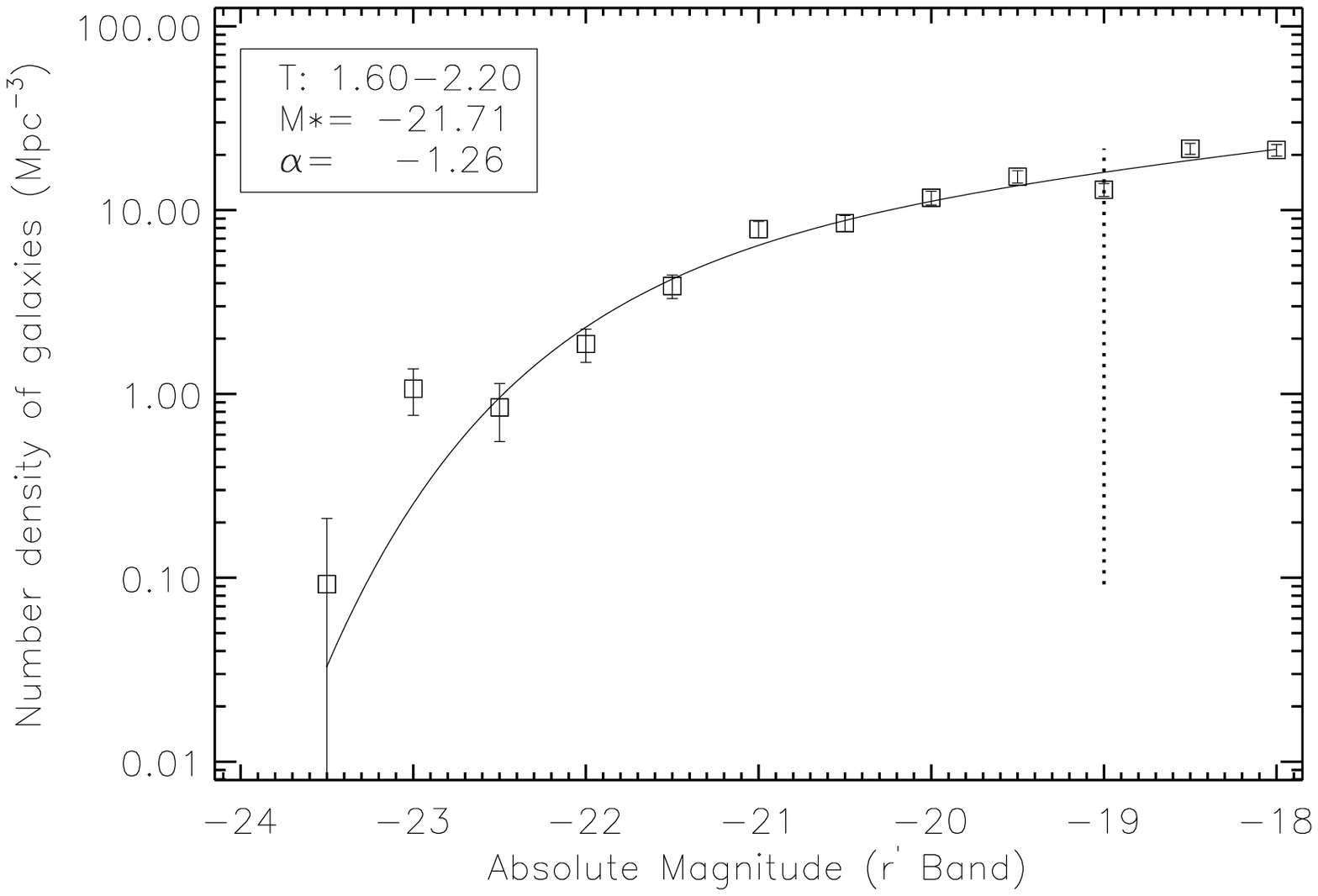,width=7.5cm,height=4.5cm}
\epsfig{file=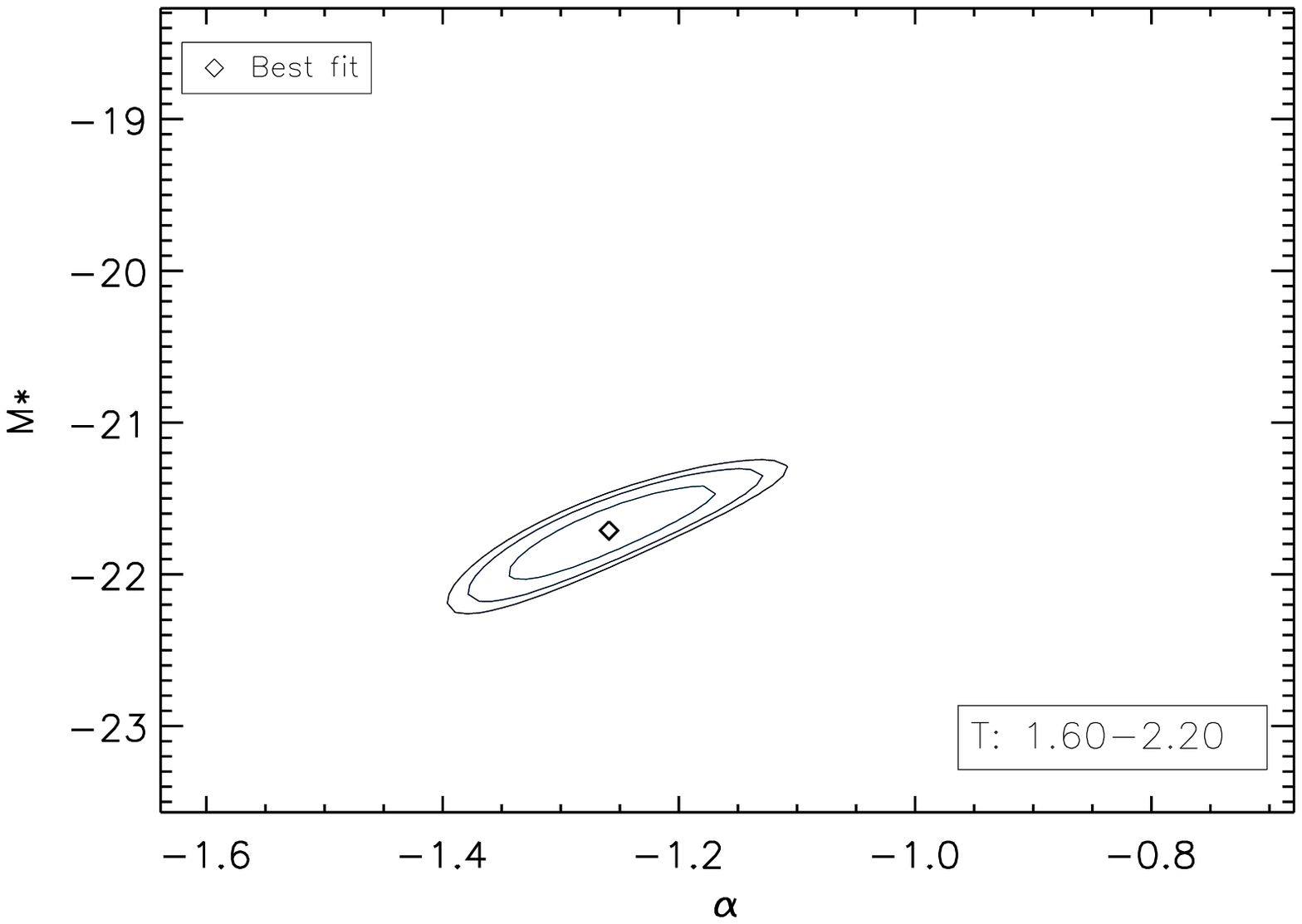,width=7.5cm,height=4.5cm}
\epsfig{file=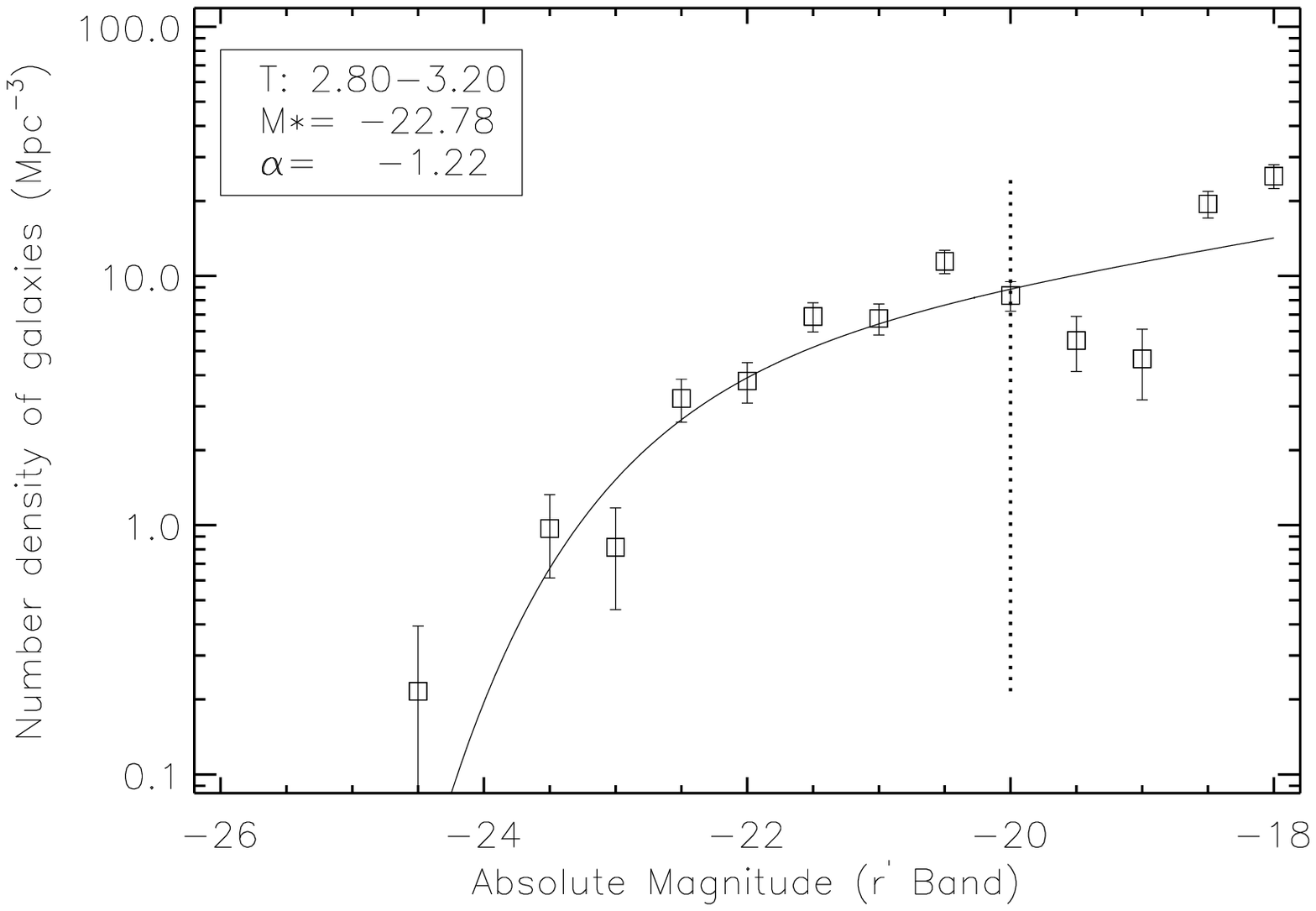,width=7.5cm,height=4.5cm}
\epsfig{file=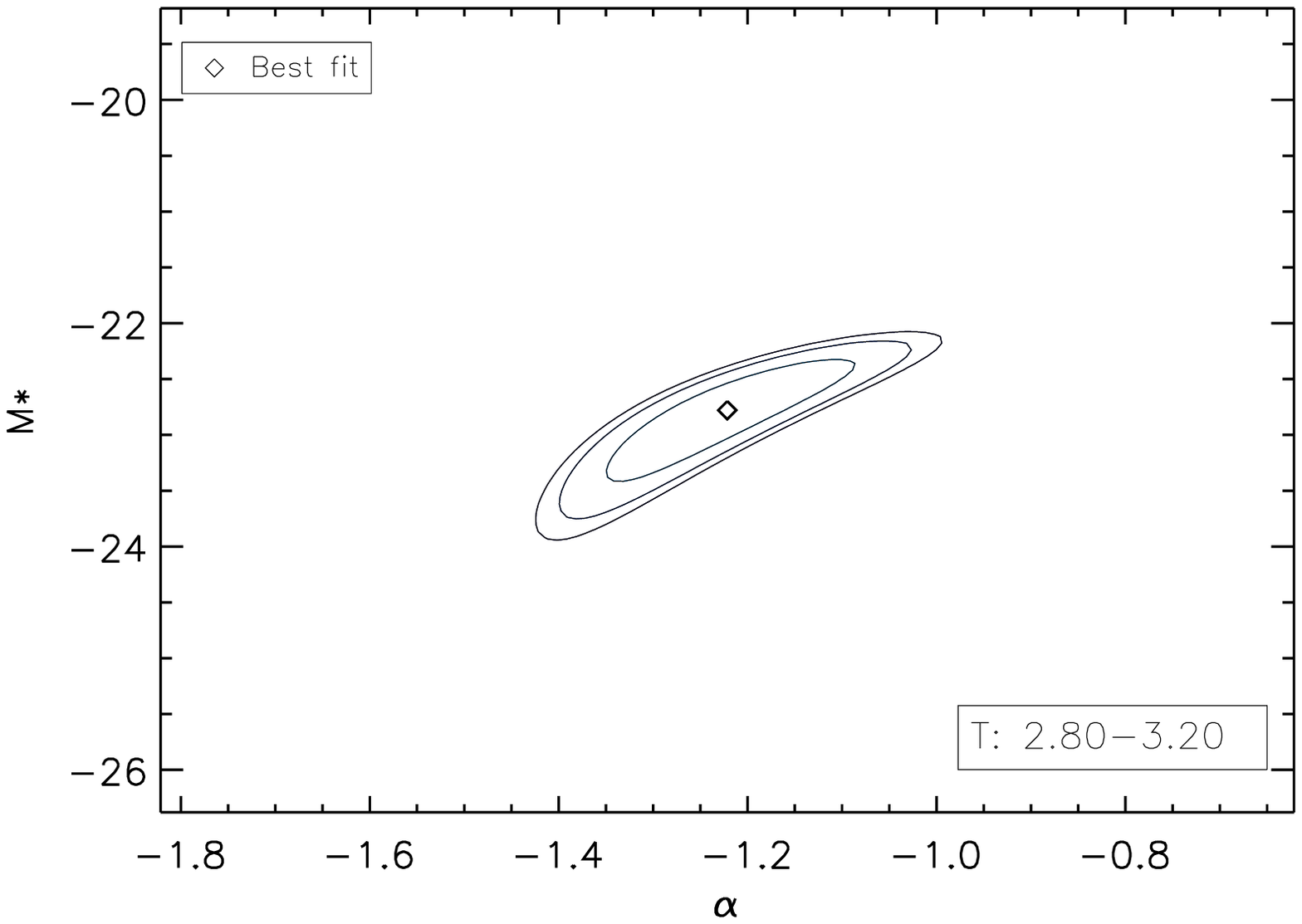,width=7.5cm,height=4.5cm}
\epsfig{file=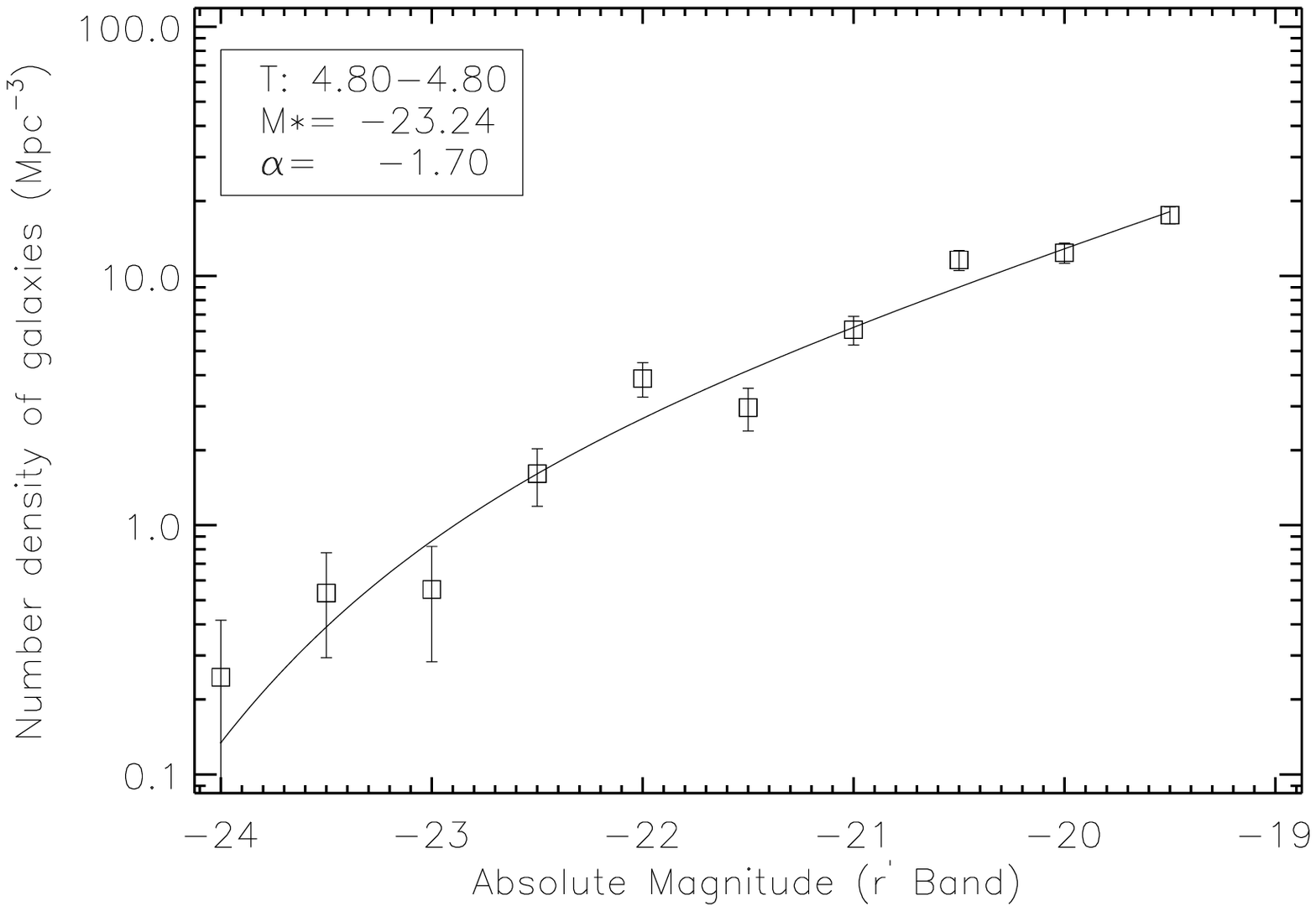,width=7.5cm,height=4.5cm}
\epsfig{file=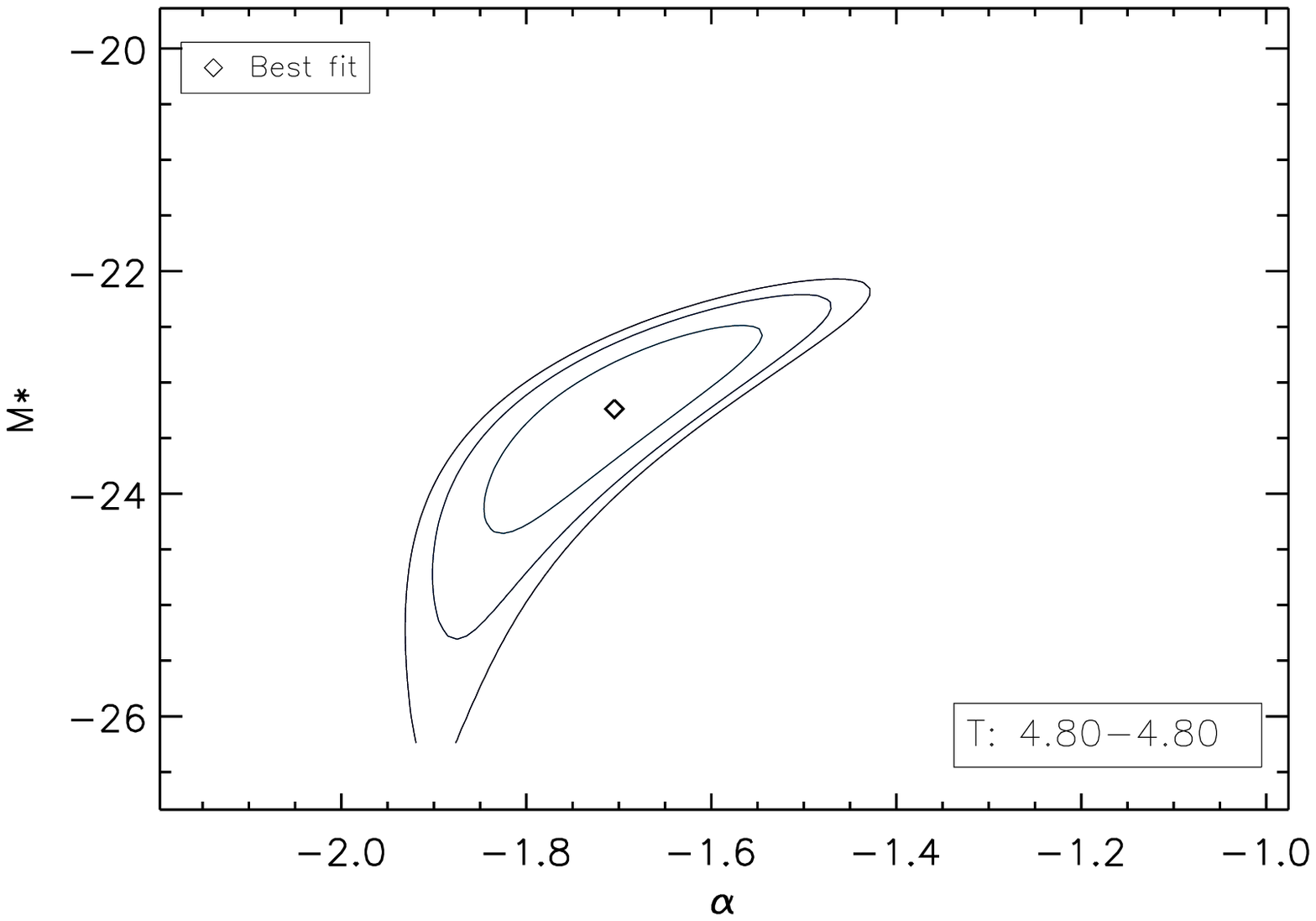,width=7.5cm,height=4.5cm}

\caption{LFs of the stacked clusters for 5 temperature ranges and the associated $1\sigma$, $2\sigma$ and $3\sigma$ contours for the $r^\prime$ band. The vertical dotted line is at the faintest common magnitude value of all stacked clusters.}
\label{t_rband_plot}
\end{figure}


\onecolumn
\begin{figure}
\center

\epsfig{file=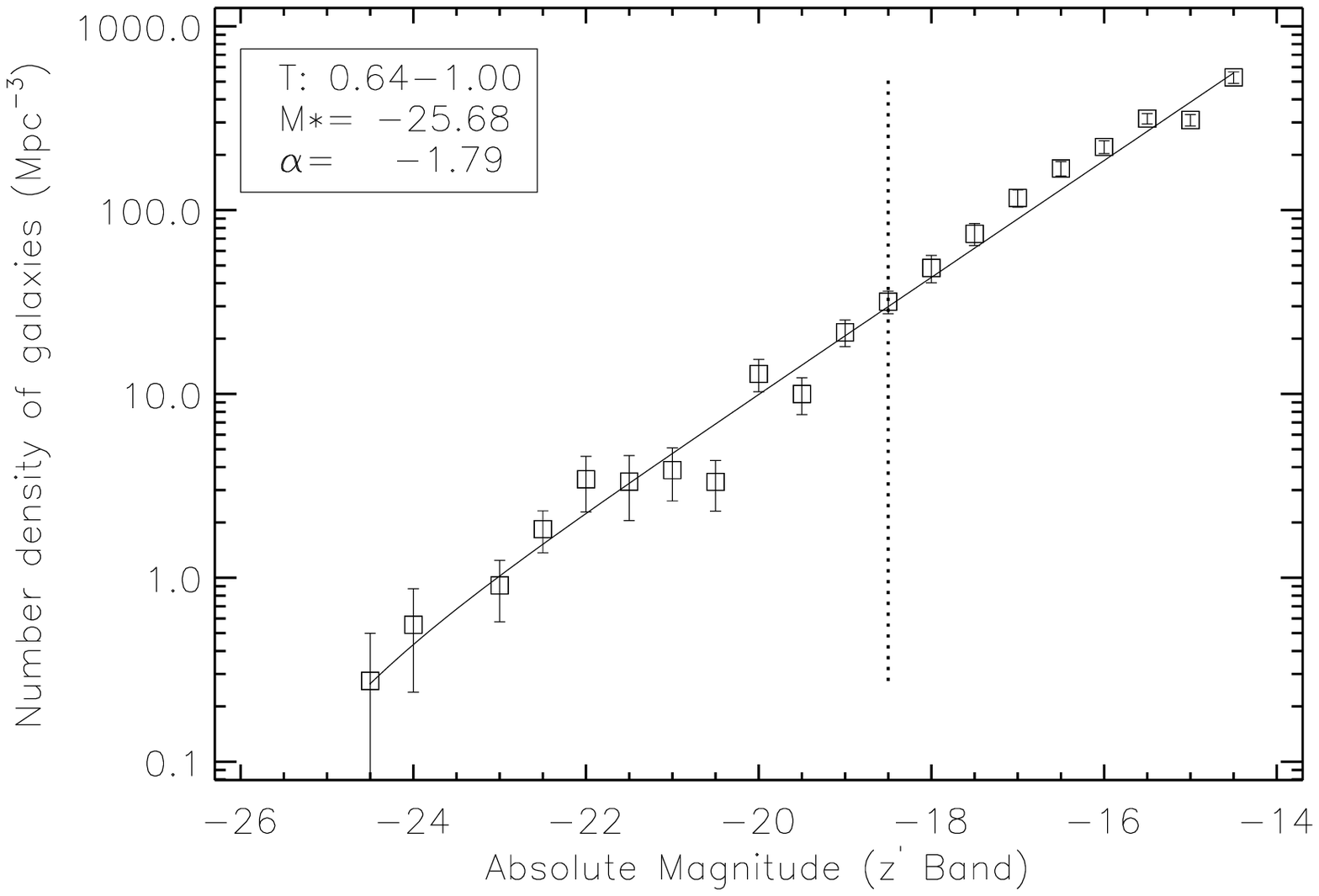,width=7.5cm,height=4.5cm}
\epsfig{file=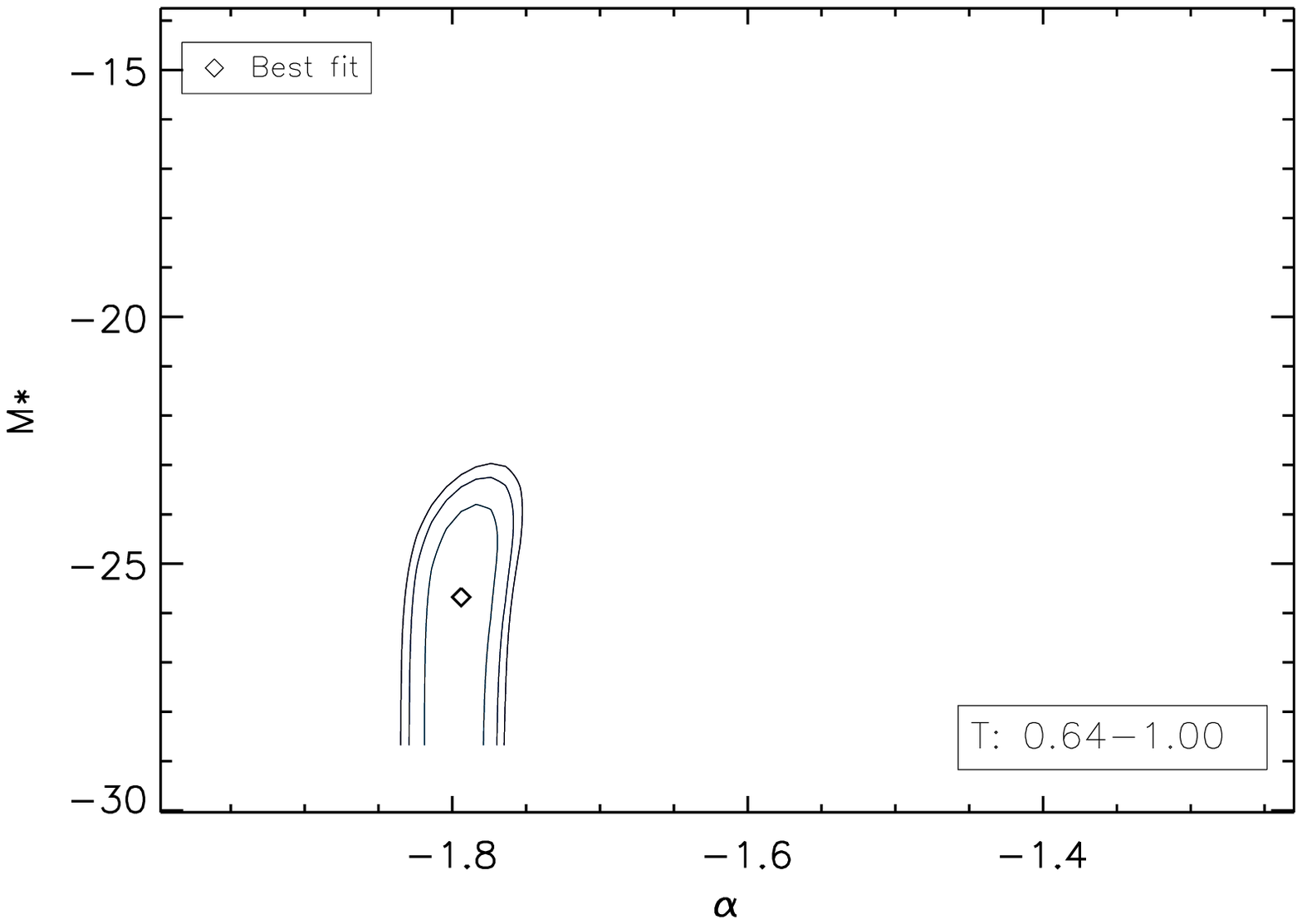,width=7.5cm,height=4.5cm}
\epsfig{file=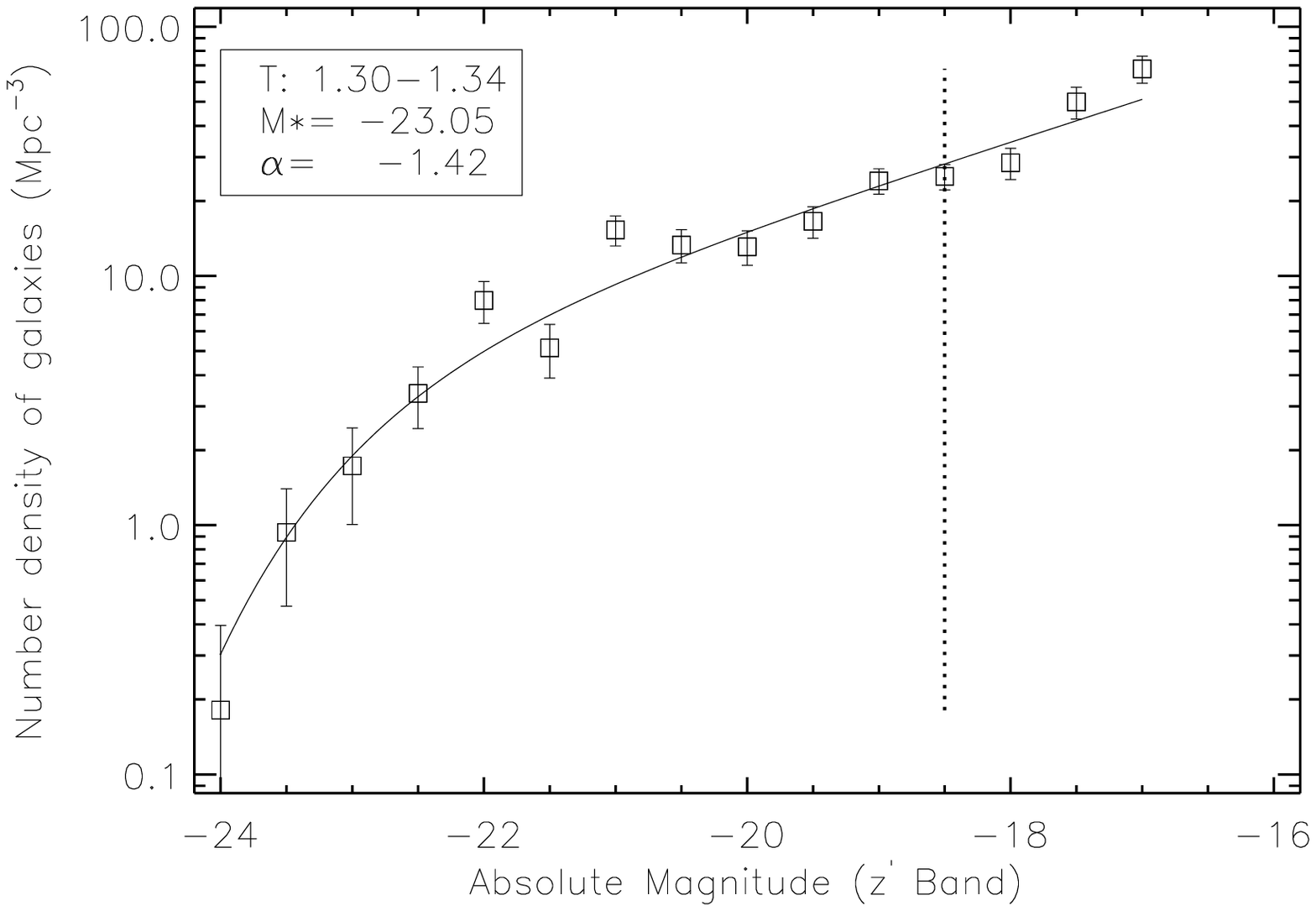,width=7.5cm,height=4.5cm}
\epsfig{file=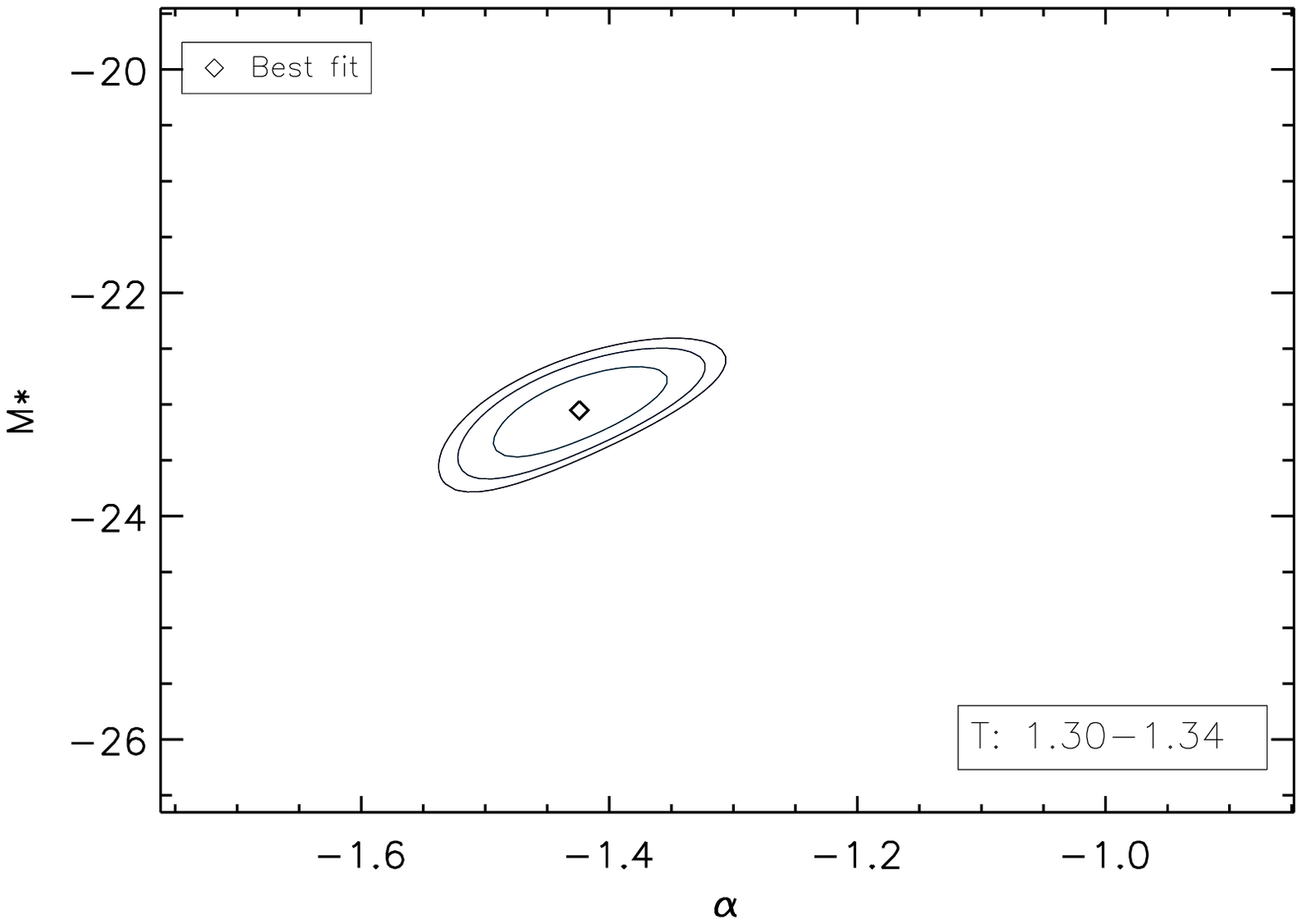,width=7.5cm,height=4.5cm}
\epsfig{file=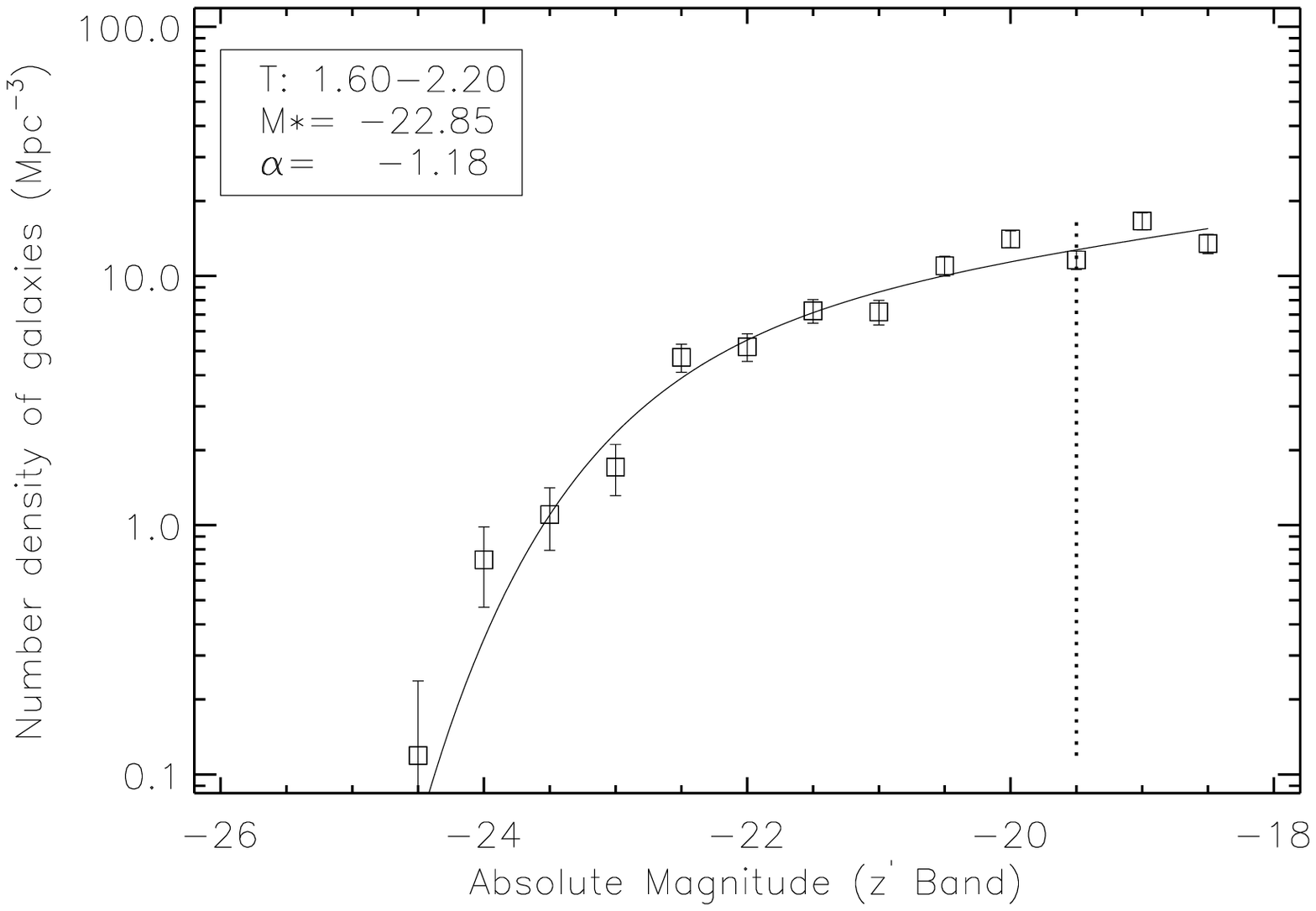,width=7.5cm,height=4.5cm}
\epsfig{file=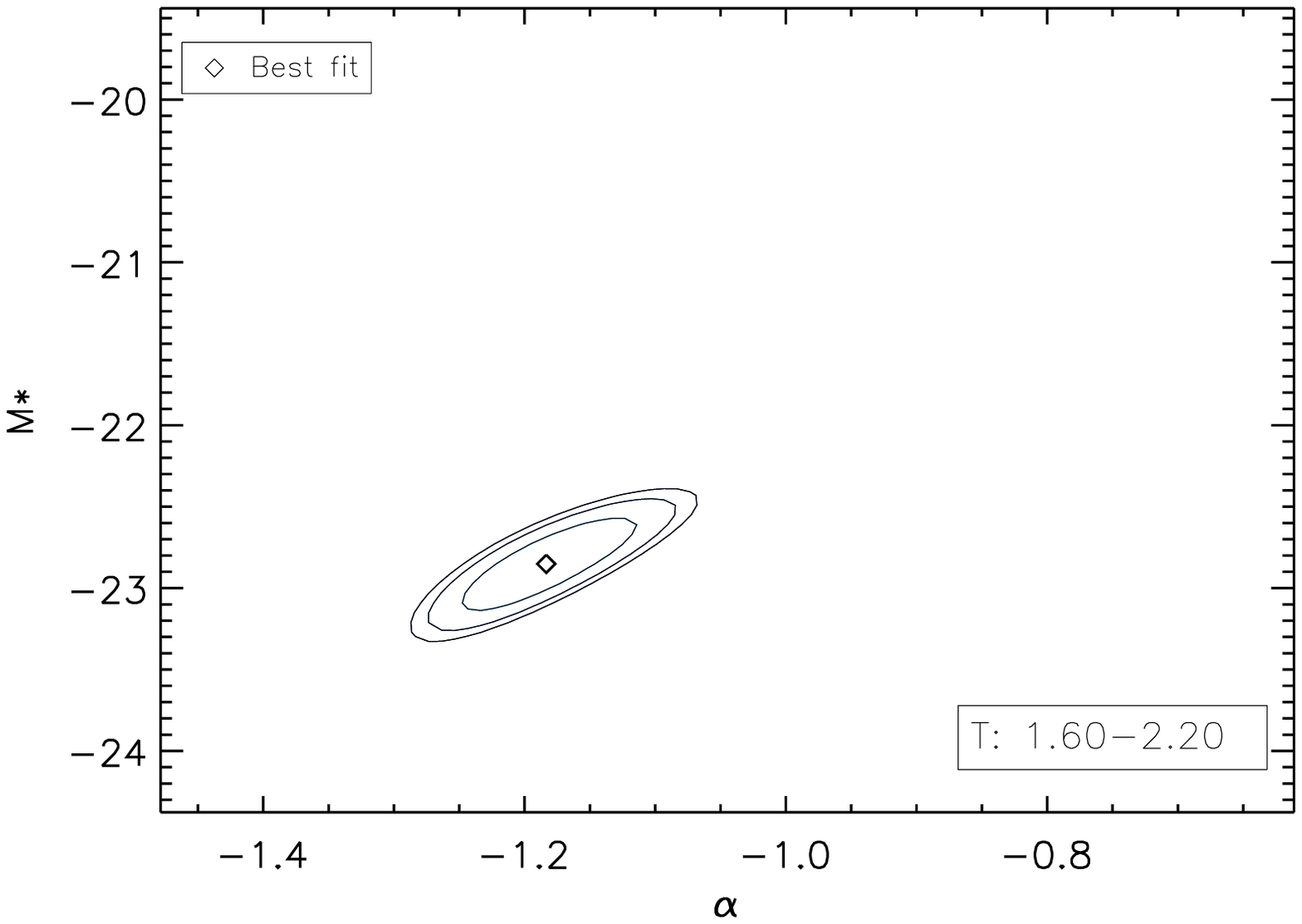,width=7.5cm,height=4.5cm}
\epsfig{file=z0.29-0.61t2.80-3.20VolLF_z.ps,width=7.5cm,height=4.5cm}
\epsfig{file=z0.29-0.61t2.80-3.20contour_z.ps,width=7.5cm,height=4.5cm}
\epsfig{file=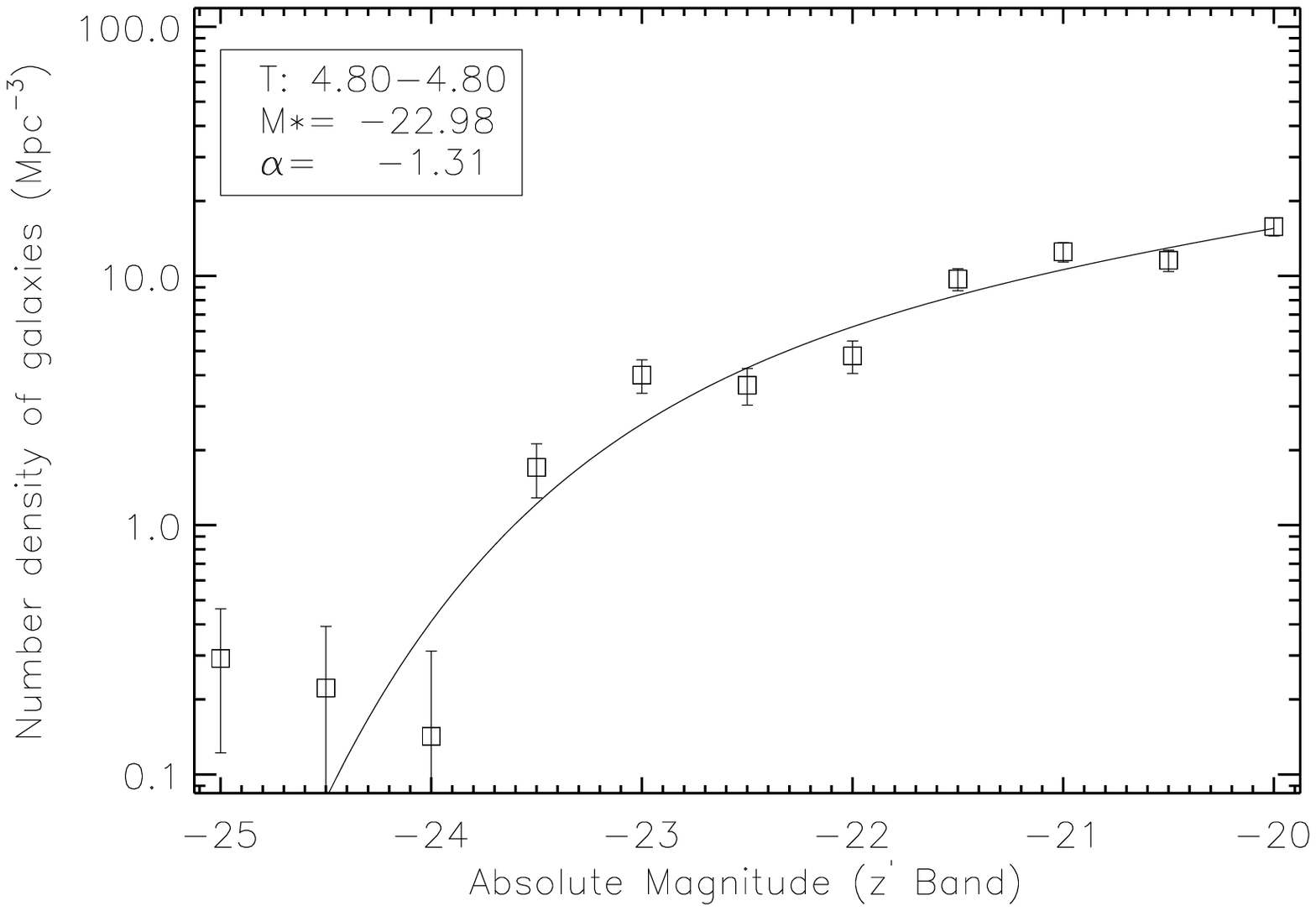,width=7.5cm,height=4.5cm}
\epsfig{file=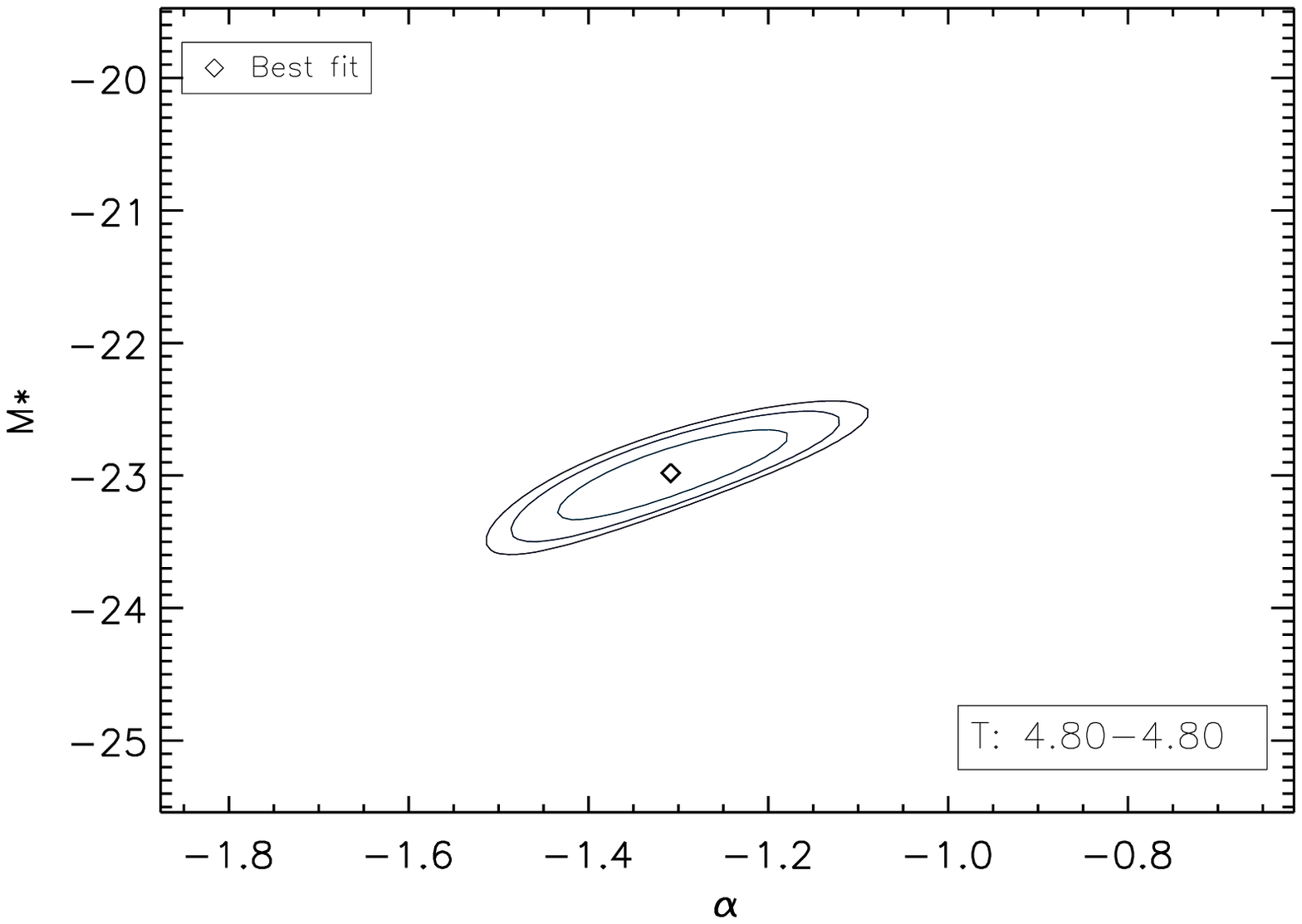,width=7.5cm,height=4.5cm}

\caption{LFs of the stacked clusters for 5 temperature ranges and the associated $1\sigma$, $2\sigma$ and $3\sigma$ contours for the $z^\prime$ zand. The vertical dotted line is at the faintest common magnitude value of all stacked clusters.}
\label{t_zband_plot}
\end{figure}

\twocolumn

\onecolumn
\appendix

\begin{center}
\section{Individual luminosity functions of C1 clusters in $\lowercase{r}^\prime$ band}
\end{center}

\begin{figure}

\center

\epsfig{file=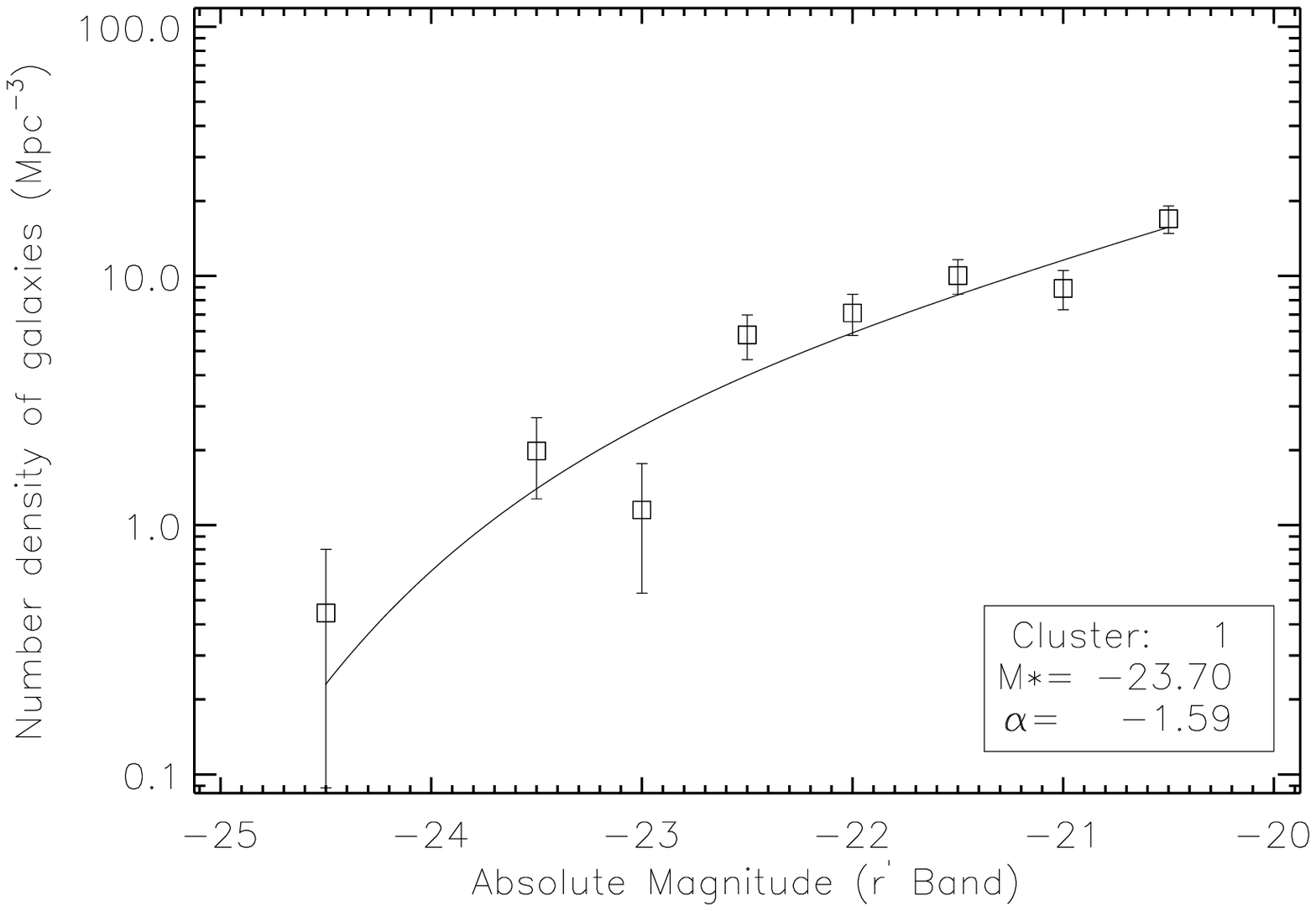,width=4cm,height=3.cm}
\epsfig{file=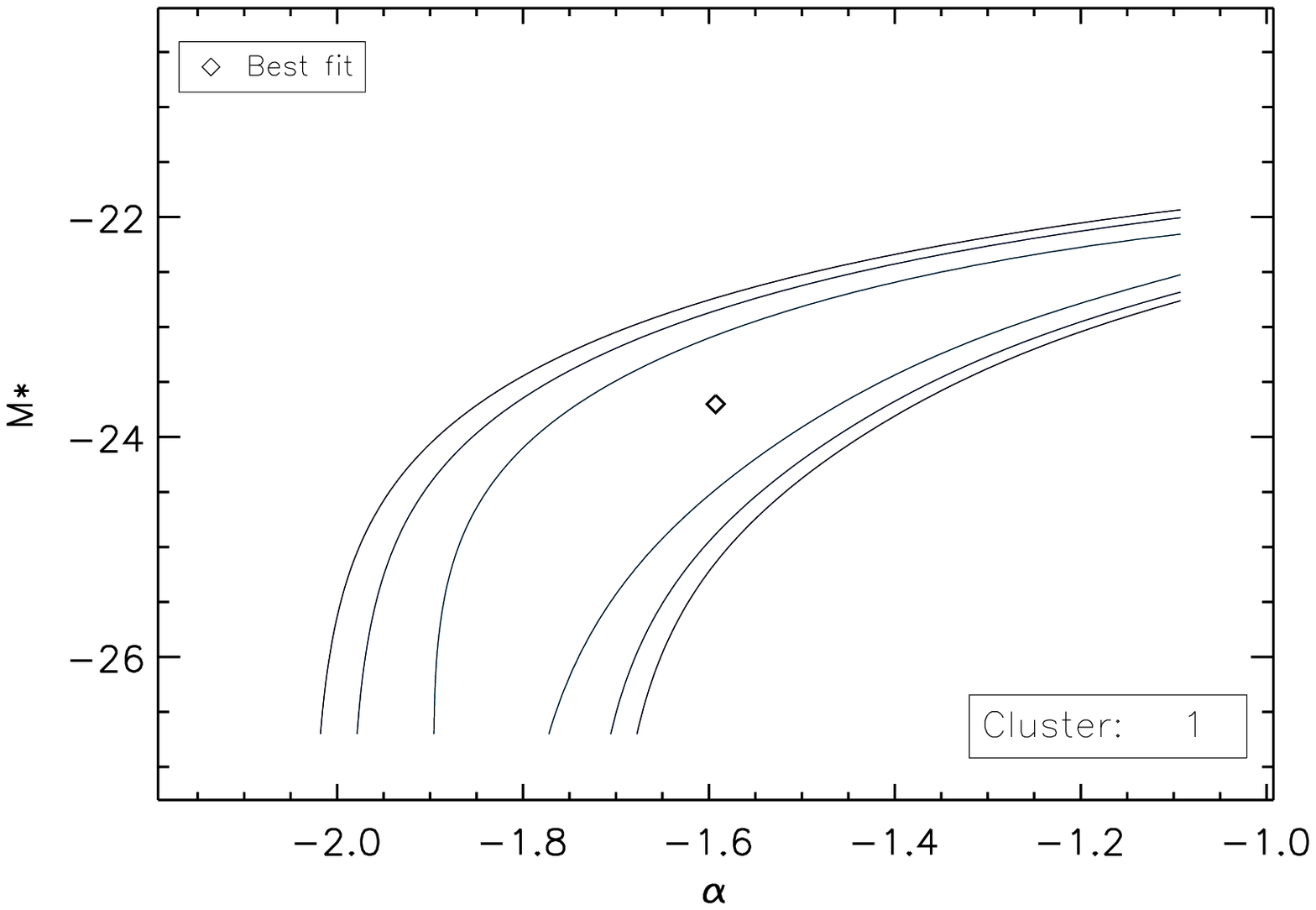,width=4cm,height=3.cm}
\epsfig{file=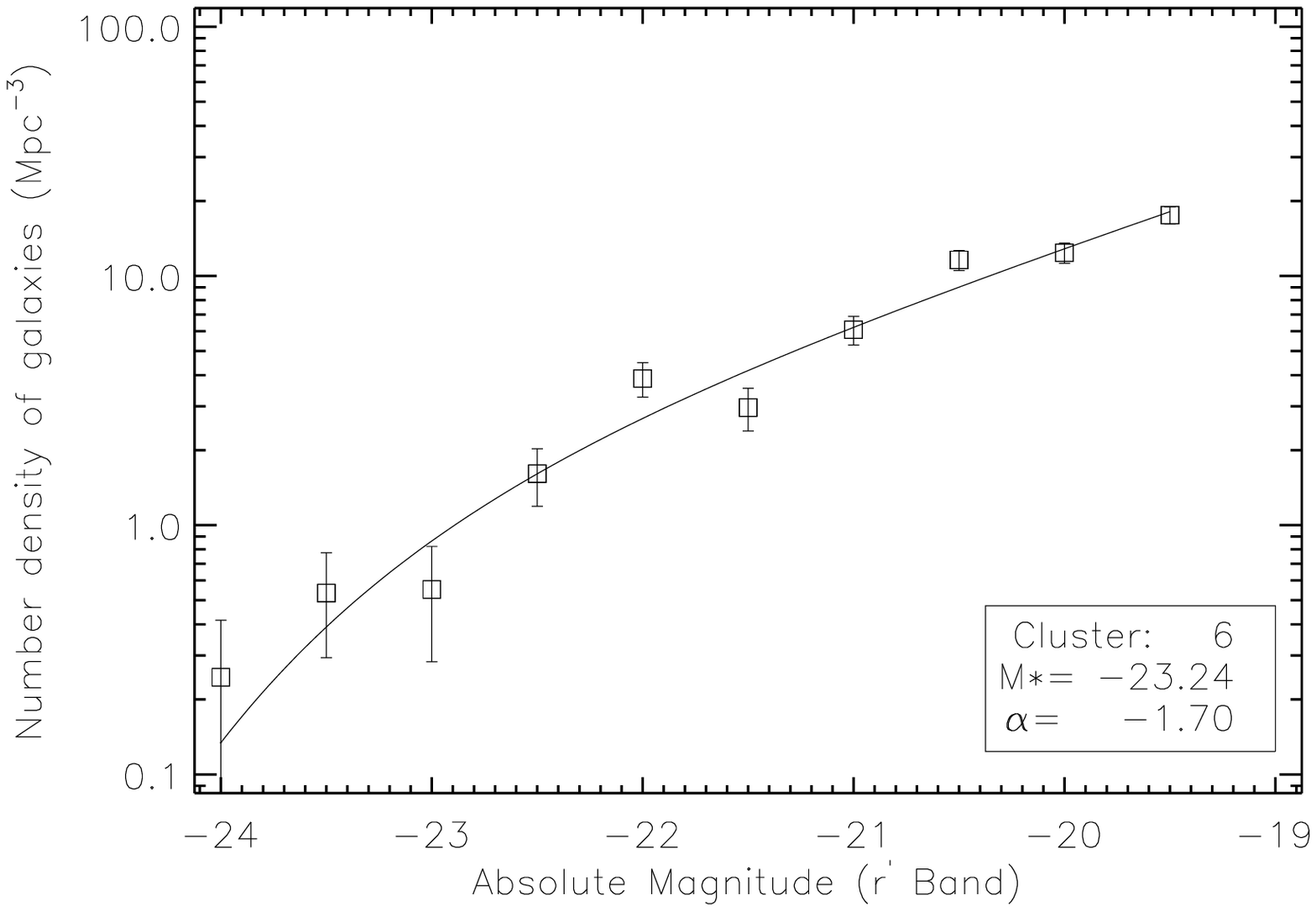,width=4cm,height=3.cm}
\epsfig{file=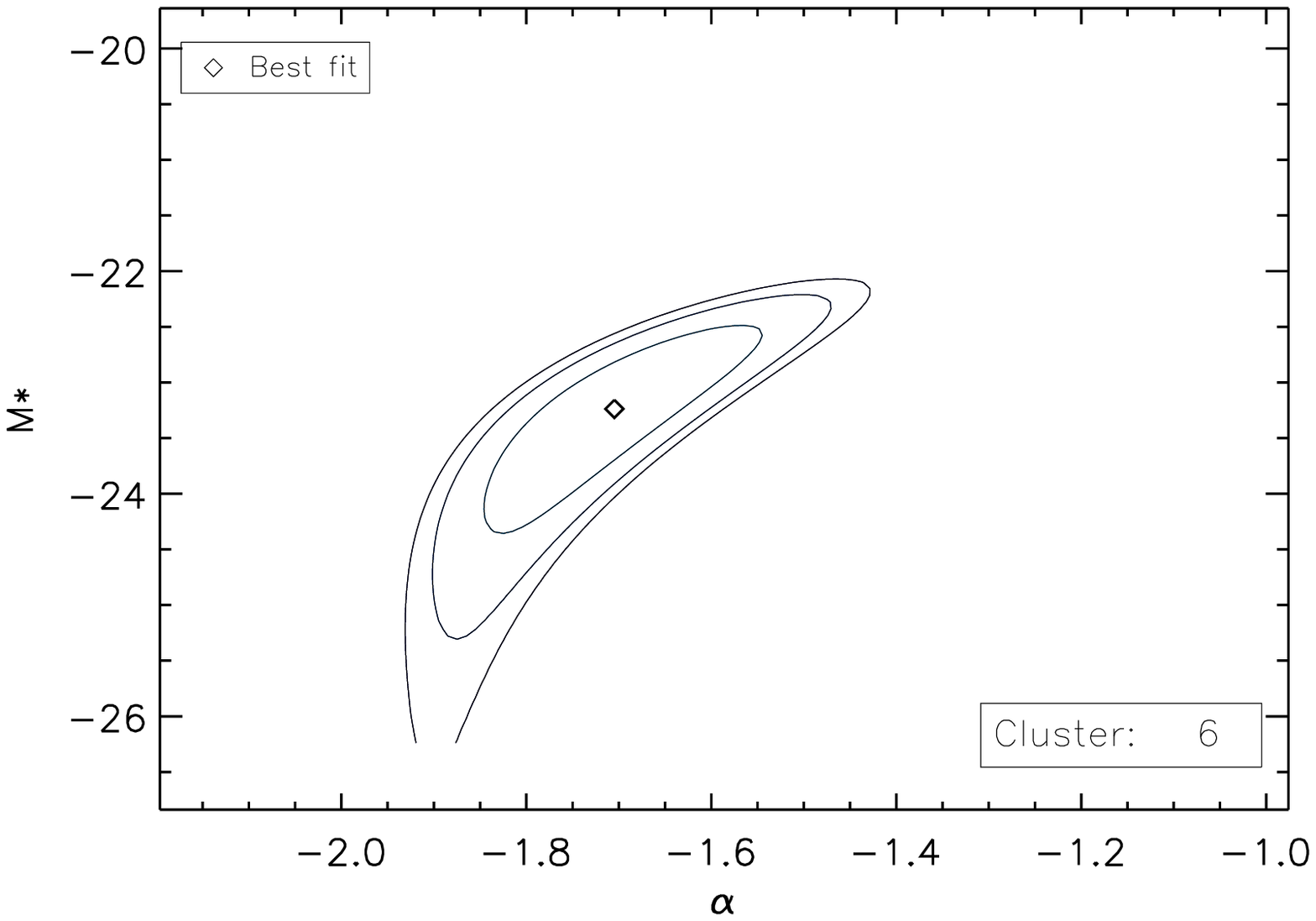,width=4cm,height=3.cm}
\epsfig{file=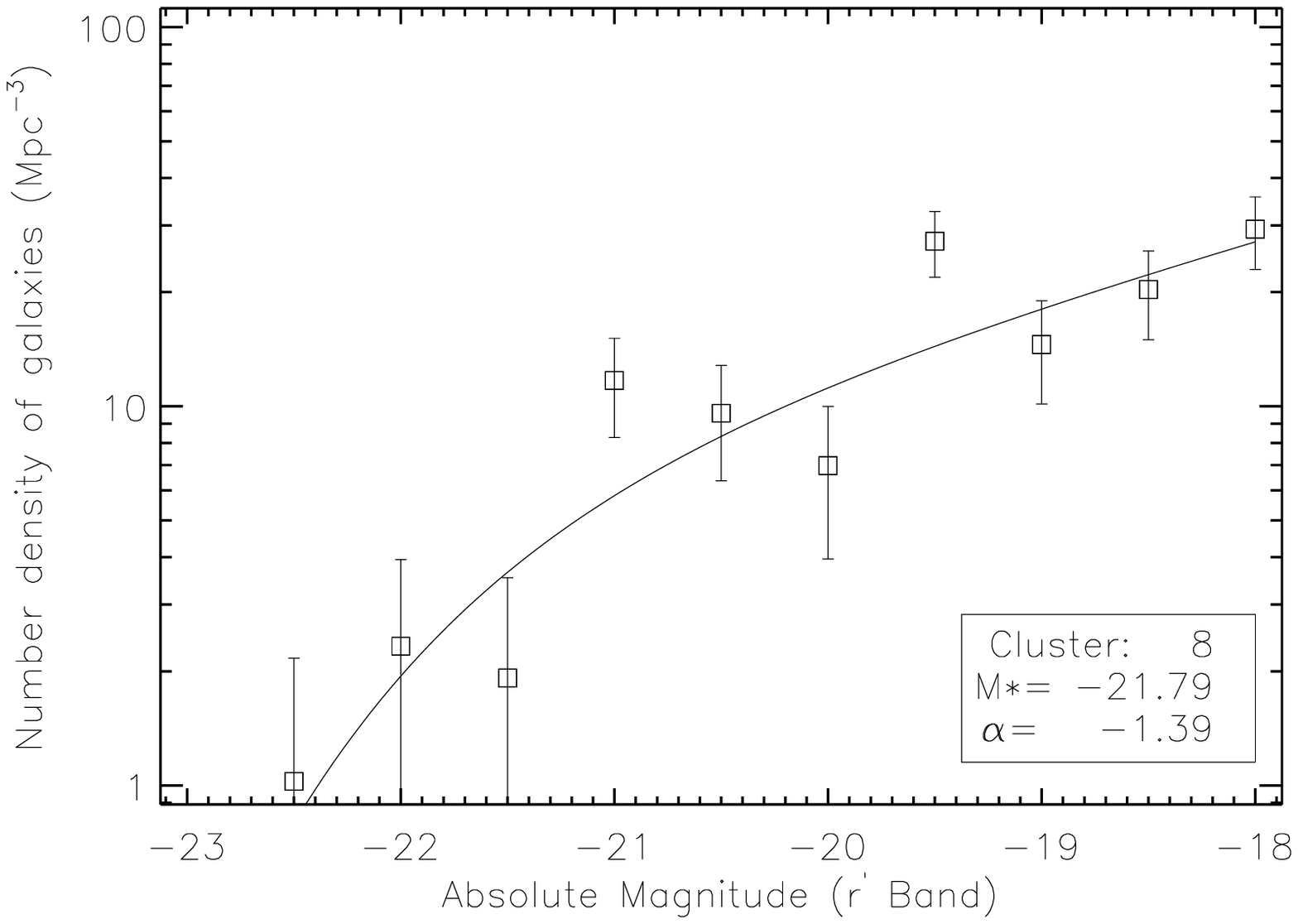,width=4cm,height=3.cm}
\epsfig{file=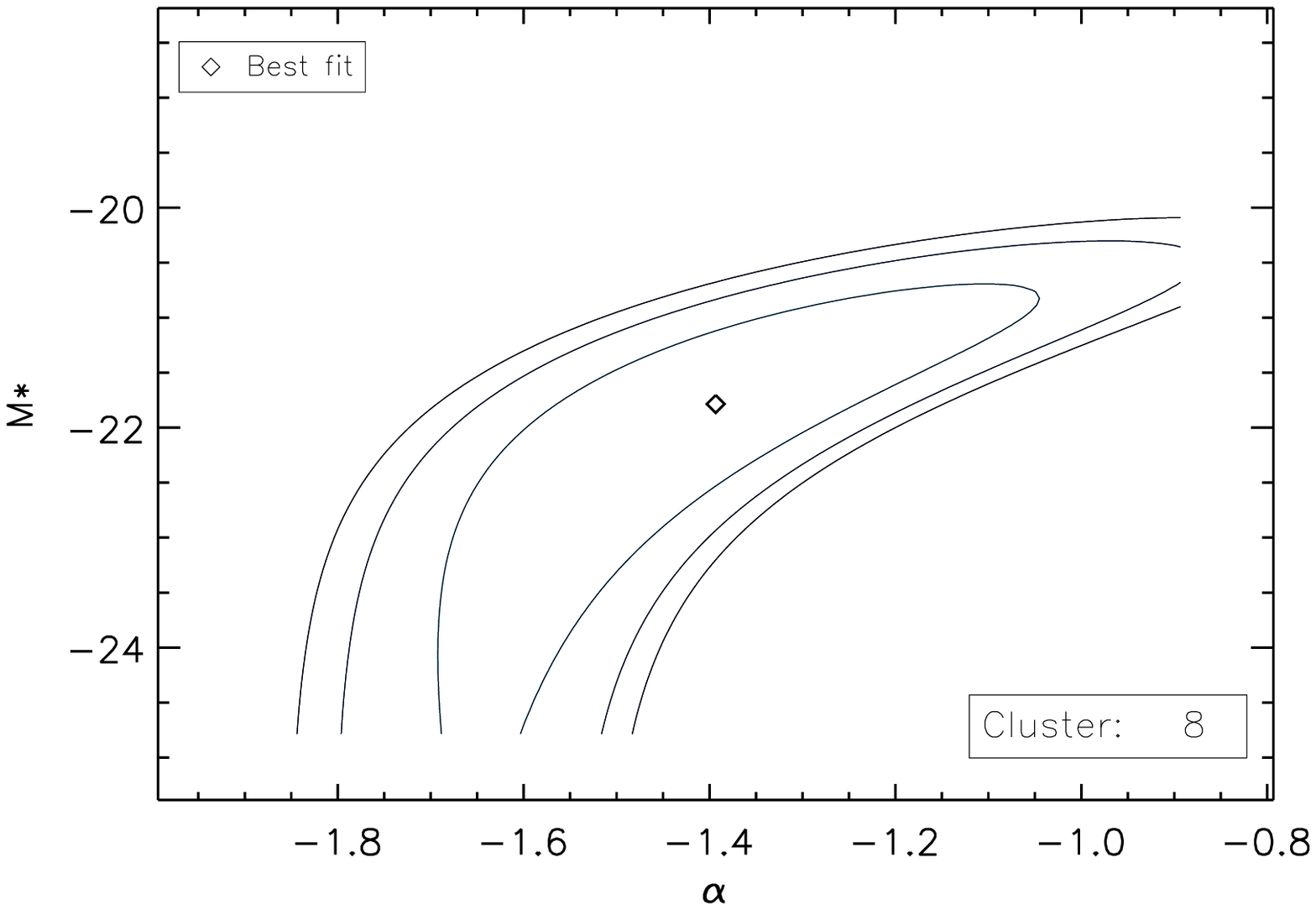,width=4cm,height=3.cm}
\epsfig{file=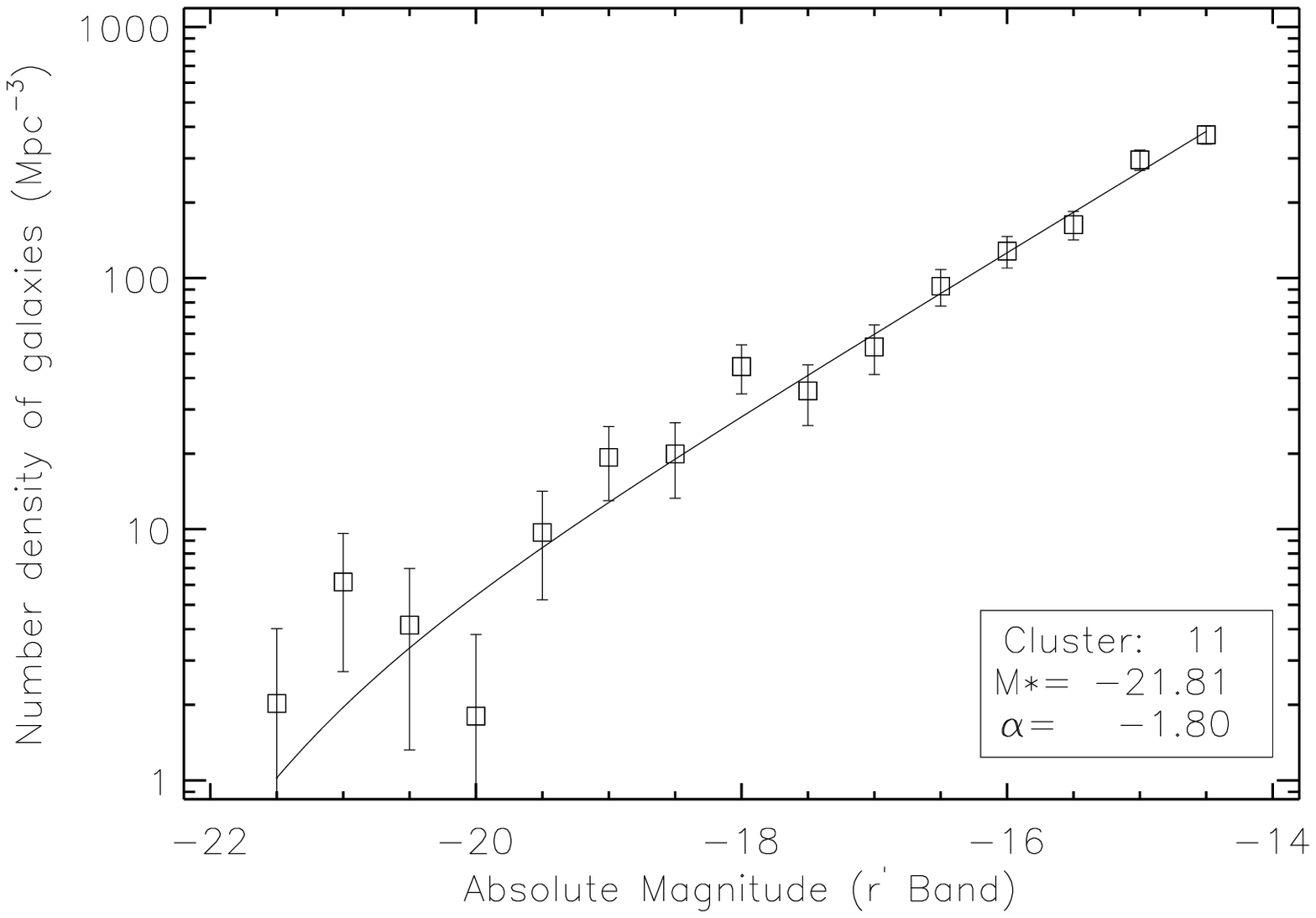,width=4cm,height=3.cm}
\epsfig{file=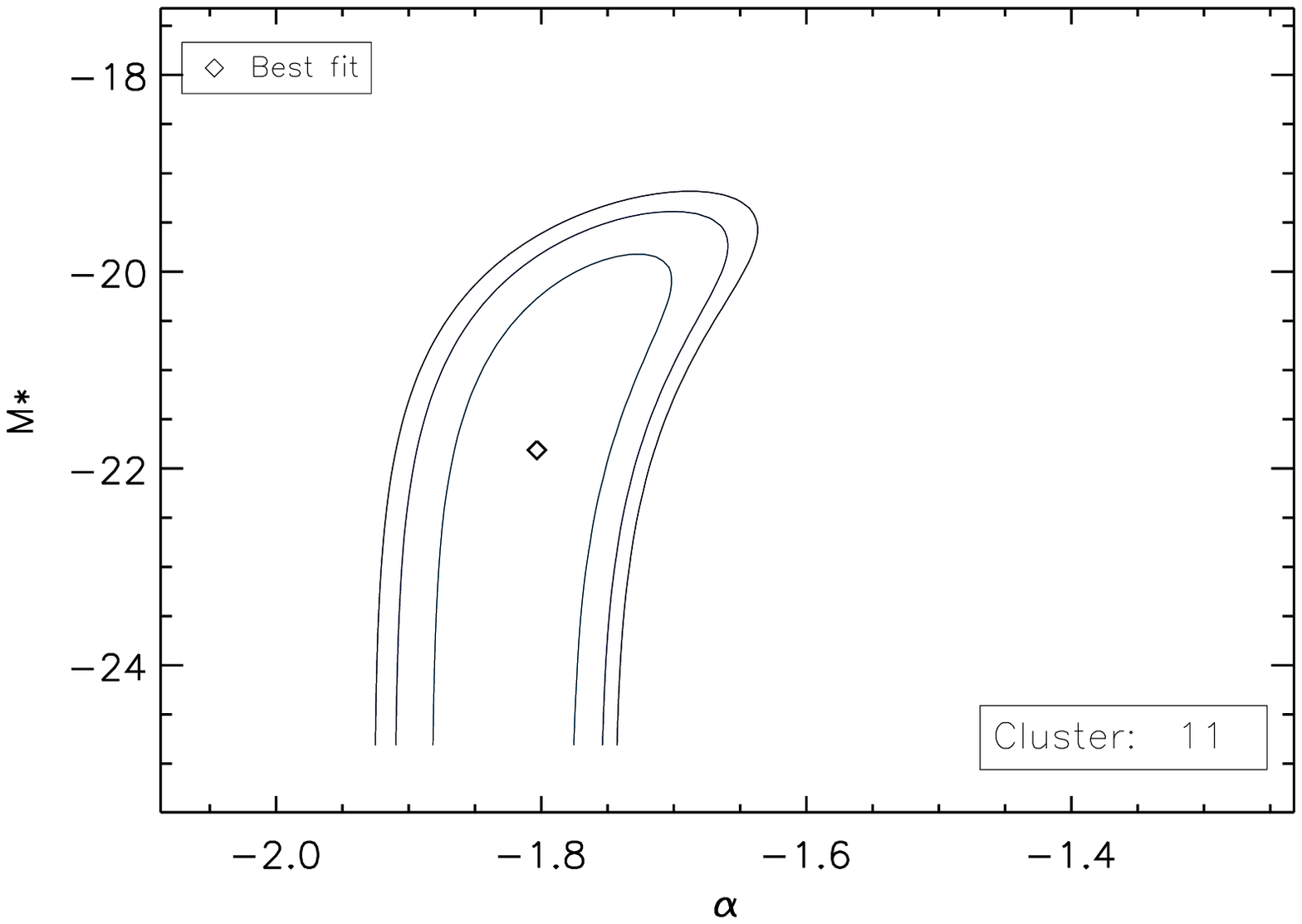,width=4cm,height=3.cm}
\epsfig{file=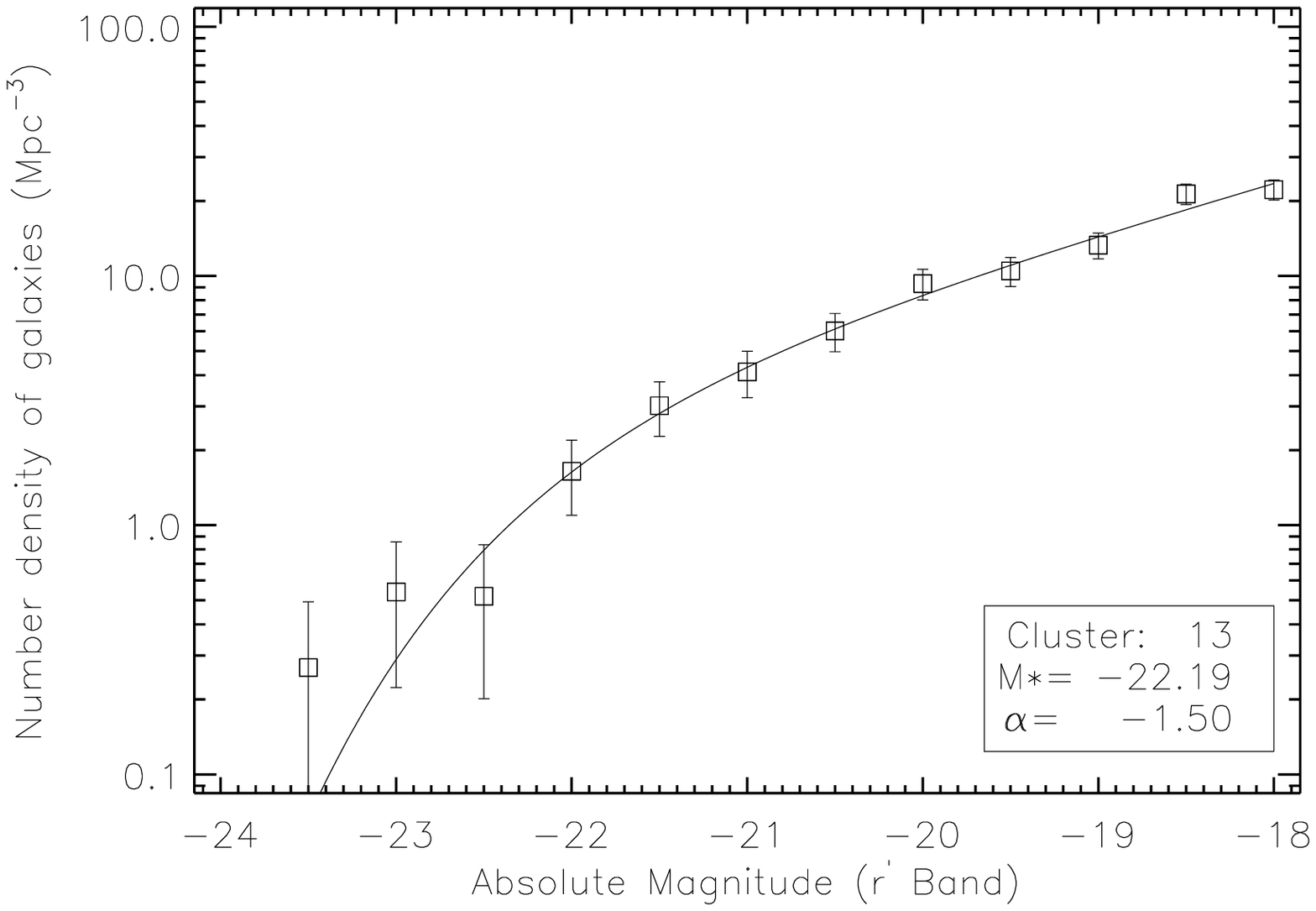,width=4cm,height=3.cm}
\epsfig{file=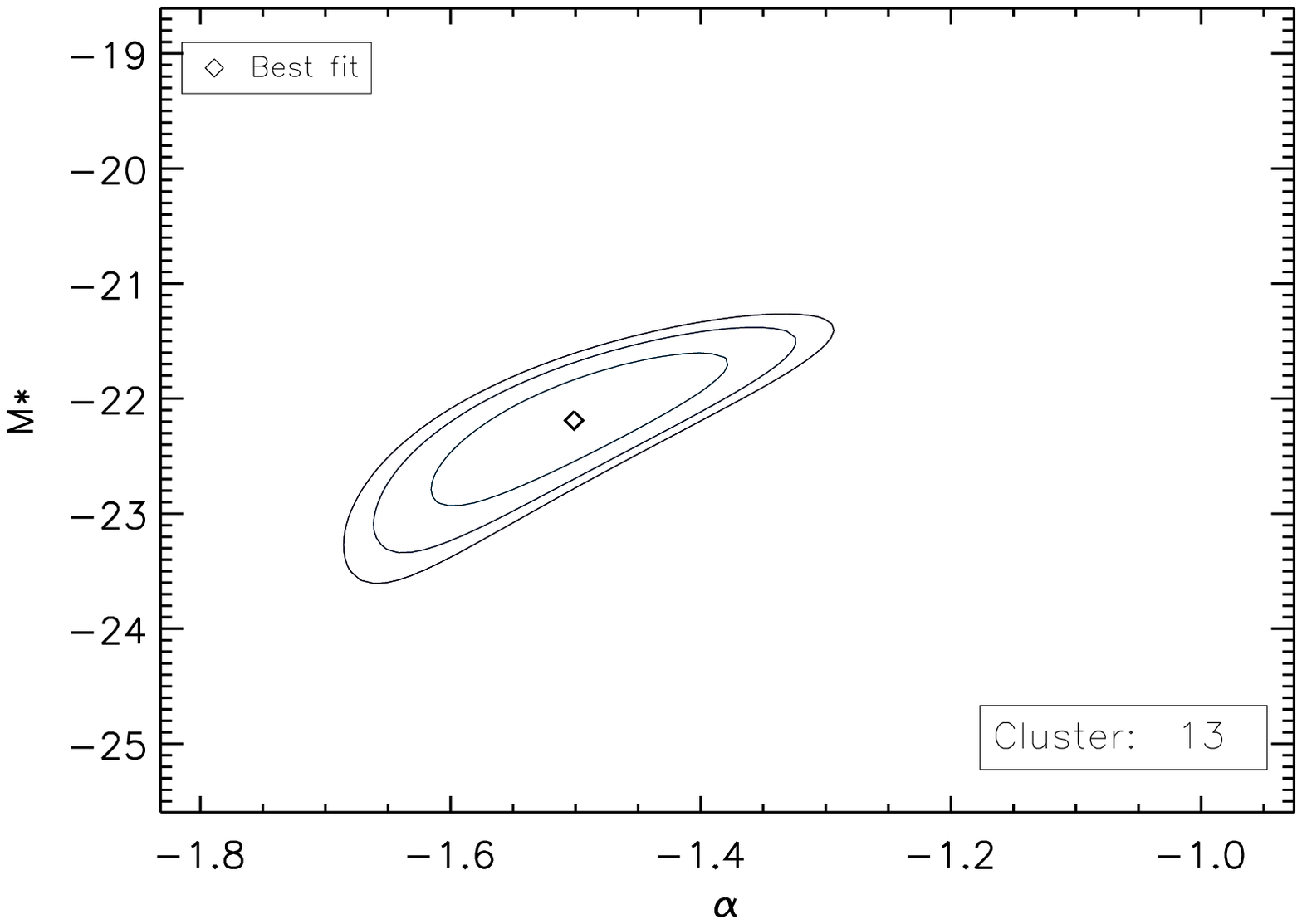,width=4cm,height=3.cm}
\epsfig{file=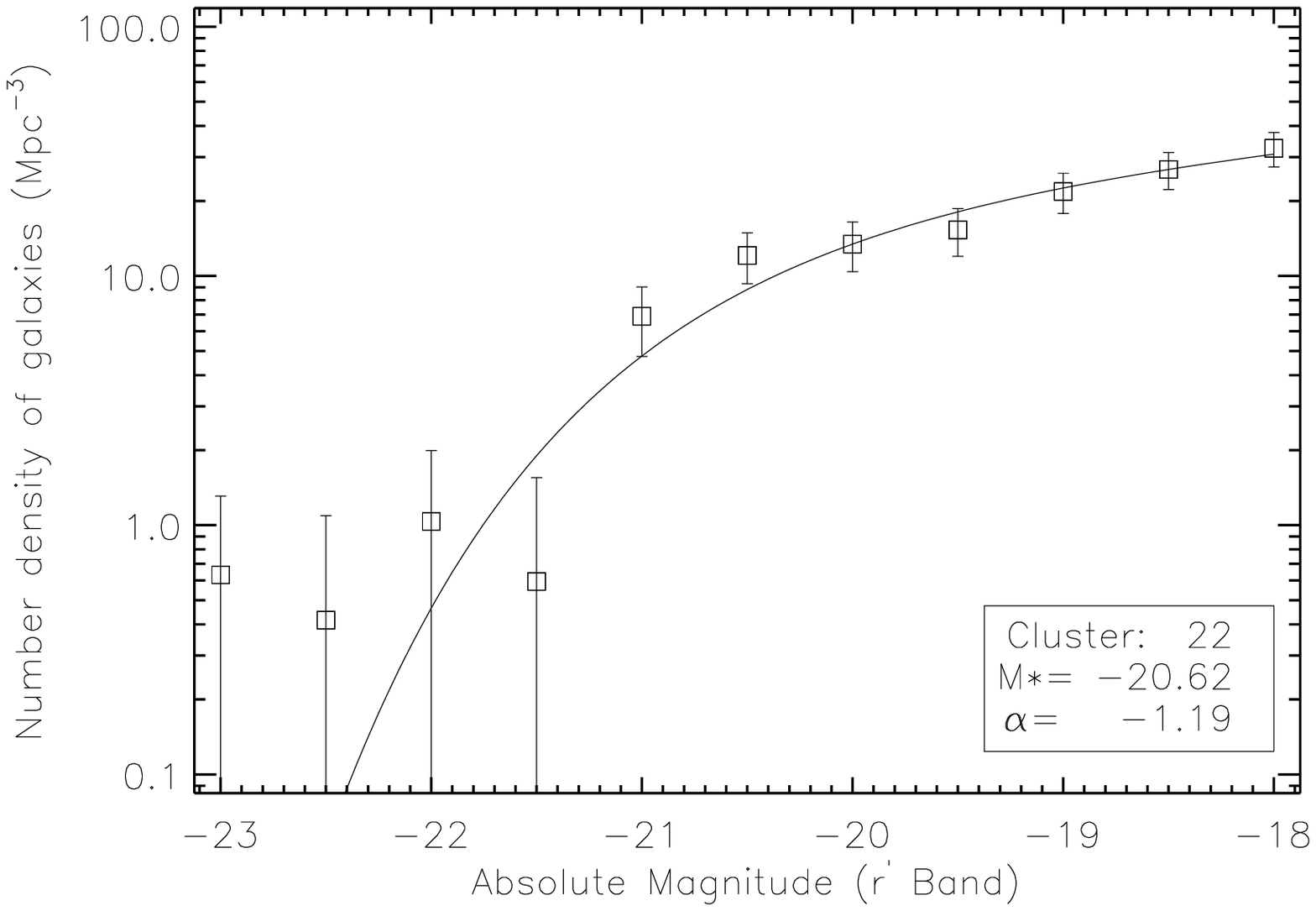,width=4cm,height=3.cm}
\epsfig{file=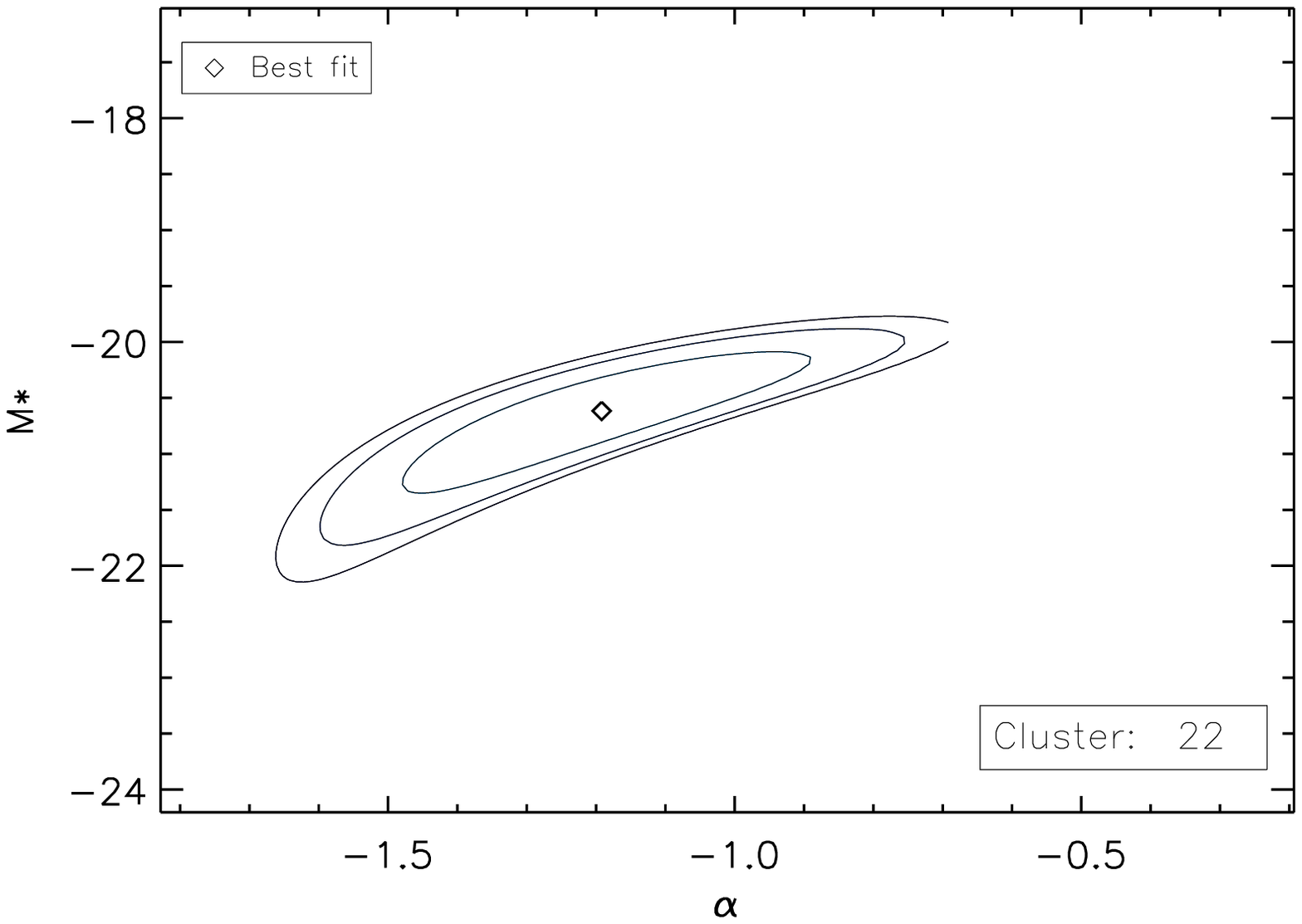,width=4cm,height=3.cm}
\epsfig{file=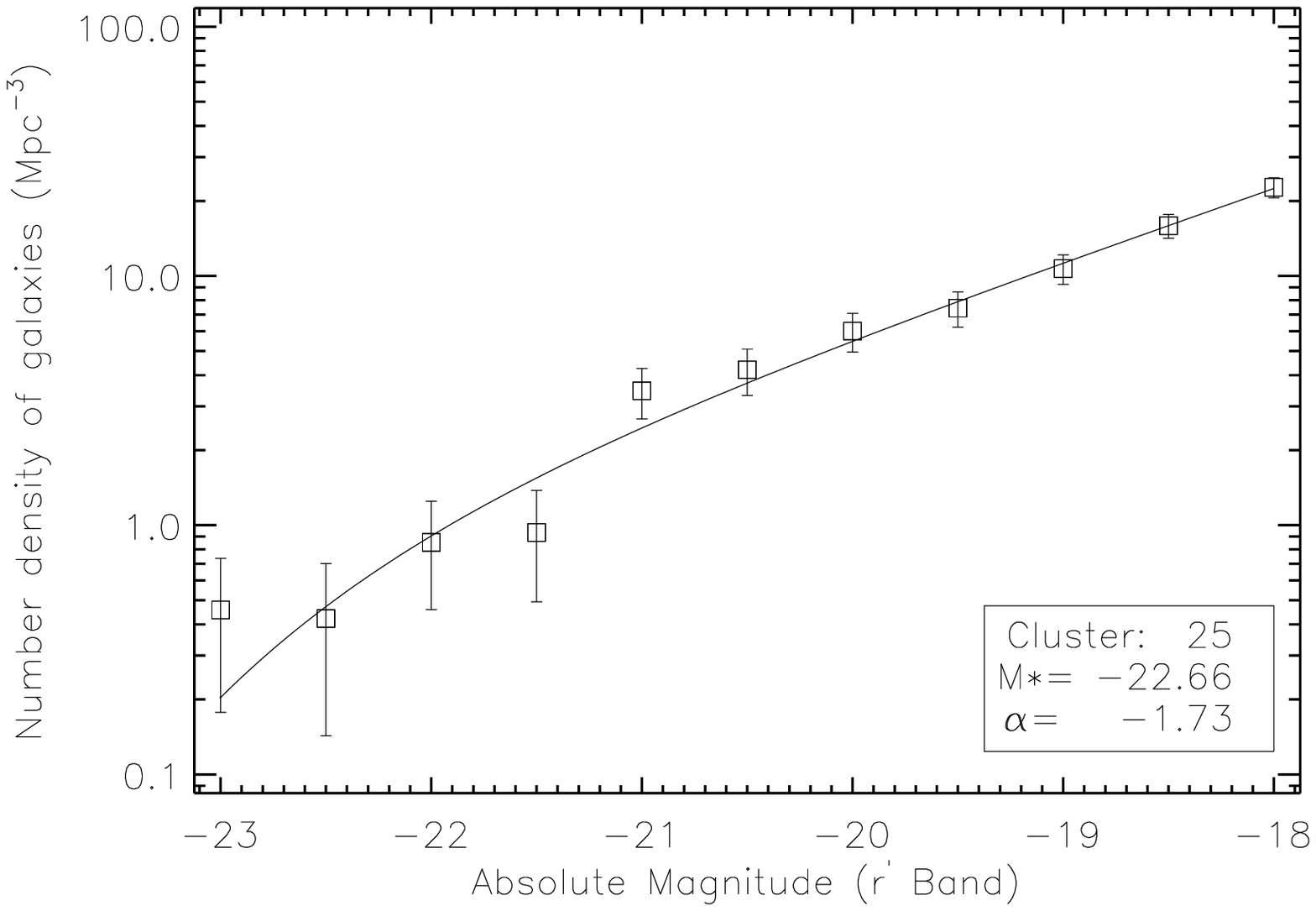,width=4cm,height=3.cm}
\epsfig{file=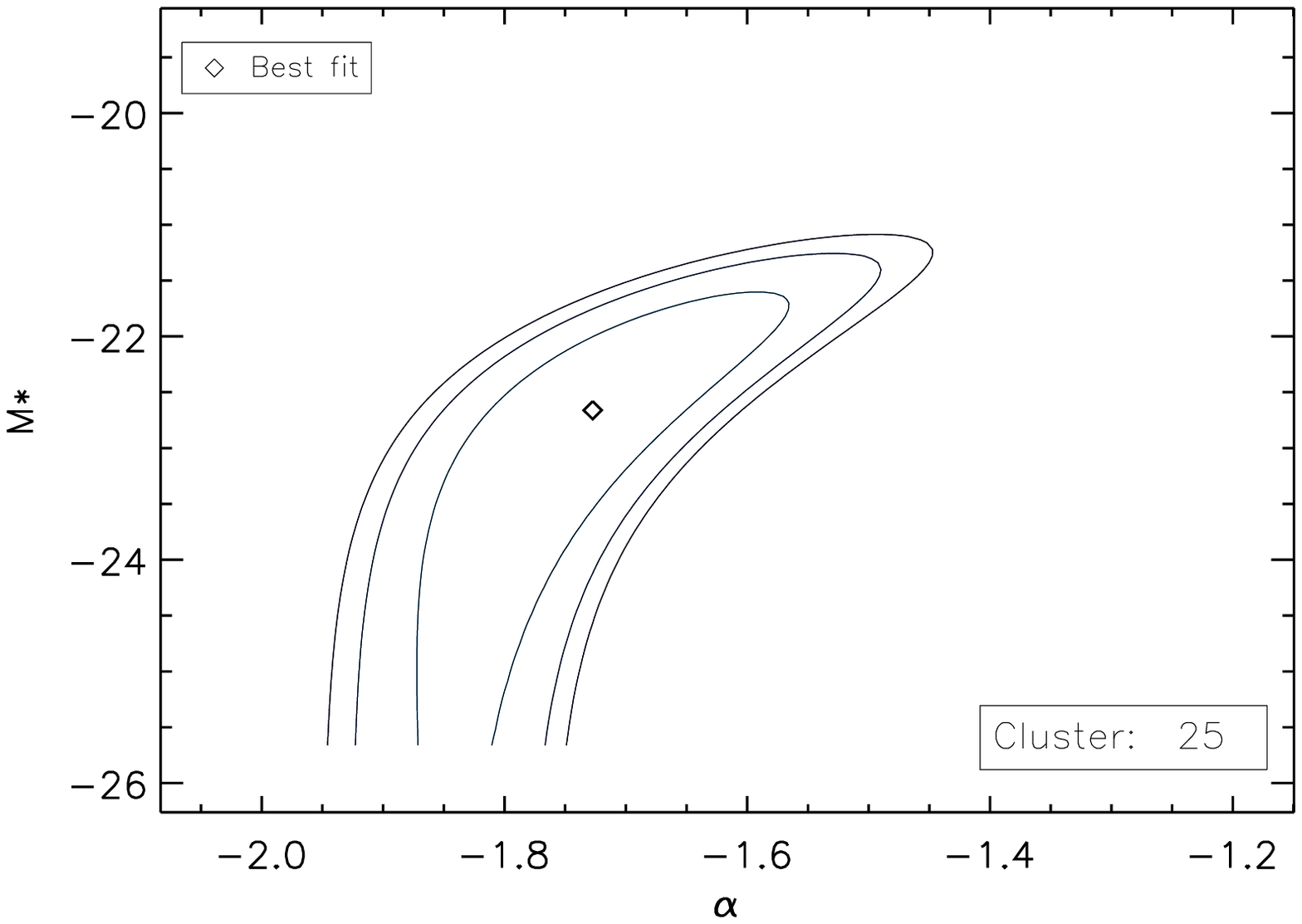,width=4cm,height=3.cm}
\epsfig{file=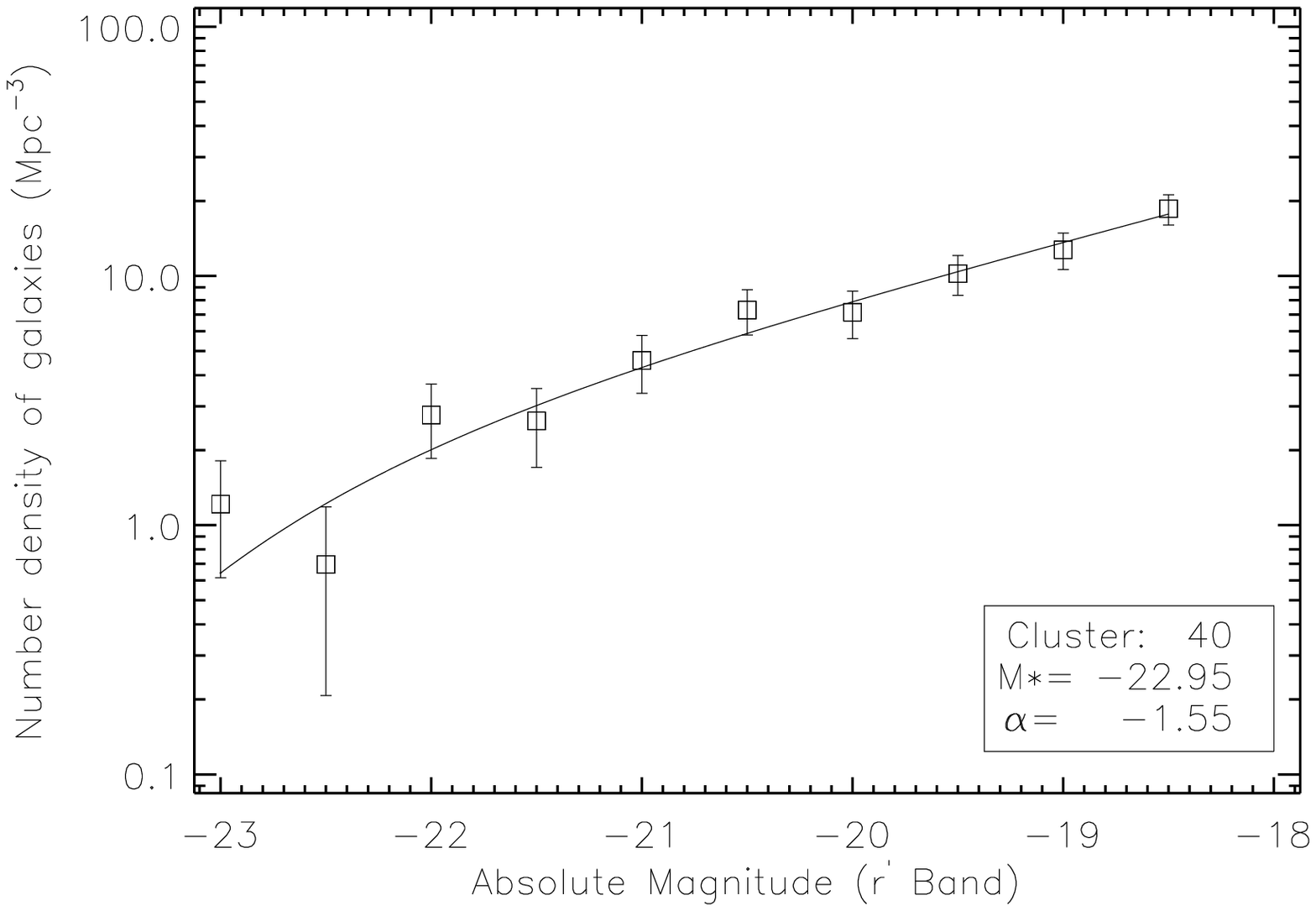,width=4cm,height=3.cm}
\epsfig{file=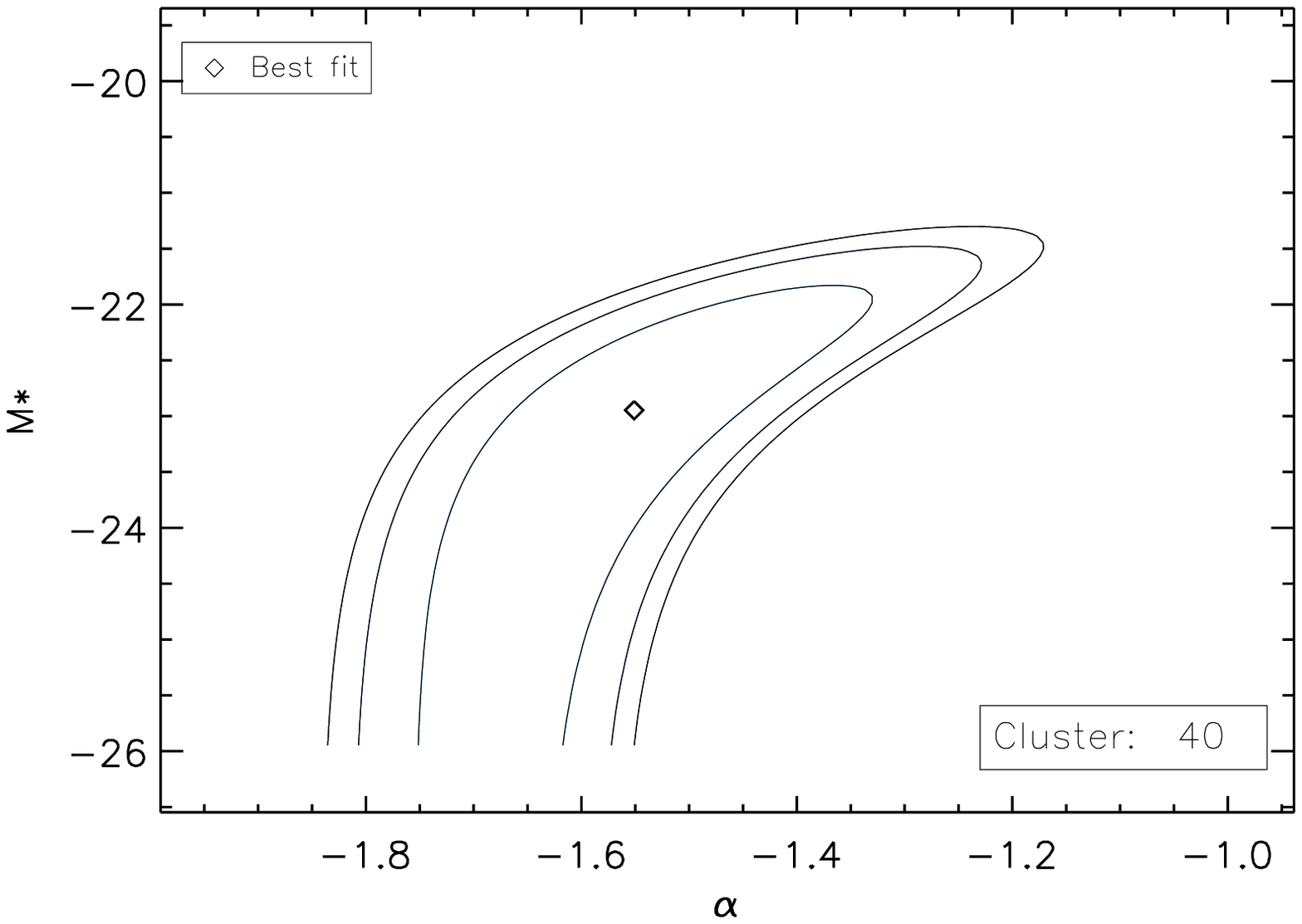,width=4cm,height=3.cm}
\epsfig{file=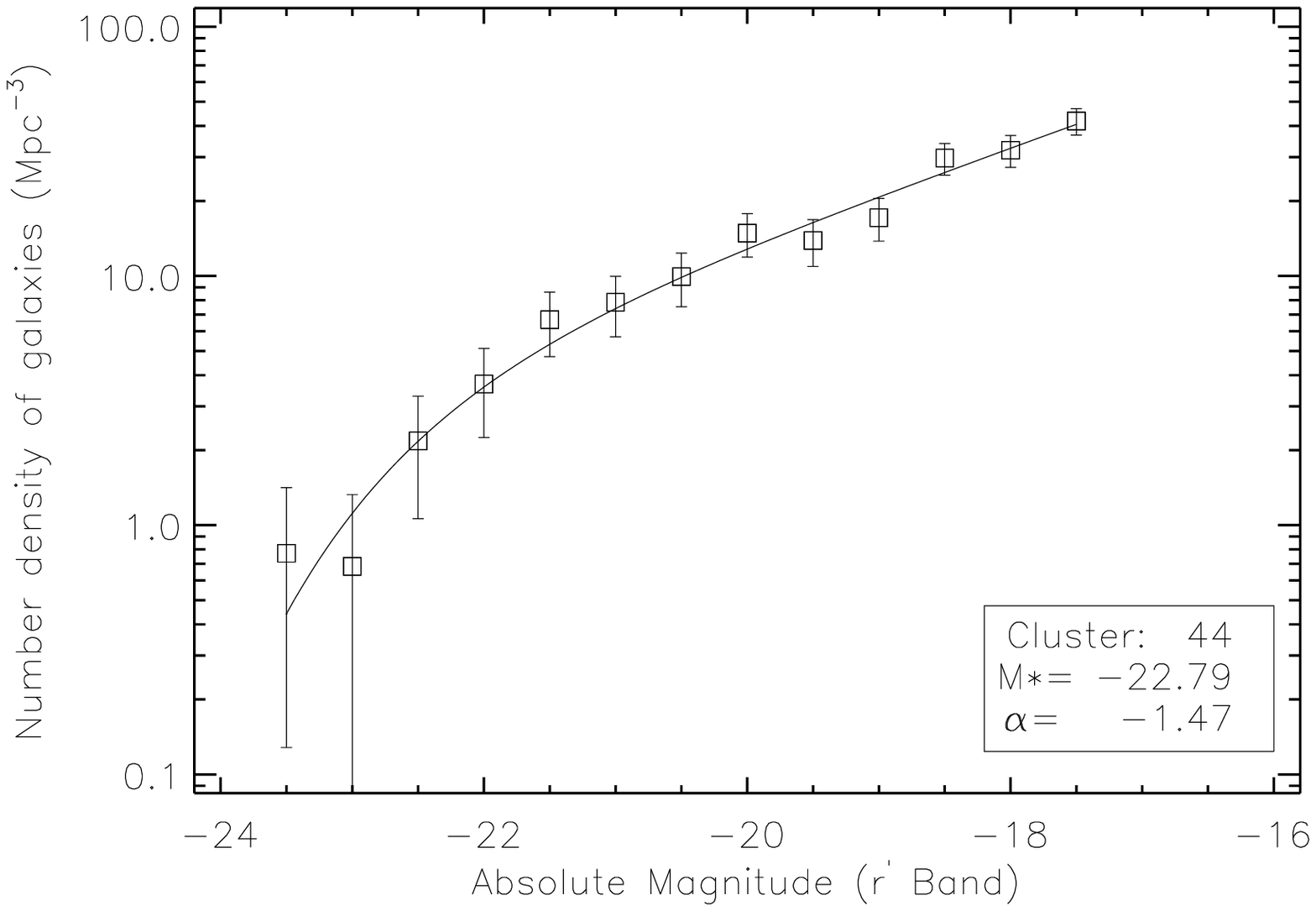,width=4cm,height=3.cm}
\epsfig{file=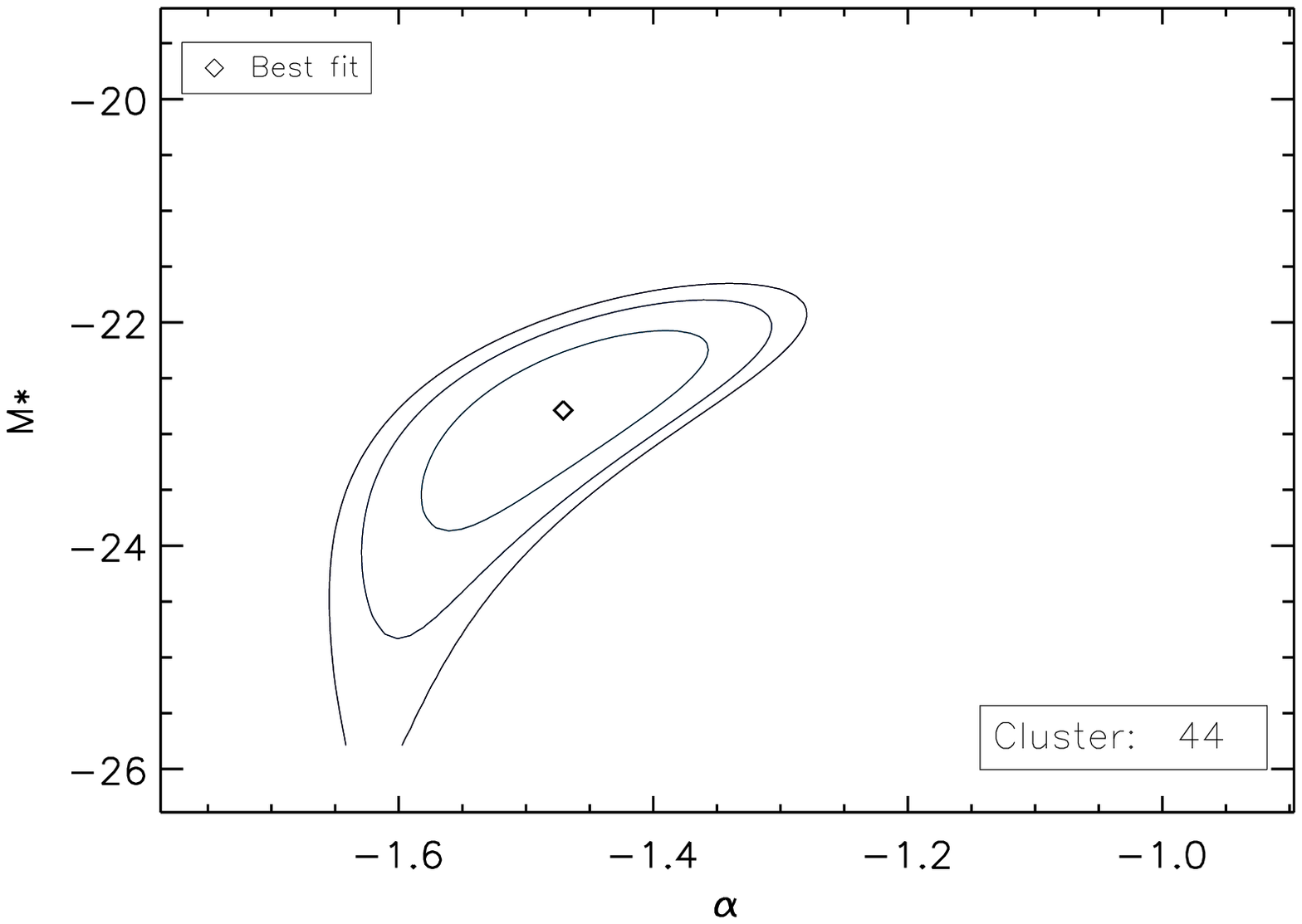,width=4cm,height=3.cm}
\epsfig{file=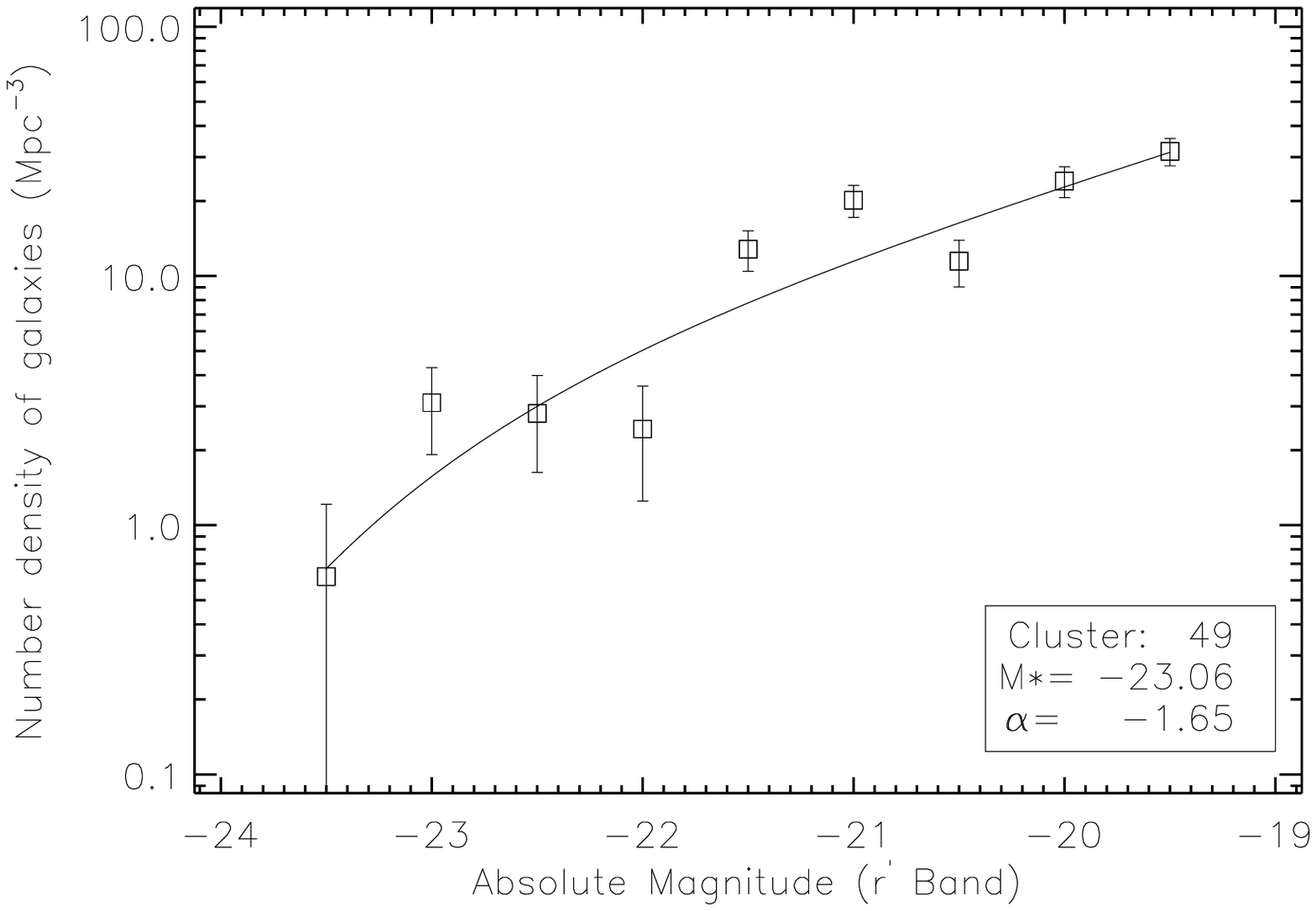,width=4cm,height=3.cm}
\epsfig{file=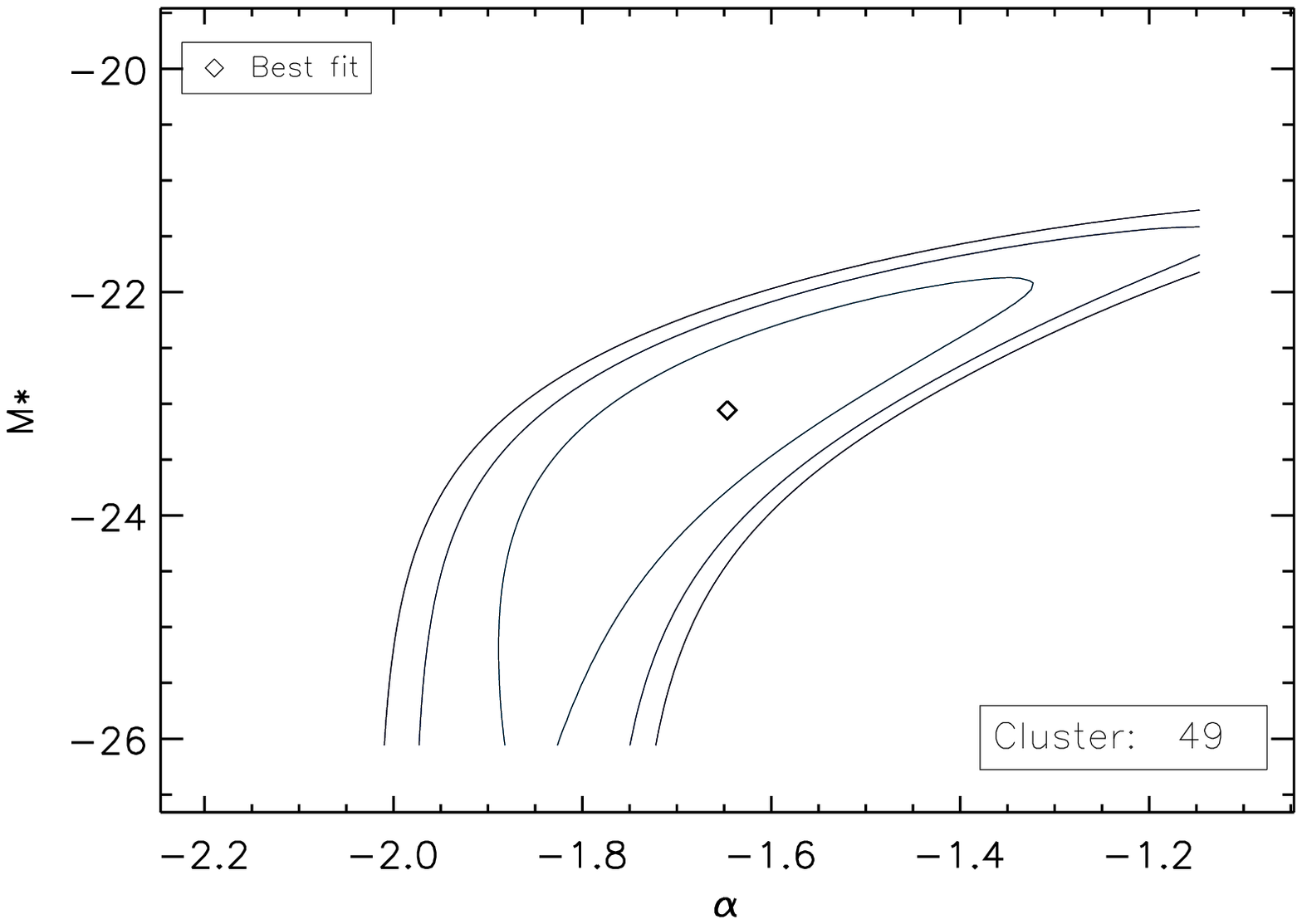,width=4cm,height=3.cm}

\epsfig{file=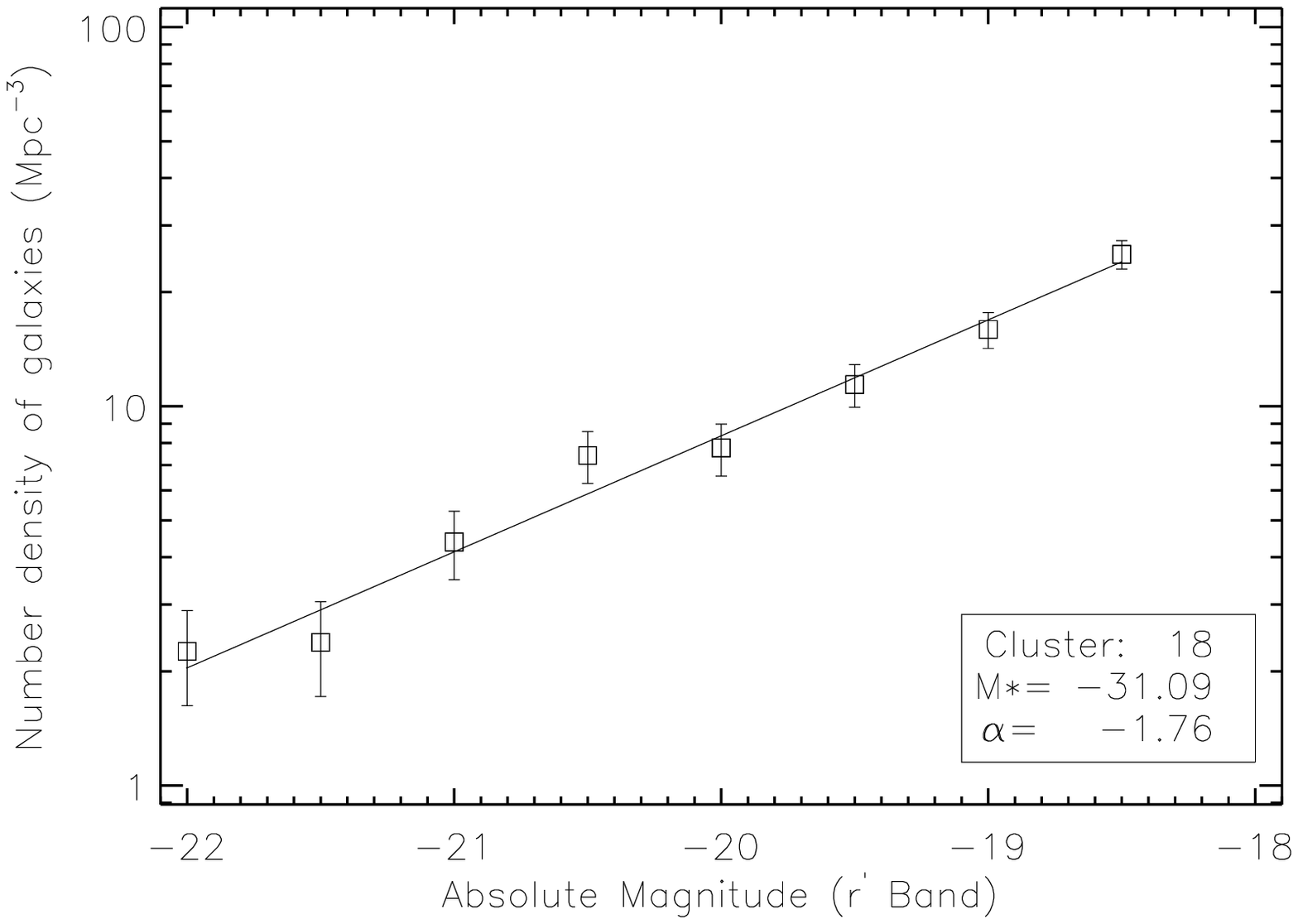,width=4cm,height=3.cm}
\epsfig{file=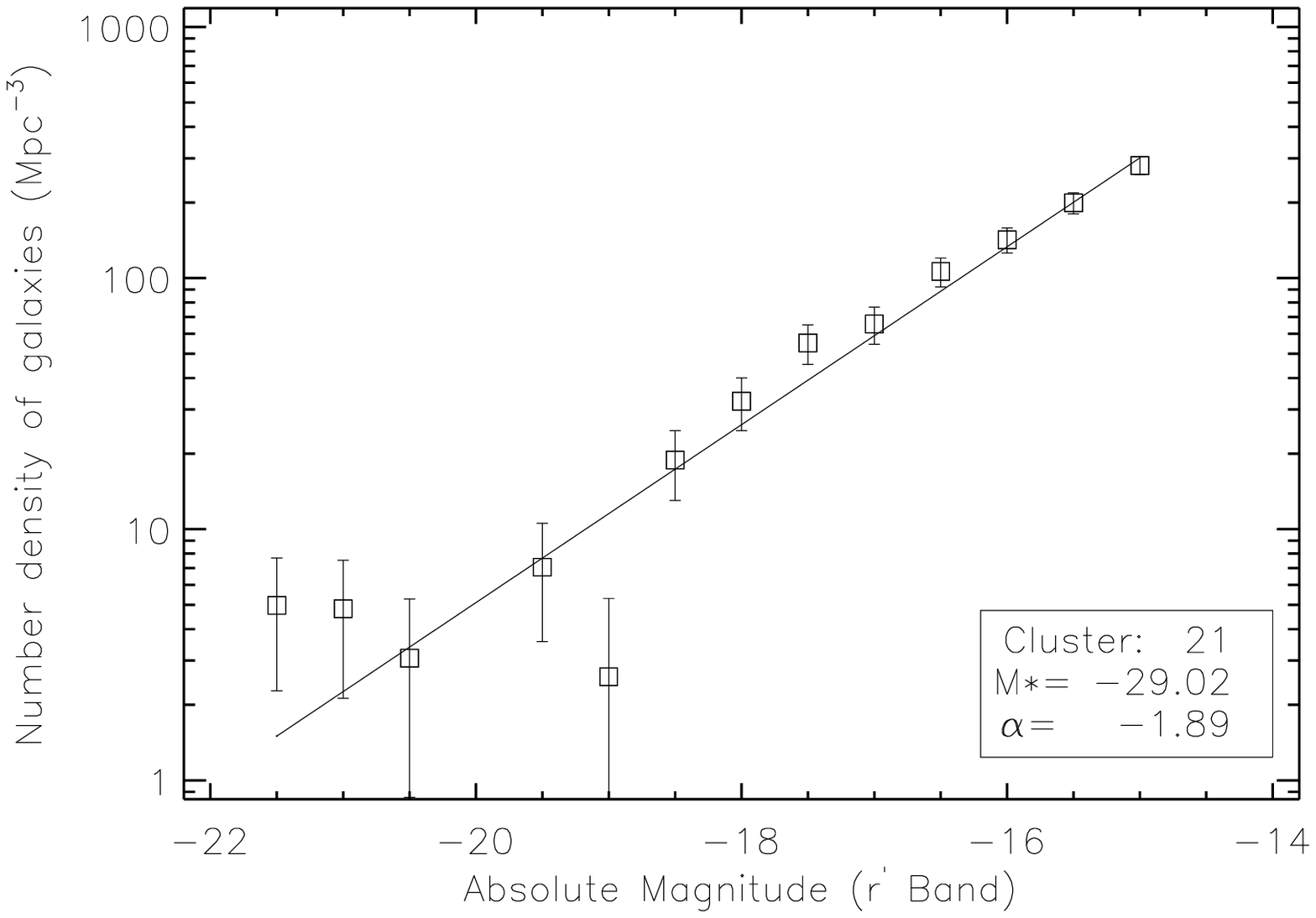,width=4cm,height=3.cm}
\epsfig{file=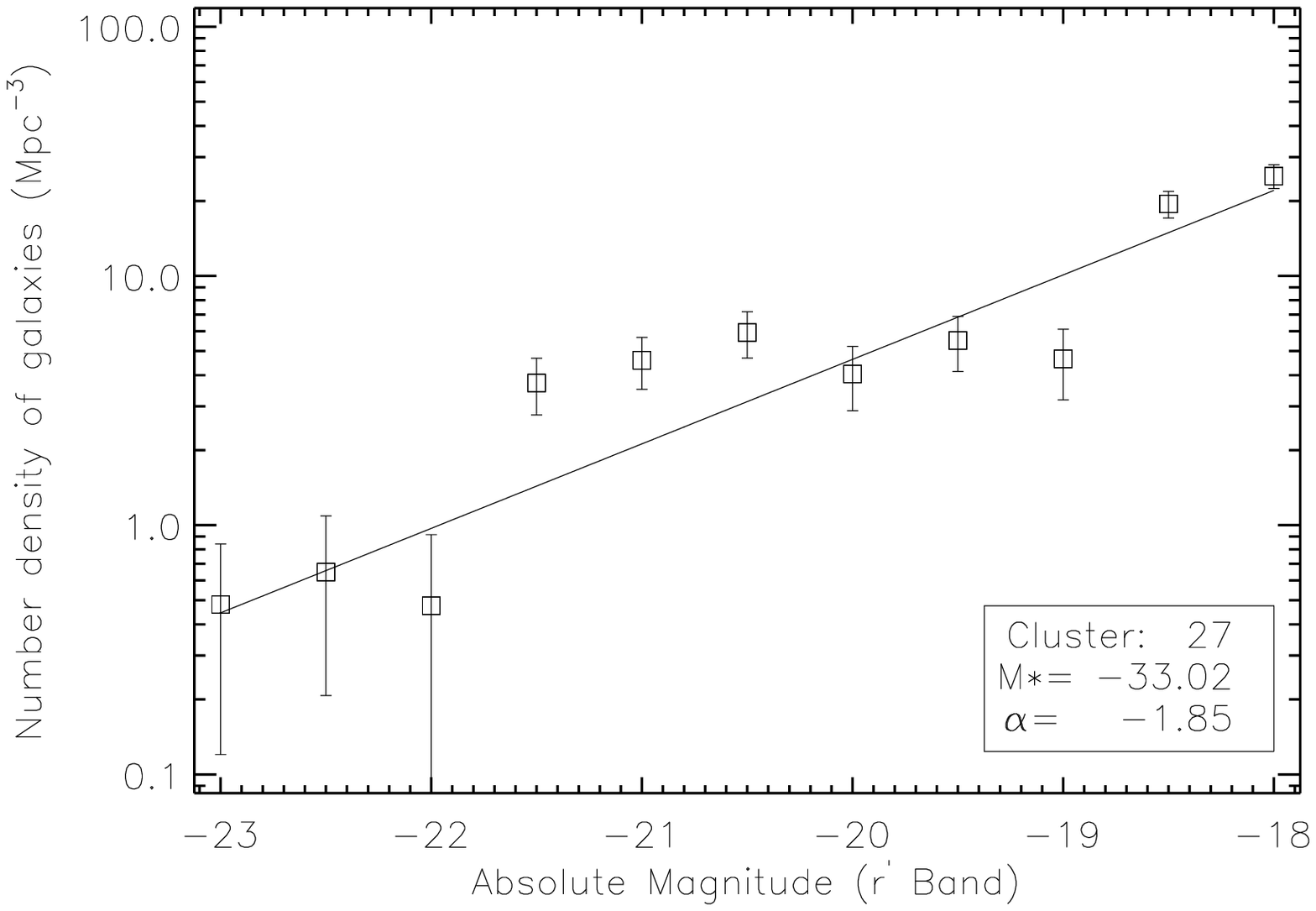,width=4cm,height=3.cm}
\epsfig{file=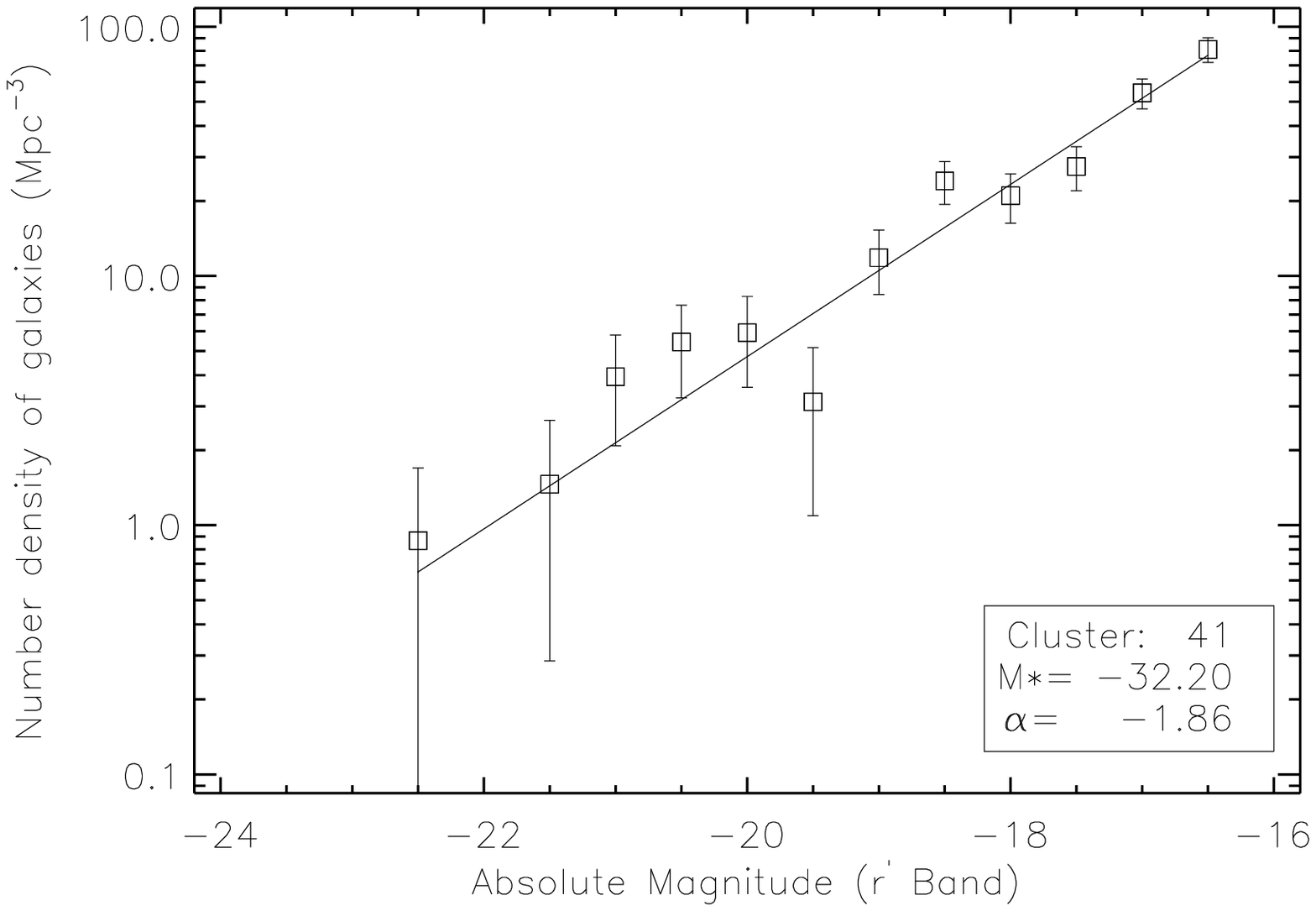,width=4cm,height=3.cm}

\caption{LFs of the 14 individual C1 clusters and contours of the well-fitted clusters for the $r^\prime$ band. Contours plots of the $1\sigma$, $2\sigma$ and $3\sigma$ confidence levels of  $\alpha$ and $M^*$ are placed next to their associated LF. Clusters with failed constrained $M^*$ (and no contours) were placed at the bottom. }
\label{Ind_r}
\end{figure}

\section{Individual luminosity functions of C1 clusters in $\lowercase{z}^\prime$ band}


\onecolumn

\begin{figure}

\center

\epsfig{file=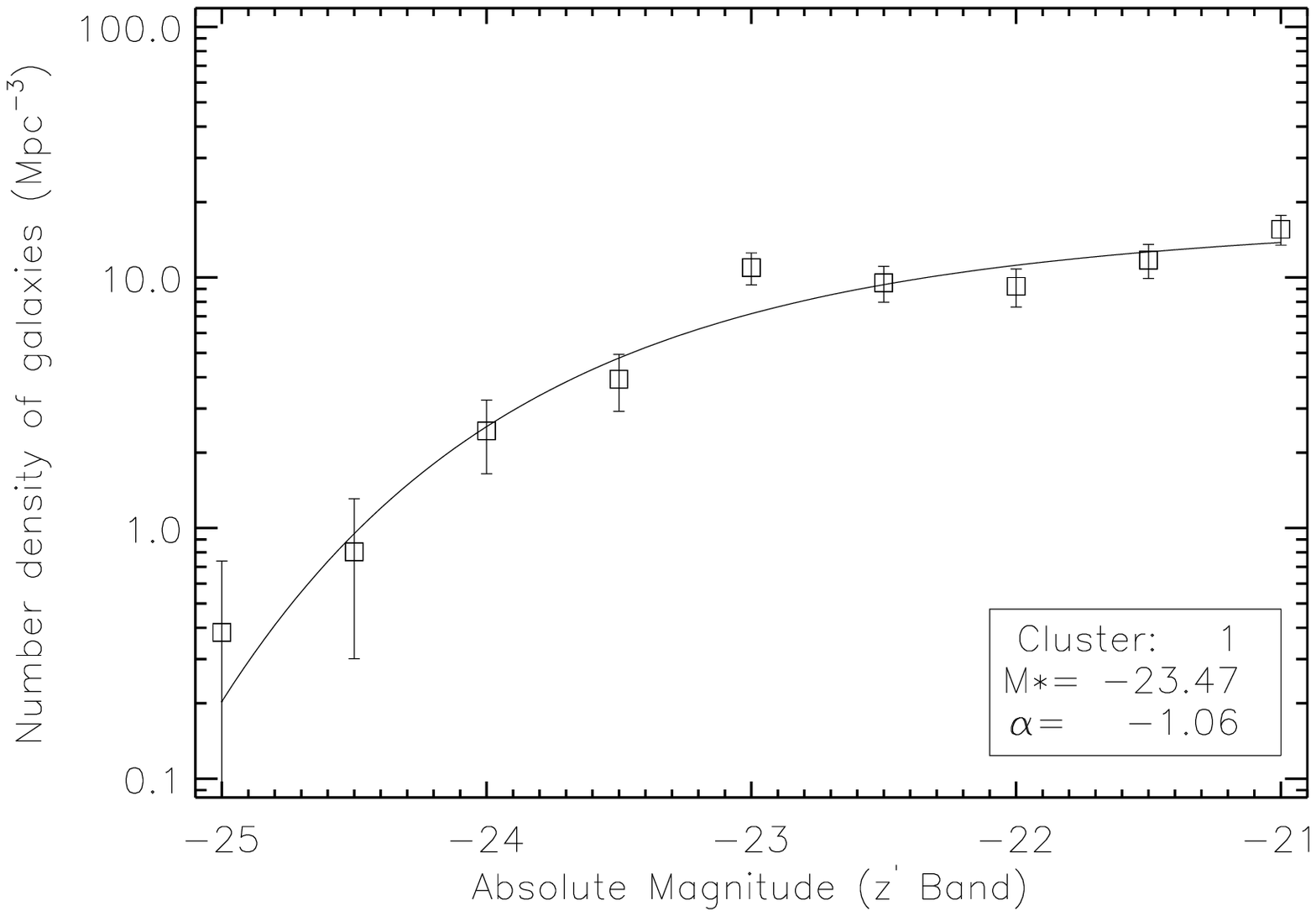,width=4cm,height=3.cm}
\epsfig{file=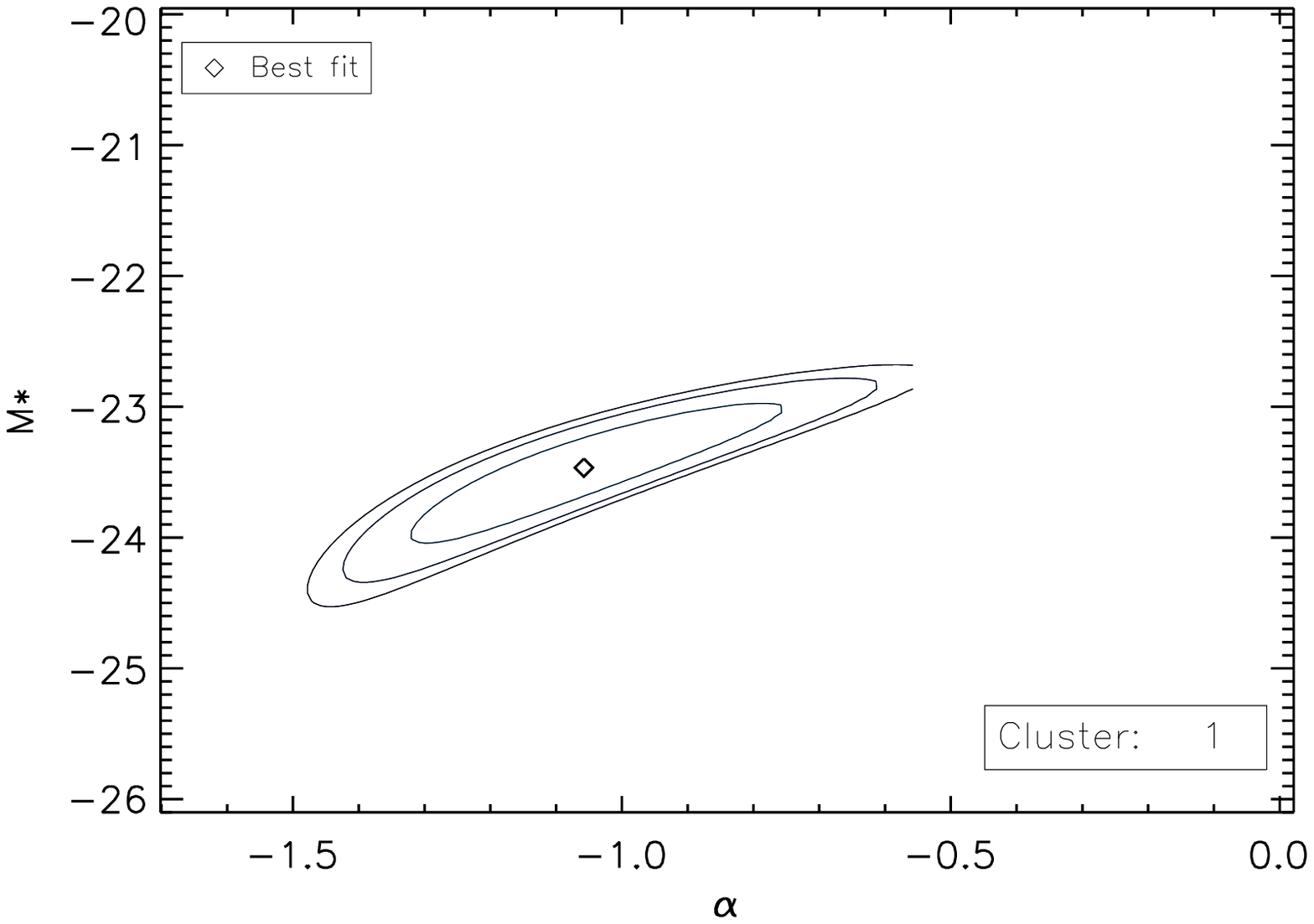,width=4cm,height=3.cm}
\epsfig{file=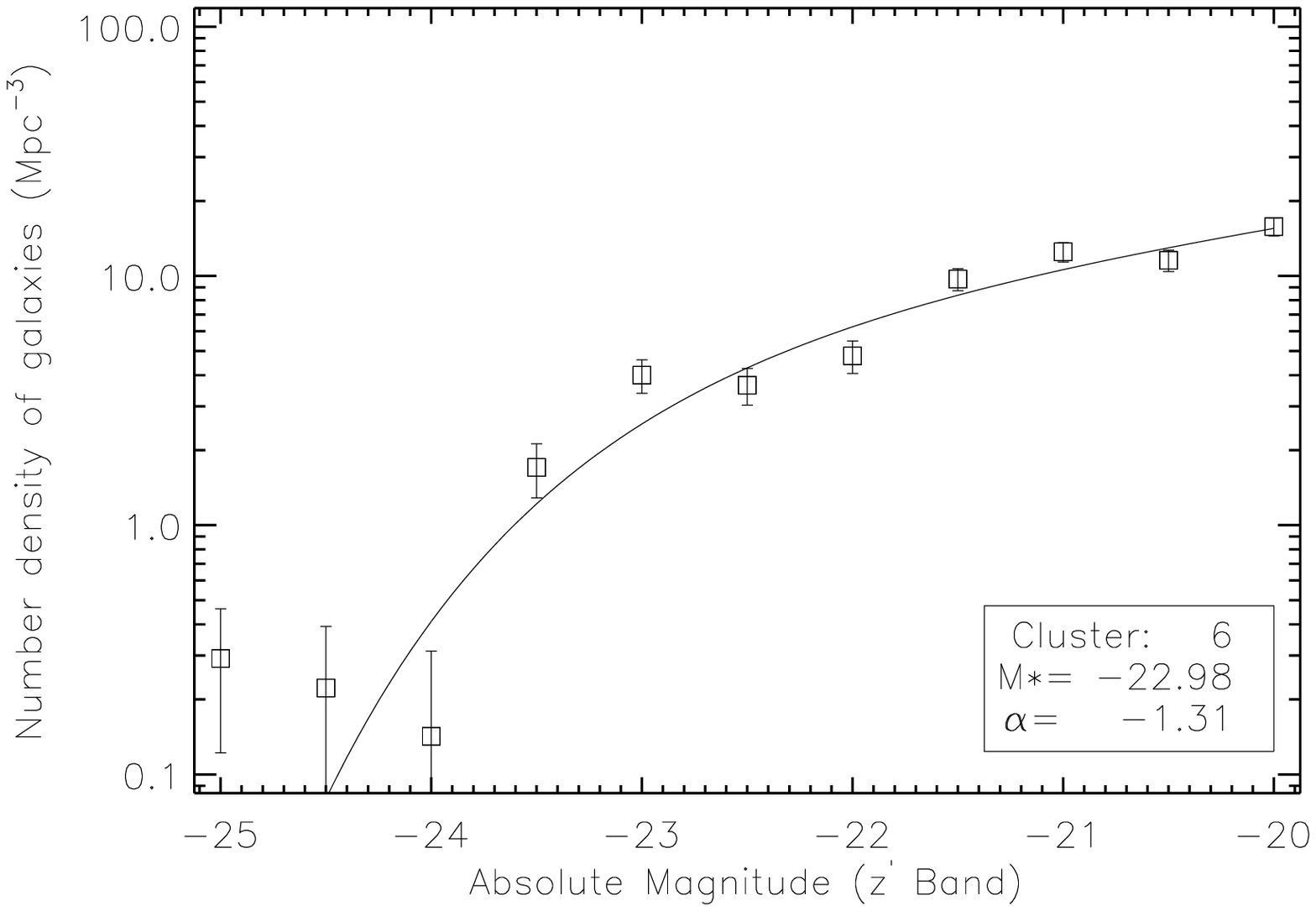,width=4cm,height=3.cm}
\epsfig{file=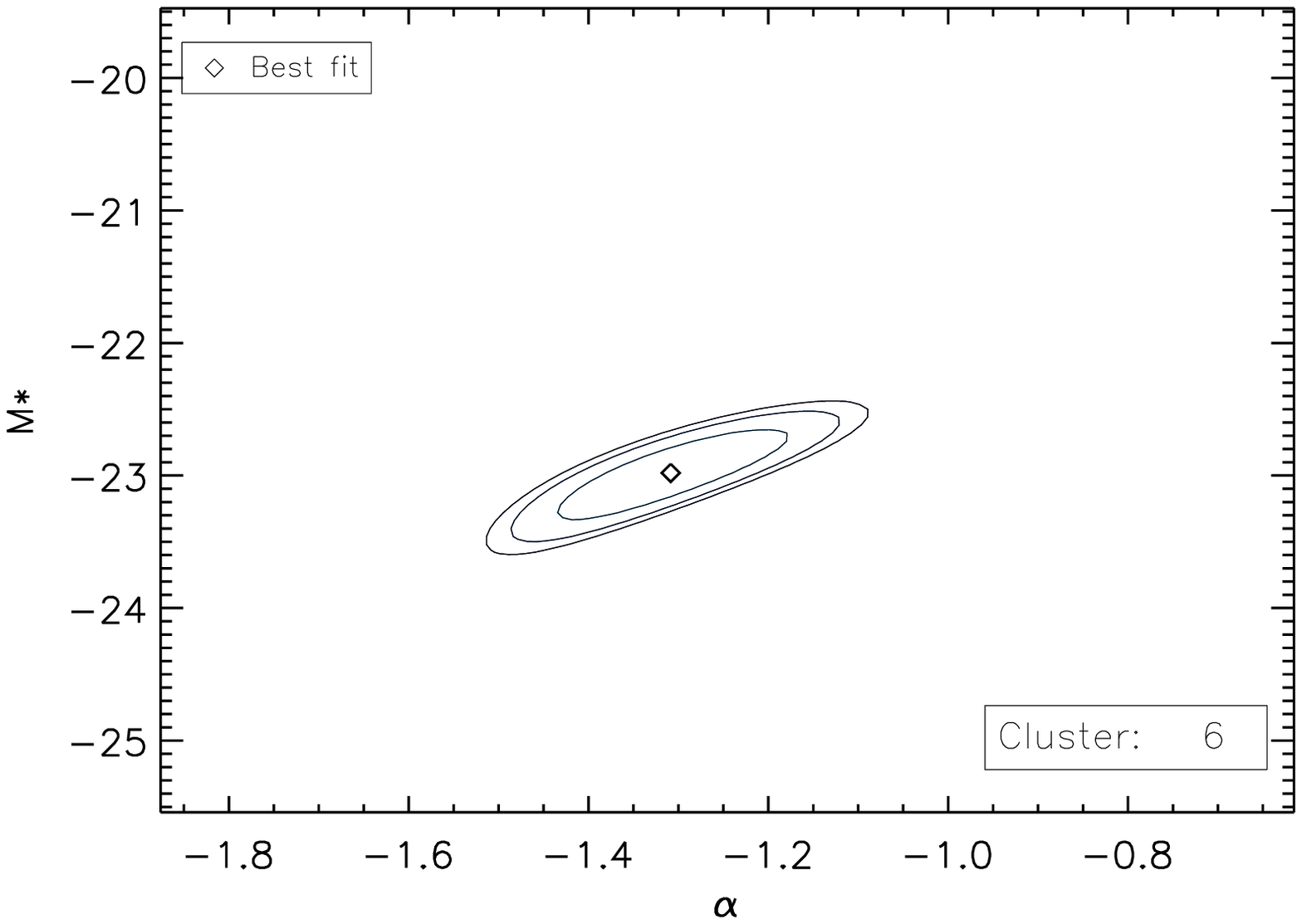,width=4cm,height=3.cm}
\epsfig{file=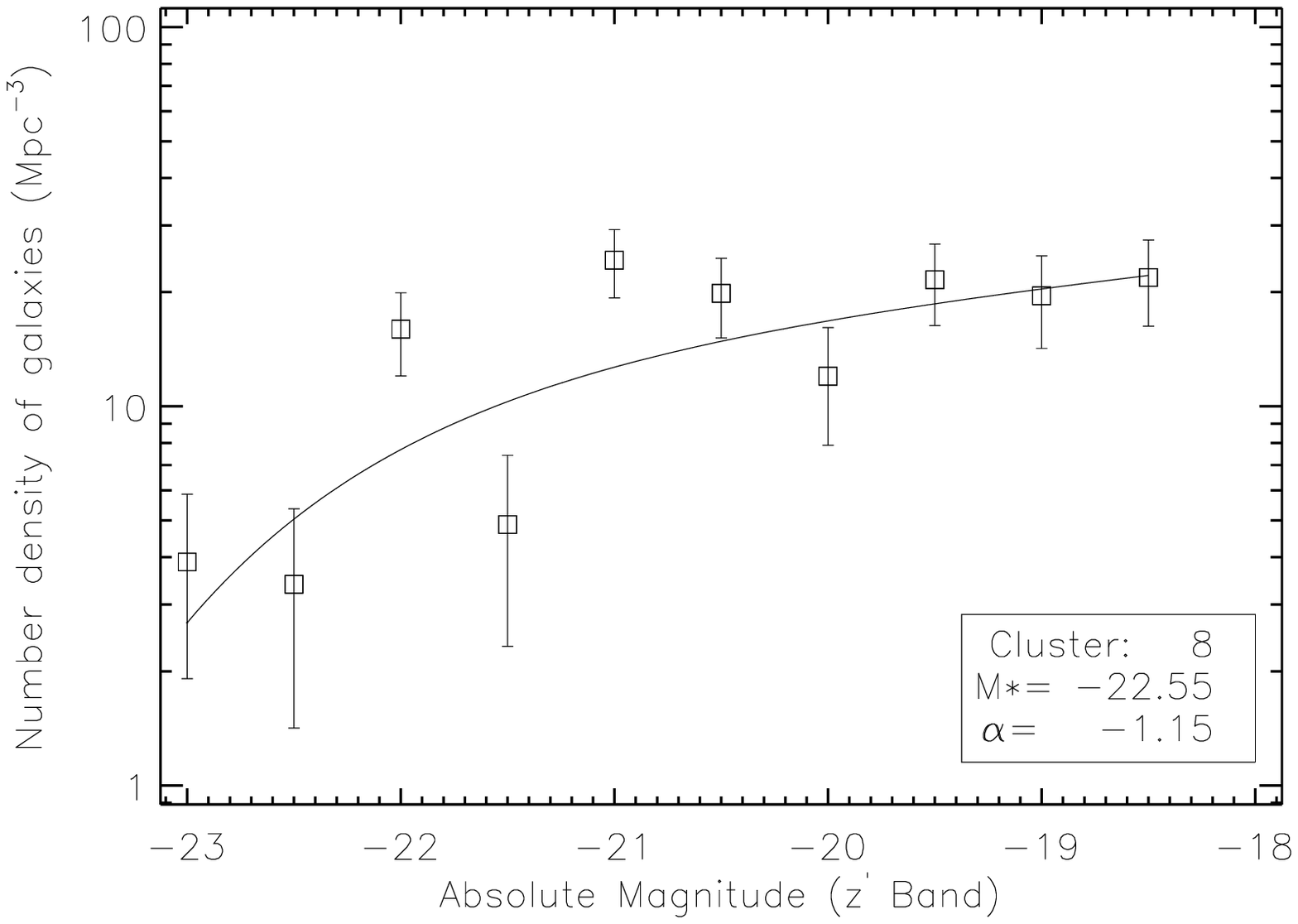,width=4cm,height=3.cm}
\epsfig{file=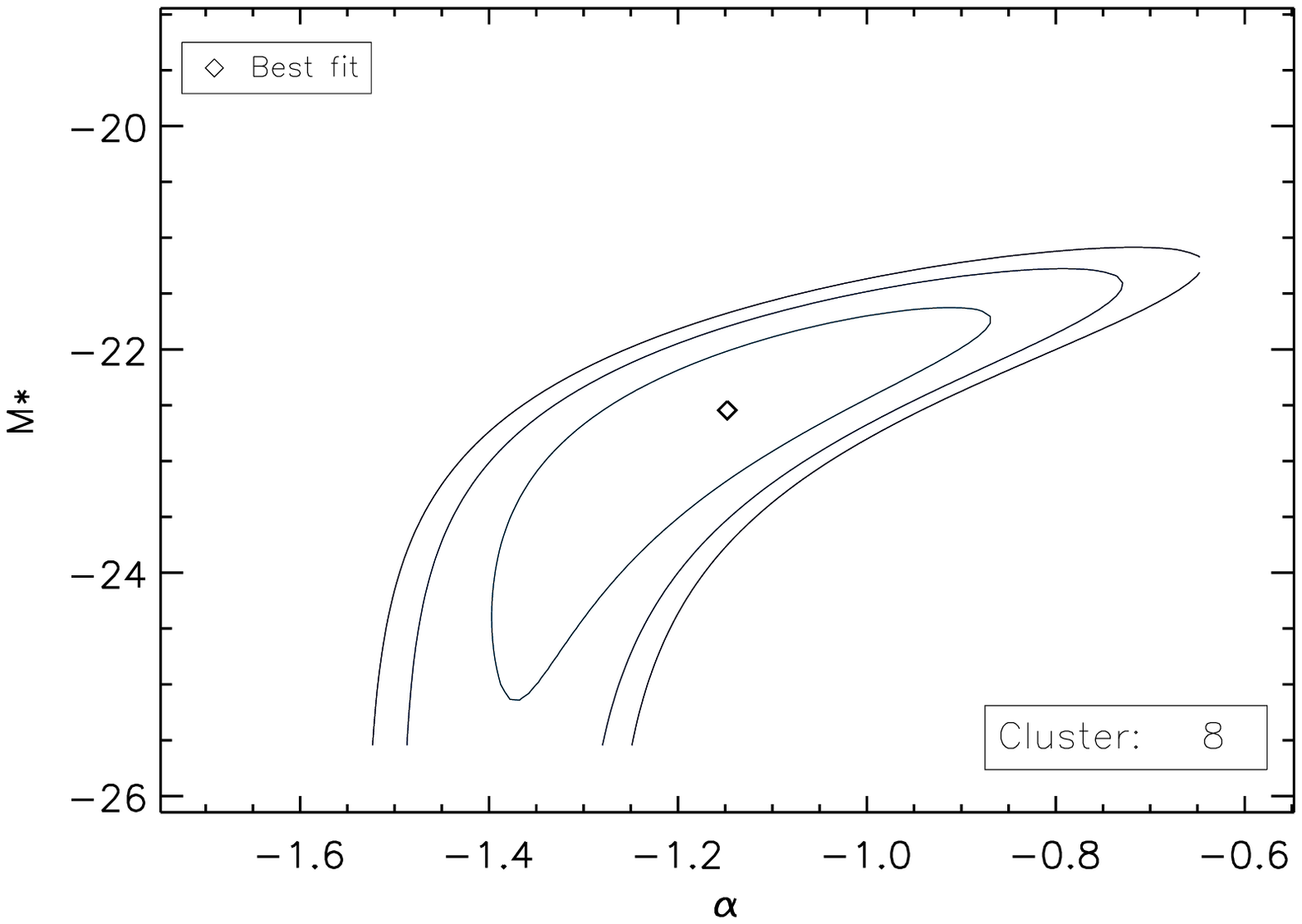,width=4cm,height=3.cm}
\epsfig{file=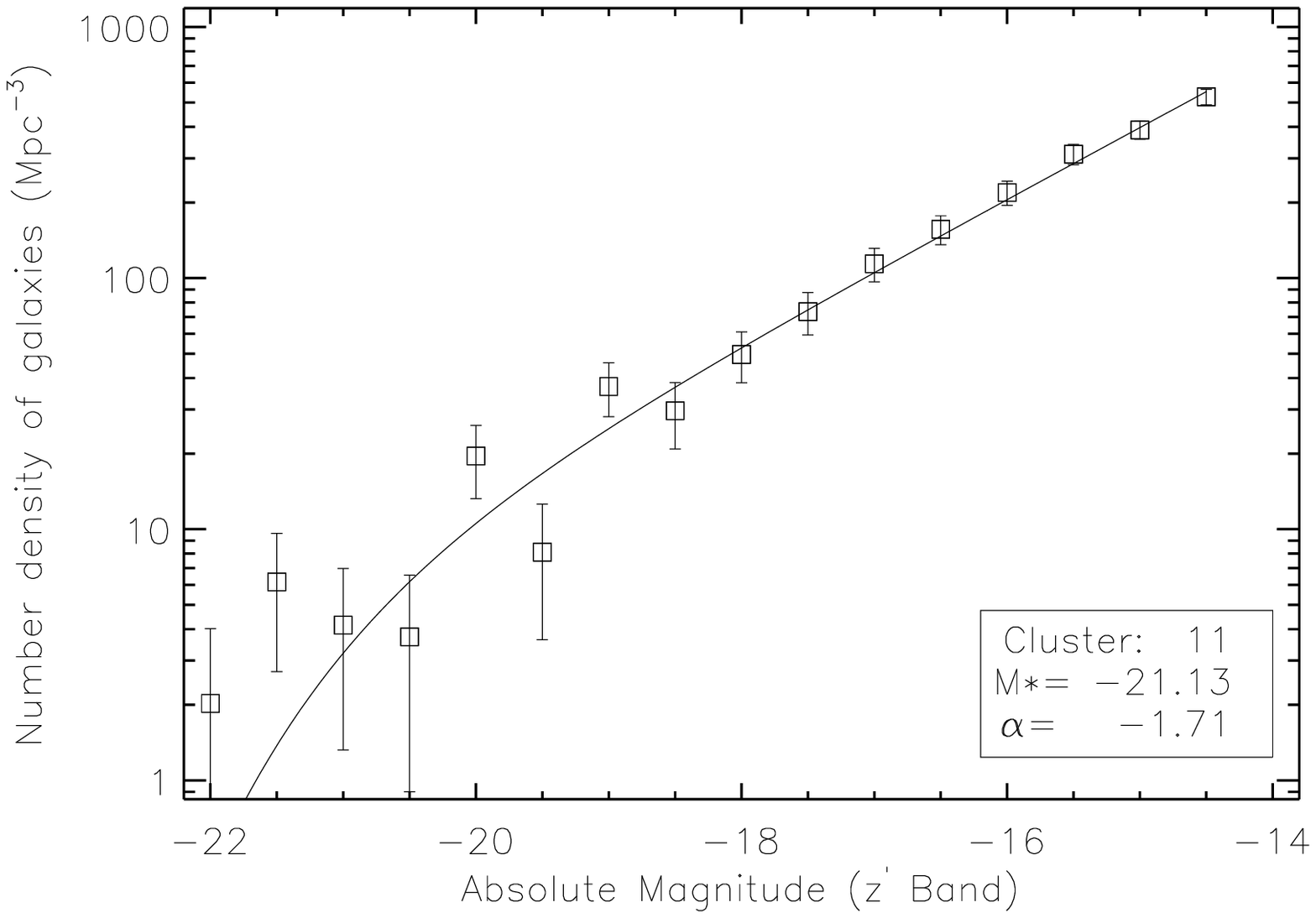,width=4cm,height=3.cm}
\epsfig{file=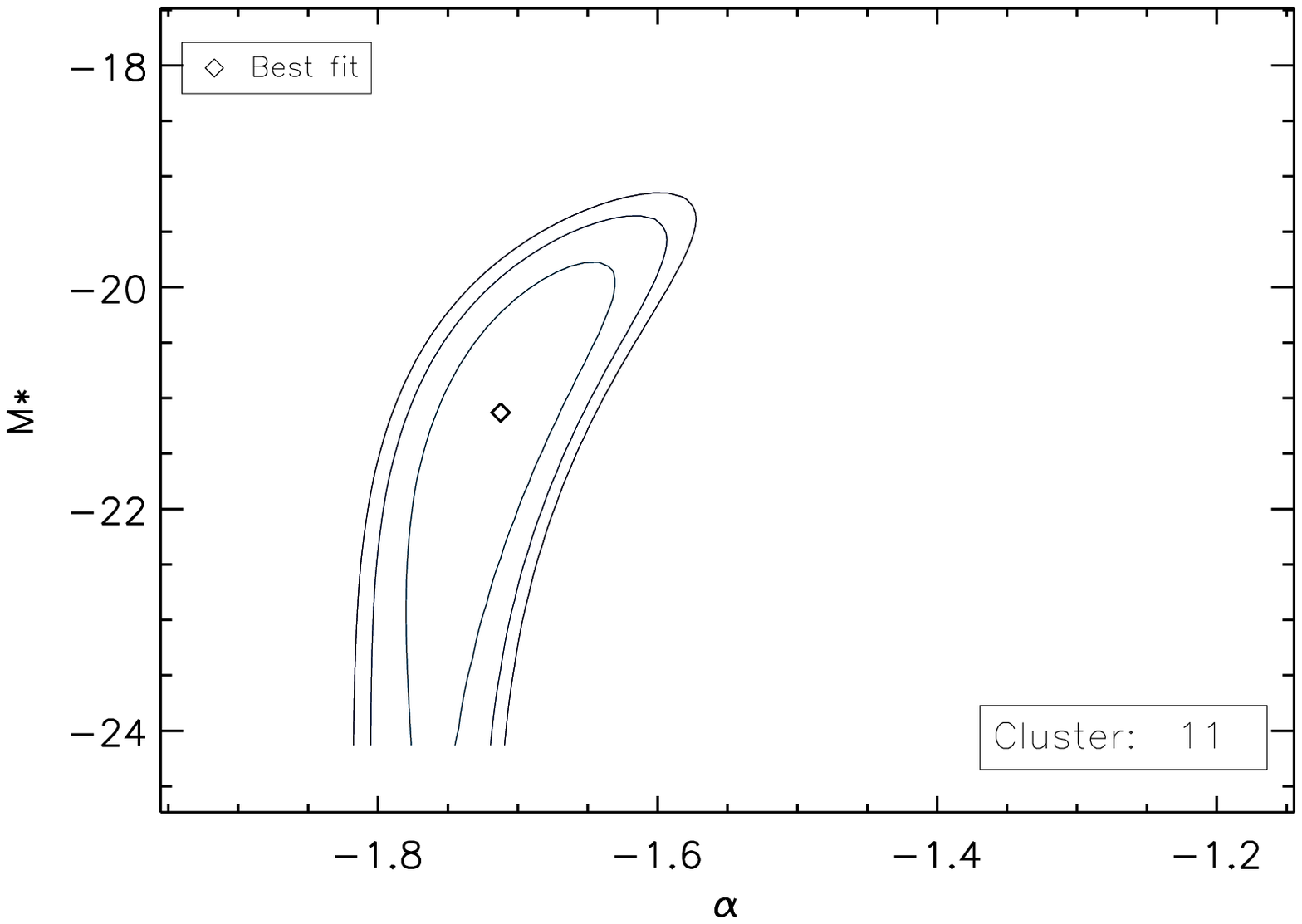,width=4cm,height=3.cm}
\epsfig{file=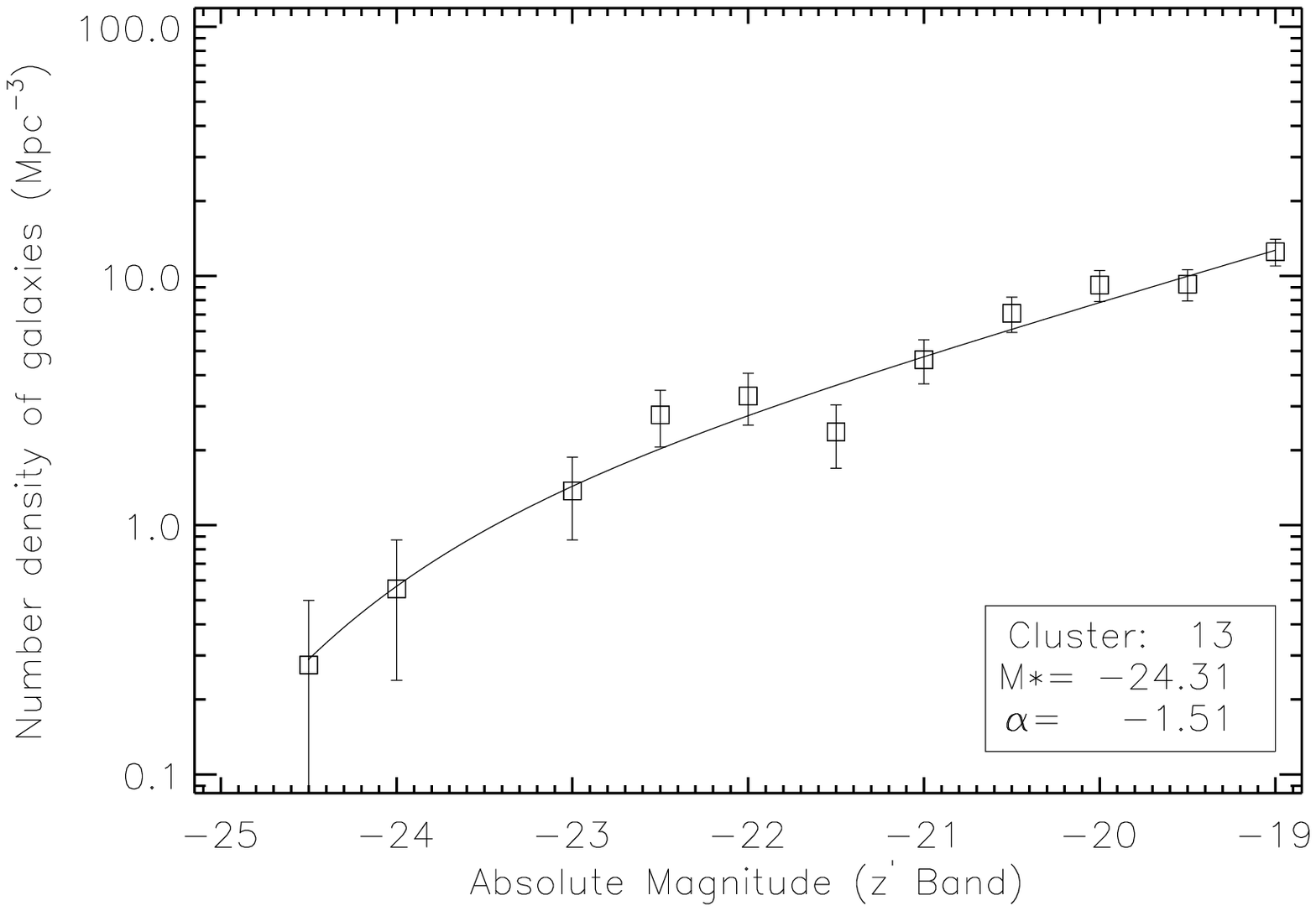,width=4cm,height=3.cm}
\epsfig{file=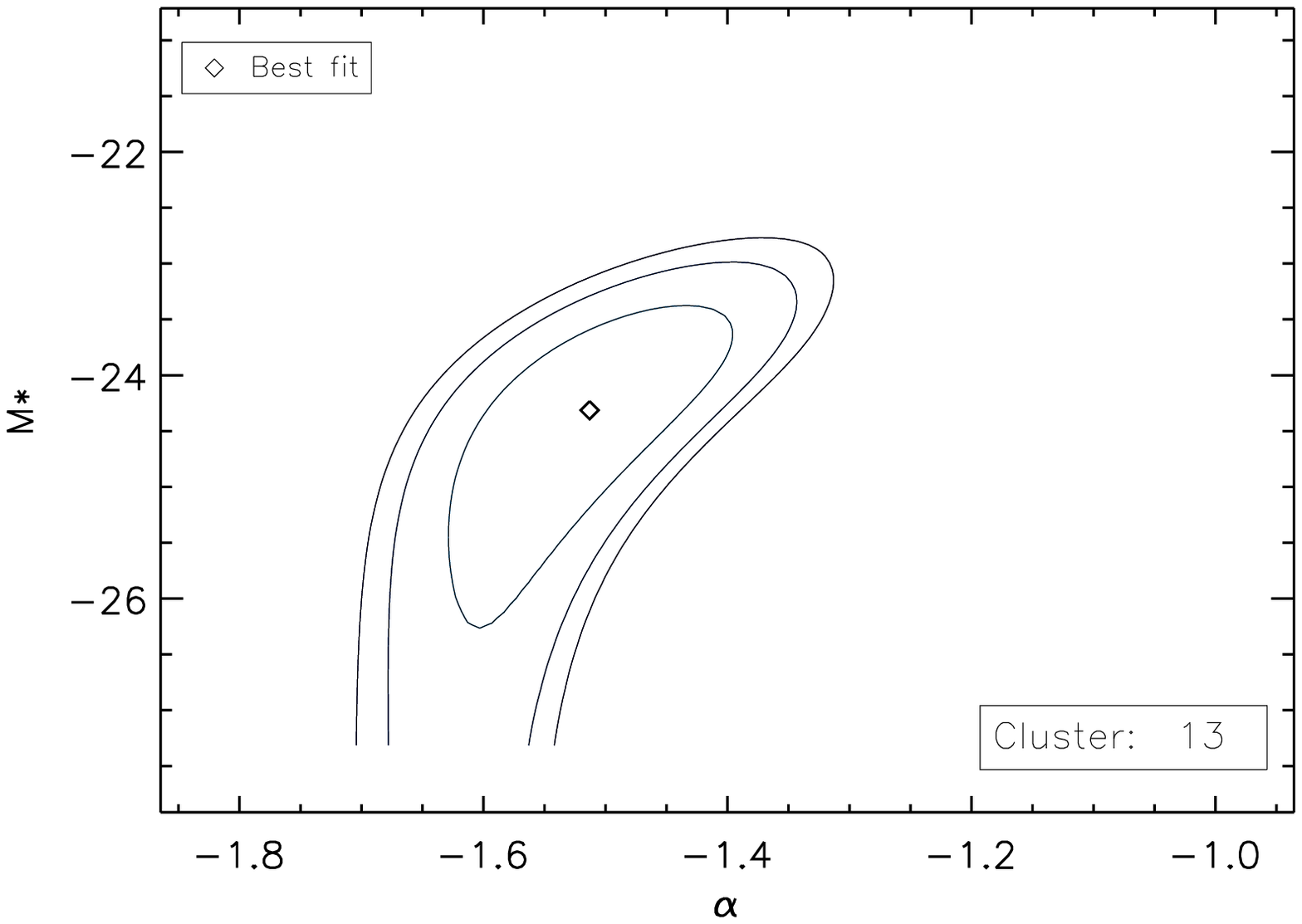,width=4cm,height=3.cm}
\epsfig{file=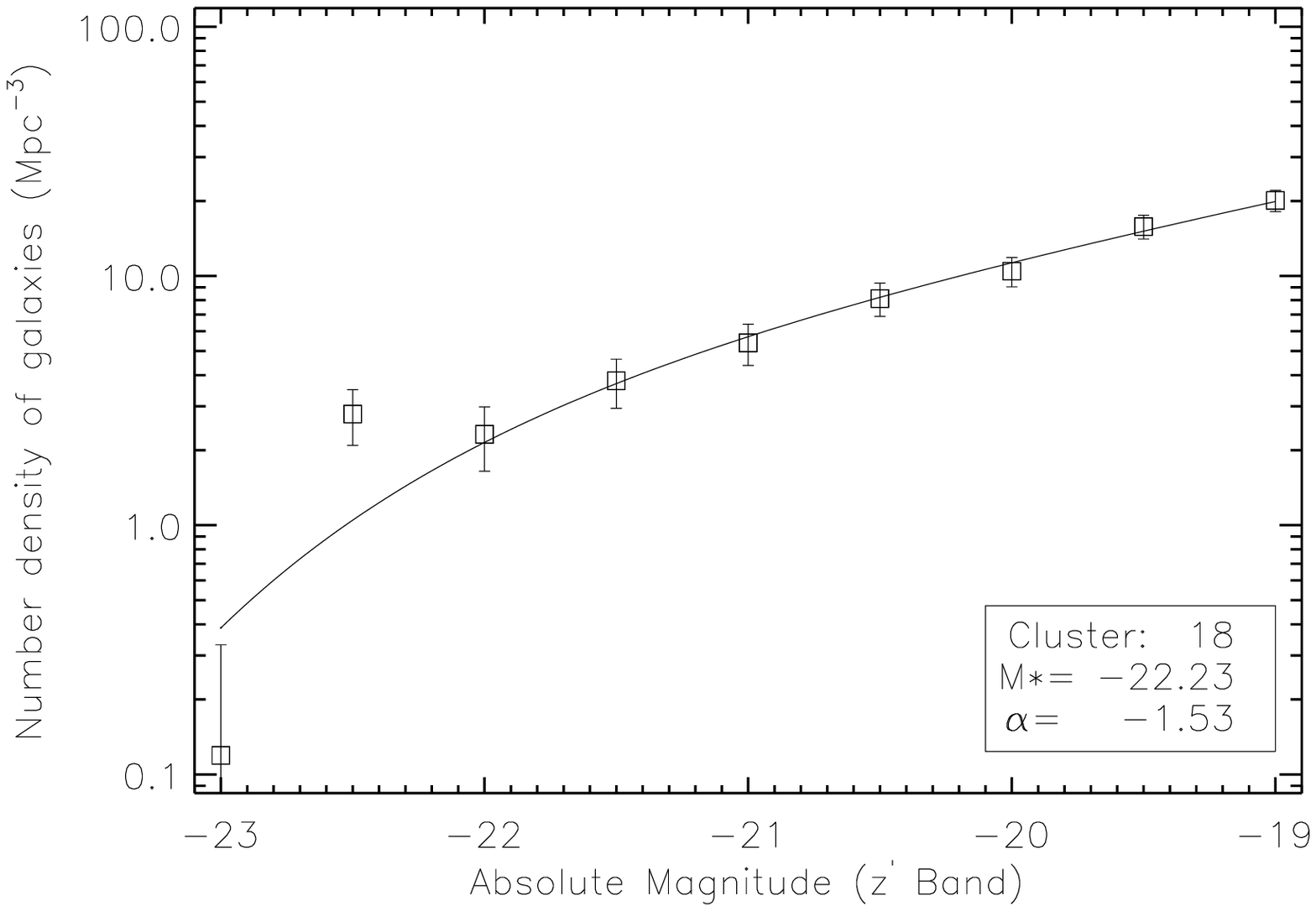,width=4cm,height=3.cm}
\epsfig{file=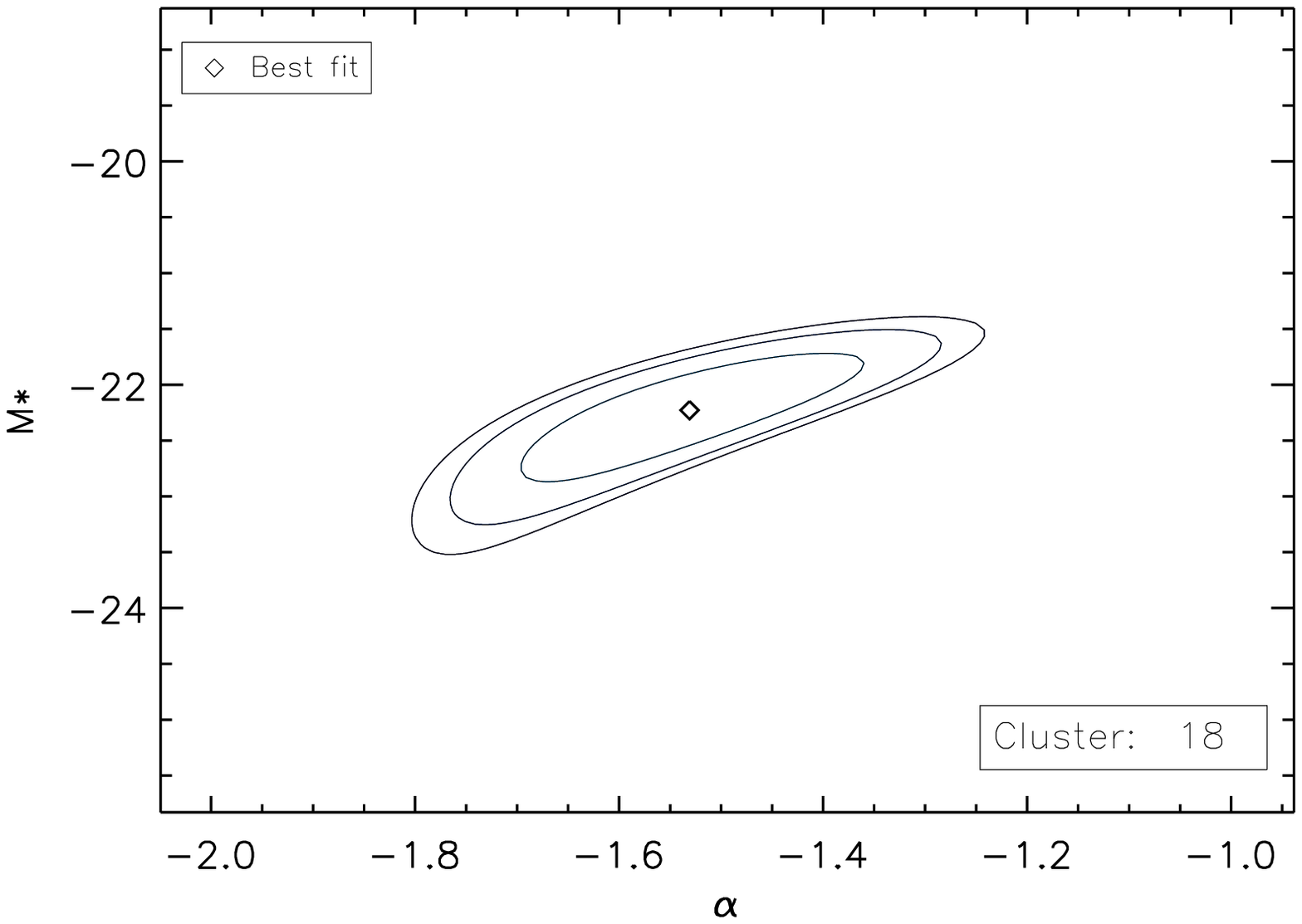,width=4cm,height=3.cm}
\epsfig{file=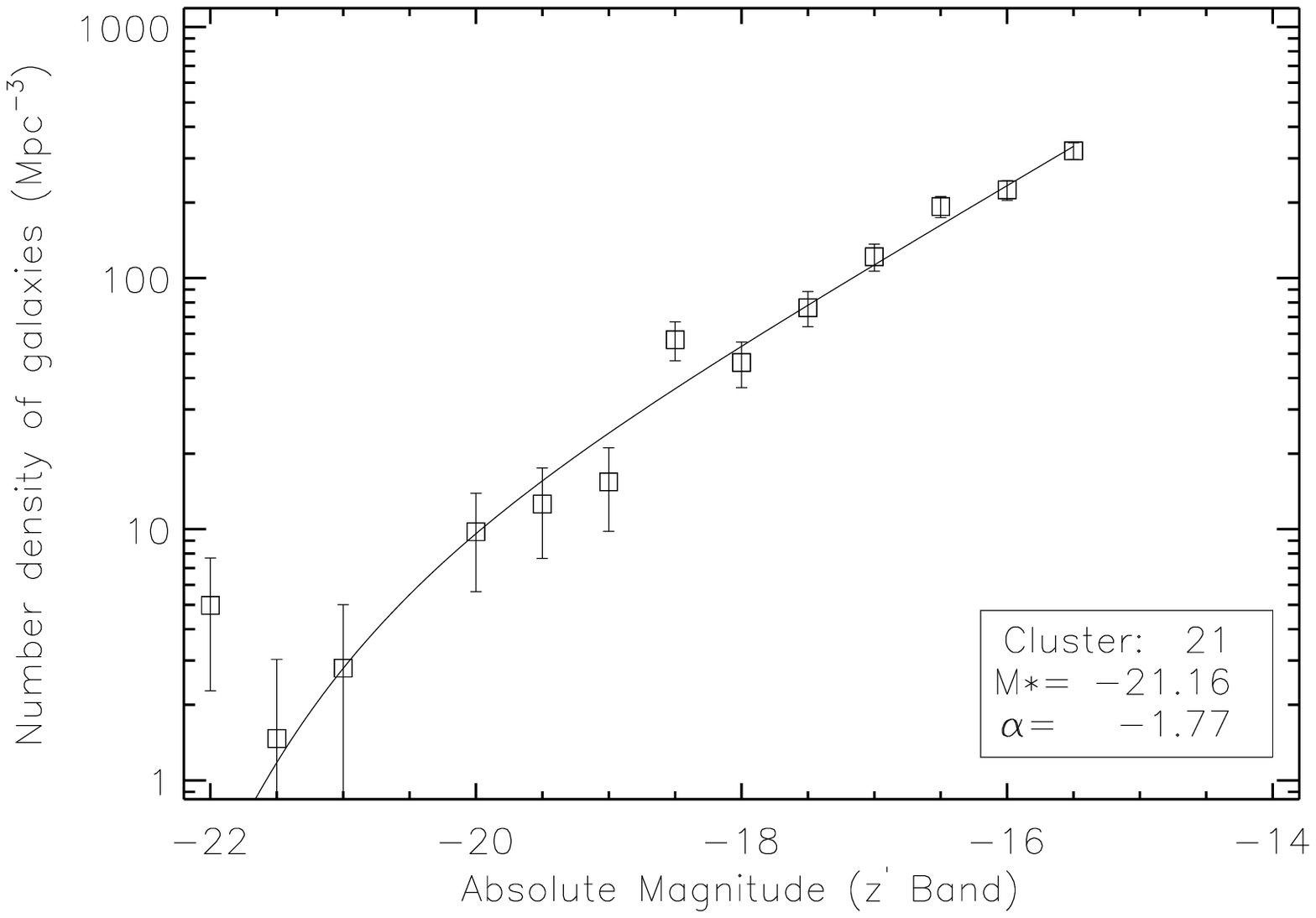,width=4cm,height=3.cm}
\epsfig{file=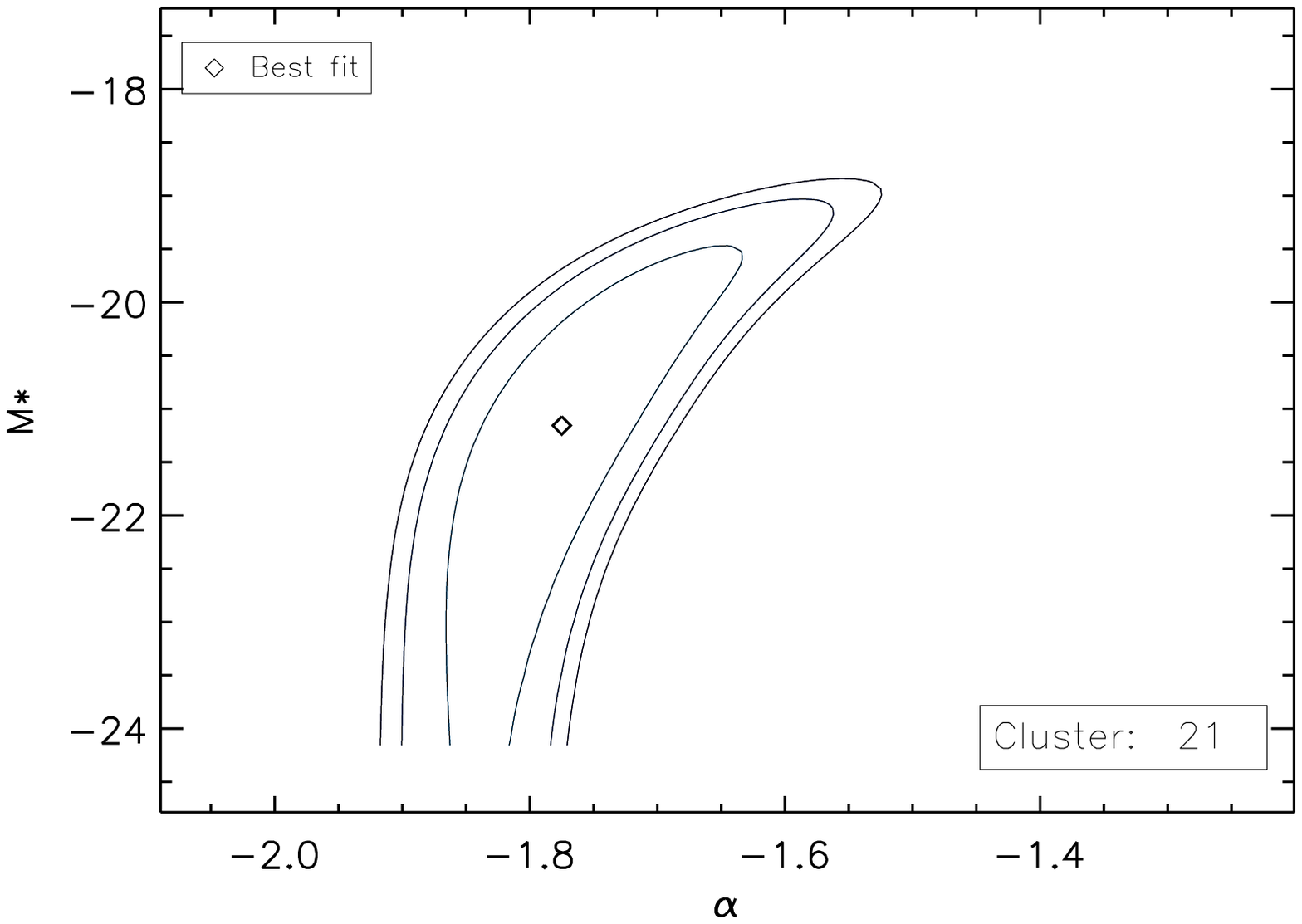,width=4cm,height=3.cm}
\epsfig{file=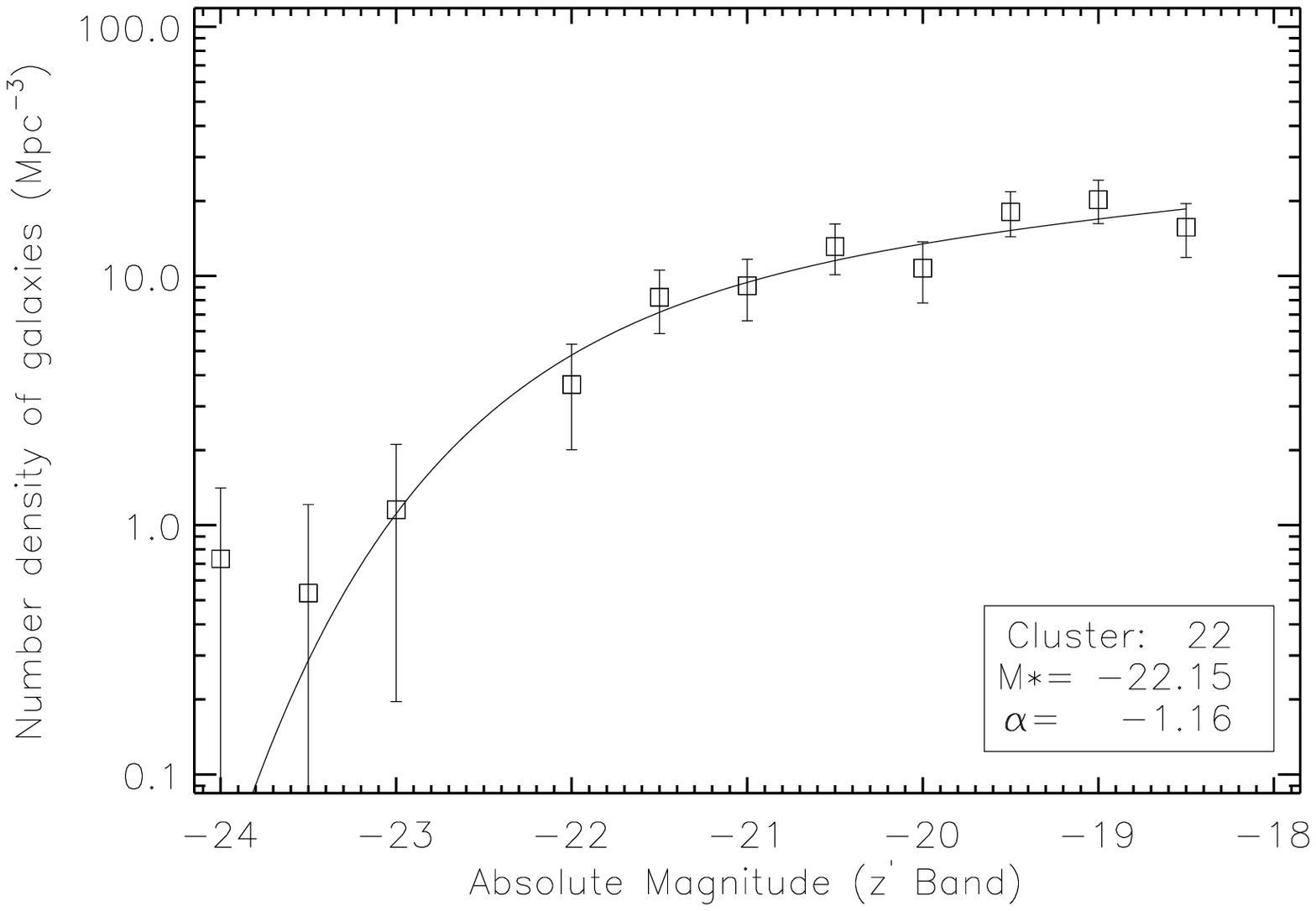,width=4cm,height=3.cm}
\epsfig{file=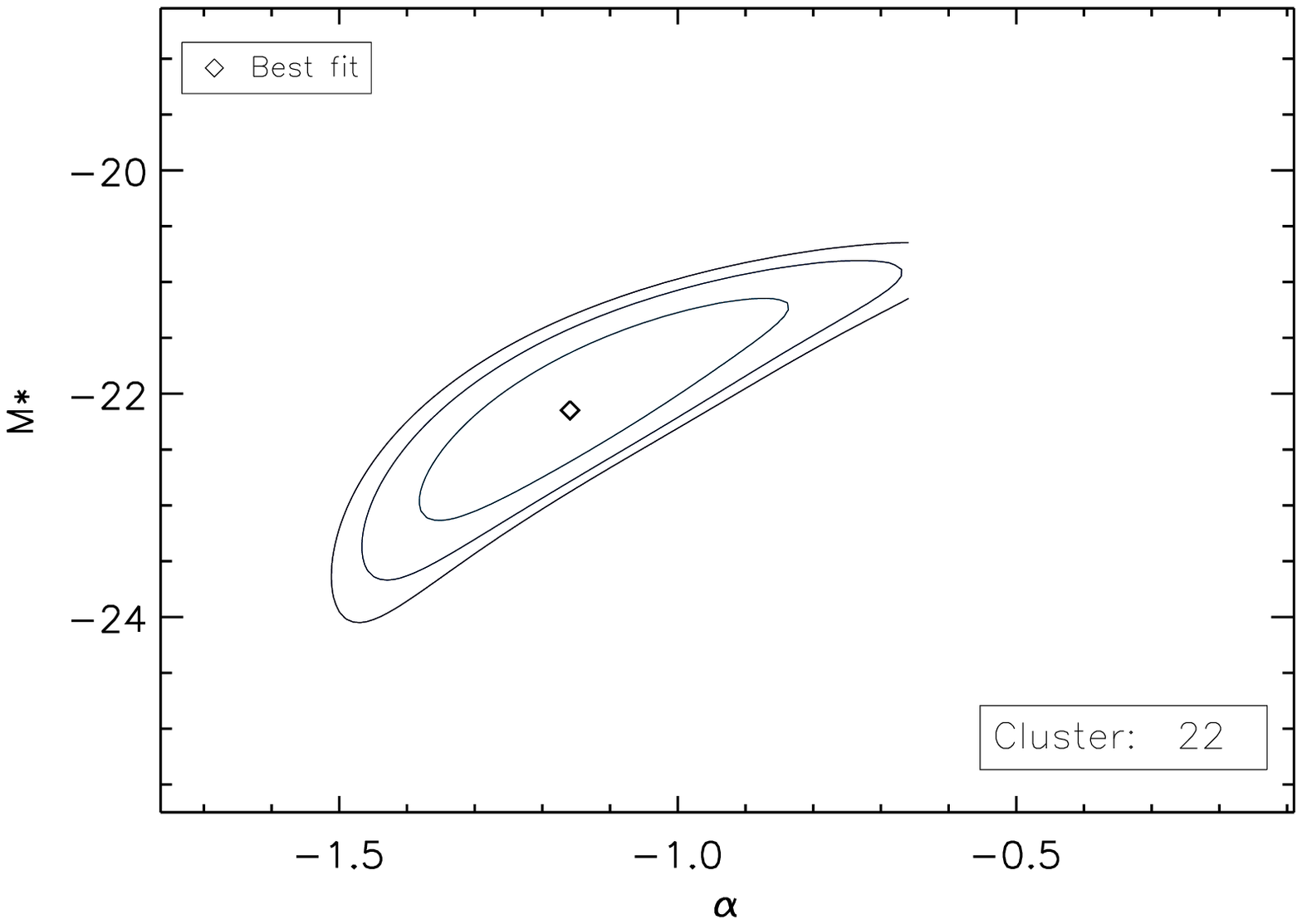,width=4cm,height=3.cm}
\epsfig{file=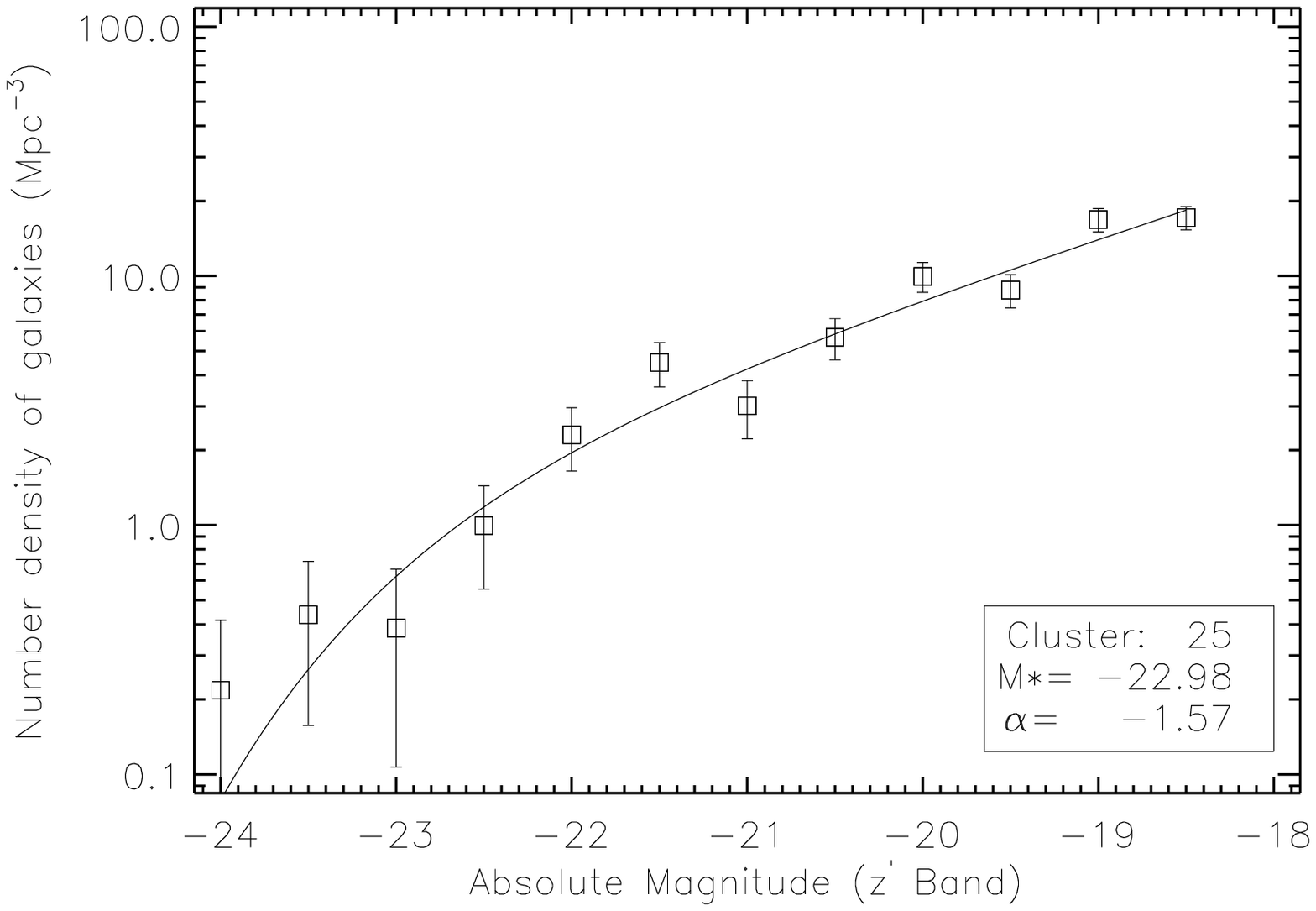,width=4cm,height=3.cm}
\epsfig{file=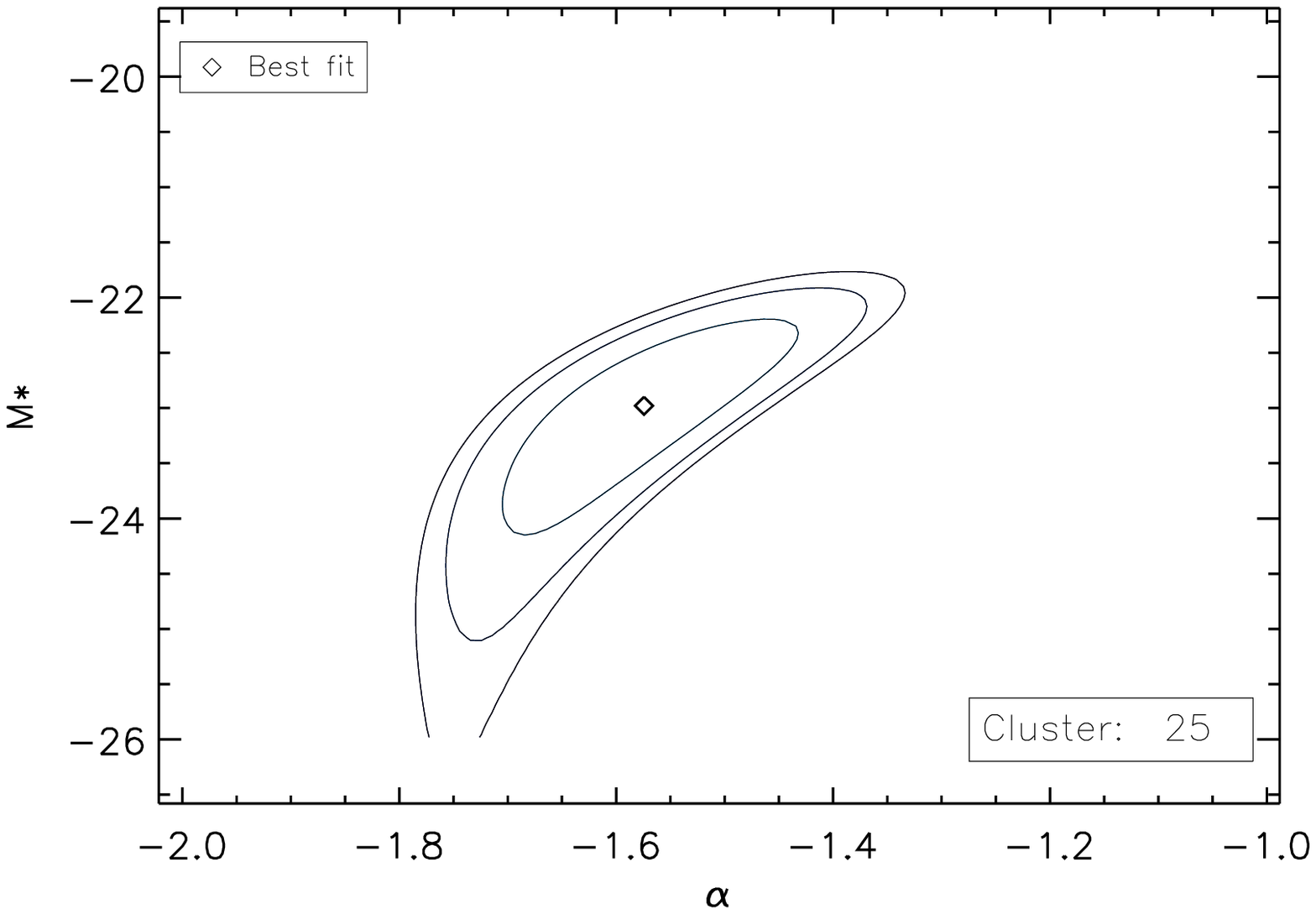,width=4cm,height=3.cm}
\epsfig{file=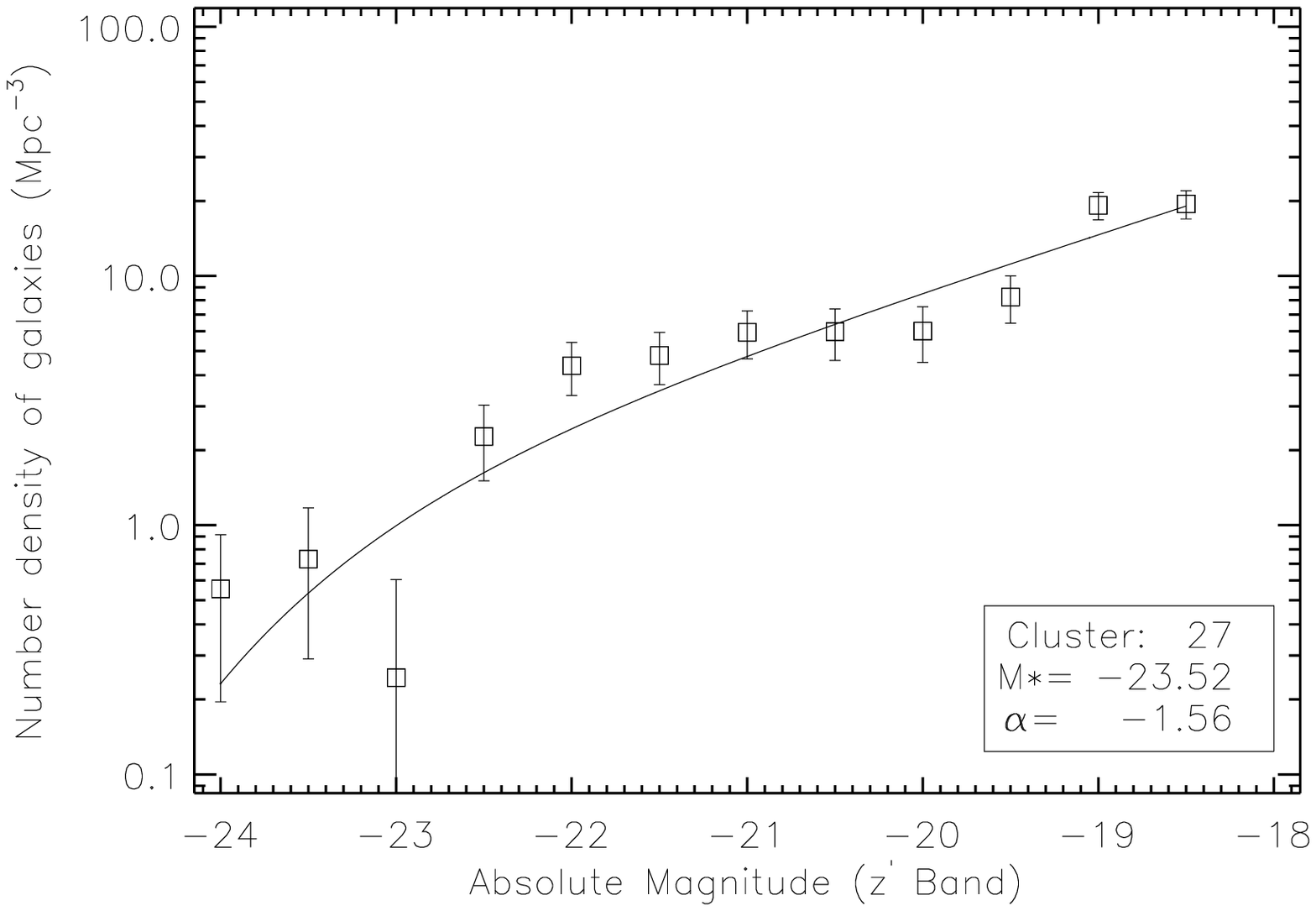,width=4cm,height=3.cm}
\epsfig{file=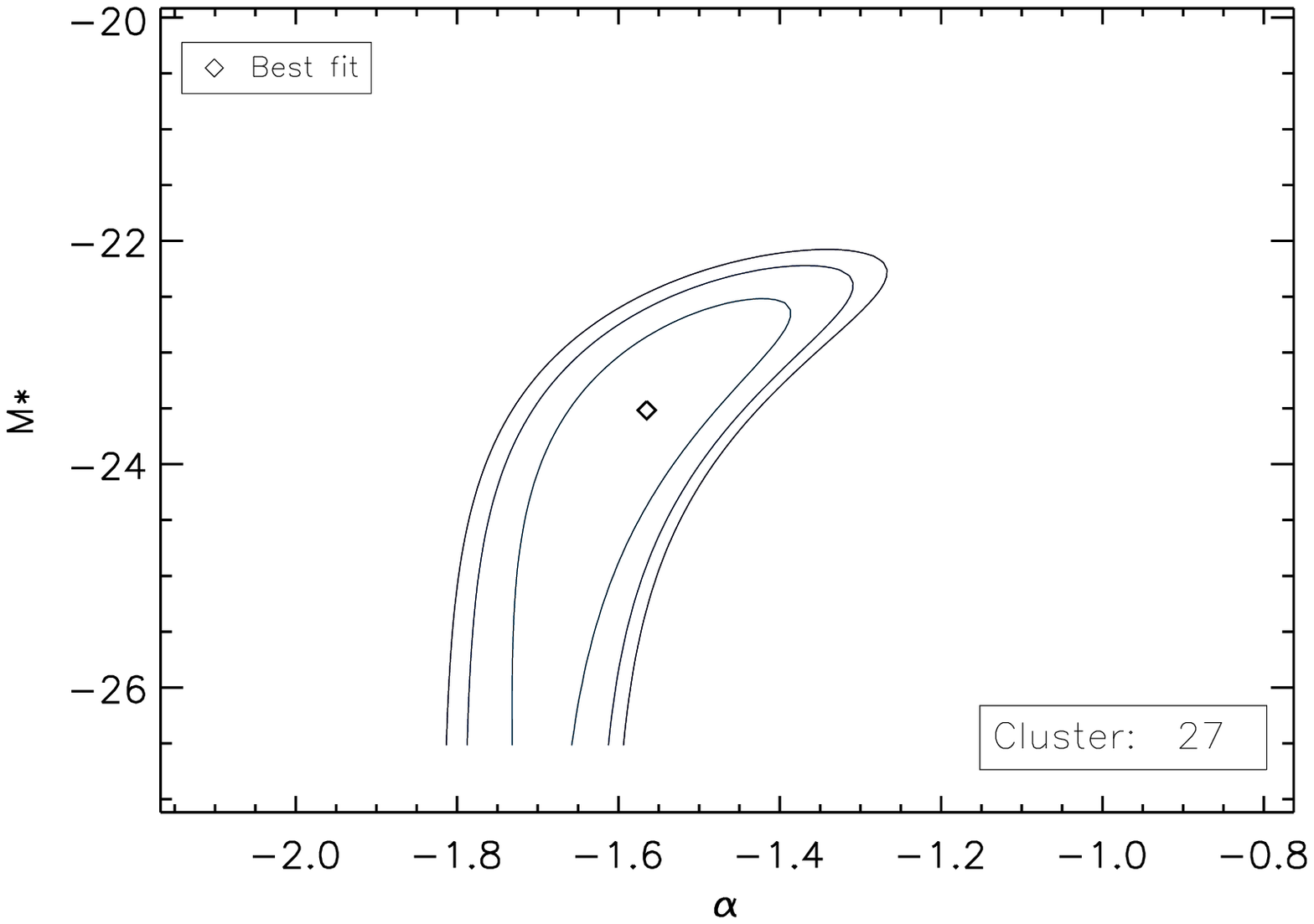,width=4cm,height=3.cm}
\epsfig{file=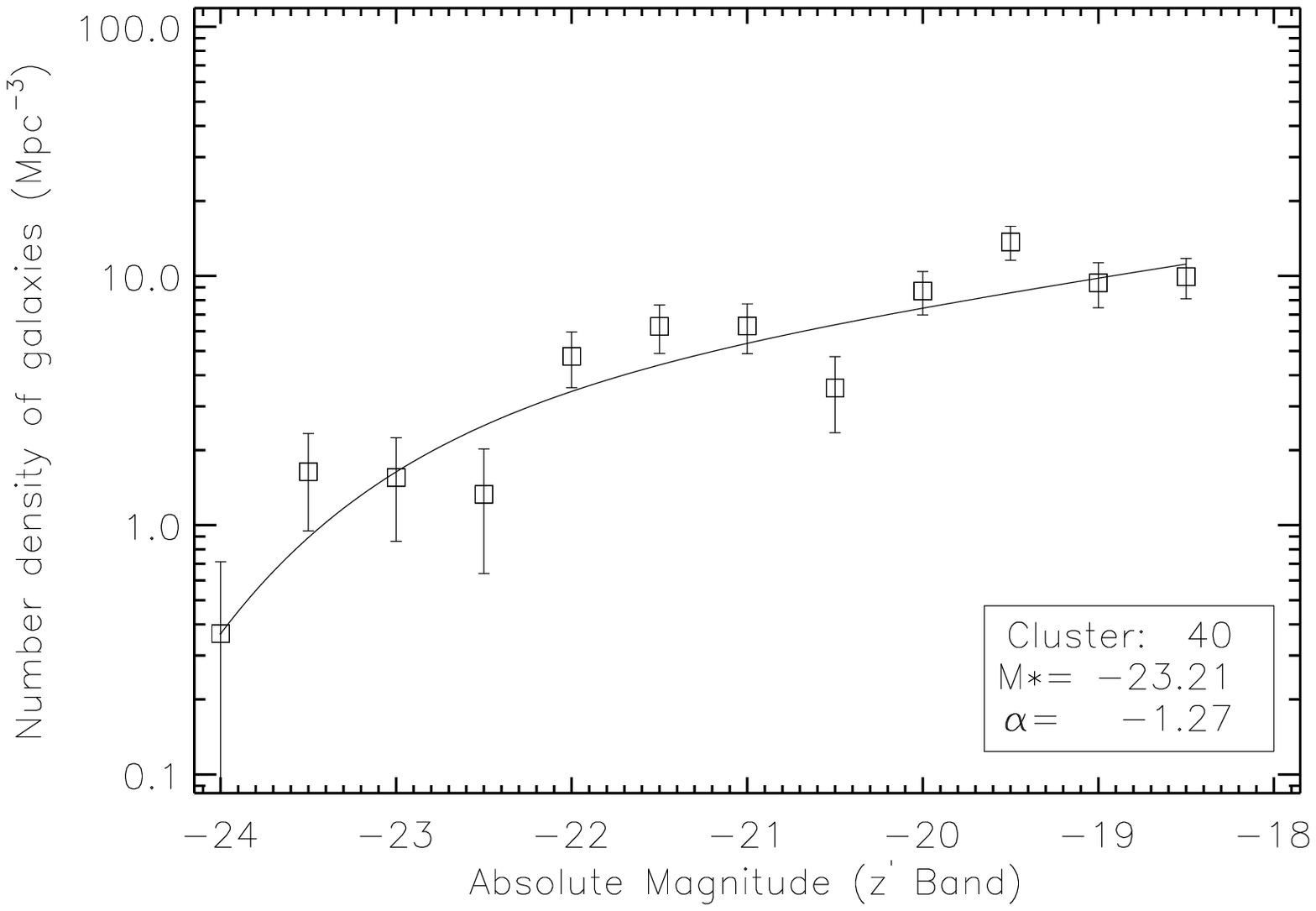,width=4cm,height=3.cm}
\epsfig{file=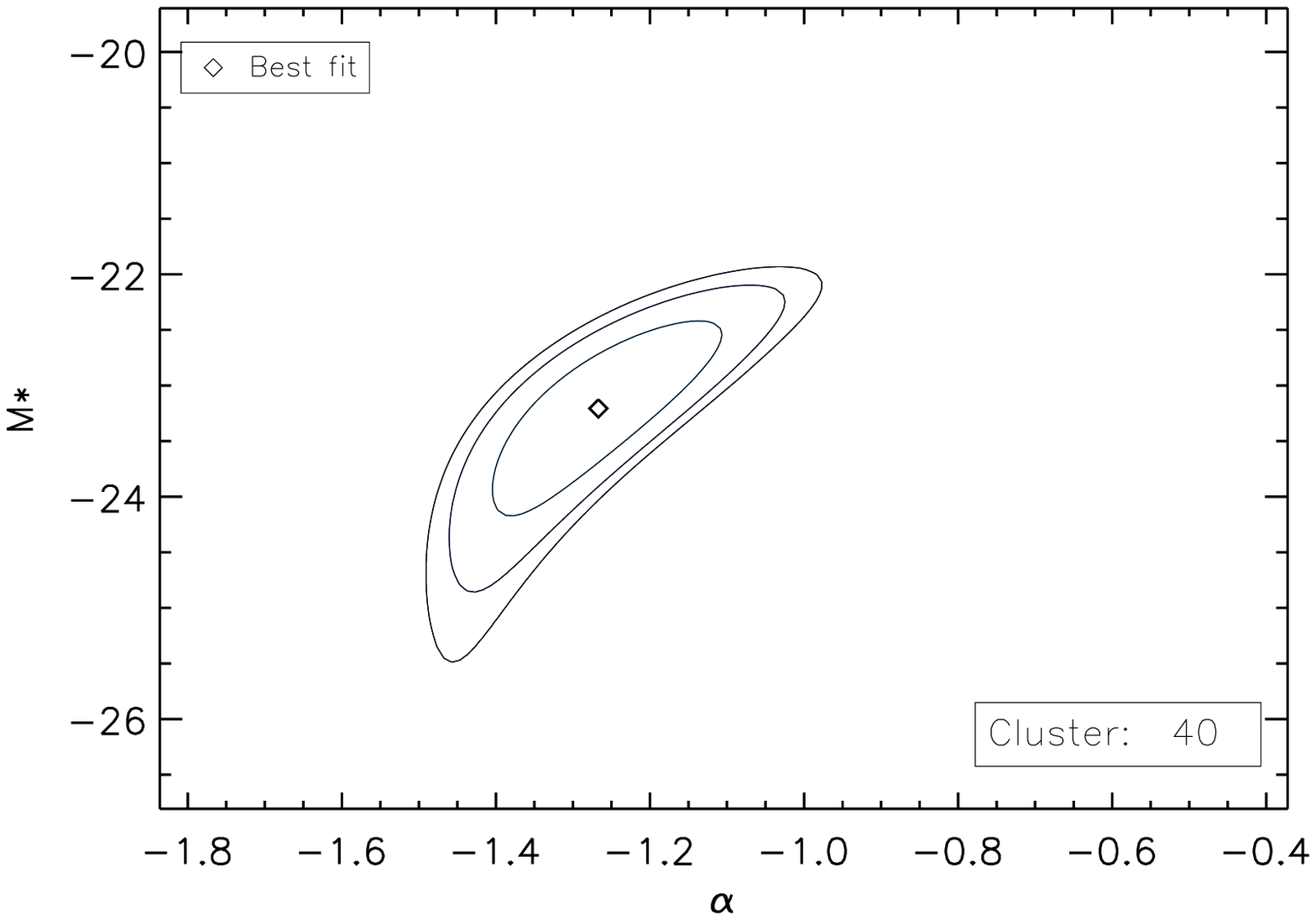,width=4cm,height=3.cm}
\epsfig{file=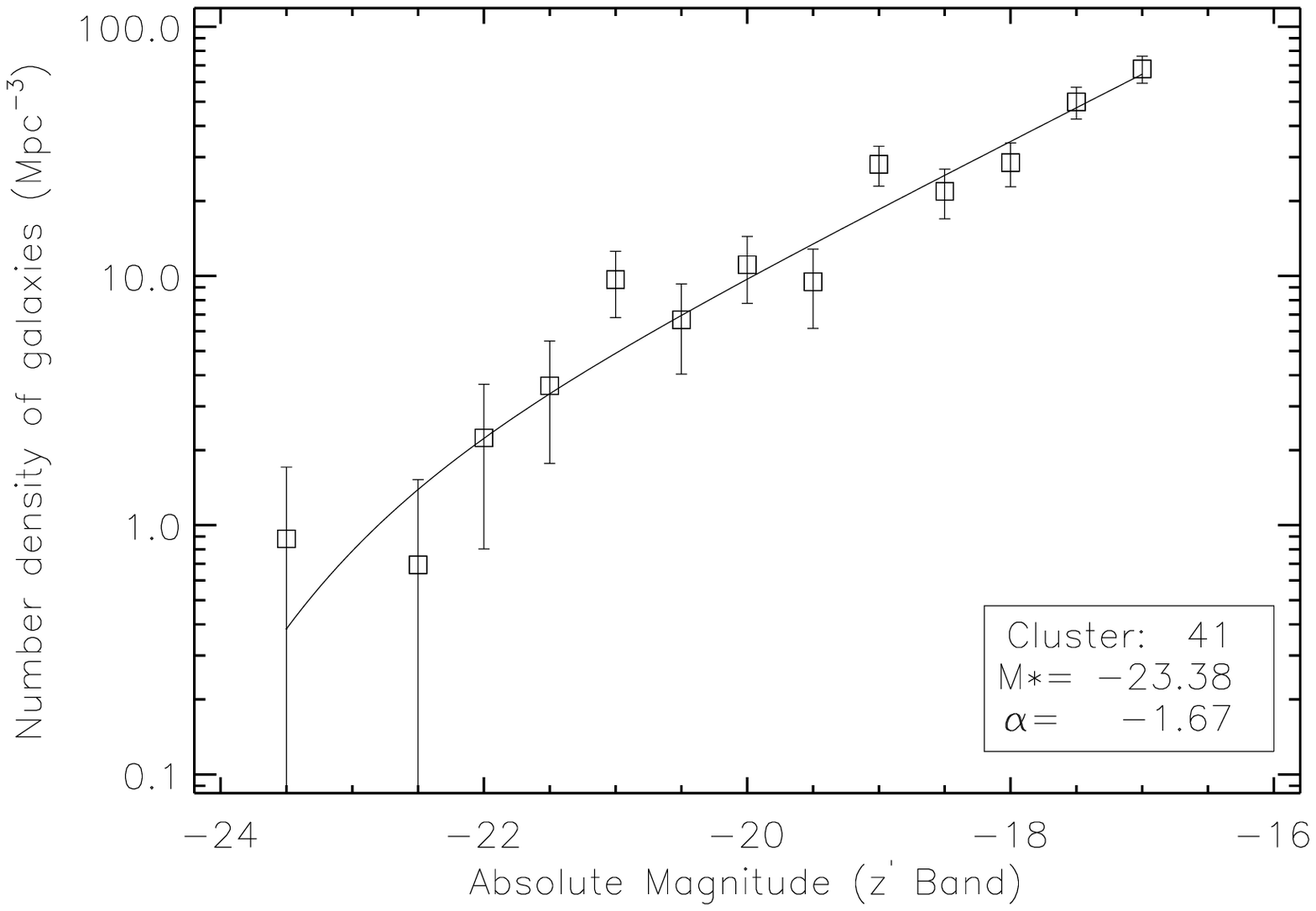,width=4cm,height=3.cm}
\epsfig{file=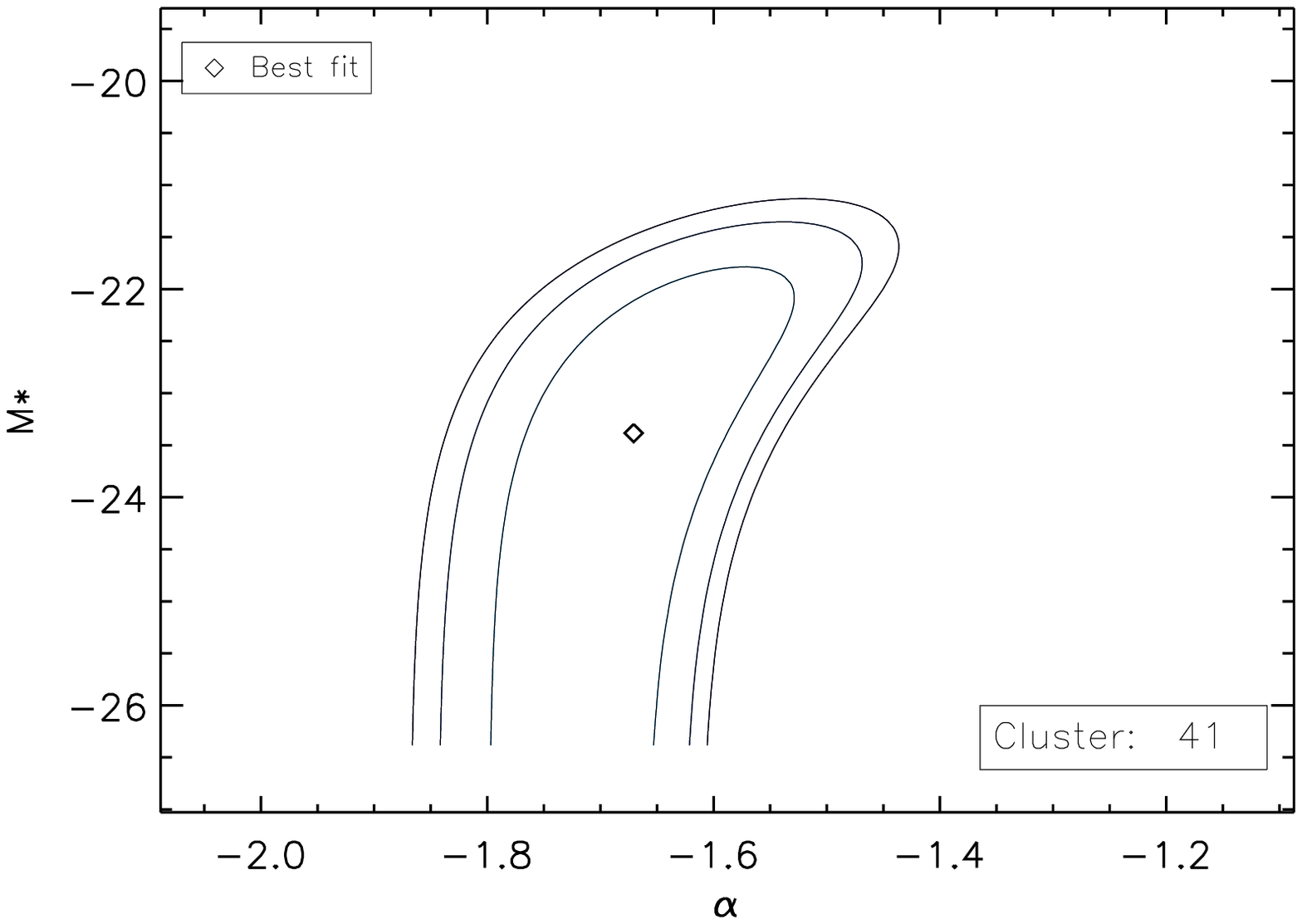,width=4cm,height=3.cm}
\epsfig{file=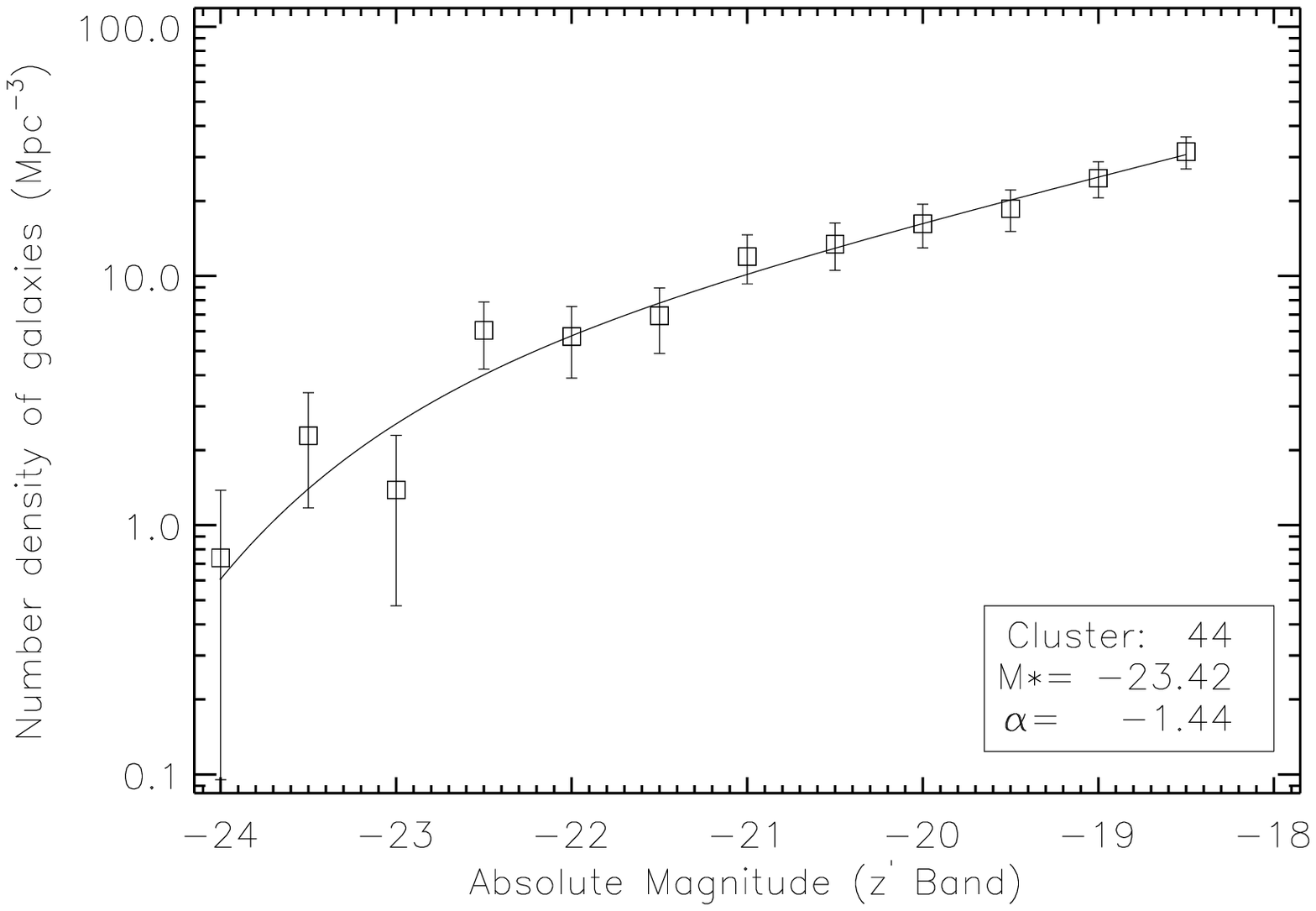,width=4cm,height=3.cm}
\epsfig{file=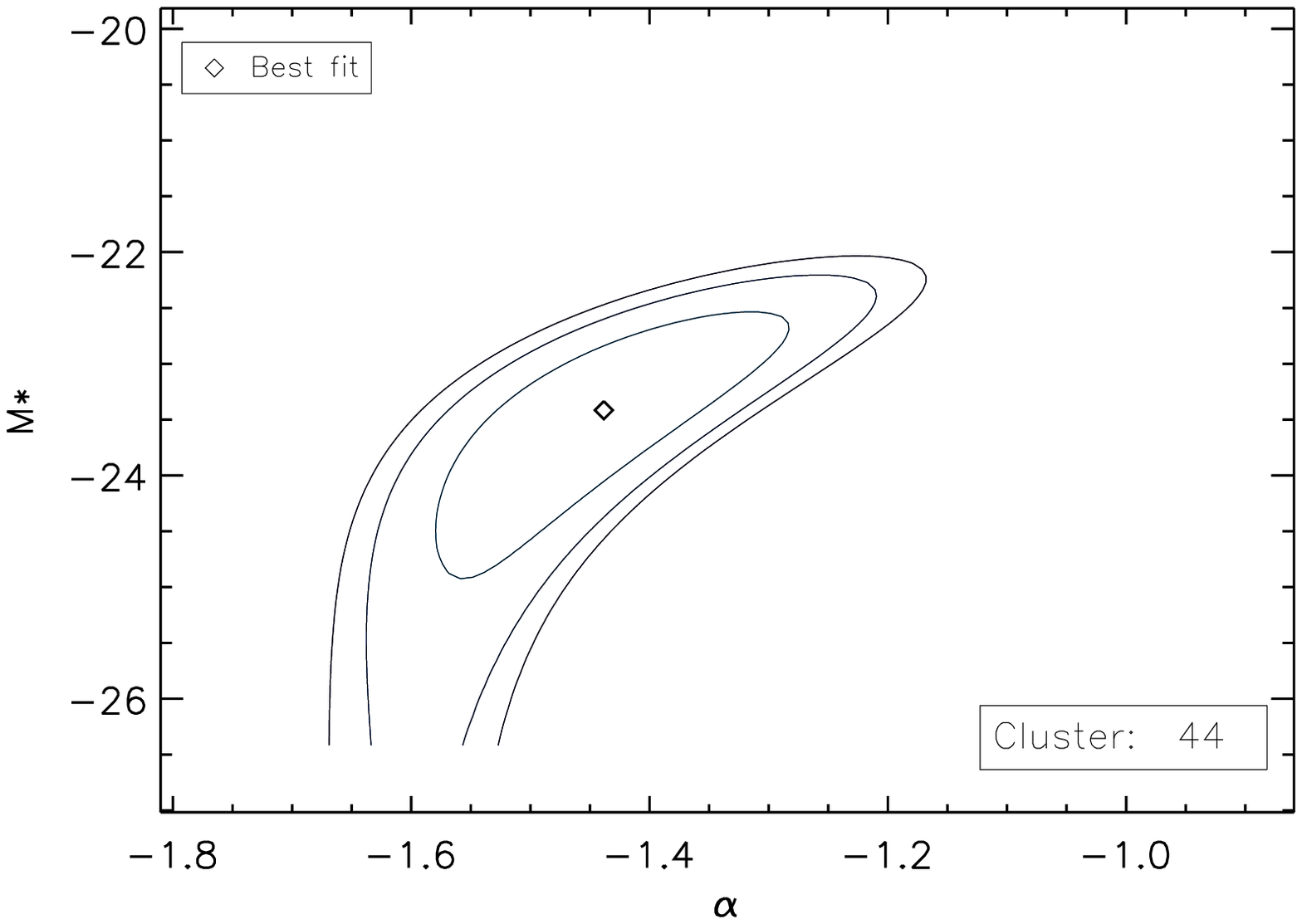,width=4cm,height=3.cm}
\epsfig{file=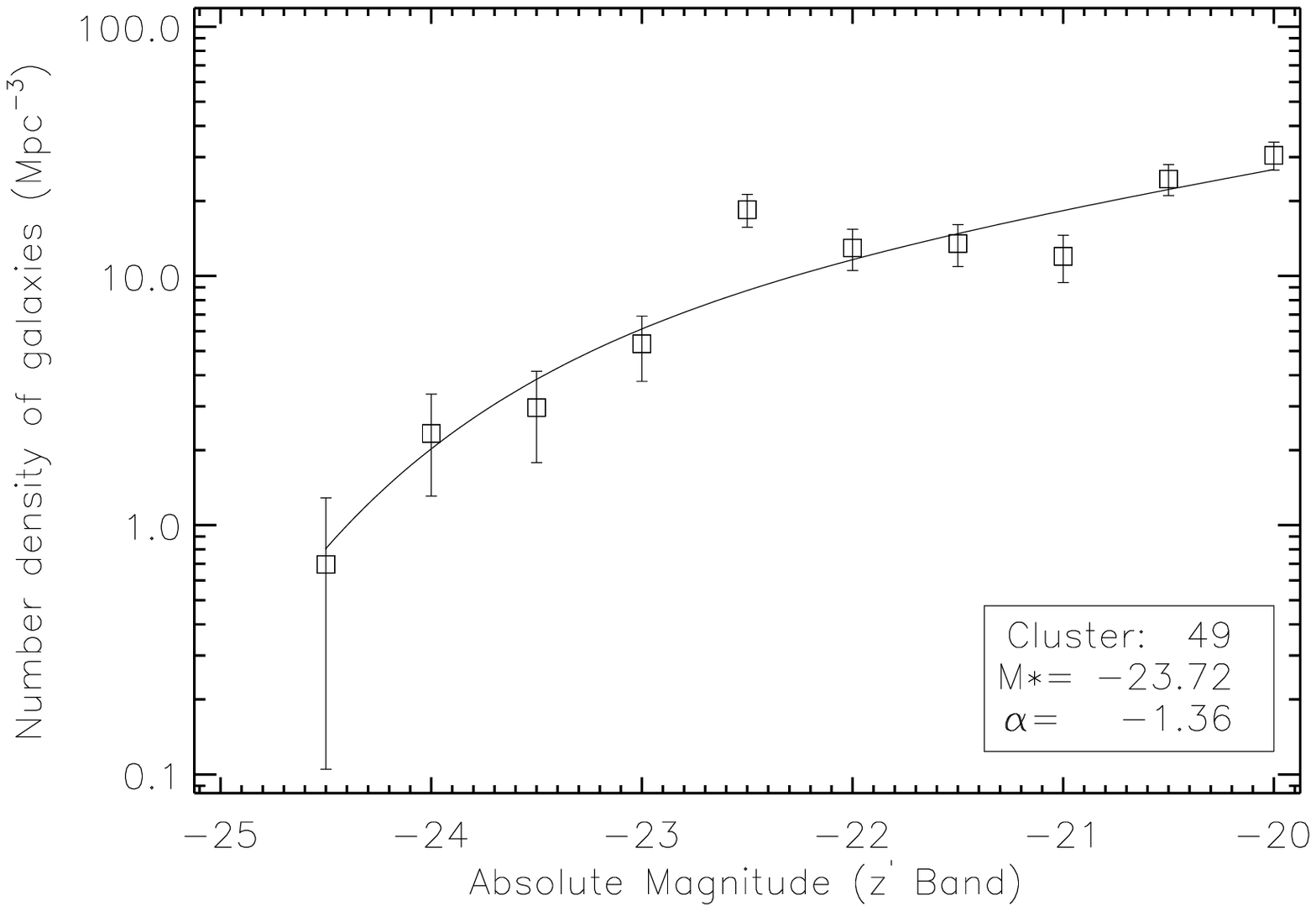,width=4cm,height=3.cm}
\epsfig{file=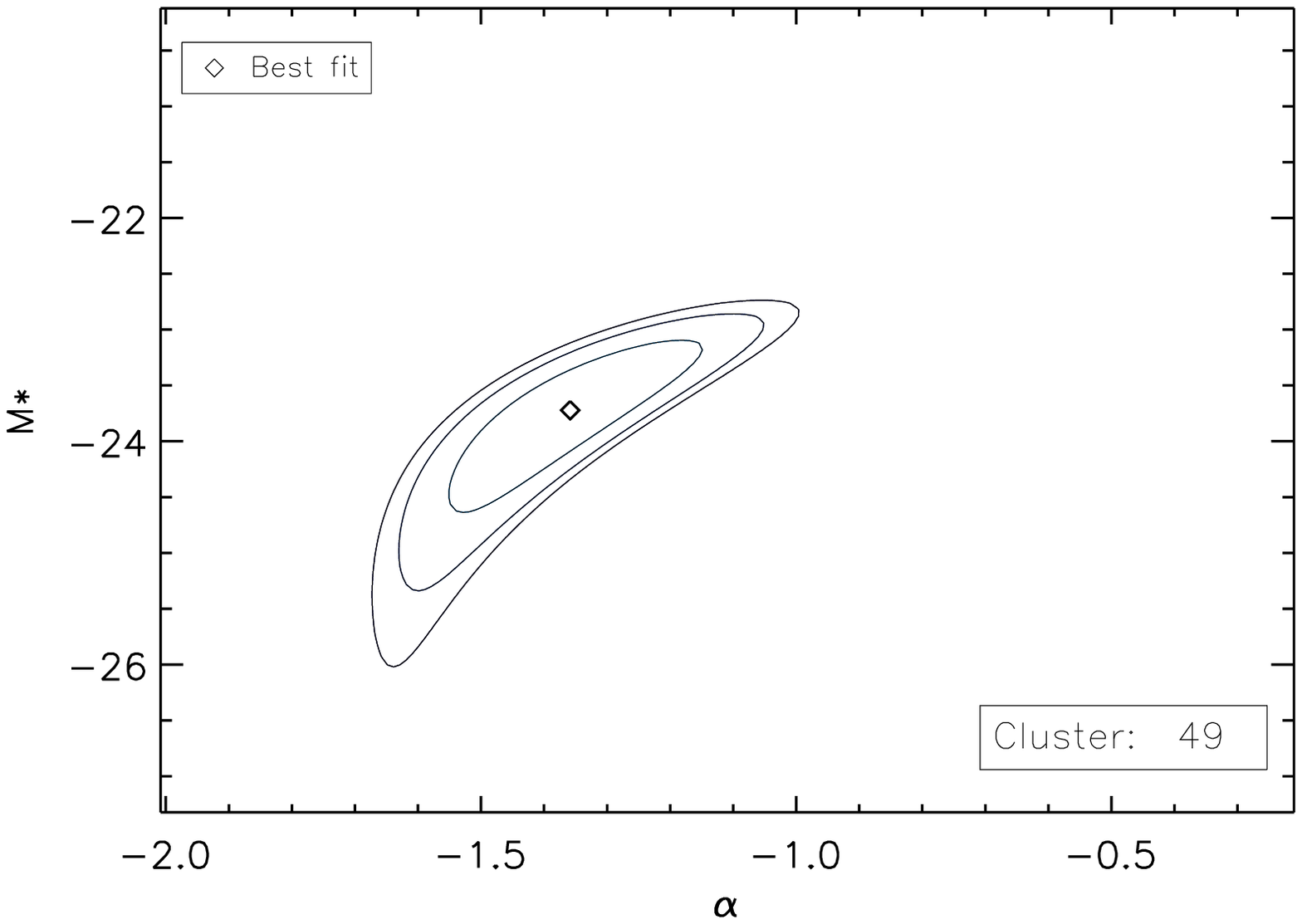,width=4cm,height=3.cm}

\caption{LFs of the 14 individual C1 clusters and the associated $1\sigma$, $2\sigma$ and $3\sigma$ contours of confidence levels of  $\alpha$ and $M^*$ for the well-fitted clusters for the $z^\prime$ band.}
\label{Ind_z}
\end{figure}

\label{lastpage}
\end{document}